\documentclass[12pt]{article}
\usepackage{latexsym,amsmath,amssymb, graphicx, subfigure}
\renewcommand{\thefootnote}{\fnsymbol{footnote}}

\usepackage{mathrsfs }
\usepackage{amsmath}
\usepackage{amsfonts}

\numberwithin{equation}{section}
	
\def\doubleset#1#2{\bgroup%
\def\doit#1#2{%
\setbox\dblsetbox=\hbox{$\cstyle #1$}%
\raise#2\ht\dblsetbox\copy\dblsetbox%
\hskip-\wd\dblsetbox%
\raise-#2\ht\dblsetbox\box\dblsetbox}%
\mathchoice%
{\def\cstyle{\displaystyle}\doit#1#2}%
{\def\cstyle{\textstyle}\doit#1#2}%
{\def\cstyle{\scriptstyle}\doit#1#2}%
{\def\cstyle{\scriptscriptstyle}\doit#1#2}\egroup}
\def\underarrow#1{\vbox{\ialign{##\crcr$\hfil\displaystyle
 {#1}\hfil$\crcr\noalign{\kern1pt\nointerlineskip}$\longrightarrow$\crcr}}}

\def\zb {{\bar{z}}}

\def\IL{\relax{\rm I\kern-.18em L}}
\def\IH{\relax{\rm I\kern-.18em H}}
\def\IB{\relax{\rm I\kern-.18em B}}
\def\ID{\relax{\rm I\kern-.18em D}}
\def\IE{\relax{\rm I\kern-.18em E}}
\def\IF{\relax{\rm I\kern-.18em F}}

\def\IG{\relax\hbox{$\inbar\kern-.3em{\rm G}$}}
\def\IGa{\relax\hbox{${\rm I}\kern-.18em\Gamma$}}
\def\IH{\relax{\rm I\kern-.18em H}}
\def\II{\relax{\rm I\kern-.18em I}}
\def\IK{\relax{\rm I\kern-.18em K}}
\def\IP{\relax{\rm I\kern-.18em P}}
\def\IQ{\relax\hbox{$\inbar\kern-.3em{\rm Q}$}}

\def\hat{\widehat}
\def\CM {{\cal M}}

\def\CR {{\cal R}}

\def\CF {{\cal F}}

\def\CL {{\cal L}}
\def\CV {{\cal V}}
\def\CO {{\cal O}}

\def\CE {{\cal E}}
\def\CG {{\cal G}}
\def\CH {{\cal H}}
\def\CC {{\cal C}}

\def\CA{{\cal A}}

\def\CQ{{\cal Q}}

\def\jb{{\bar j}}

\def\inbar{\,\vrule height1.5ex width.4pt depth0pt}
\def\Map{{\rm Map}}

\newbox\dblsetbox

\newcommand{\del}{\partial}

\newcommand{\ket}[1]{|#1\rangle}

\newcommand{\iso}{\cong}

\newcommand{\bb}{{\bar{b}}}
\newcommand{\cb}{{\bar{c}}}
\newcommand{\db}{{\bar{d}}}

\newcommand{\ib}{{\bar{\imath}}}
\newcommand{\kb}{{\bar{k}}}
\newcommand{\lb}{{\bar{l}}}

\newlength{\extraspace}
\newlength{\extraspaces}
\newcommand{\be}{\begin{equation}
\addtolength{\abovedisplayskip}{\extraspaces}
\addtolength{\belowdisplayskip}{\extraspaces}
\addtolength{\abovedisplayshortskip}{\extraspace}
\addtolength{\belowdisplayshortskip}{\extraspace}}
\newcommand{\ee}{\end{equation}}

\newcommand{\ba}{\begin{eqnarray}
\addtolength{\abovedisplayskip}{\extraspaces}
\addtolength{\belowdisplayskip}{\extraspaces}
\addtolength{\abovedisplayshortskip}{\extraspace}
\addtolength{\belowdisplayshortskip}{\extraspace}}
\newcommand{\ea}{\end{eqnarray}}

\newcommand{\bd}{\begin{displaymath}
\addtolength{\abovedisplayskip}{\extraspaces}
\addtolength{\belowdisplayskip}{\extraspaces}
\addtolength{\abovedisplayshortskip}{\extraspace}
\addtolength{\belowdisplayshortskip}{\extraspace}}
\newcommand{\ed}{\end{displaymath}}

\newcounter{saveeqn}

\newcommand{\newsection}[1]{
\vspace{12mm} \pagebreak[3] \addtocounter{section}{1}
\setcounter{equation}{0} \setcounter{subsection}{0}
\noindent{\bf \thesection. #1} \nopagebreak
\medskip
\nopagebreak
\addcontentsline{toc}{section}{\thesection. #1}}

\newcommand{\newsubsection}[1]{
\vspace{0.8cm} \pagebreak[3] \addtocounter{subsection}{1}
\setcounter{subsubsection}{0}
\noindent{ \it \thesubsection. #1} \nopagebreak \vspace{2mm}
\nopagebreak
\addcontentsline{toc}{subsection}{\thesubsection. #1}}

\newcommand{\newsubsubsection}[1]{
\medskip \pagebreak[3] \addtocounter{subsubsection}{1}
\noindent{\it \thesubsubsection. #1} \nopagebreak \vspace{2mm}
\nopagebreak
}

\begin{document}
\addtolength{\baselineskip}{1.5mm}

\thispagestyle{empty}

\vbox{} \vspace{0.0cm}

\begin{center}
\centerline{\LARGE{Quasi-Topological Gauged Sigma Models,}}
\bigskip
\centerline{\LARGE{The Geometric  Langlands Program,}} 
\bigskip
\centerline{\LARGE{And Knots}} 

\vspace{0.8cm}

{\bf{Meng-Chwan~Tan}\footnote{E-mail: tan@ias.edu}}
\\[2mm]
{\it Department of Physics, National University of Singapore\\
Singapore 119260} \\[1mm] 
\end{center}

\vspace{0.8cm}        

\centerline{\bf Abstract}\smallskip \noindent

We construct and study a closed, two-dimensional, quasi-topological $(0,2)$ gauged sigma model with target space a smooth $G$-manifold, where $G$ is any compact and connected Lie group. When the target space is a flag manifold of simple $G$, and the gauge group is a Cartan subgroup thereof,  the perturbative model describes, purely physically, the recently formulated mathematical theory of ``Twisted Chiral Differential Operators''.  This paves the way,  via a generalized $T$-duality, for a natural physical interpretation of the geometric Langlands correspondence for simply-connected, simple, complex Lie groups. In particular, the Hecke eigensheaves and Hecke operators can be described in terms of the correlation functions of certain operators that underlie the infinite-dimensional chiral algebra of the flag manifold model. Nevertheless, nonperturbative worldsheet twisted-instantons can, in some situations, trivialize the chiral algebra completely. This leads to a spontaneous breaking of supersymmetry whilst implying certain delicate conditions for the existence of Beilinson-Drinfeld $\cal D$-modules. Via supersymmetric gauged quantum mechanics on loop space, these conditions can be understood to be intimately related to a conjecture by H\"ohn-Stolz~\cite{Stolz} regarding the vanishing of the Witten genus on string manifolds with positive Ricci curvature. An interesting connection to Chern-Simons theory is also uncovered, whence we would be able to (i) relate general knot invariants of three-manifolds and Khovanov homology  to  ``quantum'' ramified $\cal D$-modules and Lagrangian intersection Floer homology; (ii)  furnish physical proofs of mathematical conjectures by Seidel-Smith~\cite{SS} and Gaitsgory~\cite{Gaitsgory-summary, Gaitsgory-Whittaker} which relate knots to symplectic geometry and Langlands duality, respectively.

\newpage

\renewcommand{\thefootnote}{\arabic{footnote}}
\setcounter{footnote}{0}

\tableofcontents

\newsection{Introduction}  

A closed, two-dimensional, quasi-topological sigma model with $(0,2)$ supersymmetry was first constructed by Witten in~\cite{TSM}. It was later studied in much greater detail in~\cite{CDO}, where an attempt to provide a physical interpretation of the mathematical theory of ``Chiral Differential Operators" (CDO's) defined in~\cite{GMS1},  was made. As the aforementioned model did not contain left-moving fermions, it could only be related to a bosonic specialization of the supersymmetric sheaf of CDO's. Shortly thereafter, a generalization of the effort in~\cite{CDO} to include left-moving fermions appeared in~\cite{MC}, wherein a quasi-topological heterotic sigma model with $(0,2)$ supersymmetry was shown to be related to the complete supersymmetric sheaf of CDO's. This generalization opened up the possibility of studying phenomenologically viable string-theoretic models via the mathematical theory of CDO's, and vice-versa. Nevertheless, the model without left-moving fermions has continued to be of great interest to us, primarily because for a certain class of nonanomalous target spaces, its chiral algebra furnishes a subset which generates (a completed enveloping algebra of) an affine algebra of a simple, simply-connected, complex Lie group ${\mathscr G}_{\mathbb C}$ at the \emph{critical} level -- a crucial ingredient in the original mathematical formulation by Beilinson and Drinfeld of the geometric Langlands program for ${\mathscr G}_{\mathbb C}$ using two-dimensional algebraic conformal field theory~\cite{BD Langlands}.   

The Langlands program has its origins in number theory~\cite{Langlands}. It relates representations of the Galois group of a number field to automorphic forms (such as ordinary modular forms of $SL(2, \mathbb Z)$). Its geometric analog, which involves complex curves of genus $g$ instead of number fields, is known as the geometric Langlands program. In 2006, the geometric Langlands program was given an elegant physical interpretation~\cite{KW} by Kapustin and Witten in terms of a four-dimensional, bounded, topologically-twisted ${\cal N} = 4$ Yang-Mills theory compactified on a complex curve of genus greater than one. In particular, they showed that the geometric Langlands correspondence which underlies the program, arises from an $S$-duality of the ${\cal N} = 4$ theory.   This gauge-theoretic interpretation, as elegant as it is, does not shed light on the utility of  two-dimensional  algebraic conformal field theory that has been ubiquitous in the mathematical literature since the seminal work of Beilinson and Drinfeld -- for a small sample, see~\cite{2a}-\cite{Rubtsov}. This is rather puzzling; afterall, the formal definition of algebraic  conformal field theory is rooted in concepts from \emph{physical} conformal field theory; one would think that it would be possible to relate any physical manifestation of the geometric Langlands correspondence to the algebraic formulation by Beilinson and Drinfeld. There have since been attempts by physicists to fill in this gap through the use of four-dimensional ${\cal N} = 2$ Yang-Mills theory~\cite{Nikita-Witten}, and two-dimensional physical conformal field theory~\cite{Gribet,Teschner}. However, these attempts are either preliminary or restricted to special examples of ${\mathscr G}_{\mathbb C}$ only. Another matter that was not addressed by Kapustin and Witten is the geometric Langlands correspondence for $g \leq 1$. In fact, Beilinson and Drinfeld's original formulation excludes complex curves of $g \leq 1$, too.  Nonetheless, the geometric Langlands correspondence has been established, at least partially, to also hold for $g \leq 1$ -- examples which generalize Beilinson and Drinfeld's formulation to include complex curves of $g= 0$ and $g=1$ can be found in~\cite{Rubtsov} and~\cite{genus 1 punctures, Nevin, genus 1 no punctures}, respectively. 

These outstanding issues prompted us to revisit the $(0,2)$ sigma model without left-moving fermions. It was quickly realized that for any physical model to be relatable to Beilinson and Drinfeld's algebraic formulation of the geometric Langlands correspondence, it ought to support, in one way or another, a \emph{family} of affine algebras of ${\mathscr G}_{\mathbb C}$ at the critical level that is parametrized by physical quantities associated with the Langlands dual group $^L{\mathscr G}_{\mathbb C}$. For a target space that is a flag manifold of ${\mathscr G}_{\mathbb C}$, the $(0,2)$ sigma model does support, in its chiral algebra, an affine algebra of ${\mathscr G}_{\mathbb C}$ at the critical level. That said, this sigma model is actually ``rigid'', in the sense that its affine algebra is unique -- i.e., we do not have a family of affine algebras  of ${\mathscr G}_{\mathbb C}$ at the critical level. This suggests that the model to consider ought not to be the $(0,2)$ model itself, but rather, a modified version of it. It soon became clear from the subsequent work~\cite{Arakawa} of Arakawa et al.~which defines a mathematical theory of ``Twisted Chiral Differential Operators'' (TCDO's), that this modified version of the $(0,2)$ model should also provide a physical interpretation of these TCDO's.  The construction of such a sigma model was the starting point of this paper.

Aside from providing a physical interpretation of (i) the mathematical theory of TCDO's, (ii) the algebraic conformal field theoretic formulation of the geometric Langlands correspondence for any ${\mathscr G}_\mathbb C$,  one of the main novelties of our purely two-dimensional sigma model approach is that it would allow us to deduce some very interesting results about the $g \leq 1$ case --  using purely physical arguments, not only can we rederive, in a much more economical fashion, various mathematical features of the  correspondence for $g \leq 1$ as established in the above references, we can also connect them to \emph{a priori} unrelated topics in algebraic and geometric topology. Furthermore,  by combining recent insights from higher-dimensional gauge theory~\cite{5-branes and knots}, we would be able to also provide physical proofs of far-reaching mathematical conjectures that relate the geometric Langlands program to symplectic geometry~\cite{SS, Kam} and knot theory~\cite{Gaitsgory-summary, Gaitsgory-Whittaker} -- an end which might prove elusive if we were to rely solely on higher-dimensional gauge theory. Let us now proceed to give a brief summary and  plan of the paper.

\smallskip\noindent{\it A Brief Summary and Plan of the Paper}  

 First, in $\S$2, we will formulate a non-dynamically $G$-gauged version of a twisted $(0,2)$ sigma model whose target space $X$ is a smooth $G$-manifold, where $G$ is any compact and connected Lie group. This gauged twisted  sigma model can be viewed as an ordinary twisted sigma model with a \emph{family} of  target spaces that are automorphic to $X$.  Then, we will focus on the relevant local operators of the gauged twisted sigma model, and study the properties of the holomorphic chiral algebra which they underlie. We will find, among other things, that the chiral algebra is infinite-dimensional -- a consequence of the fact that the twisted model is actually  \emph{quasi-topological} and not topological. We will also argue that the moduli of the chiral algebra can be interpreted in terms of the first $\check{\textrm C}$ech-cohomology of the sheaf of $G$-equivariant extended holomorphic $\partial$-closed two-forms on $X$.

In $\S$3, we will analyze the physical anomalies of the gauged twisted sigma model. The anomalies are found to be characterized by $G$-equivariant cohomology classes. This means that the anomaly-cancellation conditions are more stringent in the gauged model than in the ordinary model, as one might have expected.  
  
In $\S$4, we will introduce the notion of a sheaf of perturbative observables.  An alternative description of the chiral algebra of local operators in terms of $\check{ \textrm C}$ech cohomology will also be presented. If the gauge group is an\emph{ abelian} subgroup of $G$, the gauged twisted sigma model on any local patch of $X$ can be described in terms of a perturbed version of a free $\beta\gamma$-system. In order to obtain a complete description of the model over all  of $X$,  one will need to ``glue together'' these systems and their corresponding sheaves of chiral algebras along every pairwise intersection in $X$; this may be done using their local symmetries. In doing so, we will see that the purely mathematical obstruction to a global definition of the sheaf of chiral algebras is nothing but the physical anomaly of the model itself. As an illuminating application of our somewhat abstract discussion hitherto, we will, through a convenient example, demonstrate a novel understanding of the model's nonzero one-loop beta function solely in terms of holomorphic data. 

In $\S$5, we shall, for concreteness, consider the abelian gauge group to be a Cartan subgroup of simple $G$. One can then show that our gauged twisted sigma model  describes, purely physically, the theory of TCDO's on flag manifolds of simple, connected, complex Lie groups. If $G$ is also simply-connected, our analysis of the sheaves of TCDO's reveals that the chiral algebra $\mathscr A$ of the model would also furnish a \emph{family} of affine $G_{\mathbb C}$-algebras at the critical level, where $G_{\mathbb C}$ is a simply-connected, simple, complexified version of $G$. 

In $\S$6, we will argue that a generalized $T$-duality of the local gauged twisted sigma model over the flag manifold of $G_\mathbb C$, would imply that the family of affine $G_{\mathbb C}$-algebras at the critical level in $\S$5 is parameterized by  $^LG_{\mathbb C}$-opers on the worldsheet, where $^LG_{\mathbb C}$ is the \emph{Langlands dual} of  $G_{\mathbb C}$.  This crucial observation will allow us to furnish, in the next section, a natural physical interpretation of the geometric Langlands correspondence for $G_{\mathbb C}$.

In $\S$7, we will demonstrate, purely physically, a geometric Langlands correspondence between holomorphic $^LG_{\mathbb C}$-bundles on a complex curve $\Sigma$ and Hecke eigensheaves on the moduli space $\textrm{Bun}_{G_{\mathbb C}}$ of holomorphic $G_{\mathbb C}$-bundles on $\Sigma$.  In essence, the Hecke eigensheaves and Hecke operators of the geometric Langlands program can be described in terms of the correlation functions of certain operators which underlie the bosonic sector of the chiral algebra of the gauged twisted sigma model over the flag manifold of $G_{\mathbb C}$. Furthermore, one can also understand the uniqueness or non-uniqueness property of the Hecke eigensheaves for various $\Sigma$ as established purely mathematically, from the anomaly-cancellation conditions of the model.

In $\S$8, we will analyze the nonperturbative effects of worldsheet twisted-instantons on the chiral algebra of the gauged twisted sigma model. For certain worldsheets whereby the flag manifold model is nonanomalous and isomorphic to its untwisted counterpart, the chiral algebra is completely trivialized by such instanton effects. This results in a spontaneous breaking of supersymmetry whilst implying that there can be no Beilinson-Drinfeld $\cal D$-modules when $\Sigma$ is rational with less than three punctures (in agreement with the mathematical literature~\cite{Rubtsov}). We then go on to interpret this nonperturbative phenomenon in the context of supersymmetric gauged quantum mechanics on loop space. In doing so, we will find that (i) there can be no harmonic spinors on the loop space of flag manifolds of $G_\mathbb C$, (ii) the aforementioned condition on $\cal D$-modules is intimately related to a conjecture by H\"ohn-Stolz~\cite{Stolz} which asserts that the Witten genus must vanish on string manifolds (i.e.,~manifolds of zero first Pontraygin class) with positive Ricci curvature.  

And finally in $\S$9, we will first explain why the states of a Chern-Simons theory of a compact, simply-connected, simply-laced gauge group $G$, would be captured by certain correlation functions of the corresponding flag manifold model in the infinite-volume limit. Then, with the aid of Heegaard splittings, we will explain why knot invariants of three-manifolds ought to be related to ``quantum'' ramified $\cal D$-modules. Next, by specializing to the case where  $G = SU(2)$ and the underlying three-manifold is ${\bf S}^3$, we will (i) make contact with the Jones polynomial of an arbitrary link and its corresponding Khovanov homology, (ii) furnish a physical proof of a mathematical conjecture by Seidel-Smith~\cite{SS} which relates the latter to Lagrangian intersection Floer homology. Lastly, we will demonstrate, via a generalized $T$-duality of the flag manifold model in the infinite-volume limit,  (i) a ramified geometric Langlands correspondence for $G_\mathbb C$ (the complexification of $G$); and (ii) a correspondence  between representations of $^LG_{\mathbb C}$  and  ``classical'' ramified $\cal D$-modules on the moduli space of holomorphic parabolic $G_{\mathbb C}$-bundles on a rational curve, where $^LG_{\mathbb C}$ is the Langlands dual of $G_\mathbb C$; thereby proving physically a mathematical conjecture by Gaitsgory~\cite{Gaitsgory-summary, Gaitsgory-Whittaker}. 

\smallskip\noindent{\it A Shorter Route Through This Paper}  

This somewhat lengthy paper, though most coherent when read in its entirety, can also be approached -- depending on the reader's specific interests -- in the following ways. The reader who is solely interested in the physics of the perturbative gauged sigma model and how it can be related to the mathematical theory of TCDO's, can just read $\S$1--$\S$5. The reader who wishes to fully understand the physical interpretation of the geometric Langlands program for any $g$, will need to read all of $\S$1--$\S$7. The reader who is solely interested in instanton effects in the sigma model and how they can lead to a spontaneous breaking of supersymmetry and a physical proof of the H\"ohn-Stolz conjecture, can just  read $\S$8. That said, if the reader wishes to also understand the implications of such instanton effects in the context of the geometric Langlands program, he or she will also need to read $\S$1--$\S$7. Last but not least, the reader who is chiefly interested in the connections between the geometric Langlands program and knots as implied by the relevant physics of the sigma model, can omit $\S$8 altogether. 

\bigskip\noindent{\it Acknowledgments}

This work is supported by the NUS startup grant. I would like to thank A.~Beilinson, D.~Ben-Zvi, I.~Biswas, M.~Brion, C.W.~Chin, P.~Dalakov, D.~Gaitsgory, I.M.~Riera, C.~Sorger and D.Q.~Zhang, for various mathematical consultations. I would also like to thank T.~Pantev and J.~Yagi for useful exchanges. 

\vspace{0.2cm}

This paper is dedicated to my beloved wife, See-Hong, whose optimism and cheerfulness despite her ongoing battle with cancer, have made it possible for this work to see the light of day.

\newsection{A Quasi-Topological Gauged Sigma Model And Its Chiral Algebra}

\vspace{-0.0cm}

\newsubsection{A Non-Dynamically Gauged Version Of A Twisted $(0,2)$ Sigma Model}

\smallskip\noindent{\it The Ordinary $(0,2)$ Sigma Model}

Let us first recall the ordinary two-dimensional nonlinear sigma model
with $(0,2)$ supersymmetry on a complex manifold $X$. It governs
maps $\Phi : \Sigma \to X$, with $\Sigma$ being the worldsheet
Riemann surface. By picking local coordinates $z$, $\bar z$ on
$\Sigma$, and $\phi^{i}$, $\phi^{\bar i}$ on $X$, the map $\Phi$
can then be described locally via the functions $\phi^{i}(z, \bar
z)$ and $\phi^{\bar i}(z, \bar z)$. Let ${\overline K}$ be the
anti-canonical bundle of $\Sigma$ (the bundle of $(0,1)$-forms), such that the right-moving spinor bundle of $\Sigma$ is
given by ${\overline K}^{1/2}$; let $TX$ and $\overline {TX}$ be the
holomorphic and antiholomorphic tangent bundles of $X$; then, the
right-moving fermi fields of the model are $\psi^i$ and $\psi^{\bar i}$, and they are smooth sections of the bundles ${\overline K}^{1/2}
\otimes \Phi^*{TX}$ and ${\overline K}^{1/2} \otimes
\Phi^*{\overline {TX}}$, respectively. Here, $\psi^i$ and
$\psi^{\bar i}$ are superpartners of the scalar fields $\phi^i$
and $\phi^{\bar i}$. Let $g$ be the hermitian metric on $X$. The classical
action is then given by
\begin{eqnarray}
S& = & \int_{\Sigma} |d^2z| \  {1\over 2}g_{i{\bar j}} (\partial_z
\phi^i \partial_{\bar z}\phi^{\bar j} + \partial_{\bar z} \phi^i
\partial_z\phi^{\bar j} ) + g_{i{\bar j}} \psi^i D_z \psi^{\bar
j}, \label{action}
\end{eqnarray}
where $i, {\bar i} = 1 \dots, n={\textrm {dim}}_{\mathbb C}X$, and
$|d^2 z| = i dz \wedge d{\bar z}$. In addition,  $D_z$ is the
$\partial$ operator on ${\overline K}^{1/2} \otimes
\phi^*{\overline {TX}}$ using the pull-back of the Levi-Civita
connection on $TX$. In formulas (using a local trivialization of
${\overline K}^{1/2}$), we have \be D_z \psi^{\bar j}
= \partial_z \psi ^{\bar j} + \Gamma^{\bar j}_{\bar l \bar k}
\partial _z \phi^{\bar l} \psi^{\bar k}, \label{cov dev psi}\ee  where $\Gamma^{\bar j}_{\bar l \bar k}$ is
the affine connection of $X$.

The infinitesimal transformation  of the fields generated by the
supercharge $\overline Q_+$ under the first right-moving
supersymmetry, is given by
\begin{eqnarray}
\label{tx1}
\delta \phi^{i} = 0, & \quad & \delta \phi^{\bar i} = {{\bar \epsilon}_-} \psi ^{\bar i}, \nonumber \\
\delta \psi ^{\bar i} = 0, & \quad & \delta \psi^i = - {{\bar
\epsilon}_-}\partial_{\bar z}\phi^i;
\end{eqnarray}
while the infinitesimal transformation  of the fields generated by
the supercharge $Q_+$ under the second right-moving supersymmetry,
is given by
\begin{eqnarray}
\label{tx2}
\delta \phi^{i} = {{\epsilon}_-}\psi ^{i}, & \quad & \delta \psi^{\bar i} = -{{\epsilon}_-} \partial_{\bar z} \phi^{\bar i}, \nonumber \\
\delta \psi ^{i} = 0, & \quad & \delta \phi^{\bar i} = 0.
\end{eqnarray}
In the above, (${{\bar \epsilon}_-}$)${{\epsilon}_-}$ are
(anti)holomorphic sections of ${\overline K}^{-1/2}$.

\bigskip\noindent{\it Twisting the Ordinary $(0,2)$ Sigma Model}

Classically, the action (\ref{action}) and therefore the ordinary sigma model
that it describes, possesses a right-moving R-symmetry, giving
rise to a $U(1)_R$ global symmetry group. Denoting $q$ to be the
charge of the right-moving fermi fields under this symmetry group,
we find that $\psi^{\bar i}$ and $\psi^i$ will have charge $q =
\pm 1$, respectively. 

In order to define a twisted variant of the model, the spins of
the fermi fields need to be shifted by a multiple of their
corresponding right-moving charge $q$ under the global $U(1)_R$
symmetry group. By considering a shift in the spin $S$ via $S \to
S + {1\over 2} \left [(2{\bar s} -1)q \right]$ (where $\bar s$
is a real number), the fermi fields $\psi^i$ and $\psi^{\bar i}$ of the\emph{ twisted} model will
transform as smooth sections of the bundles ${\overline K}^{(1- \bar s)} \otimes
\Phi^*{TX}$ and ${\overline K}^{\bar s} \otimes \Phi^*{\overline{TX}}$. Notice that for $s = \bar s = {1\over 2}$, the
fermi fields transform as smooth sections of the same tensored
bundles defining the original $(0,2)$ sigma model, i.e., we get
back the untwisted model.

For our purposes in this paper, we shall consider the
case where $\bar s = 0$. Then, the fermi fields of the twisted
model will transform as smooth sections of the following bundles:
\begin{eqnarray}
\psi^i_{\bar z} \in  \Gamma \left({\overline K}^1 \otimes
\Phi^*{TX} \right), & \qquad &  \psi^{\bar i} \in  \Gamma \left(
\Phi^*{\overline{TX}}\right).
\label{ungauged section}
\end{eqnarray}
Notice that we have included additional indices in the above
fields so as to reflect their new geometrical characteristics on
$\Sigma$; the fermi field without a $\bar z$ index transform as a
worldsheet scalar, while the  fermi field with a $\bar z$ index
transform as a $(0,1)$-form on the worldsheet. In addition, as
reflected by the $i$, and $\bar i$ indices, all fields continue to
be valued in the pull-back of the corresponding bundles on $X$.
Thus, the action of the twisted variant of the ordinary $(0,2)$ sigma model
is given by
\begin{eqnarray}
S_{\mathrm twist}& = & \int_{\Sigma} |d^2z| \ {1\over 2}g_{i{\bar
j}} (\partial_z \phi^i \partial_{\bar z}\phi^{\bar j} +
\partial_{\bar z} \phi^i \partial_z\phi^{\bar j}) + g_{i{\bar j}} \psi_{\bar z}^i D_z \psi^{\bar j}.
\label{actiontwist}
\end{eqnarray}

A twisted theory is the same as an untwisted one when defined on a
$\Sigma$ which is flat. Hence, locally (where one has the freedom
to select a flat metric),  the twisting does nothing at all.
However, what happens nonlocally may be nontrivial. In
particular, note that globally, the supersymmetry parameters
$\epsilon_-$ and ${\bar \epsilon}_-$ must now be interpreted as
sections of different line bundles; in the twisted model, the
transformation laws given by (\ref{tx1}) and (\ref{tx2}) are still
valid, and because of the shift in the spins of the various
fields, we find that for the laws to remain physically consistent,
${\bar \epsilon}_-$ must now be a function on $\Sigma$ while
$\epsilon_-$ must be a section of the nontrivial bundle
${\overline K}^{-1}$. One can therefore canonically pick ${\bar
\epsilon}_-$ to be a constant and $\epsilon_-$ to vanish, i.e.,
the twisted variant of the ordinary $(0,2)$ sigma model has just $\it{one}$
canonical global fermionic symmetry generated by the supercharge
${\overline Q}_+$. Hence, the infinitesimal variations of the
(twisted) fields under this single canonical symmetry must read
(after setting ${\bar \epsilon}_-$ to 1) as
\begin{eqnarray}
\label{txtwist}
\delta \phi^{i} = 0, & \quad & \delta \phi^{\bar i} = \psi^{\bar i}, \nonumber \\
\delta \psi ^{\bar i} = 0, & \quad & \delta \psi_{\bar z}^i = -
\partial_{\bar z}\phi^i.
\end{eqnarray}
From (\ref{txtwist}), one can see that   ${{\overline Q}^2_+}  = 0$ (off-shell) on the fields. In addition, after twisting, ${{\overline Q}_+} $
transforms as a scalar. Consequently, we find that the symmetry generated by ${{\overline Q}_+} $ behaves like a BRST-symmetry.

Note at this point that the transformation laws of (\ref{txtwist})
can be expressed in terms of the BRST operator ${\overline Q}_+$,
whereby $\delta W = \{{\overline Q}_+, W]$ for any field $W$.\footnote{In a ${\mathbb Z}_2$-graded algebra, the symbol $\{A, B]$ denotes $AB - (-1)^{AB} BA$.} One
can then show that the action  (\ref{actiontwist}) can be written
as \be S_{\mathrm twist} = \int_{\Sigma} |d^2z| \{{\overline Q}_+,
V\} + S_{\mathrm top}, \label{Stwist} \ee where \be V = - g_{i \bar
j}  \psi_{\bar z}^i \partial_z \phi^{\bar j}, \label{chi} \ee
and \be S_{\mathrm top} = {1\over 2}\int_{\Sigma} |d^2z| \,  g_{i \bar j}
\left( \partial_z \phi^i
\partial_{\bar z} \phi^{\bar j} - \partial_{\bar z} \phi^i
\partial_z \phi^{\bar j} \right) 
\label{Stop}
\ee 
is $\int_{\Sigma} \Phi^*({\cal K})$,
the integral of the pull-back to $\Sigma$ of the $(1,1)$-form ${\cal K}=
{i \over 2} g_{i \bar j} d\phi^i \wedge d\phi^{\bar j}$.

In the absence of nonperturbative effects, only the degree-zero
maps of the term $\int_{\Sigma} \Phi^*({\cal K})$ contribute to the path
integral factor $e^{-S_{\mathrm twist}}$; one can therefore set $\int_{\Sigma} \Phi^*({\cal K}) = S_{\mathrm{top}} = 0$. Notice also that since ${{\overline Q}^2_+} = 0$, the term $ \{{\overline Q}_+,V\}$ in (\ref{Stwist}) is invariant under the transformation
generated by ${\overline Q}_+$. Moreover, for the transformation laws of (\ref{txtwist})
to be physically consistent, ${\overline Q}_+$ must have charge
$q = 1$ under the global $U(1)_R$ symmetry group. Since $V$ has a
corresponding charge of $q = -1$, the term $ \{{\overline Q}_+,V\}$ in (\ref{Stwist})  continues to be invariant under the $U(1)_R$
symmetry group (at the classical level). In summary, the effective
perturbative action that is both $U(1)_R$- and  ${\overline Q}_+$-invariant will be given by
\begin{eqnarray}
S_{\mathrm pert}& = & \int_{\Sigma} |d^2z| \  g_{i{\bar j}}(
 \partial_{\bar z}\phi^i  \partial_z \phi^{\bar j} +
\psi_{\bar z}^i D_z \psi^{\bar j}), \label{actionpert}
\end{eqnarray}
where it can also be written as \be S_{\mathrm pert} = \int_{\Sigma}
|d^2z| \{{\overline Q}_+, V\}. \label{Spert} \ee

\bigskip\noindent{\it A Family Of Target Spaces for the Twisted $(0,2)$ Sigma Model}

We would now like to generalize the above twisted model with action $S_{pert}$, such that it will describe a \emph{family} of target spaces which are related to $X$ via its diffeomorphism group,  whilst still being ${\overline Q}_+$- and $U(1)_R$-invariant.  To get an idea what one must do towards this end, first recall that the pair $(\phi^i (z, \bar z), \phi^{\bar i} (z, \bar z))$ can be viewed as a single-valued function which maps, in a one-to-one manner, a point in $\Sigma$ to a point in $X$; in other words,  $(\phi^i (z, \bar z), \phi^{\bar i} (z, \bar z))$ defines a section of the trivial bundle $X \times \Sigma$, where $X$ and $\Sigma$ are its fiber and base, respectively. That the model of (\ref{actionpert}) has a fixed target space is reflected in the fact that over all of $\Sigma$ is a fixed manifold $X$ in the trivial bundle $X \times \Sigma$. Therefore, if one would like to describe a family of target spaces, one ought to generalize  $(\phi^i (z, \bar z), \phi^{\bar i} (z, \bar z))$ to define a section of a \emph{nontrivial} fiber bundle $F$, where  $X \hookrightarrow F \to \Sigma$ -- indeed, one would have, in this instance, a family of complex manifolds over $\Sigma$ which are related to the fixed $X$ via the structure group of $F$. 

In the most general case, one can take the structure group of $F$ to be the noncompact diffeomorphism group of $X$. Nevertheless, if $X$ is some smooth $G$-manifold, where $G$ is any compact, connected Lie group, a natural choice for the structure group of $F$ would be $G$ itself. On such an $X$, the $G$-action is infinitesimally generated by a set of vector fields $V_a$ on $X$, where $a = 1, \dots, d={\textrm {dim}}~G$. These are holomorphic vector fields, which means that their holomorphic (antiholomorphic) components are holomorphic (antiholomorphic) functions, i.e., for  $V_a = \sum_{i =1}^{ \textrm{dim}_{\mathbb C} X} V_a^i (\partial / \partial \phi^i) + \sum_{\bar i =1}^{ \textrm{dim}_{\mathbb C} X}V_a^{\bar i}( \partial / \partial \phi^{\bar i})$, we have
\be
{{\partial V^i_a} \over {\partial \phi^{\bar j}}} ={ {\partial V^{\bar i}_a} \over {\partial \phi^{j}} } = 0.
\label{v1}
\ee
In addition, the $V_a$'s realise a $d$-dimensional Lie algebra $\mathfrak g$ of $G$, i.e., they obey
\be
[V_a, V_b] = f_{ab}{}^c V_c,
\label{v3}
\ee
where $f_{ab}{}^c$ are the structure constants of $G$. This can also be explicitly written in component form as
\begin{eqnarray}
[V_a, V_b]^i & = & V^j_a ({\partial V_b^i \over \partial \phi^j}) - V^j_b ({\partial V_a^i \over \partial \phi^j}) \nonumber \\
& = & f_{ab}{}^c V_c^i,\
\label{v4}
\end{eqnarray}
and
\begin{eqnarray}
[V_a, V_b]^{\bar i} & = & V^{\bar j}_a ({\partial V_b^{\bar i} \over \partial \phi^{\bar j}}) - V^{\bar j}_b ({\partial V_a^{\bar i} \over \partial \phi^{\bar j}}) \nonumber \\
& = & f_{ab}{}^c V_c^{\bar i}.\
\label{v5}
\end{eqnarray}
Furthermore, as the $G$-action on $X$ is supposed to leave fixed its metric, the vector fields will obey the Killing vector equations
\be
\nabla_i V_{ja} + \nabla_j V_{ia} = 0, \quad \nabla_i V_{\bar j a} + \nabla_{\bar j} V_{ia} = 0,
\label{v2}
\ee
where $\nabla$ is the covariant derivative with respect to the Levi-Civita connection on $X$, while $V_{ia} = g_{i \bar j}V^{\bar j}_a$ and $V_{\bar j a} = g_{i \bar j} V^i_a$.

With regard to our generalization of the model, the fact that  $(\phi^i (z, \bar z), \phi^{\bar i} (z, \bar z))$ now defines a section of a nontrivial bundle $F$ with structure group $G$ means that one must replace all ordinary derivatives of  $(\phi^i (z, \bar z), \phi^{\bar i} (z, \bar z))$ with covariant derivatives -- the relevant $G$-connection in this case being a local one-form gauge field $A$ on $\Sigma$ with values in $\frak g$, i.e., the $V_a$'s.  The components of $A$ obey the usual infinitesimal gauge transformation laws 
\be
\delta_{\epsilon} A^a_z =  \partial_z \epsilon^a -  f_{bc}{}^a \epsilon^b A_z^c \quad {\rm {and}} \quad   \delta_{\epsilon} A^a_{\bar z} = \partial_{\bar z} \epsilon^a + f_{bc}{}^a \epsilon^b A_{\bar z}^c,
\label{tx A}
\ee
where $\epsilon$ is a\emph{ position-dependent} zero-form on $\Sigma$ with values in $\frak g$.  Under an infinitesimal gauge transformation, the $\phi$'s, which play the role of coordinates in $X$, will change as
\be
\delta_{\epsilon} \phi^i =  \epsilon^aV_a^i \quad {{\rm and}} \quad \delta_{\epsilon} \phi^{\bar i} = -\epsilon^aV_a^{\bar i}.
\label{tx phi}
\ee
This means that a consistent generalization of the twisted model entails making the replacements
\be
\partial_{\bar z} \phi^i \rightarrow D_{\bar z} \phi^i = \partial_{\bar z} \phi^i - A^a_{\bar z}V^i_a \quad {\rm {and}} \quad \partial_z \phi^{\bar i} \rightarrow D_z \phi^{\bar i} = \partial_z \phi^{\bar i} + A^a_zV^{\bar i}_a
\label{replacement 1}
\ee 
in (i) the action (\ref{actionpert}), (ii) the field variations in (\ref{txtwist}), and (iii) $V$ in (\ref{chi}). As required of covariant derivatives, the gauge variations $\delta_{\epsilon} (D_{\bar z} \phi^i)$ and $\delta_{\epsilon} (D_z \phi^{\bar i})$ do not contain worldsheet-derivatives of the parameter $\epsilon$.  

On the other hand, under an infinitesimal gauge transformation, the $\psi$'s, which play the role of tangent vectors in $X$, will change as
\be
\delta_{\epsilon} \psi^{i}_{\bar z} =  \epsilon^a\partial_{k} V_a^{i} \psi^{k}_{\bar z}
 \quad {{\rm and}} \quad \delta_{\epsilon} \psi^{\bar i} = -\epsilon^a\partial_{\bar k} V_a^{\bar i} \psi^{\bar k}.
 \label{tx psi}
\ee
This means that a consistent generalization of the twisted model entails making the replacement
\be
D_z\psi^{\bar i}  \rightarrow {\widehat D}_z \psi^{\bar i} = D_z\psi^{\bar i} + A^a_z \nabla_{\bar k} V^{\bar i}_a \psi^{\bar k}
\label{replacement 2}
\ee
in the action (\ref{actionpert}), whereby $\nabla_{\bar k} V^{\bar i}_a = \partial_{\bar k} V_a^{\bar i} + \Gamma_{\bar j \bar k}^{\bar i}V^{\bar j}_a$. As required of covariant derivatives, the gauge variation $\delta_{\epsilon} ( {\widehat D}_z \psi^{\bar i} )$ does not contain worldsheet-derivatives of the parameter $\epsilon$ either. 

\bigskip\noindent{\it {A Non-Dynamically Gauged Version of the Twisted $(0,2)$ Sigma Model}}

Thus, the action of the generalized theory will be given by 
\begin{eqnarray}
S_{gauged}& = & \int_{\Sigma} |d^2z| \  g_{i{\bar j}}  ( D_{\bar z} \phi^i D_z\phi^{\bar j}
 + \psi^i_{\bar z} {\widehat D}_z \psi^{\bar
j}), \label{action-gen}
\end{eqnarray}
and moreover, $S_{gauged}$ will be invariant under (\ref{tx A}), (\ref{tx phi}) and (\ref{tx psi}), if the derivatives $\partial_{\bar k} V^{\bar i}_a = [\bar \partial V_a]^{\bar i}_{\bar k}$ and $\partial_{k} V^{i}_a = [\partial V_a]^{i}_{k}$  satisfy~\cite{hull}
\be
{\cal L}_a ({\partial V_b}) - {\cal L}_b ({\partial V_a})  = [\partial V_a, \partial V_b]  - f_{ab}{}^c \partial V_c \quad {\rm and} \quad {\cal L}_a ({\bar \partial V_b}) - {\cal L}_b ({\bar \partial V_a})  = [\bar \partial V_a, \bar \partial V_b]  - f_{ab}{}^c \bar \partial V_c,
\label{satisfy}
\ee
where ${\cal L}_a$ is the Lie-derivative with respect to the vector field $V_a$. As is clear from (\ref{action-gen}), one can also interpret the generalized model as a \emph{non-dynamically} $G$-gauged version of the twisted $(0,2)$ sigma model with a \emph{fixed} target space $X$. 

As in the original ungauged model, one can also write 
\be S_{gauged} = \int_{\Sigma}
|d^2z| \{ Q,  \mathscr V\}, 
\label{Sgauged}
 \ee
where
\be
\mathscr V =  - g_{i \bar j} \psi^i_{\bar z} D_z \phi^{\bar j},
\label{V}
\ee
and where the requisite field variations generated by the scalar supercharge $Q$ are 
\begin{eqnarray}
\label{txgauged}
\delta \phi^{i} = 0, & \quad & \delta \phi^{\bar i} = \psi^{\bar i}, \nonumber \\
\delta \psi ^{\bar i} = 0, & \quad & \delta \psi_{\bar z}^i = -D_{\bar z}\phi^i, \\
\delta A^a_z = 0, & \quad & \delta A^a_{\bar z} =0. \nonumber
\end{eqnarray}
Using (\ref{v1}), one can compute from (\ref{txgauged}) that $Q^2 = 0$ (off-shell) on all fields. Consequently, from (\ref{Sgauged}), one can see that $S_{gauged}$ is $Q$-invariant, as desired.  In addition, (\ref{txgauged}) implies that one can assign $(Q, \psi^{\bar i}, \psi^i_{\bar z}, \phi^i, \phi^{\bar i}, A^a_z, A^a_{\bar z})$ to have $q$-charge $(1, 1, -1, 0,0, 0, 0)$ under a global $U(1)_R$-symmetry;  one can then see from (\ref{Sgauged}) and (\ref{V}) that as desired, $S_{gauged}$ has vanishing  $q$-charge and is therefore  $U(1)_R$-invariant. Moreover, since $Q$ is nilpotent, it will mean that one can also define a $Q$-cohomology of operators in the theory. 

The above model with perturbative \emph{gauge-invariant} action $S_{gauged}$, shall be our model of interest henceforth.

\newsubsection{The Chiral Algebra Of The Gauged Twisted $(0,2)$ Sigma Model}

Classically, the gauge twisted model is scale invariant: one can compute that the trace of the stress tensor from $S_{gauged}$ vanishes, i.e., $T_{z \bar z} = 0$. The rest of the nonvanishing components of the stress tensor, at the classical level, are given by
\be
T_{zz} = g_{i \bar j} \partial_z \phi^i D_{z} \phi^{\bar j},
\label{Tzz}
\ee
and
\be
T_{\bar z \bar z} =g_{i \bar j}  \left( D_{\bar z} \phi^i \partial_{\bar z} \phi^{\bar j} +  \psi_{\bar z}^i D_{\bar z} \psi^{\bar j} \right). 
\ee 
Furthermore, one can go on to show that 
\be
T_{\bar z \bar z} = \{ Q , - g_{i \bar j} \psi_{\bar z}^i \partial_{\bar z} \phi^{\bar j} \},
\label{tZZ}
\ee
and      
\begin{eqnarray} 
\label{tzz}
[Q , T_{zz} ] & = &   g_{i \bar j} \partial_z \phi^i  {\widehat D}_z \psi^{\bar j}  =  0 \hspace{0.2cm} (\textrm {on-shell}).
\end{eqnarray}
From (\ref{tzz}) and (\ref{tZZ}), we see that $T_{zz}$ is an operator in the $Q$-cohomology while $T_{\bar z \bar z}$ is $Q$-exact and thus trivial in $Q$-cohomology. The fact that $T_{zz}$ is not $Q$-exact even at the classical level implies that the gauged twisted model is not a 2D $\it{topological}$ field theory; rather, it is a 2D $\it{quasi}$-$\it{topological}$ field theory. This is because the underlying model has $(0,2)$  and not $(2,2)$ supersymmetry. On the other hand, the fact that $T_{\bar z \bar z}$ is $Q$-exact leads to some nontrivial implications for the nature of the local operators in the $Q$-cohomology. Let us elucidate this further.

Consider a local operator $\cal O$ that is inserted at the origin. If it has
scaling dimension $(n,m)$, then, under a rescaling $z\to \lambda z$, $\bar
z\to \bar\lambda z$ (which is a symmetry of the classical theory),
it would gain a factor of $\lambda^{-n}\bar\lambda{}^{-m}$. Classically, local
operators have dimensions $(n \geq 0,m \geq 0)$.\footnote{Anomalous
dimensions under RG flow may shift the values of $n$ and $m$ quantum mechanically, but the spin given by
$(n-m)$, being an intrinsic property, remains unchanged.} However, only local operators with $m = 0$ survive in $Q$-cohomology, as the rescaling of $\bar z$ is generated by $\bar L_0=\oint d\bar z\, \bar z T_{\bar
z\bar z}$.  (Recall from the previous paragraph that $T_{\bar z\,\bar z}$
is of the form $\{Q,\dots\}$, so $\bar L_0=\{Q,V_0\} $ for some $V_0$. If $\cal O$ is to be admissible as a local physical operator, it must at least be true that $\{Q, {\cal O}]=0$. Consequently, $[\bar L_0,{\cal
O}]=\{Q,\{V_0,{\cal O}] ]$.  Since the eigenvalue of $\bar L_0$ on $\cal O$ is $m$, we have $[\bar L_0,{\cal O}]=m{\cal O}$. Therefore, if $m\not= 0$, it follows that ${\cal O}$ is $Q$-exact and thus trivial in $Q$-cohomology.)

By a similar argument, we can show that $\cal O$,  as an element of the $Q$-cohomology, varies holomorphically with $z$. Indeed, since the momentum operator (which acts on $\cal O$ as $\partial_{\bar z}$) is given by $\bar L_{-1}$, the term $\partial_{\bar z} \cal O$ will be given by the commutator $[ \bar L_{-1}, \cal O]$. Since $\bar L_{-1} = \oint d\bar z\,T_{\bar z\bar z}$, we will  have $\bar L_{-1}=\{{Q},V_{-1}\}$ for some $V_{-1}$. Thus, as $\cal O$ is physical such that ${\{Q, \cal O]} =  0$, it will be true that $\partial_{\bar z}{\cal O}=\{{Q}, \{V_{-1},{\cal O}]]$ which hence vanishes in $Q$-cohomology.

\vspace{0.4cm}{\noindent{\it The Quantum Theory}}

The observations that we have made so far are based solely on classical grounds. The question that one might then ask is whether these observations will continue to hold when we eventually consider the quantum theory. The key point to note is that if it is true classically that a cohomology vanishes, it should continue to do so in perturbation theory, whence quantum effects are small enough. Since the above observations about the local operators were made based on the classical fact that $T_{\bar z \bar z}$ vanishes in $Q$-cohomology, they will continue to hold at the quantum level, i.e., the local operators in the $Q$-cohomology of the quantum theory  $\it{continue}$ to vary holomorphically  with $z$ and have  dimension $(n,0)$.

On the other hand, $T_{zz}$, which does not vanish in $Q$-cohomology at the classical level, can potentially vanish in $Q$-cohomology at the quantum level. In fact, one-loop corrections to the action of $Q$ suggest that in the quantum theory, $[ Q, T_{zz}]  = U$, where $U$ must necessarily be a fermionic operator with dimension $(2,0)$ and $q =1$. In order to determine the explicit form of $U$, first note that from the conservation of the stress tensor, we have $\partial_{\bar z}T_{zz} = - \partial_z T_{z \bar z}$. Since $T_{z \bar z}$ in the quantum theory, while it may no longer be zero, would still be of the form $T_{z \bar z} = \{Q, G_{z \bar z}\}$ for some fermionic operator $G_{z \bar z}$,\footnote{Since perturbative quantum corrections can only annihilate cohomology classes and not create them, $T_{z \bar z}$ must remain trivial in $Q$-cohomology, i.e., even though $T_{z \bar z}$ may no longer be zero, it would still be $Q$-exact.}  for all our purposes, we can regard $\partial_z T_{z \bar z}$ to be $Q$-exact and therefore, $\partial_{\bar z}T_{zz} \sim 0$ in $Q$-cohomology. The holomorphy of $T_{zz}$ will then allow us to make a Laurent expansion  $T_{zz} (z) = \sum^{n = \infty}_{n = - \infty} L_n z^{-n -2} $, where in particular, the operator $\partial_z = L_{-1} = \oint dz \, T_{zz}$. Second, since $\delta (\partial_z {\cal O}) = \partial_z (\delta {\cal O})$, it will mean that $[Q, L_{-1} ] =  \oint dz \, [Q, T_{zz}]  = \oint dz \, U = 0$; in other words, we ought to have $U = \partial_z(\dots)$. Third, note that sigma model perturbation theory is local in $X$ and depends on an expansion of the metric tensor in a Taylor series up to some given order; i.e., corrections to the action of $Q$  can be constructed locally from the metric of $X$ appearing in the action. An example of such a correction is immediately provided by the Ricci tensor $R_{i \bar j}$ of $X$. Fourth, recall that $\psi^{\bar j}$, like $Q$, has $q=1$. And lastly, recall that all derivatives of fields must be covariant. Altogether, this means that one can write $U = \partial_z(R_{ i \bar j}D_z \phi^i    \psi^{\bar j})$, such that at the quantum level,  
\be
[Q, T_{zz}] = \partial_z(R_{ i \bar j}D_z \phi^i    \psi^{\bar j}).
\label{tzzanomaly}
\ee    
Notice that the term on the RHS of (\ref{tzzanomaly}) $\it{cannot}$ be eliminated through the equations of motion of the theory; neither can we modify $T_{zz}$ (by subtracting a total derivative term) such that it continues to be $Q$-invariant. Thus, in a ``massive''  model where $R_{i \bar j} \neq 0$, $T_{zz}$ indeed vanishes in  $Q$-cohomology at the quantum level. Moreover, because (\ref{tzzanomaly}) involves the Ricci tensor of $X$,  this vanishing of $T_{zz}$ in $Q$-cohomology can be interpreted as an effect that is associated with the one-loop beta function of the sigma model. In fact, (\ref{tzzanomaly})  is just a gauged generalization of a well-known result for the ordinary twisted $(0,2)$ sigma model.  In $\S$4.7, we will study more closely from a different viewpoint, the corrections to the action of $Q$ in a ``massive'' model; there, (\ref{tzzanomaly}) will appear in a different guise such that one can interpret it in terms of holomorphic data. 

At any rate, (\ref{tzzanomaly}) implies that for a ``massive'' model,  operators do not remain in the $Q$-cohomology after general holomorphic coordinate transformations on the worldsheet, i.e., the model is $\it{not}$ conformal at the level of the    $Q$-cohomology. Nevertheless, since $[Q, L_{-1} ] = 0$, the operators remain in the $Q$-cohomology after global translations on the worldsheet. In addition, notice that since $Q$ is a scalar with spin zero in the gauged twisted model, we ought to have $[S, Q]= 0$, where the spin operator $S= L_0 - \bar L_0$; with the condition that $\bar L_0=0$ in $Q$-cohomology, $[S, Q]= 0$ would imply that $[Q, L_0 ] = 0$; i.e., operators remain in the $Q$-cohomology after global dilatations of the worldsheet coordinates.

One can also make the following observations about the correlation functions of these local operators. Firstly, note that $\left < \{ Q, W ] \right> = 0$ for any operator $W$, and recall  that for any local physical operator ${\cal O_{\alpha}}$, we have ${\{Q, {\cal O_{\alpha}}]} = 0$. As the $\partial_{\bar z}$ operator on $\Sigma$ is given by ${\bar L}_{-1} = \oint d{\bar z}\ T_{\bar z \bar z}$, where  $T_{\bar z \bar z} = \{Q, \dots \}$, we find that ${\partial_{\bar z}\left <{\cal O}_1(z_1) {\cal O}_2(z_2) \dots {\cal O}_s(z_s)  \right >}$ is given by $ \oint d{\bar z} \left <\{Q, \dots \} \ {\cal O}_1(z_1) {\cal O}_2(z_2) \dots {\cal O}_s(z_s) \right > = \oint d{\bar z} \left < \{Q, \dots \prod_{i} {{\cal O}_i(z_i)} ] \right> = 0$, i.e., the correlation functions always vary holomorphically with $z$. Secondly, $T_{z \bar z} = \{Q, G_{z \bar z}\}$ for some $G_{z \bar z}$ in the quantum theory. Thus, the variation of the correlation functions due to a change in the scale of $\Sigma$ will be given by $\left <{{\cal O}_1(z_1)} {{\cal O}_2(z_2)} \dots {{\cal O}_s(z_s)} {\{Q, G_{z \bar z} \}} \right >= \left < \{Q, \prod_{i} {{\cal O}_i(z_i)} \cdot G_{z \bar z}] \right> = 0$, i.e., the correlation functions of local physical operators will continue to be invariant under arbitrary scalings of $\Sigma$. In other words, the correlation functions are always independent of the K\"ahler structure on $\Sigma$ and depend only on its complex structure.

\vspace{0.4cm}{\noindent{\it  A Holomorphic Chiral Algebra $\cal A$}}

Let ${\cal O} (z)$ and $\widetilde {\cal O} (z')$ be two $Q$-closed operators where their product is $Q$-closed as well. Now, consider their operator product expansion or OPE:
\be
{\cal O}(z)  {\widetilde {\cal O}}(z') \sim \sum_k f_k (z-z') {\cal O}_k (z'). 
\label{OPE}
\ee 
Here, the explicit form of the coefficients $f_k$ must be such that the scaling dimensions and $q$-charges of the operators agree on both sides of the OPE. In general, $f_k$ is not holomorphic in $z$. However, if we work modulo $Q$-exact operators in passing to the $Q$-cohomology, the $f_k$'s which are non-holomorphic and are thus not annihilated by $\partial / \partial {\bar z}$, drop out from the OPE because they multiply operators ${\cal O}_k$ which are $Q$-exact. This is true because $\partial / \partial{\bar z}$ acts on the LHS of (\ref{OPE}) to give terms which are cohomologically trivial.\footnote{Since $\{Q, {\cal O}]=0$, we have $\partial_{\bar z}{\cal O}(z)=\{Q, V(z)]$ for some operator $V(z)$, as argued before. Hence $\partial_{\bar z}{\cal O}(z)\cdot {\widetilde {\cal O}}(z')=\{Q,V(z){\widetilde {\cal O}}(z')]$.} In other words, we can take the $f_k$'s to be holomorphic coefficients in studying the $Q$-cohomology. Thus, the OPE of (\ref{OPE}) has a holomorphic structure.        

In summary, we have established that the $Q$-cohomology of holomorphic local operators defines a holomorphic chiral algebra (in the sense of the mathematical literature) which we shall henceforth call $\cal A$. It is always preserved under global translations and dilatations, though (unlike the usual physical notion of a chiral algebra) it may not be preserved under general holomorphic coordinate transformations on the Riemann surface $\Sigma$. Likewise, the OPE's of the chiral algebra of local operators are associative and invariant under translations and scalings of $z$, although they may not be invariant under arbitrary holomorphic reparameterizations of $z$. The local operators are of dimension $(n, 0)$ for $n \geq 0$, and the chiral algebra is in general only locally-defined for such operators. Nonetheless, the chiral algebra can be globally-defined, up to scaling, on a Riemann surface of genus one. To define it globally on a surface of genus $g \neq 1$ requires further analysis (which we will go into later), as the physical theory has an anomaly involving $c_1(\Sigma)$ (and the first $G$-equivariant Chern class $c^G_1(X)$) that we will compute in $\S$3. Last but not least, as is familiar for chiral algebras, the correlation functions of these operators depend on $\Sigma$ only through its complex structure.

\newsubsection{The Moduli Of The Chiral Algebra}

Let us now determine the moduli of the chiral algebra $\cal A$. To this end, first notice that the metric $g_{i \bar j}$ of the target space $X$ manifests in the classical action $S_{gauged}$ within a term of the form $\{Q, \dots \}$. Hence, in passing to the $Q$-cohomology, we find that the chiral algebra is independent of metric deformations on $X$. 

Next, note that the chiral algebra depends on the complex structure of $X$ as it is built into the definition of the fields and the fermionic symmetry transformation laws of (\ref{txgauged}). In fact, the chiral algebra has a purely holomorphic dependence on the complex structure of $X$: one can show, using the form of $S_{gauged}$ in (\ref{Sgauged}), that if $J$ denotes the complex structure of $X$, then $\partial  S_{gauged} / {\partial {\bar J}}$ is of the form $\{ Q, \dots \}$.

\bigskip\noindent{\it Adding a Modulus Term to the Action}

Now consider adding to $S_{gauged}$ a term which will serve as a modulus of the chiral algebra $\cal A$. In order to ascertain the explicit form of such a term, first note that it must preserve the classical symmetries of the theory and be $Q$- and $U(1)_R$-invariant. Second, it must be marginal with dimension $(1,1)$.  Third, it must depend on the geometry of $X$. Since $Q^2 = 0$ while $Q$ and $\psi^{i}_{\bar z}$ have $q=1$ and $-1$, respectively, the only consistent choice (distinct from  $S_{gauged}$ itself) for such a term (keeping in mind that all field derivatives must be covariant) is 
\be
{S_T=\int_{\Sigma} |d^2z| \{Q,T_{ij}\psi_{\bar z}^i D_z\phi^j\}},
\label{ST-mod}
\ee       
where $T = {1\over 2}T_{ij} d\phi^i \wedge d\phi^j$ is some two-form on $X$ that is of type $(2,0)$. 

Explicitly, we then have 
\be
 S_T = \int_{\Sigma} |d^2 z|\, T_{ij, {\bar k}}\psi^{\bar k} \psi_{\bar z}^i \partial_z \phi^j  - \int_{\Sigma} |d^2z| \, T_{ij} D_{\bar z} \phi^i D_z \phi^j,
 \label{STexp}
\ee
where $T_{ij,{\bar k}} = \partial T_{ij}/ \partial \phi^{\bar k}$.  Apart from being $Q$- and $U(1)_R$-invariant, $S_T$ -- like the action $S_{gauged}$ it is supposed to deform -- ought to be gauge-invariant as well. So, is $S_T$ gauge-invariant?  

In order to answer this question, let us first look at the first term on the RHS of (\ref{STexp}); it is manifestly \emph{not} gauge-invariant. As such, a consistent choice of $T$ would be one whereby $T_{ij,{\bar k}} = 0$, i.e., $T$ is a holomorphic $(2,0)$-form on $X$. Thus, by expanding the second term on the RHS of (\ref{STexp}), we get
\be
 S_T =  - \int_{\Sigma} |d^2z| \, T_{ij} \partial_{\bar z} \phi^i \partial_z \phi^j   + \int_{\Sigma} |d^2z| \, T_{i j} (\partial_{\bar z} \phi^i A^a_zV_a^j +  A^a_{\bar z} V_a^i \partial_{z} \phi^j - A^a_{\bar z} V_a^i A^b_z V_b^j).
\label{STex}
\ee

Next, let us look at the first term on the RHS of (\ref{STex}). Note that since $|d^2 z| = i dz \wedge d{\bar z}$, we can, assuming that $T$ is globally-defined on $X$, write this term (via Stokes' theorem) as 
\be
S^{(1)}_T = -{{i \over 2} \int_{\Sigma} T_{ij} d\phi^i \wedge d\phi^j }= -{i \int_{\Sigma} \Phi^*(T)} = -i \int_C \Phi^*({H}).
\label{H}
\ee  
Here, $C$ is some three-manifold whose boundary is $\Sigma$, such that the section $\Phi: \Sigma \to X$ extends over it, and ${H} = {dT}$ is a nonzero three-form flux. As $T$ is of type $(2,0)$ such that $T_{ij,{\bar k}} = 0$, $H$ must be of type $(3,0)$. Clearly, $d {H} = 0$, consistent with the fact $C$ cannot be the boundary of a four-manifold.\footnote{From homology theory, the boundary of a boundary is empty. Hence, since $\Sigma$ exists as the boundary of $C$, the three-manifold $C$ itself cannot be a boundary of a higher-dimensional four-manifold.} 

Since the $G$-action generates an automorphism that maps $X$ back to itself, one can choose $H$ to be $G$-invariant, i.e., ${\cal L}_a H = \{d, \iota_a\} H = d (\iota_a H) = 0$, where $\iota_a$ is a contraction with respect to the vector field $V_a$; Poincar\'e's lemma then tells us that one can write, at least locally on $X$, the relation $\iota_a H = d\theta_a$ for some ${\frak g}$-valued holomorphic $(1,0)$-form $\theta_a$.\footnote{Since $H$ is a $(3,0)$-form, $\iota_aH = d \theta_a$ will be a $(2,0)$-form -- i.e., $\theta_a$ must be a holomorphic $(1,0)$-form.} Similarly, since one can choose ${\cal L}_{a} T =  \{d, \iota_a\} T = 0$, one can easily see from $H = dT$ that $\theta_a = - \iota_a T$; in component form, this means that $\theta_{a j} = -V_a^i T_{ij}$, where $\theta_a = \theta_{a j} d \phi^j$ is globally-defined on $X$ (because $T$ and $V$ are).  By writing (\ref{STex}) in terms of $\theta_{aj}$, we have
\be
 S_T =   \int_{\Sigma} |d^2z| \ \left(T_{ij} \partial_z \phi^i \partial_{\bar z} \phi^j      -  A^a_{\bar z} \theta_{a j} \partial_{z} \phi^j + A^a_z  \theta_{a j} \partial_{\bar z} \phi^j + A^a_{\bar z} A^b_z V_b^j \theta_{aj}\right).
\label{ST-inv}
\ee
Granted that 
\be
{\cal L}_a \theta_{b i} = f_{ab}{}^c \theta_{c i} \quad {\rm and} \quad V_a^i\theta_{b i} = - V^i_b\theta_{ai},
\label{granted}
\ee 
$S_T$ will indeed be gauge-invariant~\cite{hull 2}, as desired. 

In short, if a $G$-invariant holomorphic $(2,0)$-form $T$ which obeys (\ref{granted}) exists on $X$, one can define, without any obstruction, a physically consistent modulus for the chiral algebra of the gauged twisted $(0,2)$ sigma model on $X$. 

\bigskip\noindent{\it Geometrical Description of the Modulus}

Let us now describe $S_T$ in greater detail.  Note that from (\ref{ST-inv}), one can (with the aid of (\ref{H}), and (\ref{granted}) required for gauge-invariance) also express $S_T$ as
\be
S_T = i \int_C \Phi^*({\mathscr H}), 
\label{ST-final}
\ee
where
\be
{\mathscr H} = H + d(A^a \wedge \theta_a + {1\over 2} \iota_a \theta_b A^a \wedge A^b).
\label{scr H}
\ee
Here, $d \mathscr H = 0$, and the $A^a$'s in (\ref{scr H}) are to be interpreted as ${\frak g}^*$-valued one-forms on $X$.  Because $H$ is $G$-invariant,  (\ref{scr H}) and (\ref{granted}) will then mean that $\mathscr H$ represents a ($d$-closed) $G$-equivariant extension of the three-form $H \in \Omega^3(X)$~\cite{farrill} -- in other words, $\mathscr H$ exists as a \emph{basic} (i.e.,~gauge-invariant) degree-three form in the complex $\bigoplus^{p=3}_{p=0} \, W^p({\frak g}) \otimes \Omega^{3-p} (X)$, where the Weil-algebra $W^*({\frak g})$ is generated by the $A^a$'s. What else can one say about $\mathscr H$ or $S_T$? 

Well, in the quantum theory, a shift in the (Euclidean) action $S_{gauged}$ by an integral multiple of $2\pi i$ is physically inconsequential -- the path integral factor is $e^{-S_{gauged}}$. Hence, the \emph{effective} range of the continuous modulus $\mathscr H$ would be such that $0 \leq S_T < 2\pi i$. That said, the continuous global $U(1)_R$-symmetry of the classical theory will reduce to a discrete symmetry in the quantum theory due to worldsheet twisted-instantons.\footnote{These are classical configurations defined by the relation $D_{\bar z} \phi^i = 0$.} In order for this discrete symmetry to remain anomaly-free, ${\mathscr H}\over {2 \pi}$ must be an integral cohomology class, i.e., ${1\over {2 \pi}}\int_C \Phi^*({{\mathscr H}}) \in {\mathbb Z}$. Thus, the continuous modulus of $\mathscr H$, though present in the perturbative theory, could be absent in the nonperturbative theory.              

Also, in writing $S_T$ in terms of ${\mathscr H}$ in (\ref{ST-final}), we have made the assumption that $\Phi$ extends over {\it some} three-manifold $C$ with boundary $\Sigma$. In the perturbative theory, one considers only topologically trivial sections $\Phi$ which can be extended over any chosen $C$; the assumption is therefore valid in this case. Nonperturbatively however, one must also consider the contributions coming from topologically nontrivial sections; as such, an extension of $\Phi$ over $C$ may not exist. Therefore, the present definition  of $S_T$ in (\ref{ST-final}) will not suffice. Notice too that $\mathscr T = T + A^a \wedge \theta_a + {1\over 2} \iota_a \theta_b A^a \wedge A^b$ cannot be completely determined as a degree-two form in $W({\frak g}) \otimes \Omega(X)$ by its $d$-curvature $\mathscr H = d\mathscr T$, as adding a ``flat'' (i.e.,~$d$-closed) degree-two form to $\mathscr T$ would not change $\mathscr H$ at all. This indeterminacy of $\mathscr T$ is inconsequential in the perturbative theory where $S_T$ can be made to depend solely on $\mathscr H$ via (\ref{ST-final}). Nonperturbatively on the other hand, because $C$ may not exist, $S_T$ can only be expressed in terms of $\mathscr T$ and \emph{not} $\mathscr H$, as in (\ref{ST-inv}); the explicit details of $\mathscr T$ will then be important. 

At any rate, note that the basic subcomplex   of $W(\frak g) \otimes \Omega(X)$ which $\mathscr H$ lives in, is isomorphic to the Cartan complex $C_G(\Omega(X)) = ({S} ({\frak g}^*) \otimes \Omega(X))^G$, where ${S} ({\frak g}^*)$ is the symmetric algebra on ${\frak g}^*$, and the `$G$' superscript just denotes the $G$-invariant elements of the involved complex~\cite{sternberg}.  In fact, the cohomologies of the basic and Cartan (sub)complexes are identical; they actually turn out to be the $G$-equivariant cohomology $H_G(X)$ of $X$. Consequently, one can identify $\mathscr H$ -- which is a $G$-equivariant cohomology class on $X$ -- with an element $\mathscr H_C \in C_G(\Omega(X))$, where $d_G \mathscr H_C = 0$. Here, $d_G = 1 \otimes d +F^a \otimes \iota_a$ is the differential of the complex $C_G(\Omega(X))$, and $F^a$ is a ${\frak g}^*$-valued two-form on $X$ that is the curvature of $A^a$.

\bigskip\noindent{\it Interpretation Via $H^1(X, \Omega^{2, cl}_{X, G})$}

As it will soon prove illuminating to do so, let us now attempt to give a  $\check{ \textrm C}$ech description of the $G$-equivariant cohomology class $\mathscr H_C$.  To this end, let $U_\alpha$, $\alpha=1,\dots,s$ be a collection of small open sets that provide a good cover of $X$ such that their mutual intersections are open sets as well.  

As $d_G \mathscr H_C = 0$,  it will mean that $\mathscr H_C = d_G \mathscr T_C$ \emph{locally} on $X$, where $\mathscr T_C \in C_G(X)$. Thus, on each $U_\alpha$, we will have an element $\mathscr T_{C, \alpha}$ of $C_G$, such that ${\mathscr H}_{C, \alpha} = d_G{\mathscr T}_{C, \alpha}$. On each open double intersection $U_\alpha \cap U_\beta$,  let us define $\mathscr T_{C, \alpha \beta} = \mathscr T_{C, \alpha} - \mathscr T_{C, \beta}$, where
\be
\mathscr T_{C,  \alpha \beta} = -\mathscr T_{C,  \beta \alpha}
\label{Tab}
\ee      
for each $\alpha$, $\beta$, and  
\be
\mathscr T_{C,  \alpha \beta} + \mathscr T_{C, \beta \gamma} + \mathscr T_{C, \gamma \alpha} = 0
\label{Tab...}
\ee
for each $\alpha$,  $\beta$ and $\gamma$. Since $\mathscr H_C$ is globally-defined, we have  ${\mathscr H}_{C, \alpha} = {\mathscr H}_{C, \beta}$ on the intersection $U_\alpha \cap U_\beta$ -- hence,  $d_G \mathscr T_{C,  \alpha \beta} = 0$. Because $\mathscr T_C = T_C + f_a F^a$, where $T_C \in \Omega^2(X)$ and $f_a \in \Omega^0(X)$,  the relation $d_G \mathscr T_{C,  \alpha \beta} = 0$ will mean (as $d_G F ^a= 0$)  that  $dT_{C,  \alpha \beta} + F^a \iota_a T_{C,  \alpha \beta} + (F^a df_a)_{\alpha \beta} = 0$, where $ (F^a df_a)_{\alpha \beta}  = F^a_\alpha d f_{a, \alpha} - F^a_{\beta} df_{a, \beta}$. This condition will be met if (i) $dT_{C,  \alpha \beta} = (\partial + \bar \partial)T_{C,  \alpha \beta} = 0$, (ii) $\iota_a T_{C, \alpha}  = -df_{a, \alpha}$, and (iii) $\iota_a T_{C, \beta}  = -df_{a, \beta}$. Relations (ii) and (iii) together imply that $\iota_a T_{C, \alpha \beta}  = -df_{a, \alpha \beta}$; in turn, since $d^2 =0$, this will mean that $d \iota_a T_{C, \alpha \beta} =0$. Therefore, as $dT_{C,  \alpha \beta} = 0$, we have $\{d, \iota_a\} T_{C, \alpha \beta} = {\cal L}_a (T_{C, \alpha \beta}) = 0$. In short, $T_{C,  \alpha \beta}$ is a $G$-invariant  holomorphic $\partial$-closed two-form  on $X$.  Consequently, $\mathscr T_{C,  \alpha \beta}$ must be a $G$-equivariant extension of $T_{C,  \alpha \beta}$ on $X$.

Anyhow, observe that since on each $U_\alpha$, we have ${\mathscr H}_{C, \alpha} = d_G \mathscr T_{C, \alpha}$, the shift given by $\mathscr T_{C, \alpha} \rightarrow \mathscr T_{C, \alpha} + \mathscr S_{C, \alpha}$, where $d_G \mathscr S_{C, \alpha} = 0$, leaves each ${\mathscr H}_{C, \alpha}$ invariant. In other words, in describing $\mathscr H_C$, we have an equivalence relation  
\be
\mathscr T_{C,  \alpha \beta} \sim \mathscr T'_{C,  \alpha \beta} = \mathscr T_{C,  \alpha \beta} + \mathscr S_{C,  \alpha} -\mathscr S_{C, \beta}.
\label{equiv}
\ee  
From the equivalence relation (\ref{equiv}), one can see that $\mathscr T_{C, \alpha \beta} \sim 0$ if we can express $\mathscr T_{C, \alpha \beta} =  \mathscr S_{C,  \beta} - \mathscr S_{C, \alpha}$ in $U_\alpha \cap U_\beta$. Hence, the nonvanishing $\mathscr T_{C, \alpha \beta}$'s are those which obey the identities (\ref{Tab}) and (\ref{Tab...}), modulo those that can be expressed as $\mathscr T_{C, \alpha \beta} = \mathscr S_{C,  \beta} -\mathscr S_{C, \alpha}$. This means that  $\mathscr T_{C, \alpha \beta}$ is an element of the $\check{\textrm C}$ech-cohomology group $H^1(X, \Omega^{2, cl}_{X, G})$, where $\Omega^{2, cl}_{X, G}$ is the sheaf of $G$-equivariant extended ($G$-invariant) holomorphic $\partial$-closed two-forms on $X$.

Now, if $\mathscr H_C$ is globally given by the $d_G$-exact form $\mathscr H_C = d_G \mathscr T_C$, it would mean that $\mathscr T_C$ is globally-defined and as such, $\mathscr T_{C, \alpha} = \mathscr T_{C, \beta}= \mathscr T_C$ in each $U_{\alpha} \cap U_{\beta}$, whereupon all $\mathscr T_{C, \alpha \beta}$'s must vanish. Thus, we actually have a map between the space of degree-three forms ${\mathscr H_C} \in C_G(X)$ modulo forms that can be  globally expressed as $d_G \mathscr T_C$, and the $\check{ \textrm C}$ech cohomology group $H^1(X,\Omega^{2, cl}_{X, G})$. Therefore, one can conclude that $\mathscr H$ in (\ref{ST-final}) -- which is a modulus of the chiral algebra of the gauged twisted $(0,2)$ sigma model on $X$ -- can be represented by a class in $H^1(X,\Omega^{2, cl}_{X, G})$. 

\bigskip\noindent{\it The Ordinary Case}

Last but not least, note that the gauge field $A$ can also be interpreted locally as the pull-back of a connection one-form on $P$ to an open set of $\Sigma$, where $P$ is a principal $G$-bundle over $\Sigma$. Consequently, the $\phi$'s which are minimally-coupled to $A$ can also be interpreted as sections $\phi: \Sigma \to E$ of the associated bundle $E = P \times_G X$.  Let us now take $A \to 0$ so that $F \to 0$,  whence one would just get back the ordinary \emph{ungauged} twisted $(0,2)$ sigma model on $X$ studied in~\cite{CDO}. Then, from our above discussion, $\mathscr T_{C, \alpha \beta}$ would reduce to $T_{C, \alpha \beta}$, or rather, the sheaf $\Omega^{2, cl}_{X, G}$ would reduce to the sheaf $\Omega^{2, cl}_{X}$ of \emph{ordinary} holomorphic $\partial$-closed two-forms on $X$. In other words, the corresponding modulus of the chiral algebra of the ordinary twisted model will be represented by a class in $H^1(X,\Omega^{2, cl}_{X})$ -- a conclusion that is consistent with the results in~\cite{CDO}.

\newsection{Anomalies Of The Gauged Twisted $(0,2)$ Sigma Model}

In this section, we will study the anomalies of the gauged twisted $(0,2)$ sigma model with action $S_{gauged}$ given in (\ref{action-gen}). In essence, the model will fail to exist in the quantum theory if the anomaly-cancellation  conditions are not met. We aim to determine what these conditions are. In our discussion, we shall omit the additional term $S_T$ given by (\ref{ST-final}), as the anomalies of interest do not depend on continuously varying couplings such as this one. 

\bigskip\noindent{\it A Relevant Digression}

Before we proceed further, let us make a relevant digression to describe the fermion fields $\psi^{\bar i}$ that appear in the gauged action (\ref{action-gen}). In the ungauged case, $\psi^{\bar i}$ must transform,  according to (\ref{ungauged section}),  as a smooth section of the pullback bundle $\Phi^*(\overline {TX})$. Here, $\Phi$ is a map $\Sigma \to X$, and $\overline {TX}$ is  the antiholomorphic tangent bundle over $X$.  On the other hand, in the gauged case, since $\Phi$ is a section $\Sigma \to E$ of the associated bundle $E = P \times_G X$, where $P$ is a principal $G$-bundle over $\Sigma$, it would mean that $\psi^{\bar i}$ ought to transform as a smooth section of the pullback of some bundle over $E$. Indeed, we find that $\psi^{\bar i}$ must transform as a smooth section of the pullback bundle $\Phi^*(\overline{\textrm{ker} \, d\pi_E})$, where $\textrm{ker} \, d\pi_E \to E$ is the sub-bundle of $TE \to E$ defined as the kernel of the derivative of the projection $\pi_E: E \to \Sigma$~\cite{baptista}. 

\bigskip\noindent{\it The Anomalies}

Coming back to our main discussion on the anomalies, let us first note that in sigma models, they arise because one cannot define the path integral of the worldsheet fermion fields $\psi^i_{\bar z}$ and $\psi^i$ in a physically consistent manner~\cite{moore}. Hence, it suffices for us to look at just the kinetic energy term of the fermions in the action. From $S_{\mathrm gauged}$ in (\ref{action-gen}), we see that this term is given by $(\psi, \widehat D \psi )= \int |d^2 z| g_{i \bar j} \psi^i \widehat D \psi^{\bar j}$, where $\widehat D$ is the $\partial$ operator on $\Sigma$ --  constructed using the pull-back of the connection on ${\textrm{ker} \, d\pi_E}$ --  acting on sections $\Phi^*(\overline{\textrm{ker} \, d\pi_E})$. (Notice that we have omitted the $z$ and $\bar z$ indices of the fields as they are irrelevant in the present discussion.)  By picking a spin structure on $\Sigma$, one can equivalently interpret $\widehat D$ as the Dirac operator on $\Sigma$ acting on  sections of ${\cal V} ={\overline K}^{-1/2} \otimes \Phi^*(\overline{\textrm{ker} \, d\pi_E})$,\footnote{Firstly, on a K\"ahler manifold such as $\Sigma$, the Dolbeault operator $\partial + \partial^{\dag}$ on ${\overline K}^{1/2}$ coincides with the Dirac operator. Secondly, since $\psi^{\bar i}$ is a zero-form on $\Sigma$, we have $\partial^{\dag}\psi^{\bar i} = 0$ -- in other words, $\partial + \partial^{\dag}$ is effectively $\partial$ when acting on $\psi^{\bar i}$. Altogether, this means that the action of $\partial$ on $\psi^{\bar i}\in \Gamma(\Phi^*((\overline{\textrm{ker} \, d\pi_E}))$ is equivalently to the action of the Dirac operator on sections of the bundle ${\cal V} = {\overline K}^{-1/2}\otimes \Phi^*((\overline{\textrm{ker} \, d\pi_E})$.} where $K$  is the canonical bundle of $\Sigma$ and $\overline K$ its complex-conjugate. 

Note at this point that the anomaly arises as an obstruction to defining the functional Grassmann integral of the kinetic term in $\psi^{i}$ and $\psi^{\bar i}$, as a general function on the configuration space $\cal C$ of the underlying model \cite{moore}. From our discussion in the previous paragraph, we find that the Grassmann integral is given by the determinant of $\widehat D$. As argued in \cite{moore}, one must think of the functional integral as a section of a complex determinant line bundle $\cal L$ over $\cal C$. Unless $\cal L$ is trivial, the integral would \emph{not} be a global section and therefore a function on $\cal C$. Hence, the anomaly is due to the nontriviality of $\cal L$. The bundle $\cal L$ can be characterized completely by its restriction to a nontrivial two-cycle in $\cal C$ such as a two-sphere \cite{Bott}.       

To be more precise, let us consider a family of sections $\Phi : \Sigma \to E$, parameterized by a two-sphere base which we will denote as $B$. In computing the path integral, we really want to consider the universal family of $\it{all}$ sections from $\Sigma$ to $E$. This can be represented by a $\it{single}$ section $\hat{\Phi} : {\Sigma \times B} \to E$. The quantum path integral is anomaly-free if $\cal L$, as a complex line bundle over $B$, is trivial. Conversely, if $\cal L$ can be trivialized by a local Green-Schwarz anomaly-cancellation mechanism, the quantum theory will exist. 

From the theory of determinant line bundles, we find that the basic obstruction to triviality of  $\cal L$ is its first Chern class. By an application of the family index theorem to anomalies \cite{CDO20,CDO21}, the first Chern class of $\cal L$ is given by ${\pi} ({ch_2 ( {\cal V})} )$,  where $\pi: H^4(\Sigma \times B) \to H^2(B)$. Note that the anomaly lives in $H^4(\Sigma \times B)$ and not  $H^2(B)$: it is clear that ${\pi}({ch_2({\cal V})} )$ vanishes if ${ch_2({\cal V})}$ in $H^4(\Sigma \times B)$ vanishes $\it{before}$ it is being mapped to $H^2(B)$, but if ${ch_2({\cal V})}  \neq 0$, then even if ${\pi}({ch_2({\cal V})} ) = 0$ whence $\cal L$ is trivial, it $\it{cannot}$ be trivialized by a Green-Schwarz mechanism. Thus, we need to have $ch_2({\cal V}) = 0$ for bona-fide anomaly-cancellation.

At any rate, note that we have a Chern character identity $ch({\cal E} \otimes {\cal F}) = ch({\cal E})ch({\cal F})$, where ${\cal E}$ and ${\cal F}$ are any two bundles. Also, note that $ch_2(\overline {{\cal E}}) = ch_2({\cal E})$, and by tensoring $\Phi^*(\overline{\textrm{ker} \, d\pi_E})$ with ${\overline K}^{-1/2} $ to obtain $\cal V$, we get an additional term ${1\over 2} c_1(\Sigma) c_1 ({\textrm{ker} \, d\pi_E})$ in $ch_2({\cal V})$.  Therefore, the condition  $ch_2({\cal V}) = 0$ for vanishing anomaly, can also be expressed as 
\be
 0= {1\over 2} c_1(\Sigma) c^G_1 (X) = {1\over 2} p^G_1(X),
\label{an}
\ee
where $c^G_1(X)$ and $p_1^G(X)$ are the first $G$-equivariant Chern and Pontryagin classes of $TX$, respectively. (In order to arrive at (\ref{an}), we have made use of the fact~\cite{gaio} that  $c_1 (\textrm{ker} \, d\pi_E) =  c^G_1(X)$ and $p_1({\textrm{ker} \, d\pi_E}) = p_1^G(X)$.)  The first condition in (\ref{an}) just means that we can either consider just Riemann surfaces $\Sigma$ with $c_1(\Sigma) = 0$  if $c^G_1 (X) \neq 0$, or allow $\Sigma$ to be arbitrary if $c^G_1 (X) =  0$.  Another important point to note is that the  $ c_1(\Sigma) c^G_1 (X)$ anomaly exists if and only if $K$ is \emph{nontrivial}, or equivalently, if there \emph{is} actually twisting of the model. On the other hand, the  $p^G_1(X)$ anomaly always exists, regardless of whether there is twisting or not.  The former observation regarding the $ c_1(\Sigma) c^G_1 (X)$ anomaly will be crucial when we later analyze -- within the framework of our physical interpretation of the geometric Langlands correspondence in terms of our gauged twisted $(0,2)$ sigma model -- the conditions required for the existence and uniqueness of Beilinson-Drinfeld $\cal D$-modules.

\bigskip\noindent{\it Other Potential Anomalies}

One might also wonder if the nilpotency of $Q$ would persist in the quantum theory. After all, our entire notion of a $Q$-cohomology rests upon this crucial property of $Q$. If it is really the case that $Q^2 = 0$ in the quantum theory, there would be \emph{no} nontrivial (i.e.,~nonzero at the outset) conserved charge of dimension $(0,0)$ and $q=2$ (which could therefore potentially serve as $Q^2$) at the quantum level. It suffices to show this at the classical level, as quantum corrections can only destroy and not create conservation quantities. 

Now, recall that classically, the fields of the underlying model $(\phi^i, \phi^{\bar i}, \psi^i_{\bar z}, \psi^{\bar i}, A^a_z, A^a_{\bar z})$ have $q = (0,0, -1, 1, 0,0)$. Also, all fields are of dimension $(0,0)$ except for $\psi^{i}_{\bar z}$ and $A^a_{\bar z}$ which are of dimension $(0,1)$, and $A^a_z$ which is of dimension $(1,0)$. A little thought at this point would then reveal that from the fields $\phi$, $\psi$ and $A$, one can construct a local  nontrivial (i.e.,~non total-derivative) antiholomorphic (since $Q^2$ is supposedly right-moving)  dimension-one conserved current $J( \bar z)$ with resulting charge $Q^2 = \oint  J (\bar z) d{\bar z}$ of dimension $(0,0)$ and $q=2$, in two ways. The first way involves contracting a covariantly-constant dimension $(0,0)$ tensor $\cal J$ of $X$ with two $\psi^{\bar i}$ fields and a single $D_{\bar z} \phi^i$ or $D_{\bar z} \phi^{\bar i}$ field. The second way involves contracting a covariantly-constant dimension $(0,0)$ tensor $\tilde {\cal J}$ of $X$ with one  $\psi^i_{\bar z}$ field and three  $\psi^{\bar i}$ fields.  Covariantly-constant tensors such as $\tilde {\cal J}$ or ${\cal J}$, which have a mix of $\bar i$ and $i$ indices that are fully antisymmetric or part symmetric part antisymmetric, respectively, do not, in general, exist for a generic hermitian metric on $X$.  In other words, there cannot be any nontrivial conserved current with charge $Q^2$ even at the classical level, i.e., $Q^2  = 0$ holds in the quantum theory. This claim can be further substantiated when we later show that suitable perturbative corrections to $Q$ can be found in a physically consistent manner such that the relation $Q^2 = 0$ is maintained at the quantum level.

Last but not least, note that when $\Sigma$ is curved, the Ricci scalar $R$ of $\Sigma$ is nonvanishing. As a result, the quantum expression for $T_{z \bar z}$ will in general be modified to
\be
{T_{z \bar z}} =  {\{Q, G_{z \bar z}\}+ {c \over {2 \pi}} R },
\label{tzbarzq}
\ee
where $c$ is a nonzero constant related to the central charge of the sigma model. The second term on the RHS of (\ref{tzbarzq}), given by a multiple of $R$, represents a soft conformal anomaly on the worldsheet due to a curved $\Sigma$.  $R$ scales as a $(1,1)$ operator, as required. 

There are implications for the $Q$-cohomology of operators due to this additional $R$-term. Recall from  $\S$2.2 that the holomorphy of $T_{zz}$ holds as long as $\partial_z T_{z \bar z} \sim 0$ (where `$\sim$' denotes an equivalence up to $Q$-exact terms). However, due to this additional $R$-term, we now find that $\partial_z T_{z \bar z} \nsim 0$. Hence, the invariance of the $Q$-cohomology of operators under translations on the worldsheet -- which requires $T_{zz}$ to vary holomorphically with $z$ -- no longer holds. Therefore, the local holomorphic operators apparently fail to span a chiral algebra over $\Sigma$, since one of the axioms of a chiral algebra is invariance under translations on the worldsheet. 

However, the additional $R$-term, being a $c$-number anomaly, will only affect the partition function; it will \emph{not} affect the (normalized) correlation functions that actually define our chiral algebra. Thus, assuming that (\ref{an}) holds, one will continue to have a scale-invariant chiral algebra over $\Sigma$ that, as argued in $\S$2.2, depends on $\Sigma$ only via its complex structure (as is expected of chiral algebras).

\newsection{Sheaf Of Perturbative Observables} 

In this section, we will analyze in detail, the $Q$-cohomology of local operators that define the chiral algebra of the perturbative gauged twisted $(0,2)$ sigma model. The second half of this section will be devoted to the abelian case, so that the relation of the corresponding chiral algebra to the mathematical theory of TCDO's can be elucidated in $\S$5, whence a physical interpretation of the geometric Langlands correspondence in terms of our abelian model can be furnished in $\S$7.

\newsubsection{General Considerations}

In general, a local operator which is defined up to a gauge transformation, is an operator $\cal F$ that is a function of the fields $\phi^i$, $\phi^{\bar i}$, $\psi^i_{\bar z}$, $\psi^{\bar i}$, $A^a_{z}$, $A^a_{\bar z}$, and their derivatives with respect to $z$ and $\bar z$.\footnote{Note that since we are only interested in local operators, we will work locally on an open set in $\Sigma$ -- isomorphic to an open disc in $\mathbb C$ -- with local parameters $z$ and $\bar z$. Hence, we can omit in our operators the dependence on the scalar curvature of $\Sigma$.} However, as we saw in $\S$2.2, the $Q$-cohomology is zero for operators of dimension $(n,m)$ with $m \neq 0$. Since $\psi^i_{\bar z}$, $A^a_{\bar z}$ and the derivative $\partial_{\bar z}$ both have $m=1$, and since a physical operator cannot have negative $m$ or $n$ (see $\S$2.2), $Q$-cohomology classes can be built from just $\phi^i$, $\phi^{\bar i}$, $\psi^{\bar i}$, $A^a_z$ and their derivatives with respect to $z$. Note also that the equation of motion for $\psi^{\bar i}$ is   ${\widehat D}_z \psi^{\bar i}= 0$; this means that one can ignore the $z$-derivatives of $\psi^{\bar i}$ as it can be expressed in terms of the other fields and their corresponding derivatives. Therefore, a chiral (or $Q$-invariant) operator which represents a $Q$-cohomology class can be written as
\be
{\cal F}(\phi^i,\partial_z\phi^i,\partial_z^2\phi^i,\dots;
\phi^{\bar i},\partial_z\phi^{\bar i},\partial_z^2\phi^{\bar i},\dots; A^a_z, \partial_z A^a_z, \partial_z^2 A^a_z \dots; \psi^{\bar i}),
\ee
where as indicated, $\cal F$ might depend on the $z$-derivatives of $\phi^{i}$, $\phi^{\bar i}$ and $A^a_z$ of arbitrarily high order, but not on derivatives of $\psi^{\bar i}$. If the scaling dimension of $\cal F$ is bounded, it will mean that $\cal F$ (i) depends only on the derivatives of fields up to some finite order, (ii) is a polynomial of finite degree in those, and/or (iii) is a polynomial of finite degree in $A_z^a$. Notice that $\cal F$ will always be a polynomial of finite degree in $\psi ^{\bar i}$; this is because $\psi^{\bar i}$ is fermionic whence $(\psi^{\bar j})^2 = 0$. However, the dependence of $\cal F$ on $\phi^i$, $\phi^{\bar i}$ (as opposed to their derivatives) need not have any simple form. Nevertheless, we can make the following observation: from the $U(1)_R$-charges of the fields listed below (\ref{txgauged}), we see that if $\cal F$ is of degree $k$ in $\psi^{\bar i}$, then it has $q=k$.

A general $q = k$ operator ${\cal F}(\phi^i,\partial_z\phi^i,\partial_z^2\phi^i,\dots; \phi^{\bar i},\partial_z\phi^{\bar i},\partial_z^2\phi^{\bar i},\dots; A^a_z, \partial_z A^a_z, \partial_z^2 A^a_z \dots; \psi^{\bar i})$ can be interpreted as a $(0,k)$-form on $X$ valued in a certain sum of vector bundles. In order to illustrate the general idea behind this interpretation, let us consider some explicit examples of  operators of dimension $(0,0)$ and $(1,0)$. For dimension $(0,0)$, the most general operator is ${\cal F}(\phi^i,\phi^{\bar i}; \psi^{\bar j})= f_{\bar j_1,\dots,\bar j_k}(\phi^i, \phi^{\bar i}) \psi^{\bar j_i}\dots \psi^{\bar j_k}$; i.e., $\cal F$ may depend on $\phi^i$ and $\phi^{\bar i}$ but not on their derivatives, and is $k^{th}$ order in $\psi^{\bar j}$. Since the $\psi^{\bar j}$'s anticommute, one may map $\psi^{\bar j}$ to $d\phi^{\bar j}$ whence such an operator would correspond to an ordinary $(0,k)$-form $f_{\bar j_1,\dots,\bar j_k}(\phi^i, \phi^{\bar i})d\phi^{\bar j_1}\dots d\phi^{\bar j_k}$ on $X$. For dimension $(1,0)$, there are two general cases. In the first case, we have an operator ${\cal F}(\phi^l, \phi^{\bar l}; \partial_z\phi^i; A^a_z; \psi^{\bar j})=f_{i,\bar j_1,\dots,\bar j_k}(\phi^l,\phi^{\bar l}) D_z\phi^i \psi^{\bar j_1}\dots\psi^{\bar j_k}$ that is linear in $\partial_z\phi^i$ and $A^a_z$, and is independent of any other derivatives. Notice that it can also be written as $\CF = \CF_1 + \CF_2$, where $\CF_1 = f_{i,\bar j_1,\dots,\bar j_k}(\phi^l,\phi^{\bar l}) \partial_z\phi^i \psi^{\bar j_1}\dots\psi^{\bar j_k}$ and $\CF_2 = f_{a, \bar j_1,\dots,\bar j_k}(V^l, \phi^l,\phi^{\bar l}) A^a_z \psi^{\bar j_1}\dots\psi^{\bar j_k}$.  Clearly, $\CF_1$ can be interpreted as a $(0,k)$-form on $X$ valued in the holomorphic cotangent bundle $T^*X$; alternatively, it can be interpreted as a $(1,k)$-form on $X$. On the other hand, $\CF_2$ can be interpreted as a $(0,k)$-form on $X$ valued in the bundle $E^*$ of rank $r = {\rm dim} \, \frak g$, where the local sections of the dual bundle $E$ are spanned by $A^a_z$. In the second case, we have an operator ${\cal F}(\phi^l, \phi^{\bar l}; \partial_z \phi^{\bar s}; A^a_z; \psi^{\bar j})=f^i{}_{\bar j_1,\dots,\bar j_k}(\phi^l, \phi^{\bar l}) g_{i \bar s}D_z \phi^{\bar s}\psi^{\bar j_i}\dots \psi^{\bar j_k}$ that is linear in $\partial_z \phi^{\bar s}$, and $A^a_z$ and is independent of any other derivatives. It can also be written as $\CF = \CF_1 + \CF_2$, where $\CF_1 = f^i{}_{\bar j_1,\dots,\bar j_k}(\phi^l, \phi^{\bar l}) g_{i \bar s}\partial_z \phi^{\bar s}\psi^{\bar j_i} \newline \dots \psi^{\bar j_k}$ and $\CF_2 = f_{a, \bar j_1,\dots,\bar j_k}(V_l, \phi^l, \phi^{\bar l})A^a_z\psi^{\bar j_i}\dots \psi^{\bar j_k}$. Clearly, $\CF_1$ can be interpreted as a $(0,k)$-form on $X$ valued in the holomorphic tangent bundle $TX$. On the other hand, $\CF_2$ can be interpreted as a $(0,k)$-form on $X$ valued in the bundle $E^*$. One can go on to show, in the same way, that an operator of dimension $(n > 1,0)$  and charge $q = k$ can be interpreted as a $(0,k$)-form on $X$ valued in a certain sum of vector bundles. But would this claim hold under gauge transformations? 

Notice that a local operator $\CF$ in the $Q$-cohomology is not necessarily gauge-invariant: under a gauge transformation with infinitesimal parameter $\epsilon$, we have $\CF \to \CF' = \CF + \delta_{\epsilon} \CF$, where $\delta_{\epsilon}\CF$ does not necessarily vanish. Nevertheless, since a gauge transformation commutes with the action of $Q$, $\CF'$ will still be a $Q$-cohomology class of the same dimension and $U(1)_R$-charge as $\CF$. Thus, one can interpret a gauge transformation as a change of basis in the infinite-dimensional space of $Q$-closed (modulo $Q$-exact) local operators graded by dimension and $U(1)_R$-charge in the perturbative sigma model. In fact, we will witness an explicit manifestation of this claim when we consider the canonical quantization of the sigma model in $\S$8.4.  At any rate, from (\ref{tx A}), (\ref{tx phi}) and (\ref{tx psi}), we find that $\delta_\epsilon(D_z \phi^i) = \epsilon^b A^c_z f_{bc}{}^a V^i_a - A^a_z \partial_kV_a^i \epsilon^b V^k_b$ and  $\delta_\epsilon(D_z \phi^{\bar i}) = - \epsilon^b A^c_z f_{bc}{}^a V^{\bar i}_a - A^a_z \partial_{\bar k}V_a^{\bar i} \epsilon^b V^{\bar k}_b$, along with $\delta_\epsilon \phi^i = \epsilon^aV^i_a$, $\delta_\epsilon \phi^{\bar i} = - \epsilon^aV^{\bar i}_a$, and $\delta_\epsilon \psi^{\bar i} = - \epsilon^a \partial_{\bar k} V_a^{\bar i} \psi^{\bar k}$. Note at this point that $\epsilon^a$, just like $A^a_z$, can be interpreted as a local section of a vector bundle $\CE$ of rank $r$. Hence, if $\CF$ is a dimension $(0,0)$ or $(1,0)$ operator as described in the previous paragraph,  $\CF'$ would also be a sum of operators that each have an interpretation as a $(0,k)$-form on $X$ valued in a certain vector bundle. Via a similar analysis, one can show that this would also be true if $\CF$ were to be an operator of higher dimension. Thus, the claim that an operator of dimension $(m,0)$ (for any integer $m \geq 0$) and charge $q = k$ can be interpreted as a $(0,k$)-form on $X$ valued in a certain sum of vector bundles, always holds. This structure persists in the quantum theory; however, there could be perturbative corrections to the complex structures of the bundles.

\bigskip\noindent{\it The Action of $Q$}

The classical action of $Q$ on such operators can be easily described. If we interpret $\psi^{\bar i}$ as $d\phi^{\bar i}$, then $Q$ acts on functions of $\phi^i$ and $\phi^{\bar i}$, and  is simply the $\bar\partial $ operator on $X$. This can be seen from the transformation laws $\delta\phi^{\bar i}=\psi^{\bar i}$, ${\delta\phi^i} = 0$, ${\delta \psi^{\bar i}}=0$, and $\delta A^a_{z} = \delta A^a_{\bar z} = 0$. Note that if the antiholomorphic vector fields generating the $G$-action on $X$ are covariantly-constant, i.e., $\nabla_{\bar k} V^{\bar i}_a = 0$, then $Q$ will continue to act as the $\bar\partial$ operator on a more general operator ${\cal F}(\phi^i,\partial_z\phi^i,\dots;\phi^{\bar i},\partial_z \phi^{\bar i},\dots; A^a_{z}, \partial_zA^a_z, \dots; \psi^{\bar i})$ that depends on the derivatives of $\phi^i$ and $\phi^{\bar i}$. This is because we have the equation of motion $D_z \psi^{\bar i}=0$, and this means that one can ignore the action of $Q$ on covariant derivatives $D_z^m\phi^{\bar i}$ with $m>0$. On the other hand, if $\nabla_{\bar k} V^{\bar i}_a \neq 0$, then $Q$ will only act as the $\bar \partial$ operator on physical operators that \emph{do not }contain covariant derivatives $D_z^m \phi^{\bar i}$ with $m > 0$. 

At the quantum level, there will be perturbative corrections with regard to the action of $Q$. In fact, as briefly mentioned in  $\S$2.2, eqn.~(\ref{tzzanomaly}) provides such an example: the holomorphic stress tensor $T_{zz}$ is no longer $Q$-closed because the action of $Q$ has received perturbative corrections. Let us now attempt to better understand the characteristics of such perturbative corrections. To this end, let $Q_{cl}$  denote the classical approximation to the \emph{quantum-corrected} $Q$. Then, one can write $Q = Q_{cl} + \epsilon Q' + O(\epsilon^2) $, where the parameter $\epsilon$ governs the magnitude of the perturbative quantum corrections at each order of the expansion. To ensure that we continue to have $Q^2 = 0$, we require that $\{Q_{cl}, Q' \} = 0$. That said, if $Q'=\{Q_{cl},\Lambda\}$ for some $\Lambda$, then the correction by $Q'$ can be removed via the conjugation of $Q$ with $\exp(-\epsilon\Lambda)$ (which results in a trivial change of basis in the space of $Q$-closed local operators). Hence, $Q'$  must represent a $Q_{cl}$-cohomology class. Since $Q'$ appears in sigma model perturbation theory, it ought to be constructed locally\footnote{Because sigma model perturbation theory is local in $X$ and involves a Taylor expansion of fields up to some given order, the perturbative corrections to $Q_{cl}$ will also be local in $X$, where order by order, they consist  of differential operators whose degrees depend on the underlying order.}    from the fields appearing in the sigma model action.

Note that one can ascertain $Q'$ explicitly when $\nabla_{\bar k} V^{\bar i}_a = 0$. In such a case, $Q_{cl}$ will always act as the $\bar \partial$ operator, as argued above. In other words, $Q'$ will be given by representatives of $\bar \partial$-cohomology classes on $X$. An example would be the Ricci tensor which represents a $\bar\partial$-cohomology class in $H^1(X,T^*X)$ -- it  is also constructed locally from the metric of $X$ found in the action. Hence, it satisfies the conditions required of a perturbative correction $Q'$. Another representative of a $\bar\partial$-cohomology class on $X$ which may contribute as $Q'$ would be an element of $H^1(X, \Omega^{2, cl}_{X, G})$ (since it can also be interpreted as an element of $H^1_{\bar \partial} (\Omega^{2, cl}_{X, G})$ via the $\check{ \textrm C}$ech-Dolbeault isomorphism) -- it is also constructed locally from fields found in the action $S_{gauged}$ of (\ref{Sgauged}), and is used to deform $S_{gauged}$ via $S_T$ of (\ref{ST-mod}). In fact, its contribution as $Q'$ is consistent with its interpretation as a modulus of the chiral algebra. To see this, notice that its contribution as $Q'$ means that it will parameterise a family of $Q = Q_{cl} + \epsilon Q'$ operators at the quantum level. Since the chiral algebra of local operators is defined to be closed with respect to the $Q$ operator, it will vary with the $Q$ operator  and consequently with $Q' \in H^1(X, \Omega^{2,cl}_{X, G})$, i.e., one can associate a modulus of the chiral algebra with a class in $H^1(X, \Omega^{2,cl}_{X,G})$. It is possible that these classes completely determine  $Q'$, as they are the only one-dimensional $\bar \partial$-cohomology classes on $X$ which one can construct locally from the fields found in the action.

The fact that $Q$ does not always act as the $\bar \partial$ operator even at the classical level, suggests that one needs a more general framework than just ordinary Dolbeault or $\bar \partial$-cohomology to describe the $Q$-cohomology of the gauged twisted $(0,2)$ sigma model. Indeed, as we will show shortly in $\S$4.3, the appropriate description of the $Q$-cohomology of local operators spanning the chiral algebra will be given in terms of the more abstract notion of $\check{ \textrm C}$ech cohomology.

\newsubsection{A Topological Chiral Ring}

Next, let us make an interesting and relevant observation about the ground operators in the $Q$-cohomology. Note that we had already shown in $\S$2.2, that the $Q$-cohomology of operators defines a chiral algebra with holomorphic operator product expansions. Let the local operators of the $Q$-cohomology  be given by ${\cal F}_e$, ${\cal F}_f$, $\dots$ with scaling dimensions $(h_e, 0)$, $(h_f, 0)$, $\dots$. By holomorphy, and the conservation of scaling dimensions and $U(1)_R$ charges, the OPE of  these  operators take the form 
\be
{{\cal F}_e (z) {\cal F}_f (z')} = {\sum_{q_g =  q_e + q_f} { {C_{efg}\ {\cal F}_g(z')} \over {(z- z')^{h_e + h_f -h_g}} } },
\label{OPEab}
\ee 
where we have denoted the $U(1)_R$ charges of the operators ${\cal F}_e$, ${\cal F}_f$ and ${\cal F}_g$ by $q_e$, $q_f$ and $q_g$, respectively. Here, $C_{efg}$ is a structure constant that is (anti)symmetric in the indices. If ${\widetilde {\cal F}}_e$ and ${\widetilde {\cal F}}_f$ are ground operators of dimension $(0,0)$, i.e., $h_e = h_f =0$, the OPE will then be given by
\be
{{\widetilde {\cal F}}_e (z) {\widetilde {\cal F}}_f (z')} = {\sum_{q_g =  q_e + q_f} { {C_{efg}\ {{\cal F}}_g(z')} \over {(z- z')^{-h_g}} } }.
\label{OPEc}
\ee 
Notice that the RHS of (\ref{OPEc}) is only singular if $h_g < 0$. Also recall that all physical operators in the $Q$-cohomology cannot have negative scaling dimension, i.e., $h_g \geq 0$.\footnote{As mentioned in footnote~2, for an operator of classical dimension $(n, m)$, anomalous dimensions due to RG flow may shift the values of $n$ and $m$ in the quantum theory. However, the spin $n-m$ remains unchanged. Hence, since the operators in the $Q$-cohomology of the quantum theory will continue to have $m =0$ (due to a $Q$-trivial antiholomorphic stress tensor $T_{\bar z \bar z}$ at the quantum level),  the value of $n$ is unchanged as we go from the classical to the quantum theory, i.e., $n\geq 0$ holds even at the quantum level.} Hence, the RHS of (\ref{OPEc}), given by $(z-z')^{h_g} {\cal F}_g (z')$, is nonsingular as $z \to z'$, since a pole does not exist. Note that $(z-z')^{h_g}  {\cal F}_g (z')$ must also be annihilated by $Q$ such as to live in its cohomology, since ${\widetilde {\cal F}}_e$ and ${\widetilde {\cal F}}_f$ do, too. In other words, we can write ${\widetilde {\cal F}}_g (z, z') = (z-z')^{h_g} {\cal F}_g (z')$, where ${\widetilde {\cal F}}_g (z, z')$ is a nonsingular dimension $(0,0)$ operator that represents a $Q$-cohomology class.  Thus, we can express the OPE of the ground operators as
\be
{{\widetilde {\cal F}}_e (z) {\widetilde {\cal F}}_f (z')} = {\sum_{q_g =  q_e + q_f}  C_{efg} \ {\widetilde {\cal F}}_g(z , z')}.
\label{OPEgnd}
\ee
Since the only holomorphic functions without any poles on a Riemann surface are equivalent to constants, it will mean that the operators $\widetilde {\cal F}$ are independent of the coordinate $z$ on $\Sigma$. Hence, they are completely independent of their insertion points and the metric on $\Sigma$. Therefore, we conclude that the ground operators of the $Q$-cohomology define a $\it{topological}$ chiral ring via their OPE  
\be
{{\widetilde {\cal F}}_e  {\widetilde {\cal F}}_f } = {\sum_{q_g =  q_e + q_f}  C_{efg} \ {\widetilde {\cal F}}_g}.
\label{OPEgnd1}
\ee

In any case, recall that gauge transformations preserve the grading by dimension and $U(1)_R$-charge of all $Q$-cohomology classes of the perturbative sigma model. As such, the ring structure is well-defined under infinitesimal gauge transformations. However, under \emph{finite} gauge transformations which induce a change of basis in the infinite-dimensional space of $Q$-closed (modulo $Q$-exact) local operators of the sigma model, the transformed operators ${\widetilde\CF}'_e$, ${\widetilde\CF}'_f$ and ${\widetilde\CF}'_g$ do \emph{not} necessarily satisfy (\ref{OPEgnd1}). In other words, (\ref{OPEgnd1}) will only be unambiguously-defined if ${\widetilde {\cal F}}_e$,  ${\widetilde {\cal F}}_f$ and ${\widetilde {\cal F}}_g$ are gauge-invariant ground operators.

\bigskip\noindent{\it Grading of the Ring}

In the perturbative theory, the chiral ring will have a ${\mathbb Z}$-grading by the $U(1)_R$ charge of the operators. However, the corresponding operators will either be non-Grassmannian or Grassmannian, obeying either commutators or anticommutators, depending on whether they contain an even or odd number of fermionic fields. Consequently, the ${\mathbb Z}$-grading will be reduced mod 2 to $\mathbb Z_2$, such that the ring is effectively ${\mathbb Z_2}$-graded. 

Nonperturbatively, due to worldsheet twisted-instantons, the continuous $U(1)_R$ symmetry will be reduced to a discrete subgroup. Specifically, from the relevant index theorem of the kinetic operator $\widehat D$ of the fermionic fields $\psi^i_{\bar z}$ and $\psi^{\bar i}$, (assuming, for simplicity of illustration, that there are no zero-modes for the $\psi^{i}_{\bar z}$ fields), we find that a correlation function will be nonvanishing if and only if there are exactly $p$ insertions of the $\psi^{\bar i}$ fields, where (for a nonanomalous model in which $c_1(\Sigma) = p^G_1(X) = 0$ while $c^G_1(X) \neq 0$)
\be
p = \langle c_1^G(X) , \Phi_\ast (\Sigma) \rangle.
\label{p}
\ee
(Here, $c_1^G(X)$ resides in the second $G$-equivariant cohomology $H^2_G(X)$; $\Phi_\ast(\Sigma)$ resides in the second $G$-equivariant homology $H_2^G(X)$ obtained by a push-forward by $\Phi$ of the fundamental class of $\Sigma$; the brackets in (\ref{p}) just denote the bilinear pairing $H^2_G(X) \times H_2^G(X) \to \mathbb R$.) In the presence of worldsheet twisted-instantons, $p$ is nonzero, and consequently, any nonvanishing correlation function would pick up a factor of $e^{i p {\tilde q}}$ under a $U(1)_R$ transformation of the $\psi^{\bar i}$ insertions, where $\tilde q$ is the \emph{effective} $U(1)_R$ charge of $\psi^{\bar i}$. Since the correlation function ought to remain invariant under this $U(1)_R$ transformation, we must have $\tilde q = 2\pi l/ p$, where $l \in \mathbb Z$. Therefore, the continuous $U(1)_R$ symmetry is broken down to its ${\mathbb  Z}_p$ subgroup. Thus, the initial ${\mathbb Z}$-grading by the $U(1)_R$ charges will be reduced to a $\mathbb Z_{p}$-grading. A further reduction mod 2 as discussed above, will mean that the ring is effectively $\mathbb Z_{2p}$-graded at the nonperturbative level.

\bigskip\noindent{\it The Classical Ring}

Recall that any gauge-invariant dimension $(0,0)$ operator $\widetilde {\cal F}_e$ defined in (\ref{OPEgnd1}) with $q_e = k$, can be written as $\widetilde {\cal F}_e = f_{\bar j_1,\dots,\bar j_k}(\phi^i, \phi^{\bar i}) \psi^{\bar j_i}\dots \psi^{\bar j_k}$. Under a gauge transformation with infinitesimal parameter $\epsilon$, we have, from (\ref{tx phi}) and (\ref{tx psi}), $\delta_\epsilon (d\phi^{\bar i}) = - \epsilon^a \partial_{\bar k} V^{\bar i}_a \psi^{\bar k}$ and $\delta_\epsilon \psi^{\bar i} = - \epsilon^a \partial_{\bar k} V^{\bar i}_a \psi^{\bar k}$.  This means that an operator of the form $f_{\bar j_1,\dots,\bar j_l}(\phi^i, \phi^{\bar i}) \psi^{\bar j_i}\dots \psi^{\bar j_l}$ and a $(0,l)$-form $f_{\bar j_1,\dots,\bar j_l}(\phi^i, \phi^{\bar i})d\phi^{\bar j_1}\wedge \dots \wedge d\phi^{\bar j_l}$ on $X$ gauge-transform in exactly the same way. Hence, according to our discussion in $\S$4.1,  $\widetilde {\cal F}_e$ would correspond to a gauge-invariant $(0,k)$-form $f_{\bar j_1,\dots,\bar j_k}(\phi^i, \phi^{\bar i})d\phi^{\bar j_1}\wedge \dots \wedge d\phi^{\bar j_k}$ on $X$.  

At the classical level (where perturbative corrections are absent),  it was also argued in $\S$4.1 that $Q$ will act on a dimension $(0,0)$ operator such as  $\widetilde {\cal F}_e$  (which does not contain the covariant derivatives $D^m_z \phi^i$ or $D_z^m\phi^{\bar i}$ with $m>0$) as the $\bar \partial$ operator. Moreover, as elaborated in $\S$2.3,  gauge-invariant forms are also $G$-equivariant forms. Altogether, this means that the classical ring is just the graded $G$-equivariant Dolbeault ring $H^{0, *}_G(X)$. This ring, for compact $X$, is also finite-dimensional.

\newsubsection{A Sheaf of Chiral Algebras}

Let us now explain the idea of a ``sheaf of chiral algebras'' on $X$. To this end, note that both the $Q$-cohomology of local (on the worldsheet $\Sigma$) operators and the supersymmetry generator $Q$, can be described locally on $X$. Hence, one is free to restrict the local operators to be well-defined only on a given open set $U \subset X$. Since in the perturbative theory, we are considering sections $\Phi :\Sigma \to X$ with no multiplicities, operator product expansions between local operators will make sense in $U$. From here, one can naturally proceed to restrict the definition of the operators to smaller open sets, such that  a global definition of the operators can be obtained by gluing together the open sets on their unions and intersections. From this description, in which one associates a chiral algebra, its OPE's, and chiral ring to every open set $U \subset X$, we arrive at what is mathematically understood as a ``sheaf of chiral algebras''. We will call this sheaf $\widehat {\cal A}$.

\bigskip \noindent{\it Description of $\cal A$ Via $\check{ \textrm C}$ech Cohomology}

In the perturbative theory, one can also describe  the  $Q$-cohomology  classes as a kind of $\check{ \textrm C}$ech cohomology. Specifically, we will show that the chiral algebra $\cal A$ of $Q$-cohomology classses of the gauged twisted $(0,2)$ sigma model on $X$, can be described by the classes of the $\check{ \textrm C}$ech cohomology  of the sheaf $\widehat {\cal A}$ of locally-defined chiral operators. To this end, we shall generalize the argument in $\S$2.3 -- which provides a $\check{ \textrm C}$ech cohomological description of a $\bar \partial$-cohomology -- to demonstrate an isomorphism  between the $Q$-cohomology classes and the classes of the $\check{ \textrm C}$ech cohomology of $\widehat {\cal A}$.

Let us start by considering a contractible open set $U \subset X$ that is homeomorphic to an open ball in $\mathbb C^n$, where $n = {\textrm {dim}}_{\mathbb C} (X)$. As explained at the start of $\S$2.3, the $Q$-cohomology is independent of metric deformations of $X$. Hence, let us, for convenience, pick a metric on $U$ whereby $\nabla_{\bar k} V^{\bar i}_a = 0$. Then, according to our discussion in $\S$4.1,  $Q$ will act classically as the $\bar \partial$ operator on any local operator $\cal F$ in $U$. In other words, $\cal F$ can be interpreted as a $\bar\partial$-closed $(0,k)$-form valued in a certain vector bundle $\widehat F$ over $U$ -- that is, in the absence of perturbative corrections at the classical level, any operator ${\cal F}$ in the $Q$-cohomology will be a class of $H^{0,k}_{\bar \partial}(U, \widehat {F})$ on $U$. Since $U$ is contractible, $\widehat F$ will be a trivial bundle over $U$. This means that $\widehat {F}$ will always possess a global section, i.e., it corresponds to a soft sheaf.  Since the higher $\check{ \textrm C}$ech cohomologies of a soft sheaf are trivial \cite{Wells}, we will have $ {H_{\textrm{$\check{ \textrm C}$ech}}^{k} (U, {\widehat {F}} )} = 0$ for $k > 0$. Mapping this back to Dolbeault cohomology via the $\check{ \textrm C}$ech-Dolbeault isomorphism, we find that $H^{0,k}_{\bar \partial}(U, \widehat {F}) = 0$ for $k > 0$.  Recalling that small quantum corrections in the perturbative limit can only annihilate and not create cohomology classes, we conclude that local operators ${\cal F}$  with $q > 0$ will  necessarily vanish in  $Q$-cohomology on $U$.

Now define a good cover of  $X$ by open sets $\{U_e \}$.  Since the intersection of open sets $\{U_e \}$ also give open sets (homeomorphic to open balls in $\mathbb C^n$),  $\{U_e \}$, as well as all of their intersections, have the same feature as $U$    described above: $Q$-cohomology vanishes for $q > 0$ on $\{U_e \}$ and their intersections.  

Let the operator ${\cal F}_1$ on $X$ be a $Q$-cohomology class with $q = 1$. It is here that we shall import the usual arguments relating a $\bar\partial$- to a $\check{ \textrm C}$ech cohomology, to demonstrate an isomorphism between the $Q$-cohomology and a $\check{ \textrm C}$ech cohomology.  When restricted to an open set        $U_e$, the operator ${\cal F}_1$ must be trivial in $Q$-cohomology since $q > 0$, i.e., ${\cal F}_1 =[Q,{\cal C}_e]$, where ${\cal C}_e$ is an operator that is well-defined in $U_e$ with $q=0$. 

Now, since $Q$-cohomology classes such as ${\cal F}_1$ -- albeit locally-defined on $\Sigma$ -- can be globally-defined on $X$, we have ${\cal F}_1 =[Q,{\cal C}_e]= [Q,{{\cal C}_f}]$ over the intersection  $U_e\cap U_f$, so $[Q,{{\cal C}_e}- {{\cal C}_f}]=0$. Let
${\cal C}_{ef}= {{\cal C}_e}- {{\cal C}_f}$, where ${\cal C}_{ef}$ is defined in $U_e\cap U_f$. Then, $[Q, {\cal C}_{ef} ]=0$, and over any triple intersection $U_e \cap U_f \cap U_g$, we have
\be
{\cal C}_{ef}=     -{\cal C}_{fe}, \quad {{\cal C}_{ef}}+ {{\cal C}_{fg}} + {{\cal C}_{ge}} =0.
\label{cab}
\ee
Moreover, for ($q =0$) operators ${\cal K}_e$ and ${\cal K}_f$ whereby $[Q, {{\cal K}_e} ] = [Q, {{\cal K}_f} ]= 0$, we have an equivalence relation 
\be
{\cal C}_{ef} \sim  {{\cal C}'_{ef} = {{\cal C}_{ef} + {\cal K}_e - {\cal K}_f}}.
\label{cab1}
\ee
Note that since  $[Q, {\cal C}_{ef}]=0$ and ${\cal C}_{ef} \neq \{Q, \dots]$, the collection $\{\CC_{ef} \}$ are operators in the $Q$-cohomology with well-defined operator product expansions, and whose dimension $(0,0)$ subset furnishes a topological chiral ring with $q =0$.  
 
Since the local operators with $q > 0$ vanish in $Q$-cohomology on an arbitrary open set $U$, the sheaf $\widehat {\cal A}$ of the chiral algebra of operators has as its local sections  the $\psi^{\bar i}$-independent (i.e.,~$q =0$) operators  ${\widehat {\cal F}} (\phi^i,\partial_z\phi^i,\dots; \phi^{\bar i},\partial_z\phi^{\bar i},\dots; A^a_z, \partial_z A^a_z, \dots)$ 
that are annihilated by $Q$. Each $\CC_{ef}$ with $q =0$ is thus a section of
$\widehat{\cal A}$ over the intersection $U_e\cap U_f$. From (\ref{cab}) and (\ref{cab1}), we see that the collection $\{\CC_{ef} \}$ defines the elements  of the first $\check{ \textrm C}$ech cohomology group $H_{{\rm \check{C}ech}}^1(X, \widehat{\cal A})$.

Recall that the $Q$-cohomology classes are defined as those operators which are $Q$-closed, modulo those which can be globally written as $\{ Q, \dots]$ on $X$. In other words, ${\cal F}_1$ vanishes in $Q$-cohomology if we can write it as ${\cal F}_1 = [Q ,{\cal C}_e]=[Q,{{\cal C}_f}] =  [Q ,{\cal C}]$, i.e., ${\cal C}_e = {\cal C}_f$ and hence ${\cal C}_{ef} = 0$. Therefore, a vanishing $Q$-cohomology with $q =1$ corresponds to a vanishing first $\check{ \textrm C}$ech cohomology. Thus, we have obtained a map from the $Q$-cohomology with $q =1$ to a first $\check{ \textrm C}$ech cohomology.      

Similar to the case of relating a $\bar \partial$- to a $\check{ \textrm C}$ech cohomology,  one can also reverse our arguments and construct an inverse of this map. Suppose we are given a family $\{ {\cal C}_{ef} \}$ of sections of $\widehat {\cal A}$ over the corresponding intersections $\{U_e \cap U_f \}$, and they obey (\ref{cab}) and (\ref{cab1}) so that they define the elements of $H_{{\rm \check{C}ech}}^1(X, \widehat{\cal A})$. We can then proceed as follows.  Let the set $\{f_a \}$ be partition
of unity subordinates to the open cover of $X$ provided by $\{U_e\}$. This means that the elements of $\{f_e \}$ are continuous functions on $X$, and they  vanish outside the corresponding elements in  $\{U_e \}$ whilst obeying $\sum_e f_e=1$. Let ${\cal F}_{1,e}$ be a chiral operator defined in $U_e$ by ${\cal F}_{1,e}= \sum_g[ Q, f_g]  {\cal C}_{eg}$.\footnote{Normal ordering of the operator product
between $[Q, f_c(\phi^i,\phi^{\bar i})]$ and ${\cal C}_{ef}$ is needed for regularization purposes.} ${\cal F}_{1,e}$ is well-defined throughout $U_e$, since in $U_e$, $[Q, f_g]$ vanishes wherever ${\cal C}_{eg}$ is not defined. Clearly, ${\cal F}_{1,e}$ has $q =1$, since ${\cal C}_{eg}$ has $q =0$ and $Q$ has $q=1$. Moreover, since ${\cal F}_{1,e}$ is a chiral operator defined in $U_e$, it will mean that $\{ Q, {\cal F}_{1,e}\} = 0$  over $U_e$. For any $e$ and $f$, we have ${\cal F}_{1,e}- {\cal F}_{1,f} = \sum_g [Q , f_g]  ({\cal C}_{eg}- {\cal C}_{fg})$. Using (\ref{cab}), this is $\sum_g [Q , f_g] {\cal C}_{ef} = [Q, \sum_g f_g ] {\cal C}_{ef}$. This vanishes since $\sum_g f_g = 1$. Hence, ${\cal F}_{1,e} = {\cal F}_{1,f}$ on $U_e \cap U_f$, for $\it{all}$ $e$ and $f$. In other words, we have uncovered a globally-defined $q =1$ operator ${\cal F}_1$ where $\{Q, {\cal F}_1 \} = 0$ on $X$. Notice that ${\cal F}_{1,e}$ and thus ${\cal F}_1$ is not defined to be of the form $\{Q, \dots \}$. Therefore, we have arrived at a map from the $\check{ \textrm C}$ech cohomology group $H_{{\rm \check{C}ech}}^1(X, \widehat{\cal A})$ to the $Q$-cohomology group with $q =1$, i.e., $Q$-closed $q =1 $ operators modulo those that can be globally written as $\{Q, \dots ]$. The fact that this map is an inverse of the first map can indeed be verified.      

Since there is nothing unique about the $q =1$ case, we can repeat the above procedure for operators with $q > 1$. In doing so, we find that  the $Q$-cohomology coincides with the $\check{ \textrm C}$ech cohomology of $\widehat {\cal A}$ for all $q$. Hence, the chiral algebra $\cal A$ of the gauged twisted $(0,2)$ sigma model will be given by  $\bigoplus_{q} H^{q}_{\textrm {$\check{ \textrm C}$ech}} (X, {\widehat{\cal A}})$ as a vector space. As there will be no ambiguity, we shall henceforth omit the label `$\check{ \textrm C}$ech' when referring to the cohomology of $\widehat {\cal A}$. 

\bigskip\noindent{\it A Sheaf of Vertex Algebras}  

Note that mathematically, the sheaf $\widehat {\cal A}$ would be known as a sheaf of vertex algebras. It would be described purely from the $\check{ \textrm C}$ech viewpoint: the field $\psi^{\bar i}$ would be omitted and locally on $X$, one would just consider operators constructed from $\phi^i$, $\phi^{\bar i}$, $A^a_z$ and their  $z$-derivatives as generators of $\widehat {\cal A}$. The chiral algebra $\cal A$ of $Q$-cohomology classes with $q > 0$ would then be constructed as $\check{ \textrm C}$ech $q$-cocycles. Notice that in this framework, one would not need to resort to any computation involving the  path integral. Instead, one would utilize the abstraction of $\check{ \textrm C}$ech cohomology to define the spectrum of operators in the quantum sigma model. In this sense, the study of the sigma model can be made mathematically rigorous. 

Unlike its close cousins the sheaf of CDO's and CDR~\cite{GMS1, MSV1} -- which have been shown to be relevant to the ordinary (heterotic) twisted $(0,2)$ sigma model and the half-twisted $(2,2)$ model, respectively~\cite{CDO, MC, Ka} -- a mathematical interpretation of the sheaf $\widehat {\CA}$ is currently unavailable.  Nonetheless, in the case where the gauge group is an abelian subgroup $T \subset G$, we will show in $\S$5 that $\widehat {\cal A}$ for a certain class of $X$'s is just the sheaf of TCDO's recently formulated by Arakawa et al.~in~\cite{Arakawa}.

\newsubsection{Relation To A Perturbed Version Of A Free $\beta\gamma$ System}

Let us now provide a useful physical description of the local structure of the sheaf $\widehat {\cal A}$ for when the gauge group is an abelian subgroup $T \subset G$. To this end, we will describe in a novel way the $Q$-cohomology of operators which are regular in a small open set $U \subset X$. As before, we assume that $U$ is homeomorphic to an open ball in $\mathbb C^n$ and is thus contractible. 

Now notice from $S_{\mathrm gauged}$ in (\ref{Sgauged}) and $\mathscr V$ in (\ref{V}), that the hermitian metric on $X$ appears  within a $\{Q, \dots \}$-term in the action. Thus, any shift in the metric will also appear within $Q$-exact (i.e.,~$Q$-trivial) terms. Consequently, for our present purposes, we can arbitrarily redefine the values of the hermitian metric on $X$, since they do not affect the analysis of the $Q$-cohomology. As such, to describe the local structure, we can always select a flat hermitian metric on $U$. As explained in $\S$2.3, the action also contains terms derived from $H^1(X,\Omega^{2,cl}_{X,T})$. However, from (\ref{ST-mod}), we see that these terms are likewise $Q$-exact and therefore, can be ignored in our analysis of the local structure in $U$. Thus, noting that $\partial_k V^{i}_a = \partial_{\bar k} V^{\bar i}_a = 0$ (whence (\ref{satisfy}) is trivially satisfied)  since the gauge group $T$ is abelian,\footnote{As the gauge group $T$ is abelian, the structure constants $f_{ab}{}^c$ must vanish for all $a, b, c = 1, 2, \dots, d$, where $d$ is the dimension of the group. From (\ref{v4}) and (\ref{v5}), we then have  the conditions $V^j_a  \partial_j V_b^i = V^j_b \partial_j V_a^i$ and $V^{\bar j}_a \partial_{\bar j} V_b^{\bar i} = V^{\bar j}_b \partial_{\bar j} V_a^{\bar i}$. Since the nonzero vector fields that generate the abelian action on $X$ are linearly-independent in all indices, i.e., $V^i_a \neq V^j_a$, $V^{\bar i}_a \neq V^{\bar j}_a$, $V^j_a \neq V^j_b$ and $V^{\bar j}_a \neq V^{\bar j}_b$, these conditions will be met by $\partial_k V^l_c = \partial_{\bar k} V^{\bar l}_c = 0$ for any component.} one can write the local action of the abelian twisted $(0,2)$ sigma model on $U$ as 
\be
I = {1 \over 2 \pi} \int_{\Sigma} |d^2 z| \, \sum_{i, \bar j, a, b} \delta_{i \bar j} \left ( \partial_z \phi^{\bar j} \partial_{\bar z}\phi^i +   \psi ^i_{\bar z} \partial_z \psi^{\bar j} \right ) + \delta_{i \bar j} \left(\partial_{\bar z} \phi^i A^a_z V_a^{\bar j}  - A^a_{\bar z} V^i_a \partial_z \phi^{\bar j}  - A^a_{\bar z} V_a^{i} A^b_z V_b^{\bar j} \right).
\label{Suold}
\ee 
From (\ref{Suold}), we have, from the equation of motions,  the constraints  $\partial_{\bar z} (\partial_z \phi^{\bar j}) = -(\partial_{\bar z} A^a_z) V^{\bar j}_a$ and $\partial_{\bar z} (\partial_{z} \phi^{i}) = (\partial_{z} A^a_{\bar z}) V^{i}_a$. These constraints will be satisfied if $\partial_z \phi^{\bar j}$, $A^a_z$ and $\partial_z \phi^i$ vary holomorphically with $z$, while $A^a_{\bar z}$ varies antiholomorphically with $\bar z$. Then, via integration by parts, and the holomorphy and antiholomorphy of $A^a_z$ and $A^a_{\bar z}$,  one can simplify (\ref{Suold}) to
\be
I = {1 \over 2 \pi} \int_{\Sigma} |d^2 z| \, \sum_{i, \bar j, a, b} \delta_{i \bar j} \left ( \partial_z \phi^{\bar j} \partial_{\bar z}\phi^i +   \psi ^i_{\bar z} \partial_z \psi^{\bar j}  -  A^a_{\bar z} V_a^{i} A^b_z V_b^{\bar j} \right).
\label{Su}
\ee 

Let us now describe the $Q$-cohomology classes of operators regular in $U$.  As explained earlier, these are operators of dimension $(n,0)$ which vary holomorphically with $z$ and are independent of $\psi^{\bar i}$. Such operators  are of the form ${\widehat {\cal F}} (\phi^i,\partial_z\phi^i,\dots; \phi^{\bar i},\partial_z \phi^{\bar i},\dots; A^a_z, \partial_z A^a_z, \dots)$. Note that because $\nabla_{\bar k} V^{\bar i}_a = 0$ over $U$ (since the Levi-connection $\Gamma_{\bar j \bar k}^{\bar i}$ vanishes for a flat metric on $U$, in addition to having the condition $\partial_{\bar k} V^{\bar i}_a = 0$), from our discussion in $\S$4.1, we find that $Q$ will act  as the $\bar \partial$ operator at the classical level. In this case, the $Q$ operator may receive perturbative corrections from $\bar\partial$-cohomology classes such as the Ricci tensor and classes in $H^1(X,\Omega^{2,cl}_{X, T})$. However, note that since we have picked a flat hermitian metric on $U$, the corresponding Ricci tensor on $U$ is zero. Moreover, as explained above, classes from $H^1(X, \Omega^{2,cl}_{X, T})$ do not contribute  when analyzing the $Q$-cohomology on $U$. Hence, we can ignore the perturbative corrections to $Q$ in our present discussion. Therefore, on the classes of operators in $U$, $Q$ acts as $\bar \partial =\psi^{\bar i}\partial/\partial\phi^{\bar i}$, and if $\widehat {\cal F}$ is to be annihilated by $Q$, it would mean that as a function of $\phi^i$, $\phi^{\bar i}$, $A^a_z$ and their $z$-derivatives, $\widehat {\cal F}$ must be independent of $\phi^{\bar i}$ (but not its derivatives); in other words,  $\widehat {\cal F}$  should depend only on the other variables  $\phi^i$ and $A^a_z$, and the derivatives of $\phi^i$, $\phi^{\bar i}$ and $A^a_z$.\footnote{Once again, we can disregard the action of $Q$ on $z$-derivatives of $\phi^{\bar i}$ as $\partial_z\psi^{\bar i}=0$ and $\delta \phi^{\bar i} = \psi^{\bar i}$.} Hence, the $Q$-invariant operators take form ${\widehat {\cal F}}(\phi^i,\partial_z\phi^i,\dots;\partial_z\phi^{\bar i},\partial_z^2\phi^{\bar i},\dots ; A^a_z, \partial_z A^a_z, \partial_z^2 A^a_z, \dots)$. In other words, the operators, in their dependence on $\phi^{k, \bar k}$, the center of mass coordinates of the string whose worldsheet theory is the abelian twisted $(0,2)$ sigma model, is holomorphic. The local sections of $\widehat {\cal A}$ are just given by the operators in the $Q$-cohomology  of the local abelian twisted $(0,2)$ sigma model with action (\ref{Su}).

Let us set $\beta_i  =  \delta_{i \bar j} \partial_z \phi^{\bar j}$ and $\gamma^i = \phi^i$, where $\beta_i$ and $\gamma^i$ are bosonic operators of dimension $(1,0)$ and $(0,0)$, respectively. Then, the $Q$-cohomology of operators that are regular in $U$ can be represented by arbitrary local functions of $\beta$, $\gamma$ and $A^a_z$ of the form ${\widehat {\cal F}} (\gamma, \partial_z \gamma, \partial_z^2 \gamma, \dots, \beta, \partial_z \beta, \partial_z^2 \beta, \dots A^a_z, \partial_z  A^a_z, \partial_z^2 A^a_z, \dots)$. From the flat action (\ref{Su}), one can deduce, via standard methods in quantum field theory, that the operator products $A^a_z \cdot \beta$ and $A^a_z \cdot \gamma$ are trivial;\footnote{Note that the $V^i_a$'s and $V^{\bar i}_a$'s are, in the abelian case at hand, independent of the $\phi^i$ and $\phi^{\bar i}$ fields. Hence, there are \emph{no} nontrivial propagators arising from their presence in the action (\ref{Su}).} on the other hand, the products  $\gamma\cdot\gamma$ and $\beta\cdot\beta$ are nonsingular, while
\be
\beta_i(z)\gamma^j(z')=-{\delta_{ij}\over z-z'}+{\rm regular}.
\label{beta-gamma OPE}
\ee
We can construct an action for the fields $\beta$, $\gamma$ and $A^a_z$, viewed as free elementary fields, which leads to these OPE's.  It is the following action of a perturbed version of a free $\beta\gamma$ system: 
\be
I_{\beta\gamma}= {1\over
2\pi} \int |d^2z|  \sum_{i,a,b}   \left( \beta_i \partial_{\bar z}\gamma^i -   A^a_{\bar z} V_a^{i} A^b_z V_{i b} \right).
\label{bcaction}
\ee
Hence, we find that the perturbed free $\beta\gamma$ system above reproduces the $Q$-cohomology of $\psi^{\bar i}$-independent operators of the abelian twisted $(0,2)$ sigma model on $U$, i.e., the local sections of the sheaf $\widehat{\cal A}$.

At this juncture, one can make another important observation about the relationship between the local abelian twisted $(0,2)$ sigma model with action (\ref{Su}) and the local (in the sense of the target space) version of the perturbed free $\beta\gamma$ system of (\ref{bcaction}). To begin with, note that the holomorphic stress tensor ${\widehat T}(z) = -2 \pi T_{zz}$ of the local sigma model is given by
\be
{\widehat T}(z) =  - \delta_{i \bar j} \partial_z \phi^{\bar j}\partial_z \phi^i. 
\label{T(z)}
\ee
(Here and below, normal ordering is understood for ${\widehat T}(z)$). Via the identification of the fields $\beta$ and $\gamma$ with $\partial_z \phi$ and $\phi$, respectively, we find that ${\widehat T}(z)$ can also be written  as
\be
{\widehat T}(z) = - \beta_i \partial_z \gamma^i.
\label{Tz}
\ee 
This coincides with the holomorphic stress tensor of the local perturbed free $\beta \gamma$ system. Simply put, the abelian twisted $(0,2)$ sigma model and the perturbed free $\beta\gamma$ system have the same $\it{local}$ holomorphic stress tensor. This means that locally on $X$, the sigma model and the $\beta\gamma$ system have the same generators of general holomorphic coordinate transformations on the worldsheet.

One may now ask the following question: does the $\beta\gamma$ system reproduce the $Q$-cohomology of $\psi^{\bar i}$-independent operators globally on $X$, or just within a small open set $U$? Well, the $\beta\gamma$ system will certainly reproduce the $Q$-cohomology of $\psi^{\bar i}$-independent operators globally on $X$ if there is no obstruction to defining the system globally on $X$ -- i.e., one finds, after making global sense of the action (\ref{bcaction}), that the corresponding theory remains anomaly-free at the quantum level. Let's study this aspect more closely.  

First and foremost, the classical action (\ref{bcaction}) is globally sensible if we interpret the bosonic fields $\beta$ and $\gamma$ correctly.  Since $\gamma$ is not minimally-coupled to the gauge field $A$, it ought to be interpreted as a map $\gamma:\Sigma\to X$.  As for $\beta$, it ought to be interpreted as a $(1,0)$-form on $\Sigma$ valued in the pull-back $\gamma^*(T^*X)$. As always, the $V^i_a$'s and $V_{ib}$'s are to be interpreted as (co)vector fields, but now, on a (target) space that is not necessarily flat. With this interpretation, (\ref{bcaction}) becomes the action of what one might call a nonlinear perturbed $\beta\gamma$ system. Nevertheless, by choosing $\gamma^i$ to be local coordinates on a small open set $U\subset X$, one can make the action linear. In other words, a local version of (\ref{bcaction}) represents the action of a linear perturbed $\beta \gamma$ system. 

Now that we have made global sense of the action of the $\beta\gamma$ system at the classical level, we move on to discuss what happens at the quantum level.  To this end, let us first integrate out from the path integral, the non-propagating fields $A^a_z$ and $A^a_{\bar z}$ (which do not affect the anomalies) via their equations of motion. Next, let us perform an expansion around a classical solution of the nonlinear (i.e.,~global) $\beta\gamma$ system given by a holomorphic map $\gamma_0:\Sigma \to X$, i.e., let ${\gamma} =\gamma_0 +\gamma'$. Then, the action is $(1/2\pi) (\beta , {\overline D \gamma'})$ to quadratic order about $\gamma_0$. The field $\gamma'$, being a deformation of the coordinate $\gamma_0$ on $X$, is a section of the pull-back $\gamma_0^* (TX)$. Thus, the kinetic operator of the $\beta$ and $\gamma$ fields which characterizes the anomaly of the $\beta\gamma$ system, is the $\overline D$ operator on sections of $ \gamma_0^*(TX)$. Note that $\overline D$ is the complex conjugate of the $D$ operator in $S_{pert}$ of (\ref{actionpert}). Complex conjugation flips the sign of the anomalies. That said, the fields involved here are bosonic while in $S_{pert}$ they are fermionic -- this results in a second sign flip. Hence, the anomalies of the $\beta\gamma$ system can be computed as the anomalies of the model with action $S_{pert}$.\footnote{Notice that the $D$ operator in (\ref{actionpert}) acts on sections of the pull-back of the antiholomorphic bundle $\overline {TX}$ instead of the holomorphic bundle $TX$. However, this difference is irrelevant with regard to anomalies since $ch_2(\overline E) = ch_2(E)$ for any holomorphic vector bundle $E$.}  The anomalies for the model with action $S_{pert}$ have been computed in~\cite{CDO}. The computation is similar to that furnished in $\S$3; one just replaces the bundle $\overline{\textrm{ker} \, d\pi_E}$ therein with the bundle $\overline{TX}$, and proceed with the same calculation (where $\Phi : \Sigma \to X$ is now a map).  This gives us the anomaly-cancellation conditions for the $\beta\gamma$ system as $c_1(\Sigma) c_1(X)/2 = p_1(X)/2 = 0$.

That being said,  note that since the compact, connected $T$-action is supposed to leave fixed the metric on $X$ -- i.e., the $T$-action generates an automorphism that maps $X$ back to itself -- the standard de Rham cohomology groups will coincide with the cohomology groups defined by $T$-invariant  forms on $X$.\footnote{Since the $G$-action $t_g$ is an automorphism of $X$, it would induce an automorphism $t^*_g$ on the de Rham cohomology groups $H^k(X; \mathbb R)$. Hence, $G$ acts as a group of automorphisms on  $H^k(X; \mathbb R)$. Let  $H^k(X; \mathbb R)^G$ be the fixed point set of this action. Also, if $[\Omega^k(X)]^G$ denotes a $G$-invariant $k$-form on $X$, the inclusion $I: [\Omega^k(X)]^G \to \Omega^k(X)$ induces an isomorphism $I^* : H^k([\Omega(X)]^G) \cong H^k(X, \mathbb R)^G$.  Since $G$ is connected, $t_g$ and the identity map $1_X$ are homotopic, i.e.,  $t_g \cong 1_X$ for every $g \in G$. This implies that the induced automorphisms are the same, i.e.,  $t^*_g = {\textrm {Id}}$, where Id is the identity on $H^k(X; \mathbb R)$. Hence, $H^k(X; \mathbb R)^G$ is the whole $H^k(X;\mathbb R)$, and we have an isomorphism $I^* : H^k([\Omega(X)]^G) \cong H^k(X, \mathbb R)$.} In other words, $c_1(X) = [c_1(X)]^T$ and $p_1(X) = [p_1(X)]^T$, where the `$T$' superscript just indicates that the class is $T$-invariant. In turn, because $T$ is abelian, we have $[c_1(X)]^T = c^T_1(X)$ and $ [p_1(X)]^T = p_1^T(X)$, whereby if $H^m_T(X)$ is the degree-$m$ part of the $T$-equivariant cohomology of $X$, $c^T_1(X) \in H^2_T(X)$ and $p^T_1(X) \in H^4_T(X)$.\footnote{This is -- in the context of our sigma model -- an implicit consequence of the fact that $T$-invariant forms coincide with $T$-equivariant forms (see remark~14 of~\cite{libine}). To understand this explicitly, note that any $G'$-equivariant cohomology class $\mathscr O_d$ of degree $d$ can -- in the Cartan model of equivariant cohomology -- be written as $\mathscr O_d = \CO_d + \sum_{i=1}^p {1 \over i!} F^{a_1} \dots F^{a_i} \Delta_{a_1 \dots a_i}$, where the forms $\Delta_{a_1 \dots a_p} \in \Omega^{d - 2p}(X)$ are totally symmetric in their indices, $F = dA + A\wedge A$, and $\CO_d$ is a $G'$-invariant form on $X$ of degree $d$~\cite{farrill}. In our abelian case of $G' = T$, we have, from our discussion below (\ref{Suold}), the constraints $\partial_z A^a_{\bar z} = 0 = \partial_{\bar z} A^a_z$. In turn, this means that $F^a = 0$ and hence, $\mathscr O_d = \CO_d$ and $d_C = d$, where $d_C$ is the differential of the equivariant cohomology complex. In other words, where our sigma model is concerned, a $T$-equivariant cohomology class and a $T$-invariant de Rham cohomology class can be regarded as one and the same thing.}  Altogether, this means that one can actually write $c_1(X) =  [c_1(X)]^T = c^T_1(X)$ and $ p_1(X) = [p_1(X)]^T =  p^T_1(X)$, whence the anomaly-cancellation conditions for the $\beta\gamma$ system are
\be
0= {1\over 2} c_1(\Sigma) c^T_1 (X) = {1\over 2} p^T_1(X).
\label{beta anomaly}
\ee   
Notice that (\ref{beta anomaly}) is exactly (\ref{an}) for gauge group $T$ -- i.e., the anomalies of the abelian twisted $(0,2)$ sigma model also manifest in the nonlinear perturbed $\beta\gamma$ system. Furthermore, if (\ref{beta anomaly}) is satisfied whence the anomalies vanish, the  nonlinear perturbed $\beta\gamma$ system will reproduce the $Q$-cohomology of $\psi^{\bar i}$-independent operators globally on $X$. This means that  one would be able to find, without any obstruction, a global section of $\widehat {\cal A}$.    


\bigskip\noindent{\it Describing $\CA$ Via the $\beta\gamma$ System and Other Relevant Issues}        

At any rate, notice that the perturbed $\beta\gamma$ system lacks the presence of right-moving fermions and thus, the $U(1)_R$ charge $q$ carried by the fields $\psi^i_{\bar z}$ and $\psi^{\bar i}$ of the underlying abelian twisted $(0,2)$ sigma model. Locally, the $Q$-cohomology of the sigma model is nonvanishing only for $q =0$. Globally and generically however, there can be  higher degrees in cohomology. Since the chiral algebra of operators furnished by the linear perturbed $\beta\gamma$ system gives the correct description of the   $Q$-cohomology of $\psi^{\bar i}$-independent operators on $U$, one can then expect the globally-defined chiral algebra of operators furnished by the nonlinear perturbed $\beta\gamma$ system to correctly describe the $Q$-cohomology classes of zero degree (i.e.,~$q =0$) on $X$. How then can one utilize the nonlinear perturbed $\beta\gamma$ system to describe the higher cohomology? The answer lies in the analysis carried out in $\S$4.3. In the $\beta\gamma$ description, we lack a close analog of $\bar \partial$-cohomology at our disposal. Nonetheless, we can exploit the more abstract notion  of $\check{ \textrm C}$ech cohomology. As before, we begin with a good cover of $X$ by small open sets $\{U_e \}$, and, as explained in $\S$4.3, we can then describe the         $Q$-cohomology classes of positive degree (i.e.,~$q > 0$) by $\check{ \textrm C}$ech $q$-cocycles, i.e., they can be described by the $q^{th}$ $\check{ \textrm C}$ech cohomology of the sheaf $\widehat {\cal A}$ of the chiral algebra of the linear perturbed $\beta\gamma$ system with action being a linearized version of (\ref{bcaction}).

Another issue that remains to be elucidated is the appearance of the various moduli of the sigma model in the nonlinear perturbed $\beta\gamma$ system. Recall from $\S$2.3  that the moduli of the chiral algebra of the sigma model consist of the complex  structure of $X$  and a class in $H^1(X, \Omega^{2, cl}_{X, T})$. The complex structure, via the definition of the fields themselves, is automatically built into the classical action (\ref{bcaction}). However, one cannot incorporate a class from $H^1(X, \Omega^{2,cl}_{X,T})$ within the action in this framework. Nevertheless, as we will explain in $\S$4.6, the modulus represented by a class in $H^1(X, \Omega^{2, cl}_{X,T})$ can be built into the definition of certain $\check{ \textrm C}$ech cocycles through which one can define a family of sheaves of chiral algebras.                      

A final remark to be made is that in the study of quantum field theory, one would like to be able to go beyond just defining the $Q$-cohomology classes or a sheaf of chiral algebras. One would also like to be able to compute physically meaningful quantities such as the correlation functions of these cohomology classes of local operators. In the sigma model, the correlation functions can be computed from standard methods in quantum field theory. But at first sight, there seems to be an obstacle in doing likewise for the nonlinear perturbed $\beta\gamma$ system. This can be seen as follows. Let the correlation function of $s$ local operators ${\cal O}_1$, ${\cal O}_2$, $\dots$, ${\cal O}_s$ on a genus $g$ Riemann surface $\Sigma$ be given by $\left < {\cal O}_1(z_1) \dots {\cal O}_s(z_s) \right>_g$, where ${\cal O}_i(z_i)$ has $U(1)_R$ charge $q_i$. Note that due to a $U(1)_R$ anomaly, the correlation functions of our perturbative model will be nonvanishing if and only if $\sum_i q_i = n (1-g)$, where $n ={\textrm dim}_{\mathbb C}X$. Thus, generic nonzero correlation functions require that not all the $q_i$'s be zero. In particular, correlation functions at string tree level vanish unless $\sum_i q_i = n$. However, the operators of $q_i \neq 0$ cannot be expressed in a standard manner in the nonlinear perturbed $\beta\gamma$ system. They are instead expressed in terms of $\check{ \textrm C}$ech $q_i$-cocycles. This means that in order for one to compute the corresponding correlation functions using the nonlinear perturbed $\beta\gamma$ system, one must translate the usual quantum field theory recipe employed in the sigma model into a $\check{ \textrm C}$ech language. The computation in the $\check{ \textrm C}$ech language will involve products of $\check{ \textrm C}$ech cohomology groups and their maps into complex numbers. 

\bigskip\noindent{\it An Illuminating Example}

A straightforward but illuminating example would be the following computation of a correlation function involving the gauge-invariant local operator $\CO_i = g_{k \bar j}(\phi^l, \phi^{\bar l}) D_z \phi^{k} \psi^{\bar j}$  on the sphere (i.e.,~at string tree level). To this end, first note that one can write $\CO_i = \CO_{i, \phi} + \CO_{i, A}$, where $\CO_{i, \phi} =  g_{k \bar j}(\phi^l, \phi^{\bar l}) \partial_z \phi^{k} \psi^{\bar j}$ and $\CO_{i, A} =  g_{k \bar j}(\phi^l, \phi^{\bar l}) A^a_z V^{k}_a \psi^{\bar j}$. Second, recall from $\S$4.1 that a dimension $(1,0)$ operator $\CO_{i, \phi}$ with $U(1)_R$ charge $q_i =1$, can be interpreted as a $(0, 1)$-form valued in the holomorphic cotangent bundle $T^*X$; on the other hand, a dimension $(1,0)$ operator $\CO_{i, A}$ with $U(1)_R$ charge $q_i =1$, can be interpreted as a $(0,1)$-form valued in the bundle $E^*$ of rank $r = {\rm dim} \, \frak t$, where $\frak t$ is the Lie algebra of the gauge group $T$.  Despite the nontrivial geometrical interpretation of $\CO_i$, one can nevertheless interpret it as a class in the $\check{ \textrm C}$ech cohomology group $H^{1}( X, {\widehat \CA}_1)$, where ${\widehat \CA}_1$ is a sheaf whose local sections are generated by dimension $(1,0)$ functions of $\partial_z \gamma^k$, $A^a_z$ and $\gamma^l$.  Third, note that the path integral would localize onto supersymmetric configurations characterized by setting the fermionic field variations in (\ref{txgauged}) to zero~\cite{mirror manifolds}; this means that  the path integral would boil down to an integral over the moduli space of solutions to $D_{\bar z} \phi^{l} = 0$. In this particular case of the worldsheet being a sphere, a further simplification occurs; one can set $A_z = 0$ everywhere on $\Sigma$ via a gauge transformation\footnote{Recall that from our discussion following (\ref{Suold}),  we have the constraints $\partial_z A_{\bar z} = \partial_{\bar z}A_z = 0$; this implies that $F = d A = 0$. Since in this case, $\Sigma = {\bf S}^2$ is simply-connected, a vanishing field strength means that one can write the corresponding holomorphic component of the connection one-form $A$ in pure gauge, i.e, $A_z = i \partial_z (U^{\dagger})^{-1} \cdot U^{\dagger}$,  where $U \in G$ and $G$ is abelian. This expression shows that $A_z$ can be set to zero \emph{everywhere} on $\Sigma$ by a gauge transformation~\cite{GSW2}.} and write $\CO_i$ as $\CO_{i, \phi}$ and $D_{\bar z} \phi^{l} = 0$ as $\partial_{\bar z} \phi^l = 0$; i.e., the path integral would reduce to an integral over the moduli space of holomorphic maps  $\phi^l: \Sigma \to X$. That being said, since we are considering degree-zero maps in the perturbative theory, the moduli space of holomorphic maps is just $X$ itself; i.e., the path integral would really be an integral over the target space $X$. Altogether therefore, since $\sum q_i = n$ is required for $U(1)_R$ anomaly-cancellation, we find that a nonvanishing $\it{perturbative}$ correlation function on the sphere involving just the dimension $(1,0)$ operators ${\cal O}_i$, can be computed as 
\be
\left < {\cal O}_1(z_1) \dots {\cal O}_n(z_n) \right>_0 = \int_X W^{n,n}.
\label{corr1}
\ee                             
Here, $W^{n,n}$ is a top-degree $(n,n)$-form on $X$ which can be interpreted as a class in the $\check{ \textrm C}$ech cohomology group $H^n(X, K_X)$, and $K_X$ is the canonical sheaf of $(n,0)$-forms on $X$.  Explicitly, $W^{n,n}$ is obtained via the following antisymmetric product of $\check{ \textrm C}$ech cohomology classes $H^{1}( X, {\widehat \CA}_1)_1, \dots, H^{1}( X, {\widehat \CA}_1)_s$ which represent the anticommuting fermionic operators $\CO_1, \dots, \CO_s$: 
\be
  H^1 (X, {\widehat \CA}_1)_1\wedge \dots \wedge H^{1} (X, {\widehat \CA}_1)_n  \tilde{\rightarrow}    H^1 (X, \Omega^1_X)_1\wedge \dots \wedge H^{1} (X, \Omega^1_X)_n  \rightarrow  H^n(X, K_X).
\ee          
In the above, we have made use of the fact that because ${\widehat \CA}_1$ (whose local sections are independent of $\beta$, $A$ and their $z$-derivatives) can be identified with the sheaf of CDO's whose local sections are $\beta$-independent and of dimension $(1,0)$, it is in turn isomorphic to the sheaf $\Omega^1_X$ of holomorphic one-forms on $X$ (see $\S$2 of~\cite{GMS1}). In this framework, the integral over $X$ in (\ref{corr1}) just defines a map $H^n(X, K_X) \to \mathbb C$. But can one say more? Most certainly. 

In fact, since (even-dimensional) $T$ acts freely on $X$, one can regard $X$ as a principal bundle $X \to X / T$ with fiber $T$ and smooth base $X/T$, and by integrating first over the fiber of $X$, one can reduce the integral over $X$ to an integral over $X / T$. Because the $T$-action generates an automorphism of $X$, $W^{n,n}$ -- being a representative of a $\check{ \textrm C}$ech cohomology class -- will be a $T$-invariant form on $X$.\footnote{The $\check{\rm C}$ech-Dolbeault isomorphism implies that $W^{n,n}$ can be interpreted as a Dolbeault cohomology class. In turn, footnote~17 (which also applies to Dolbeault cohomology) implies that $W^{n,n}$ can be interpreted as a $T$-invariant form.} Hence, the integration over the fiber will just give a factor proportional to the volume of the group $T$. In short, one can also write (\ref{corr1}) as
\be
\left < {\cal O}_1(z_1) \dots {\cal O}_n(z_n) \right>_0  = \int_{X/T} W^{d,d}
\label{fpt}
\ee
(up to some normalization factor), where $d = \textrm{dim}_{\mathbb C} \, (X/T)$, and $W^{d,d}$ is a top-degree $(d,d)$-form on $X/T$ that can be interpreted as a class in the $\check{ \textrm C}$ech cohomology group $H^d(X, K_{X/T})$.

Although the above procedure involving products of $\check{ \textrm C}$ech cohomology groups and their maps into complex numbers is unusual for a physicist, it has been utilized in \cite{KS, Sharpe, Donagi} as a powerful means to compute certain quantum (i.e.,~nonperturbative) correlation functions in heterotic string theory. Analogous procedures follow for the computation of correlation functions involving other types of local operators.

\bigskip\noindent{\it Chiral Equivariant Quantum Cohomology}

Last but not least, note that in a $\it{non}$-$\it{perturbative}$ computation of any correlation function of the abelian twisted $(0,2)$ sigma model, the local operators will  be represented by $\check{ \textrm C}$ech cohomology classes in the moduli space of worldsheet twisted-instantons. The current procedure would then serve as a basis for a chiral generalization of $(0,2)$ quantum $T$-equivariant cohomology involving operators of dimension zero\emph{ and} greater.

\newsubsection{Local Symmetries}

So far, we have obtained an understanding of the local structure of the $Q$-cohomology. We shall now proceed towards our real objective of gaining an understanding of its global structure. In order to do so, we will need to glue the local descriptions that we have studied above, together.

To this end, we must first cover $X$ by small open sets $\{U_e\}$.  Recall here that in each $U_e$, the $Q$-cohomology is described by the chiral algebra of local operators of a perturbed free $\beta\gamma$ system on $U_e$. We will need to glue these local descriptions together over the intersections $\{ U_e \cap U_f \}$, so that the global structure of the $Q$-cohomology can be described via a globally-defined sheaf of chiral algebras on the \emph{entire} manifold $X$.

Note that the gluing has to be carried out using the automorphisms of the perturbed free $\beta\gamma$ system. Thus, one must first ascertain the underlying symmetries of the system, which are in turn divided  into  geometrical and non-geometrical symmetries. The geometrical symmetries are used in gluing together the local sets $\{{U_e} \}$ into the entire target space $X$. The non-geometrical symmetries on the other hand, are used in gluing the local descriptions at the algebraic level.

As usual, the generators of these symmetries will be given by the charges of the conserved currents of the perturbed free $\beta\gamma$ system. In turn, these generators will furnish the Lie algebra $\mathfrak s$ of the symmetry group. Let the elements of $\mathfrak s$ which generate the geometrical and non-geometrical symmetries be  written as ${\mathfrak v}$ and ${\mathfrak c} = ({\frak f}, {\frak e})$, where as we will explain shortly, $\frak v$, $\frak f$, and $\frak e$  are associated with a vector field $V$, a $\frak t$-valued one-form $F_a$ (where $\frak t$ is the Lie algebra of $T$), and a $\partial$-closed $T$-equivariant two-form $E$ on $X$, respectively.   Since the conserved charges must also be conformally-invariant, it will mean that an element of $\mathfrak s$ must be given by an integral of a dimension-one current, modulo total derivatives. In addition, these currents ought to be gauge-invariant and covariant under coordinate transformations of $X$. 

\bigskip\noindent{\it A Relevant Digression}

Before we proceed to construct the dimension-one gauge-invariant and covariant currents of the perturbed free $\beta \gamma$ system, it would be useful to discuss the following issues. 

Firstly, notice from (\ref{tx psi}) and the fact that $\partial_k V^i_a = \partial_{\bar k} V^{\bar i} _a = 0$ in our abelian theory, that components of holomorphic and antiholomorphic vectors are invariant under gauge transformations. As a result, in each $U_e$ where one has picked the flat metric $\delta_{i \bar j}$, components of holomorphic and antiholomorphic one-forms are also invariant under gauge transformations.  

Secondly, recall from our discussion in $\S$2.3 that gauge-invariant forms are also $T$-equivariant forms.  However, since $T$ is abelian, these forms are simply $T$-invariant (see remark~14 of~\cite{libine}).  Therefore, ${{\cal L}_a} (d \gamma^i) = \{d, \iota_a\} (d \gamma^i) = 0$ (as $dV^i_a = 0$), and because $(d\gamma^i, \partial / \partial \gamma^k)  = \delta^i_k$, it would mean that basis holomorphic one-forms $d\gamma^i$\emph{ and} tangent vectors $\partial / \partial \gamma^i$ are gauge-invariant. 

Thirdly, note that since $\beta_k$ is a dimension-one elementary field, one  can define its charge as $Q_{\beta_k} = \oint \beta_k dz$. Then, from the $\beta$-$\gamma$ OPE in (\ref{beta-gamma OPE}), we have $[Q_{\beta_k}, \gamma^i] = \delta^i_k$. This implies that $Q_{\beta_k}$ acts as the $\partial / \partial \gamma^k$ operator. Hence, one can regard $\beta_k$ as a basis holomorphic tangent vector $\partial / \partial \gamma^k$ on $X$.\footnote{This has also been the mathematical viewpoint on the subject~\cite{GMS1}.} 

Last but not least, since the structure constants of our abelian gauge group $T$ vanish, one can see from (\ref{tx A})-(\ref{tx phi}) that $D_z \gamma^i = \partial_z \gamma^i - A^a_z V^i_a$ is gauge-invariant. With these facts in mind, we are now ready to construct our currents.

\bigskip\noindent{\it The Geometrical Symmetries}

Let us now describe the current which is associated with the geometrical symmetries. Firstly, if we have a holomorphic vector field $V$ on $X$ where $V = V^i (\gamma) {\partial \over {\partial \gamma^i}}$, one can construct a gauge-invariant and covariant  
dimension-one current $J_V=-V^i \beta_i$. The corresponding
conserved charge is then given by $K_V=\oint J_V dz  $. One can compute that the operator product expansion of $J_V$ with the elementary fields $A^a_z$ is trivial, but that with the elementary fields $\gamma$ is 
\be
J_V(z)\gamma^k(z')\sim {V^k(z')\over z-z'}.
\label{jv}
\ee 
Under the symmetry transformation generated by $K_V$, we have $\delta_\epsilon \gamma^k = i \epsilon [ K_V, \gamma^k ]$, where $\epsilon$ is an infinitesinal transformation parameter. Thus, from (\ref{jv}), we see that $K_V$ generates an infinitesimal diffeomorphism $\delta_\epsilon\gamma^k=i \epsilon V^k$ of $U$. In other words, $K_V$ generates the holomorphic diffeomorphisms of the target space $X$.  Therefore, $K_V$ spans the purely geometrical subset $\mathfrak v$ of $\mathfrak s$, as claimed. For finite diffeomorphisms, we would have a coordinate transformation ${\tilde \gamma}^k = g^k (\gamma)$, where each $g^k (\gamma)$ is a holomorphic function in the $\gamma^k$s. Since we are using the symmetries of the $\beta \gamma$ system to glue the local descriptions over the intersections $\{U_e \cap U_f\}$, on an arbitrary intersection $U_e \cap U_f$, $\gamma^k$ and ${\tilde \gamma}^k$ will be defined in $U_e$ and $U_f$, respectively.      

One can also compute the operator product expansion of $J_V$ with the elementary fields $\beta$ to be
\be
J_V(z)\beta_k(z')\sim -{\partial_k V^i \beta_i (z')\over z-z'}.
\label{jv-beta}
\ee
Under the symmetry transformation generated by $K_V$, we have $\delta_\epsilon \beta_k = i \epsilon [ K_V, \beta_k ]$, where $\epsilon$ is an infinitesinal transformation parameter. Thus, from (\ref{jv-beta}), we see that $K_V$ generates  an infinitesimal change $\delta_\epsilon\beta_k= - i \epsilon \partial_k V^i \beta_i$ of the elementary fields $\beta$.  For finite changes, we would have the transformation ${\tilde \beta}_i   =    D_i{}^k \beta_k$, where $D$ is an $ N \times N$ (where $N = {\rm dim}_{\mathbb C} X$) matrix such that $[D] = [\partial g]^{-1}$, i.e., $[D^{-1}]_i{}^k = \partial_i g^k$. This transformation is purely geometrical, and it just tells us how the basis holomorphic tangent vector $\beta_k$ transforms into $\tilde\beta_i$ as one goes from the coordinate system in $U_e$ to that in $U_f$. Again, this just affirms the fact that $K_V$ spans the  subset  $\mathfrak v$ of $\mathfrak s$.  

Clearly, the symmetry transformations generated by $K_V$ are \emph{all} there are to the purely geometrical transformations of the holomorphic coordinates $\gamma^k$ and basis tangent vectors $\beta_k$ of $X$.  Hence, one can use the geometrical symmetries generated by $K_V$ to glue the local sets $\{ U_e \}$ together on  intersections of small open sets to form the entire target space $X$. Note however, that ${\mathfrak v}$ is {\it not} a Lie subalgebra of $\mathfrak s$, but only a linear subspace. This is because $\mathfrak v$ does not close upon itself as a Lie algebra.  This leads to nontrivial consequences for $\mathfrak s$. In fact,  this property of $\mathfrak v$ is related to the physical anomalies of the underlying sigma model.  We will explain this shortly.

\bigskip\noindent{\it The Non-Geometrical Symmetries}

Let us now describe the current that is associated with the non-geometrical symmetries. Such a current, by virtue that its charge should not generate any transformation of the coordinates $\gamma^k$, ought to consist only of the $\gamma$ and $A$ fields. A little thought  would then reveal that if $B = B_i (\gamma) d{\gamma^i}$ is a holomorphic $(1,0)$-form on $X$, one can define a gauge-invariant and covariant dimension-one current as  $J_C=B_i D_z\gamma^i$, where the conserved charge is $K_C = \oint J_C dz$. Notice that one can also write $J_C =J_F + J_E$ and $K_C = K_F + K_E$, where $J_F = - B_iA^a_zV^i_a$, $J_E = B_i \partial_z \gamma^i$, $K_F = \oint J_F dz$ and $K_E = \oint J_E dz$.

Let us study $J_F$ and $K_F$ first. One can compute that the operator product expansion of $J_F$ with the elementary fields $\beta$ is
\be
J_F(z) \beta_k (z') \sim -{  {V^i_a\partial_k B_i A^a_z}(z')  \over z-z'}.
\label{jf1}
\ee 
Under the symmetry transformation generated by $K_F$, we have $\delta_\epsilon \beta_k = i \epsilon [ K_F, \beta_k ]$, where $\epsilon$ is an infinitesinal transformation parameter. Thus, from (\ref{jf1}), we see that $K_F$ generates an infinitesimal change  $\delta_\epsilon\beta_k= -i \epsilon  {V^i_a \partial_k B_i A^a_z}$ of the elementary fields $\beta$. Notice at this point that one can interpret $(B_iV^i_a)$ as a $\frak t$-valued holomorphic function $f_a$ on $X$. In turn, one can  regard $V^i_a \partial_k B_i = \partial_k (B_iV^i_a) = \partial_k f_a$ as the $F_{k, a}$-component of a $\frak t$-valued holomorphic one-form $F_a = F_{l, a} \, d\gamma^l = (\partial_l f_a) d \gamma^l$ on $X$. In addition, since $[V_a, V_b]^i =0$, the $V^i_a$'s are only defined up to scaling by a constant.  These last two observations together imply that  one can promote $\delta_\epsilon\beta_k$ to a finite change $\delta \beta_k = -F_{k, a} A^a_z$. In this sense, $K_F$ can be understood to span the purely non-geometrical subset $\frak f$ of $\frak s$.

Let us now study $J_E$ and $K_E$. One can compute that the operator product expansion of $J_E$ with the elementary fields $\beta$ is
\be
J_E(z) \beta_k (z') \sim {  {\partial_k B_i \partial_z \gamma^i}(z')  \over z-z'} - {\beta_k(z') \over {(z-z')^2}}.
\label{jf2}
\ee 
Under the symmetry transformation generated by $K_E$, we have $\delta_\epsilon \beta_k = i \epsilon [ K_E, \beta_k ]$, where $\epsilon$ is an infinitesinal transformation parameter. Thus, from (\ref{jf2}), we see that $K_E$ generates an infinitesimal change  $\delta_\epsilon\beta_k= i \epsilon \partial_k B_i \partial_z \gamma^i$ of the elementary fields $\beta$. Since one is free to rescale the $B_i$'s by a constant, one can promote  $\delta_\epsilon\beta_k$ to a finite change $\delta\beta_k = E_{ki}  \partial_z \gamma^i$, where $E = \partial B = E_{ki} \, d\gamma^k \wedge d\gamma^i$ is a holomorphic $(2,0)$-form on $X$ and $E_{ki} = \partial_k B_i$.

If $B$ is an exact form in $U_e$, it would mean that $B_i = \partial_i  H$ in $U_e$  for some locally-defined function $H$ that is holomorphic in $\gamma$. In such a case, $\oint J_E dz  = \oint \partial_i H \partial_z\gamma^i dz$. From the action (\ref{bcaction}), we have the equation of motion  $\partial_{\bar z} \gamma^i = 0$. Hence, $\oint J_E dz  = \oint \partial_i H d\gamma^i  = \oint dH  = 0$ by Stoke's theorem. In other words, the conserved charge $K_E$ vanishes if $B$ is exact in $U_e$ and vice-versa. Via Poincare's lemma, $B$ is locally exact if and only if $B$ is a closed form on $X$, i.e., $\partial B = \partial_i B_j - \partial_j B_i = 0$. Thus, for every nonvanishing holomorphic $(2,0)$-form $E = \partial B$ on $X$, we will have a nonvanishing conserved charge $K_E = \oint J_E dz $. At any rate, notice that $E$ is annihilated by $\partial$ since $\partial^2 = 0$. Moreover, as stated earlier, $B$ and therefore $E$ in $U_e$ are gauge-invariant; i.e., $E$ when restricted to $U_e$ is $T$-equivariant. As such, for \emph{every} local holomorphic section of the sheaf $\Omega^{2,cl}_{X, T}$ of $T$-equivariant  $\partial$-closed two-forms $E$ on $X$, we have a \emph{nonvanishing} conserved charge $K_E$. In this sense, $K_E$ can be understood to span the purely non-geometrical subset $\mathfrak e$ of $\mathfrak s$.

\bigskip\noindent{\it Local Field Transformations} 

A summary of how the different fields of the perturbed  free $\beta \gamma$ system on $U$ transform locally under its geometrical and  non-geometrical symmetries generated by $K_V$ and $K_C = K_F + K_E$, respectively, is as follows:      
\begin{eqnarray}
\label{auto1}
{\tilde \gamma}^i & = & g^i (\gamma) ,\\
\label{auto2}
\tilde A^a_z &= & A^a_z, \\
\label{autobeta}
{\tilde \beta}_i  & = &    D_i{}^k  \beta_k + \partial_z \gamma^j E_{i j} -   F_{i, a} A^a_z,  
\end{eqnarray} 
where $i,j,k = 1, 2, \dots, N={\textrm{dim}_{\mathbb C} X}$. As explained, $D$ and $E$ are $N \times N$ matrices such that $[D] = [\partial g]^{-1}$ and $[E] = \partial B$, i.e., $[D^{-1}]_i{}^k = \partial_i g^k$ and $[E]_{ij} = \partial_i B_j$; $F_{i,a}$ is the $i^{th}$-component of a $\frak t$-valued holomorphic one-form $F_a = F_{l, a} \, d\gamma^l = (\partial_l f_a)  d \gamma^l$ (where $f_a = V^k_a B_k$) on $X$. It can be verified that $\tilde \beta$, $\tilde \gamma$ and $\tilde A$ obey the correct OPE's amongst themselves. We thus conclude that the fields must undergo the above transformations (\ref{auto1})-(\ref{autobeta}) when we glue a local description (in a small open set) to another local description (in another small open set) over the mutual intersection of open sets using the automorphisms of the perturbed free $\beta\gamma$ system.

\bigskip\noindent{\it A Nontrivial Extension of Lie Algebras and Groups} 

Last but not least, let us now study the properties of the symmetry algebra $\mathfrak s$ of the perturbed free $\beta\gamma$ system on $U$. From the analysis thus far, we find that we can write $\mathfrak s = {\mathfrak c} \oplus {\mathfrak v}$ as a linear space, where $\mathfrak c = ({\mathfrak f}, {\mathfrak e})$. Note that $\mathfrak c$ is a trivial abelian subalgebra of $\mathfrak s$ because the commutator of $K_C$ with itself vanishes: the OPE of $J_C$ with itself is nonsingular since the current is constructed from $\gamma^k$, $\partial_z \gamma^k$ and $A^a_z$ only. Hence, $\mathfrak s$ can be expressed in an extension of Lie algebras as follows: 
\be
0 \to {\mathfrak c} \to {\mathfrak s} \to {\mathfrak v} \to 0.
\label{extension}
\ee   
In fact, (\ref{extension}) is an exact sequence of Lie algebras -- in other words, $\mathfrak c$ is ``forgotten'' when we project $\mathfrak s$ onto $\mathfrak v$. This is true because $[{\mathfrak v}, {\mathfrak c}] \subset {\mathfrak c}$. One can verify this claim as follows. 

The action of $\mathfrak v$ on $\mathfrak c$ can be ascertained from the $J_V (z) J_C(z')$ OPE
\begin{eqnarray}
\label{OPE1}
-V^i\beta_i(z)  \cdot B_j D_{z'}\gamma^j(z') & \sim & {1\over
z-z'} \left[V^i\partial_i B_k + B_i \partial_kV^i \right] \partial_{z'}\gamma^k  -  {1\over z-z'} [V^i \partial_i B_k] A^a_{z'} V_a^k  \nonumber \\
&&   + {1\over (z-z')^2}V^iB_i. 
\end{eqnarray} 
The commutator of $K_V$ with $K_C$, and thus $[{\mathfrak v}, {\mathfrak c}]$, is just the residue of the simple poles on the RHS of (\ref{OPE1}). The numerator of the first term on the RHS of (\ref{OPE1}), given by $V^i\partial_i B_k + B_i \partial_kV^i$, is equivalent to $({\cal L}_V(B))_k$, the $k^{th}$ component of the one-form that results from the action of a Lie derivative of $V$ on $B$; as such, this term takes values in $\frak e$. The numerator of the second term on the RHS of (\ref{OPE1}), given by $V^i \partial_i B_k$, can be interpreted as the $k^{th}$ component $P_k$ of a one-form $P = P_l d \gamma^l = (V^i \partial_i B_l) d \gamma^l$; as such, this term takes values in $\frak f$. Therefore, we find that $[{\mathfrak v}, {\mathfrak c}]$ takes values in ${\frak e}$ and $\frak f$; in other words, $[{\mathfrak v}, {\mathfrak c}] \subset \frak c$, as claimed. 


Let us now compute the commutator between two elements of $\mathfrak v$. To this end, let $\{ V, W \}$ be vector fields on $U$ that are holomorphic in $\gamma$.  Let $V$ and $W$ be associated with the currents $J_V(z)$ and $J_W (z')$, respectively.  The $J_V (z)  J_W (z')$ OPE is then computed to be
\be
\label{OPE2}
J_V (z) J_W (z')   \sim  - {{(V^i\partial_iW^j-W^i\partial_iV^j)\beta_j} \over z-z'} -{{(\partial_k\partial_jV^i)(\partial_iW^j\partial_{z'}\gamma^k)}\over z-z'}   - {{\partial_jV^i\partial_iW^j(z')} \over (z-z')^2}.
\ee  
The last term on the RHS of (\ref{OPE2}) is a double pole, i.e., it does not contribute to the commutator; thus, we shall ignore it. From the mathematical relation $[V,W]^j= ({\cal L}_V(W))^j = V^i\partial_iW^k-W^i\partial_iV^j$, we see that the first term on the RHS of (\ref{OPE2}) takes values in $\mathfrak v$. This term results from a single contraction of elementary fields in the OPE, and it corresponds to the Poisson bracket between $J_V$ and $J_W$ in the classical $\beta \gamma$ theory. On the other hand, the second term on the RHS of (\ref{OPE2}) takes values in $\mathfrak c$, and is the reason why  $[\mathfrak v, \mathfrak v] \nsubseteq {\mathfrak v}$. Note that this term results from a multiple contraction of elementary fields, just like the anomalies of conformal field theory. Hence, since $\mathfrak v$ does not close upon itself as a Lie algebra, $\mathfrak s$ is not a semi-direct product of $\mathfrak v$ and $\mathfrak c$. Consequently, the extension of Lie algebras in (\ref{extension}) is nontrivial. Is the nontriviality of the extension of Lie algebras of the symmetries of the $\beta\gamma$ system on $U$, then related to the physical anomalies of the underlying sigma model? Let us see. 

The exact sequence of Lie algebras in (\ref{extension}) will result in the following group extension when we exponentiate the elements of $\mathfrak s$:
\be
 1 \to \widetilde C \to \widetilde S \to \widetilde V \to 1.
\label{group extension}
\ee  
Here, $\widetilde S$ is the symmetry group of all admissible automorphisms of the $\beta\gamma$ system, $\widetilde C$ is the symmetry group of the non-geometrical automorphisms, and $\widetilde V$ is the symmetry group of the geometrical automorphisms. Just as in (\ref{extension}), (\ref{group extension}) is an exact sequence of groups, i.e., the kernel of the map $\widetilde S \to \widetilde V$ is given by $\widetilde C$. This means that the non-geometrical symmetries are ``forgotten'' when we project the full symmetries onto the geometrical symmetries. Since (\ref{group extension}) is derived from a nontrivial extension of Lie algebras in (\ref{extension}), it will be a nontrivial group extension. In fact, since $[{\frak v}, {\frak c}] \subset {\frak c}$ means that $\widetilde C$ is a $\widetilde V$-module, the nontriviality of  the group extension will be captured by a class in the second $\check{\rm C}$ech cohomology $H^2({\widetilde V}, {\widetilde C})$ (see Corollary~9.24 of~\cite{rothman}). Let us now ascertain what this class is. 

To this end, first recall that for every local holomorphic section of the sheaf $\Omega^{2,cl}_{X, T}$ of $T$-equivariant  $\partial$-closed two-forms $E = \partial B$ on $X$, we have a nonvanishing element of $\frak e$. Next, recall that under the \emph{only} nontrivial infinitesimal symmetry generated by $\frak f$, we have, in $U_e$, the variation $\delta_{\epsilon} \beta_k = -i \epsilon V^i_a \partial_k B_i A^a_z$. Since $B$ in $U_e$ is gauge-invariant (i.e.,~$T$-equivariant) and is therefore just $T$-invariant because $T$ is abelian (see remark~14 of~\cite{libine}), we have,  in $U_e$,  the condition $(\CL_{V_a} (B))_k = V^i_a \partial_i B_k = 0$. Thus, \emph{if} $E = \partial B = \partial_i B_k - \partial_k B_i = 0$, it would mean that  $V^i_a \partial_k B_i = 0$ and therefore, $\delta_{\epsilon} \beta_k = 0$. In other words, for \emph{every} local holomorphic section of the sheaf $\Omega^{2,cl}_{X, T}$, we have a \emph{nontrivial} symmetry generated by an element of $\frak f$.  Altogether, one can conclude that there is a one-to-one correspondence between nonvanishing local holomorphic sections of the sheaf $\Omega^{2,cl}_{X, T}$ and elements of $\widetilde C$. Hence, the class which captures the nontriviality of the group extension ought to come from $H^2({\widetilde V}, \Omega^{2,cl}_{X, T})$.

\bigskip\noindent{\it Relation to the Anomaly-Cancellation Condition of the Sigma Model}

Looking back at (\ref{OPE2}), one can see that the second term on the RHS -- which is solely responsible for the nontriviality of the group extension -- takes values \emph{purely} in $\frak e \subset {\frak c}$;  in particular, it will remain \emph{unchanged} as one varies  $A^a_z$. Therefore, one can conveniently set $A^a_z = 0$ via a gauge transformation (which can always be done since we are working locally on $\Sigma$) when computing the sought-after class in  $H^2({\widetilde V},\Omega^{2,cl}_{X, T})$. At $A^a_z = 0$, the automorphism relations given by (\ref{auto1}) and (\ref{autobeta}) coincide with the automorphism relations (4.4a)-(4.4b) in~\cite{GMS1} of the sheaf of CDO's. Moreover, as in $\S$2-6 of~\cite{GMS1}, our only nontrivial OPE (\ref{beta-gamma OPE}) is that of a free $\beta\gamma$ system. Consequently, one can repeat the computations that follow (4.4a)-(4.4b) in~\cite{GMS1} which eventually lead to Theorem~6.2 in \emph{loc.~cit.}, whence one would find that the class characterizing the nontriviality of the group extension is $p_1^T \in H^2(\widetilde V, \Omega^{2,cl}_{X, T})$, where $p_1^T$ is the $T$-equivariant universal first Pontryagin class.\footnote{Note that the standard $\check{\rm C}$ech cocycles which represent the universal classes in $H^2(\widetilde V, \Omega^{2,cl}_{X, T})$ take values in $\Omega^{2,cl}_{X, T}$. Hence, the universal classes -- like  $\Omega^{2,cl}_{X, T}$ -- will be $T$-equivariant.}  In turn, this implies  that there is \emph{no} obstruction to a globally-defined sheaf of chiral algebras $\widehat A$ if 
\be
{1\over 2}  p^T_1(X) = 0. 
 \label{group anomaly}
\ee 
This is exactly the anomaly-cancellation condition of the underlying $T$-gauged twisted $(0,2)$ sigma model computed in (\ref{an})! (We do not see the other anomaly involving $c_1(\Sigma)$ because we are working locally on $\Sigma$.) In hindsight, this `coincidence' should not be entirely surprising: note that a physically valid, anomaly-free sigma model must be defined over \emph{all} of $X$ (and $\Sigma$), and since a nonzero $p^T_1(X)$ captures the obstruction to gluing the local descriptions together to arrive at a global description, it would mean that the sigma model -- which is described locally by the perturbed free $\beta\gamma$ system on $U$ -- cannot be globally-defined over all of $X$ unless (\ref{group anomaly}) holds. Thus, the nontriviality of the extension of Lie algebras of the symmetries of the $\beta\gamma$ system on $U$, is indeed related to the physical anomaly of the underlying sigma model.

\newsubsection{Gluing the Local Descriptions Together}

Now, we will describe explicitly, how one can glue the local descriptions together using the automorphisms of the perturbed free $\beta\gamma$ system on $U$ to obtain a globally-defined sheaf of chiral algebras. In the process, we will see how the cohomology class in (\ref{group anomaly}) emerges as an obstruction to gluing the locally-defined sheaves of  chiral algebras globally on $X$. Moreover, we can also obtain the other anomaly in (\ref{an}) -- which is not captured in (\ref{group anomaly}) -- when we consider gluing the sheaves of chiral algebras globally over $X$ $\it{and}$ $\Sigma$. In addition, we will see that a modulus of the resulting sheaf emerges as a $\check{ \textrm C}$ech cohomology class generated by a relevant $\check{ \textrm C}$ech cocycle.

As a start, let's take a suitable collection of small open patches $U_e\subset \Bbb{C}^n$, where $n=\textrm{dim}_{\mathbb C}X$. Next, consider the corresponding set of  patches $\{U_e \}$. The idea is to glue these patches together to arrive at a good cover of  $X$. On every $U_e$ is a perturbed free $\beta\gamma$ system which defines a sheaf $\widehat{\cal A}$ of chiral algebras. In gluing these free conformal field theories together, we will obtain a globally-defined sheaf of chiral algebras. 

It will be convenient for us to first describe how we can geometrically glue the set of patches $\{U_e\}$ together to form $X$. For each $e,f$, let us pick a patch $U_{ef}\subset U_{e}$, and likewise another patch $U_{fe}\subset U_f$. Let us define a geometrical symmetry ${\hat v}_{ef}$ (characterized by holomorphic diffeomorphisms on $U$) between these patches as  
\be
{\hat v}_{ef}: U_{ef}\cong U_{fe}.
\label{g symmetry}
\ee
Note that ${\hat v}$ can be viewed as a geometrical gluing operator corresponding to an element of the geometrical symmetry group $\widetilde V$. From the above definition, we see that ${\hat v}_{fe}={\hat v}_{ef}^{-1}$. We want to identify an arbitrary point $P \in  U_{ef}$ with an arbitrary point $Q \in U_{fe}$, if $Q={\hat v}_{ef}(P)$. This identification will be consistent if for any $U_e$, $U_f$, and $U_g$, we have
\be
{\hat v}_{ge} {\hat v}_{fg} {\hat v}_{ef}=1
\label{triple}
\ee
in any triple intersection $U_{efg}$ over which all the maps ${\hat v}_{ge}$,  ${\hat v}_{fg}$ and ${\hat v}_{ef}$ are defined. The relation in (\ref{triple}) tells us that the different pieces $U_e$ can be glued together via the set of maps $\{{\hat v}_{ef} \}$ to make $X$. The complex structure moduli of $X$ will then manifest as parameters in the ${\hat v}_{ef}$'s. 

Say we now have a sheaf of chiral algebras on each $U_e$; to get a sheaf of chiral
algebras on $X$, we will need to glue them together on overlaps. The gluing must be done using the automorphisms of the conformal field theories. Thus, for each pair $U_e$ and $U_f$, we select a conformal field theory symmetry $\hat s_{ef}$ that maps the perturbed free $\beta\gamma$ system on $U_e \cap U_{ef}$, to the perturbed free $\beta\gamma$ system on $U_f \cap U_{fe}$. If 
\be
\hat s_{ge}\hat s_{fg}\hat s_{ef}=1,
\label{consistent}
\ee
i.e., if the gluing is consistent, we get a globally-defined sheaf of chiral algebras. Note that $\hat s$ can be viewed as a gluing operator corresponding to an element of the full symmetry group $\widetilde S$. As usual, we have ${\hat s}_{fe} = {\hat s}^{-1}_{ef}$. Moreover, recall at this point that from the exact sequence of groups in (\ref{group extension}), we have a map $\widetilde S \to \widetilde V$ which ``forgets'' the non-geometrical symmetry group $\widetilde C \subset \widetilde S$. As such, for $\it{any}$ arbitrary set of $\hat s$'s which obey (\ref{consistent}), the geometrical condition (\ref{triple}) will be automatically satisfied, regardless of  what the non-geometrical gluing operator $\hat c$ corresponding to an element of $\widetilde C$, is. Hence, every possible way to glue the conformal field theories together via $\hat s$, determines the same way to geometrically glue the set of patches $\{U_e \}$ together to form $X$ over which the resulting conformal field theory is defined.      

The above discussion translates to the fact that for a given set of ${\hat v} _{ef}$'s which obey (\ref{triple}), the corresponding set of  $\hat s_{ef}$'s which obey (\ref{consistent}) is not uniquely determined: for each $U_{ef}$, we can still pick an element ${\cal C}_{ef}\in H^0(U_{ef},\Omega^{2,cl}_{T})$ which represents an element of $\mathfrak{c}$ (as discussed in $\S$4.5), so that $\textrm {exp}({\cal C}_{ef})$ represents an element of $\widetilde C$; one can then transform $\hat s_{ef}\to {\hat
s}'_{ef}=  \textrm{exp}({\cal C}_{ef})\hat s_{ef}$, where ${\hat s}'_{ef}$ is another physically valid gluing operator. The condition that $\hat s'$ obeys the gluing identity (\ref{consistent}), i.e., $\hat s'_{ge}\hat s'_{fg}\hat s'_{ef}=1$, is that in each triple
intersection $U_{efg}$, we should have
\be
{\cal C}_{ge}+ {\cal C}_{fg} + {\cal C}_{ef}=0. 
\label{cocycle}
\ee
From ${\hat s}'_{fe} = ({\hat s}'_{ef})^{-1}$, we have ${\cal C}_{ef} = -{\cal C}_{fe} $. Moreover, ${\widetilde {\cal C}}_{ef} \sim {\cal C}_{ef} + {\cal S}_e - {\cal S}_f$ for some $\cal S$, in the sense that the $\widetilde{\cal C}$'s will obey (\ref{cocycle}) as well. In other words, the $\cal C$'s in (\ref{cocycle}) must define an element of the $\check{ \textrm C}$ech cohomology group $H^1(X,\Omega^{2,cl}_{X,T})$.  Also, in projecting from ${\hat s}_{ef}'$ to the geometrical gluing operator ${\hat v}_{ef}$, ${\textrm{exp}}({\cal C}_{ef})$  is ``forgotten''. Therefore, in going from $\hat s$ to ${\hat s}'$, the symmetry $\hat v$, and consequently $X$, remain unchanged.  Now, let us use a specific $\hat s$ operator to define the specific symmetries of a perturbed free $\beta\gamma$ system, which in turn will define a unique sheaf of chiral algebras. In this sense, with any sheaf and an element ${\cal C} \in
H^1(X,\Omega^{2,cl}_{X, T})$, one can define a new sheaf by going from $\hat s \to
\exp({\cal C}) \hat s$. So, via the action of $H^1(X,\Omega^{2,cl}_{X,T})$, we get a family of sheaves of chiral algebras on $X$. Hence, the modulus of the sheaf of chiral algebras is represented by a class in $H^1(X,\Omega^{2,cl}_{X,T})$, in agreement with our analysis in $\S$2.3.

\bigskip\noindent{\it The Anomaly}

We now move on to discuss the case when there is an obstruction to the gluing. Essentially, the obstruction occurs when (\ref{consistent}) is not satisfied by $\hat s$. In such a case, one generally has, on triple intersections $U_{efg}$, the following relation
\be
\hat s_{ge} \hat s_{fg}\hat s_{ef}= \textrm{exp}({\cal C}_{efg})
\label{obstruction sheaf}
\ee
 for some ${\cal C}_{efg}\in H^0(U_{efg},\Omega^{2,cl}_T)$. The reason for (\ref{obstruction sheaf}) is as follows. First, note that the LHS of (\ref{obstruction sheaf}) projects purely to the group of geometrical symmetries associated with $\hat v$. If $X$ is to exist mathematically, there will be no obstruction to its construction, i.e., the LHS of (\ref{obstruction sheaf}) will map to the identity under the projection. Hence, the RHS of (\ref{obstruction sheaf}) must represent an element of the abelian group $\widetilde C$ (generated by $\mathfrak c$) whose action on the coordinates $\gamma^i$ of the $U_e$'s is necessarily trivial.     

 Recall that the choice of ${\hat s}_{ef}$ was not unique. If we transform $\hat s_{ef}\to \exp({\cal C}_{ef})\hat s_{ef}$ via a (non-geometrical) symmetry of the system, we get 
\be
{\cal C}_{efg}\to {\cal C}'_{efg} = {\cal C}_{efg}+ {\cal C}_{ge}+ {\cal C}_{fg}+ {\cal C}_{ef}.
\label{cocycle 3}
\ee
 If one can choose the ${\cal C}_{ef}$'s to set $\it{all}$ ${\cal C}_{efg}'=0$, then there is no obstruction to gluing and one can obtain a globally-defined sheaf of chiral algebras. 

In any case, in quadruple overlaps $U_e\cap U_f\cap U_g\cap U_h$,
the $\cal C$'s obey
\be
{\cal C}_{efg}- {\cal C}_{fgh}+ {\cal C}_{ghe} - {\cal C}_{hef} =0.
\label{cocycle 4}
\ee
Together with the equivalence relation (\ref{cocycle 3}), this means that the
$\cal C$'s in (\ref{cocycle 4}) must define an element of the $\check{ \textrm C}$ech cohomology group $H^2(X, \Omega^{2,cl}_{X,T})$. In other words, the obstruction to gluing the locally-defined sheaves of chiral algebras is captured by a nonvanishing cohomology class  $H^2(X, \Omega^{2,cl}_{X,T})$. As discussed at the end of $\S$4.5, this class can be represented by $ p^T_1(X)$. Thus, we have obtained an interpretation of the anomaly in the abelian twisted $(0,2)$ sigma model in terms of an obstruction to a global definition of the sheaf of chiral algebras derived from a perturbed free $\beta\gamma$ system that describes the model locally on $X$.

\bigskip\noindent{\it The Other Anomaly} 

According to our analysis in $\S$3, the abelian twisted $(0,2)$ sigma model ought to possess two anomalies; one involving $p^T_1(X)$, and the other involving $c_1(\Sigma)c^T_1(X)$. We have already seen how the first anomaly arises from the $\check{ \textrm C}$ech perspective. How then can one observe the second anomaly in the current framework? 

We have hitherto constructed a sheaf of chiral algebras globally on $X$ but only locally on the worldsheet $\Sigma$. Since the canonical bundle $K$ can be trivialized when one works locally on $\Sigma$, one would not see the second anomaly involving $c_1(\Sigma)c^T_1(X)$. This was explained in $\S$3.  In any case, note that the perturbed free $\beta\gamma$ system is conformally invariant -- in other words, it can be defined globally on any Riemann surface $\Sigma$. But, notice that the anomaly which we are looking for is given by $c_1(\Sigma) c^T_1(X)$. Thus,  it will vanish even if we use a perturbed free $\beta\gamma$ system that can be globally-defined on $\Sigma$,  if we continue to work locally on $X$. Therefore, the only way to see the second anomaly is to work globally on both $X$ $\it{and}$   $\Sigma$. (In fact, recall that the underlying sigma model is to be defined on \emph{all }of $\Sigma$ and $X$.) We shall describe how to do this now.  

Let us cover $\Sigma$ and $X$ with small open sets $\{P_\tau \}$ and $\{U_e \}$, respectively. This will allow us to cover $X \times \Sigma$ with open sets $W_{e \tau} = U_e \times P_{\tau}$.  On each $P_\tau$,
define a perturbed free $\beta\gamma$ system with target $U_e$. In other words, on each open set $W_{e \tau}$, define a perturbed free $\beta\gamma$ system which in turn furnishes us with a sheaf of chiral algebras. What we want to do is to glue the sheaves of chiral algebras on the $W_{e \tau}$'s together on overlaps, to get a globally-defined sheaf of chiral algebras that is valid over all of $X$ and $\Sigma$. As before, the gluing ought to be done using the admissible automorphisms of the perturbed free $\beta\gamma$ system. 

Recall from $\S$4.5 that the admissible automorphisms are given by the symmetry group $\widetilde S$. Note that the set of geometrical symmetries $\widetilde V \subset \widetilde S$ considered in $\S$4.5 can be extended to include holomorphic diffeomorphisms of the worldsheet $\Sigma$ -- as mentioned above, the perturbed free $\beta\gamma$ system, being conformally invariant, is also invariant under arbitrary holomorphic reparameterizations of the coordinates on $\Sigma$. Previously in $\S$4.5, there was no requirement to consider and exploit this additional geometrical symmetry along $\Sigma$ in gluing the local descriptions together simply because we were working locally on $\Sigma$. There, gluing of the local descriptions at the purely geometrical level was carried out using $\widetilde V$, where $\widetilde V$ consists of the group of holomorphic diffeomorphisms of $X$. Now that we want to work globally on $\Sigma$ as well, one will need to use the symmetry of the free conformal field theory under holomorphic diffeomorphisms of $\Sigma$ to glue the $P_{\tau}$'s together to form $\Sigma$. In other words, gluing of the local descriptions at the purely geometrical level must now be carried out using the geometrical symmetry group ${\widetilde V}'$, where ${\widetilde V}'$ consists of the group of holomorphic diffeomorphisms on $\Sigma$ $\it{and}$ $X$. Let the conformal field theory gluing map from $W_{e\tau}$ to $W_{f\nu}$ be given by    ${\hat s}_{e\tau,f\nu}$.  Let the corresponding geometrical and non-geometrical gluing maps from $W_{e\tau}$ to $W_{f\nu}$ be given by ${\hat v}'_{e\tau,f\nu}$ and ${\hat c}'_{e\tau,f\nu}$, respectively. Since we have a sensible concept of a holomorphic map $\gamma : \Sigma \to X$, and since $X$ and $\Sigma$ are pre-defined to exist mathematically, there is no obstruction to gluing at the purely geometrical level, i.e., 
\be
{\hat v}'_{g\sigma,e\tau}{\hat v}'_{f\nu,g\sigma} {\hat v}'_{e\tau,f\nu} =1
\ee
in triple intersections. There will be no obstruction to gluing at \emph{all} levels if one has the relation
\be
{\hat s}_{g\sigma,e\tau} {\hat s}_{f\nu,g\sigma} {\hat s}_{e\tau,f\nu} =1. 
\label{consistency}
\ee
However, (\ref{consistency}) may not always be satisfied. Similar to our previous arguments concerning the anomaly $p^T_1(X) \in H^2(X, \Omega^{2,cl}_{X,T})$, since one has a map ${\hat s}_{e\tau,f\nu} \to {\hat v}'_{e\tau,f\nu}$ in which ${\hat c}'_{e\tau,f\nu}$ is ``forgotten'', in general, we will  have
\be
{\hat s}_{g\sigma,e\tau} {\hat s}_{f\nu,g\sigma} {\hat s}_{e\tau,f\nu} = \textrm{exp}({\cal C}_{e \tau f \nu g \sigma}),   
\ee                    
where the ${\cal C}_{e \tau  f\nu  g\sigma}$'s on any triple overlap defines a class in the two-dimensional $\check{ \textrm C}$ech cohomology group $H^2(X \times {\Sigma}, {\mathscr S})$. $\mathscr S$ is a sheaf associated with the non-geometrical symmetries of the perturbed free $\beta\gamma$ system. Being non-geometrical in nature, these symmetries ought to act trivially on the $\gamma^i$  and $z$ coordinates of $X \times \Sigma$.  Let us now determine what $\mathscr S$ is. 

Earlier on in our discussion, when we worked locally on $\Sigma$ but globally on $X$, we constructed a gauge-invariant and covariant dimension-one current $J_C$ from a holomorphic $(1,0)$-form $B$ on $X$; its conformally-invariant conserved charge $K_C = \oint J_C dz$ was shown to generate the non-geometrical symmetries of the perturbed free $\beta\gamma$ conformal field theory. Therefore, if one wants to work globally on both $\Sigma$ and $X$, one would need to construct an analogous gauge-invariant and covariant dimension-one current $J_{C'}$ from a $(1,0)$-form $B'$ on $X \times \Sigma$; its conformally-invariant conserved charge $K_{C'} = \oint J_{C'} dz$ would then generate the non-geometrical symmetries in this particular extended case. To this end, first note that since the dimension-one current $J_{C'}$ should have nonsingular OPE's with the $\gamma$ fields, it can only depend linearly on $\partial_z \gamma$ or $A^a_z$, and be holomorphic in  $\gamma$ and $z$. Next, note that gauge transformations do not act on the coordinates $z$ of $\Sigma$. Last but not least, note that a holomorphic function which is obtained by contracting a holomorphic vector with a holomorphic one-form is gauge-invariant on $U_e$.\footnote{Recall from our discussion in $\S$4.5  that holomorphic vectors and one-forms  on $U_e$ are automatically and separately gauge-invariant under $T$.} Bearing these points in mind, we find that $J_{C'}$ ought to be given by
\be
J_{C'} = {B'_i}(\gamma, z) D_z \gamma^i + B_{z}(\gamma, z).
\ee                       
Here, ${B'_i}$ and $B_{z}$ are components of a holomorphic $(1,0)$-form $B' = {B'_i} d \gamma^i + B_{z} dz$ on $X \times \Sigma$, where ${B'_i}$ and $B_{z}$ have scaling dimension zero and one, respectively. The $\gamma$-dependence of $B_{z}(\gamma, z)$ just reflects the fact that it transforms as a holomorphic function of the above-mentioned kind on $X$ whilst being a component of a $(1,0)$-form on $\Sigma$. The $z$-dependence of ${B'_i}(\gamma, z)$ just reflects the fact that it transforms as a holomorphic function of the above-mentioned kind on $\Sigma$ whilst being a component of a $(1,0)$-form on $X$.

Notice that one can also write $J_{C'} = J_{E'} + J_{F'} + J_{B}$ and $K_{C'} = K_{E'} + K_{F'} + K_{B}$, whereby $J_{E'} = {B'_i} \partial_z \gamma^i$, $J_{F'} = - {B'_i}A^a_zV^i_a$, $J_{B} = B_z$, $K_{E'} = \oint J_{E'} dz$, $K_{F'} = \oint J_{F'} dz$ and $K_B = \oint J_B dz$.  If $B'$ is exact, i.e, $B'=\partial H'$ for some local function $H' (\gamma,z)$ on $X\times \Sigma$ holomorphic in $\gamma$ and $z$, we will have ${B'_i} = \partial_i H'$ and $B_{z} = \partial_z H'$. As a result, the conserved charge $K_{E'} + K_B = \oint (J_{E'} + J_B) dz = \oint (\partial_i H') d\gamma^i + (\partial_z H') dz = \oint dH' = 0$ by Stoke's theorem. In other words, the conserved charge $K_{E'} + K_B$ vanishes if $B'$ is exact in $W_{e\tau}$ and vice-versa. Via Poincare's lemma, $B'$ is locally exact if and only if $B'$ is a closed form on $X \times \Sigma$. Thus, for every nonvanishing holomorphic $(2,0)$-form $C' = \partial B'$ on $X \times \Sigma$, we will have a nonvanishing conserved charge $K_{E'} + K_B = \oint (J_{E'} + J_B) dz$. At any rate, notice that $C'$ is annihilated by $\partial$ since $\partial^2 = 0$. Moreover, $B'$ and therefore $C'$ in  $W_{e\tau}$ are gauge-invariant; i.e., $C'$ when restricted to  $W_{e\tau}$ is $T$-equivariant. Hence, for \emph{every} local holomorphic section of the sheaf $\Omega^{2,cl}_{X \times \Sigma, T}$ of $T$-equivariant  $\partial$-closed two-forms $C'$ on $X \times \Sigma$, we have a \emph{nonvanishing} conserved charge $K_{E'} + K_B$. 

So what about $J_{F'}$ and $K_{F'}$? Well, note that under the \emph{only} nontrivial infinitesimal symmetry generated by $K_{F'}$, we have,  in $W_{e \tau}$, the variation  $\delta_{\epsilon} \beta_k = -i \epsilon V^i_a \partial_k {B'_i} A^a_z$. According to our discussion in $\S$4.5, since $B'$ in $W_{e \tau}$ is gauge-invariant (i.e.,~$T$-equivariant) and is therefore just $T$-invariant because $T$ is abelian (see remark~14 of~\cite{libine}), we have, in $W_{e \tau}$, the conditions $(\CL_{V_a} (B'))_k = V^i_a \partial_i {B'_k} = 0$ and $(\CL_{V_a} (B'))_z = V^i_a \partial_i B_z = 0$. Thus, \emph{if} $C' = \partial B' =  0$ -- which implies that $\partial_i {B'_k}- \partial_k {B'_i} = 0$ -- it would mean that  $V^i_a \partial_k {B'_i} = 0$ and hence, $\delta_{\epsilon} \beta_k = 0$. In other words, for \emph{every} local holomorphic section of the sheaf $\Omega^{2,cl}_{X\times \Sigma, T}$, we have a \emph{nontrivial} symmetry generated by an element of $K_{F'}$.

Therefore, we find that the sheaf $\mathscr S$ associated with the non-geometrical symmetries that act trivially on the coordinates $(\gamma, z)$ can be identified as $\Omega^{2,cl}_{X\times\Sigma, T}$. Hence, the obstruction to a globally-defined sheaf of chiral algebras -- with target space $X$ and defined on all of $\Sigma$ -- will be captured by a class in the $\check{ \textrm C}$ech cohomology group $H^2(X\times \Sigma, \Omega^{2,cl}_{X\times\Sigma, T})$. As such, the physical anomalies of the underlying sigma model ought to be captured by the $T$-equivariant cohomology classes which take values in $H^2(X\times \Sigma, \Omega^{2,cl}_{X\times\Sigma, T})$. 

Note at this juncture that one can, on $X \times \Sigma$,  write $C' =  {C'_{ij}} d\gamma^i \wedge d \gamma^j + {C'_{i z}} d\gamma^i \wedge dz$, where ${C'_{ij}} (\gamma, z)  = {1\over 2} (\partial_i {B'_j} - \partial_j {B'_i})$ and ${C'_{i z}} (\gamma, z)  = (\partial_i B_z - \partial_z {B'_i})$. Since gauge transformations \emph{never} act on the $z$ coordinates, one can, according to the explicit form of $C'$ given, write $\Omega^{2,cl}_{X \times\Sigma, T} = (\Omega^{2,cl}_{X,T} \otimes {\cal O}_\Sigma) \oplus (\Omega^{1,cl}_{X,T} \otimes \Omega^{1,cl}_\Sigma)$, where ${\cal O}_{\Sigma}$ is a sheaf of holomorphic functions in $z$ only on $\Sigma$, and $\Omega^{1,cl}_\Sigma$ is the sheaf of holomorphic one-forms in $z$ only on $\Sigma$ that are annihilated by the operator $dz \wedge \partial / \partial z$. Similarly, $\Omega^{i,cl}_{X,T}$ is the sheaf of $T$-equivariant $i$-forms which are holomorphic in $\gamma$ only on $X$ that are annihilated by the operator $d \gamma^{i} \wedge \partial / \partial \gamma^{i}$. Thus, on a compact Riemann surface $\Sigma$, where the only globally-defined holomorphic functions in $z$ are equivalent to constants, i.e.,  $H^0(\Sigma,{\cal O}_\Sigma)\cong \mathbb{C}$,  we have the expansion 
\be
H^2(X\times \Sigma,\Omega^{2,cl}_{X\times
\Sigma, T})= H^2 (X,\Omega^{2,cl}_{X,T}) \oplus
( H^1(X,\Omega^{1,cl}_{X,T}) \otimes
H^1(\Sigma,\Omega^{1,cl}_{\Sigma}) )\oplus \dots.
\label{sheaf anomaly}
\ee

As $c_1(\Sigma) \in H^1(\Sigma, \Omega^{1,cl}_{\Sigma})$ and $c^T_1(X) \in H^1(X, \Omega^{1,cl}_{X,T})$, the two physical anomalies $p^T_1(X)$ and $ c_1(\Sigma) c^T_1(X)$ take values in the first and second term on the RHS of (\ref{sheaf anomaly}), respectively.\footnote{If $M$ is a K\"ahler manifold, $p_1(M)$ would be a $(2,2)$-form that is annihilated by both $\partial$ and $\bar \partial$; this just reflects the fact that  $p_1(M)$ represents an element of $H^2(M, \Omega^{2,cl}_{M})$. One can also show the latter statement to be true for any complex manifold $X$. To this end, choose any connection on the holomorphic tangent bundle $TX$ where its $(0,1)$ part is the natural $\bar\partial$ operator of this bundle; since $\bar\partial^2=0$, its curvature $c_1(TX)$ (abbreviated in this paper as $c_1(X)$), would be a two-form of type $(2,0)\oplus (1,1)$. In turn, this means that $c_k(X)$ for $k \geq 0$ would be a $2k$-form of type $(k,k) \oplus (k+1, k-1) \oplus \dots (2k,0)$, which therefore implies that it must represent an element of $H^k (X, \Omega^{k,cl}_{X})$ -- indeed, like the sheaf $\Omega^{k,cl}_{X}$ which generates $H^k (X, \Omega^{k,cl}_{X})$, $c_k(X)$ will be annihilated by both $\partial$ and $\bar \partial$. Therefore, $c_1(X)$ must represent an element of $H^1(X,\Omega^{1,cl}_X)$, while $p_1(X) = c^2_1(X) - 2c_2(X)$  must represent an element of $H^2(X,\Omega^{2,cl}_X)$.  Similarly, $c^T_1(X)$ and $p_1^T(X)$ ought to represent elements of $H^1(X,\Omega^{1,cl}_{X,T})$ and  $H^2(X,\Omega^{2,cl}_{X,T})$, respectively.}  Note that the terms on the RHS of (\ref{sheaf anomaly}) must independently vanish for $H^2(X\times \Sigma, \Omega^{2,cl}_{X\times\Sigma, T})$ to be zero. In other words, we have obtained a consistent, alternative interpretation of the physical anomalies  which arise due to a nontriviality of the determinant line bundles over the configuration space of the sigma model that are associated with the Dirac operators of the underlying Lagrangian,  in terms of an obstruction to the gluing of sheaves of chiral algebras. 

By extending the arguments surrounding (\ref{cocycle}) to the present context,  we find that for a vanishing  anomaly, (apart from the geometrical moduli encoded in the complex structure of $X$), the moduli of the globally-defined sheaf of chiral algebras on $\Sigma$, with target space $X$, must be parameterized by classes in $H^1(X\times\Sigma,\Omega^{2,cl}_{X\times\Sigma, T})$.

\newsubsection{The Conformal Anomaly}

To end this section, we shall now present an illuminating application of our discussion on the sheaves of chiral algebras which has hitherto been somewhat abstract. In the process, we will be able to obtain a solely physical interpretation of a purely mathematical result, and vice-versa.

From (\ref{Tzz}), we see that the holomorphic stress tensor $T(z) \sim T_{zz}$ of the abelian twisted $(0,2)$ sigma model lacks the $\psi^{\bar i}$ fields. This means that it is an operator with $q = 0$. Hence, from the $Q$-$\check{ \textrm C}$ech cohomology dictionary established in $\S$4.3, if $T(z)$ is to be nontrivial in $Q$-cohomology, such that the model and its chiral algebra are conformally-invariant, it will be given by an element of $H^0(X, \widehat{\cal A})$ -- a global section of the sheaf of chiral algebras $\widehat{\cal A}$. Recall that the local sections of $\widehat{\cal A}$ are furnished by the physical operators in the chiral algebra of the perturbed linear $\beta\gamma$ system. Since the perturbed linear $\beta\gamma$ system  describes a local version of the underlying abelian twisted $(0,2)$ sigma model, one can write the local holomorphic stress tensor of the model as the local holomorphic stress tensor of the perturbed linear $\beta\gamma$ system, which in turn is given by     
\be
{\cal T} (z) =  - : \beta_i \partial_z \gamma^i: \hspace{-0.1cm}(z).
\ee 
(See $\S$4.4).  Under an automorphism of the $\beta\gamma$ system, ${\cal T}(z)$ will become 
\be
{\widetilde {\cal T}} (z) = - :{\tilde\beta}_i \partial_z {\tilde \gamma}^i: \hspace{-0.1cm}(z),
\label{tildefraktz}
\ee 
where the fields $\tilde\beta$ and $\tilde \gamma$ are defined in the automorphism relations of (\ref{auto1})-(\ref{autobeta}). It is clear that on an overlap $U_e \cap U_f$ in $X$, ${{\cal T}(z)}$ will be regular in $U_e$ while ${\widetilde{\cal T}(z)}$ will be regular in $U_f$. Note that both ${\cal T}(z)$ and ${\widetilde{\cal T}(z)}$ are at least local sections of $\widehat {\cal A}$. And, if there is no obstruction to ${\cal T}(z)$ or ${\widetilde{\cal T}(z)}$ being a global section of $\widehat {\cal A}$, it will mean that $T(z)$ is nontrivial in $Q$-cohomology, i.e., $T(z) \neq \{Q, \dots \}$ and $[Q, T(z) ] =0$, and the abelian twisted $(0,2)$ sigma model will be conformally-invariant. For ${\cal T}(z)$ or ${\widetilde{\cal T}(z)}$  to be a global section of $\widehat{\cal A}$, it must be true that ${\cal T}(z) = {\widetilde{\cal T}(z)}$ on any overlap $U_e \cap U_f$ in $X$. Let us examine this further with an example. 

\bigskip\noindent{\it An Example and the One-Loop Beta Function in Terms of Holomorphic Data}

For ease of illustration, let us consider $X = \mathbb{CP}^1$. This example is nonanomalous as $p^T_1(X) = 0$; it also has a \emph{nonvanishing} Ricci tensor. Since $\mathbb{CP}^1$ can be considered as the complex $\gamma$-plane plus a point at infinity, we can cover it with two open sets, $U_1$ and $U_2$, where $U_1$ is the complex $\gamma$-plane, $U_2$ is the complex $\tilde\gamma$-plane, and $\tilde \gamma = 1 /\gamma$. Thus, we can write the (normal ordered) stress tensors in $U_1$ and $U_2$, respectively, as
\be
{{\cal T} (z)} = - \textrm{lim}_{z' \to z}  \big( \beta (z') \partial_z \gamma (z) - \beta(z') \cdot  \partial_z \gamma (z) \big)
\ee 
and 
\be
{\widetilde {\cal T}} (z) =  - \textrm{lim}_{z' \to z} \left(  \tilde\beta (z') \partial_z \tilde\gamma (z) - \tilde\beta(z') \cdot  \partial_z \tilde\gamma (z)       \right).
\ee
By substituting the definitions of $\tilde\beta$ and $\tilde \gamma$ from (\ref{auto1})-(\ref{autobeta}) into $\widetilde{\cal T}(z)$,   a short computation will give\footnote{Note that in our computation, we have, for convenience, chosen (i) a point $z'$ such that $\gamma(z') \to 0$, (ii)  the arbitrary local (1,0)-form $B(\gamma)d\gamma$ on $\mathbb{CP}^1$ (associated with the current $J_E$ of  $\S$4.5) to be such that $B(\gamma) = - \gamma$.}   
\be
{\widetilde{\cal T}(z)} - {\cal T}(z)  = \partial_z \left ( {{\partial_z \gamma} \over{\gamma}} - {V_a A^a_z \over {\gamma}} \right)(z). 
\label{fraktzanomaly}         
\ee
The only consistent way to modify $\cal T \to {\cal T}'$ and $\widetilde {\cal T} \to \widetilde {\cal T}'$ so that $ {\cal T}' =\widetilde {\cal T}'$ over $U_1\cap U_2$ while preserving the stress tensor OPE's such that ${\cal T}' \cdot \gamma = {\cal T} \cdot \gamma$ and $\widetilde {\cal T}' \cdot \gamma = \widetilde {\cal T} \cdot \gamma$, is to add to  ${\cal T}$ and  $\widetilde {\cal T}$ dimension-two terms involving $\gamma$, $A^a_z$,  $\tilde \gamma$ or ${\tilde A}^a_z$ only. The RHS of (\ref{fraktzanomaly}) is invariant under $\gamma \to \zeta \gamma$ and $A^a_z \to \zeta A^a_z$, where $\zeta \in \mathbb C^*$. Therefore, the modification ought to preserve this invariance too; i.e., the dimension-two terms we add must also be invariant under  $\gamma \to \zeta \gamma$ and $A^a_z \to \zeta A^a_z$. Such terms will be given by $\partial^2_z \gamma / \gamma$, $(\partial_z \gamma)^2 / \gamma^2$, $(V_a\partial_z A^a_z) / \gamma$ and $(V_aA^a_z \partial_z \gamma) / \gamma^2$. To cancel the quantity on the RHS of (\ref{fraktzanomaly}), one will need to add a linear combination of \emph{all }such terms to ${\widetilde{\cal T}(z)} - {\cal T}(z)$ on the LHS of (\ref{fraktzanomaly}). Because ${(\partial_z \gamma)^2 / \gamma^2} = {(\partial_z \tilde \gamma)^2 / \tilde \gamma^2}$ and $\partial^2_z \gamma / \gamma = - \partial^2_z \tilde \gamma / \tilde \gamma + 2(\partial_z \gamma)^2 / {\tilde \gamma}^2$, this linear combination would have a pole at \emph{both} $\gamma=0$ and $\tilde\gamma=0$. Thus, it cannot be used to redefine $\cal T$ and/or $\widetilde {\cal T} $ (which has to be regular in $U_1$ and/or $U_2$, respectively). Hence, we find that neither ${\cal T}(z)$ nor ${\widetilde{\cal T}(z)}$ can be a global section of $\widehat{\cal A}$, i.e., ${{\cal T}(z), {\widetilde{\cal T}(z)}} \notin H^0( \mathbb{CP}^1, \widehat {\cal A})$. In other words, $T(z)$ is not in the $Q$-cohomology of the abelian twisted $(0,2)$ sigma model -- there is a conformal anomaly.      This is consistent with the observation made in $\S$2.2 via (\ref{tzzanomaly}),    where $[Q, T_{zz}] \neq 0$ in general but
\be
[Q, T_{zz}] = \partial_z \left (R_{ i \bar j} \partial_z \phi^i \psi^{\bar j} - R_{ i \bar j}V^i_a A^a_z  \psi^{\bar j}\right).
\label{tanomaly}
\ee    
Since $Q$ generates a (BRST-like) symmetry (i.e.,~an automorphism) of the abelian twisted model via the field transformations (\ref{txgauged}), (\ref{fraktzanomaly}) can be viewed as an analog in $\check{ \textrm C}$ech cohomology of (\ref{tanomaly}). Indeed, the operator $R_{ i \bar j} \partial_z \phi^i \psi^{\bar j}$ can be shown to correspond exactly to ${\partial_z \gamma} / {\gamma}$, as follows. Apart from an obvious comparison of (\ref{tanomaly}) and (\ref{fraktzanomaly}), note that ${\partial_z \gamma} / {\gamma} = - {\partial_z \tilde\gamma} / {\tilde \gamma}$, i.e., ${\partial_z \gamma} /{\gamma}$ is a holomorphic operator over $U_1 \cap U_2$. Moreover, it cannot be expressed as a difference between an operator that is holomorphic in $U_1$ and an operator that is holomorphic in $U_2$. Thus, it is  a dimension-one class in the first $\check{ \textrm C}$ech cohomology group $H^1(\mathbb{CP}^1, \widehat{\cal A})$. Hence, from our $Q$-$\check{ \textrm C}$ech cohomology dictionary,   ${\partial_z \gamma} /{\gamma}$ will correspond to a dimension-one operator in the $Q$-cohomology of the sigma model with $q =1$ -- namely, $R_{ i \bar j} \partial_z \phi^i \psi^{\bar j}$ (which,  according to our discussion in $\S$4.1, indeed takes the correct form of a $Q$-invariant, dimension $(1,0)$ operator with $q=1$). As a corollary, we also find that the operator $R_{i \bar j} V_a^i \psi^{\bar j}$ ought to correspond to $V_a / \gamma$, so that the operator $R_{i \bar j} V_a^i A^a_z \psi^{\bar j}$ ought to correspond to $V_a A^a_z / \gamma$, as one would expect. Since the Ricci tensor $R_{ i \bar j}$ is proportional to the one-loop beta-function of the sigma model,  this correspondence allows one to interpret the one-loop beta-function purely in terms of holomorphic data.  By setting $A^a_z \to 0$ so that one is in the ordinary limit, (\ref{fraktzanomaly}) and (\ref{tanomaly}) indeed agree exactly with (3.2) and (5.17) of~\cite{CDO} where the interpretation of the one-loop beta function in terms of holomorphic data was first elucidated for the ordinary model. 

\bigskip\noindent{\it Higher-Dimensional Examples}
  
One can certainly consider other higher-dimensional examples in a similar fashion. It should be possible to show that ${\widetilde{\cal T}(z)} \neq {\cal T}(z)$ for any $X$ whose Ricci tensor is \emph{nonvanishing}. However, for brevity, we shall not pursue this matter further. In summary, we have found that the obstruction to a globally-defined $T(z)$ operator on $X$ -- which is characterized by the nonvanishing Ricci tensor of $X$ -- translates to a lack of invariance under arbitrary, holomorphic reparameterizations on the worldsheet $\Sigma$ of the $Q$-cohomology of the underlying  abelian twisted $(0,2)$ sigma model with target space $X$.

\newsection{Relation To The Mathematical Theory Of TCDO's}

In preparation of our physical interpretation in $\S$7 of the geometric Langlands correspondence for simply-connected, simple, complex Lie groups via our abelian twisted $(0,2)$ sigma model, we will devote this section to elucidating the relation of the abelian model to the mathematical theory of TCDO's. In doing so, we will, among other things, come across a crucial hint to a purely physical nonperturbative effect -- which we will discuss in detail in $\S$8 -- that would imply certain delicate conditions for the existence of Beilinson-Drinfeld $\cal D$-modules of the geometric Langlands program.

\newsubsection{Making Contact With The Sheaf Of TCDO's}

Note that if $X$ is a smooth flag manifold of a compact, connected, simple Lie group $G$, we have  $p^T_1(X) =0$ (although $c^T_1(X) \neq 0$), where $T \subset G$ is an abelian subgroup. In this case, (\ref{an}) tells us that the $T$-gauged twisted $(0,2)$ sigma model is anomaly-free (assuming $c_1(\Sigma) = 0$). Moreover, if the gauge group $T = T_C$, where $T_C$ is the Cartan subgroup of $G$ with Lie algebra $\frak h$ whence ${\rm dim} ({\frak h}) = r = {\rm rank}(\frak g)$, one can regard $A^a_z$ (where $a = 1, 2, \dots, r$) as an element of (a complexification of) $\frak h^*$. In turn, since  $X = G/ T_C = G_{\mathbb C} / B$, where $G_{\mathbb C}$ is the complexification of $G$, and $B \subset G_{\mathbb C}$ is a Borel subgroup, the $A^a_z$'s will be isomorphic to the elements of $H^1(X, \Omega^{1, cl}_X)$, where $ \Omega^{1, cl}_X$ is the sheaf of holomorphic $\partial$-closed $(1,0)$-forms on $X$  (see (4.20) of~\cite{Arakawa}). \emph{We shall consider this particular model henceforth}.


Let us now see what the general automorphism relations of (\ref{auto1})-(\ref{autobeta}) imply for such a model. To this end, cover $X$ with a set of open patches $\{U_e\}$, where $U_e \subset \mathbb C^n$, and $n = {\rm dim}_{\mathbb C} X$. Now, consider a holomorphic vector field $\xi = \xi^k \partial/ \partial \gamma^k$ on $X$.  Then,  over an intersection $U_e \cap U_f$, the change in $\beta_k$ (and hence $\partial / \partial \gamma^k$, as explained in $\S$4.5) under the symmetry generated by $K_F$, would, from (\ref{autobeta}),  imply the change $\delta\xi = \xi^kF_{k, a} A^a_z = (\iota_\xi \lambda^{(1)}_a) A^a_z$ in $\xi$, where $\iota_\xi  \lambda^{(1)}_a$ is an antiderivation by $\xi$ of a (complex) $\frak h$-valued holomorphic one-form $\lambda^{(1)}_a$ on $X$. Similarly, the changes in $\beta$ and hence the corresponding changes in $\xi$ under the symmetries generated by the rest of the conserved charges $K_V$ and $K_B$, can be obtained from the first two terms on the RHS of (\ref{autobeta}). Since these two terms coincide with the changes in the beta-field of the sheaf of CDO's given by the RHS of (4.4b) in~\cite{GMS1}, we conclude that the changes in $\xi$ under $K_V + K_E$ can be obtained solely from the automorphism relations of the sheaf of CDO's. For our present purpose, there is no need for us to detail what these changes in $\xi$ are. The interested reader is invited to do so him or herself.

Next, consider a holomorphic function $\cal O$ of the coordinates $\gamma$ of $X$. Then, over an intersection $U_e \cap U_f$, the changes in $\cal O$ under the full symmetry generated by $K_V + K_F + K_E$ can be obtained from the RHS of (\ref{auto1}), which is a purely geometrical coordinate transformation of $\gamma$.  Since the RHS of (\ref{auto1}) coincides with the changes in the gamma-field of the sheaf of CDO's given by the RHS of (4.4a) in~\cite{GMS1}, the changes in $\gamma$ under the full symmetry group can be obtained solely from the automorphism relations of the sheaf of CDO's.

Last but not least, consider the gauge field $A^a_z = \lambda^*_a$. Then, over an intersection $U_e \cap U_f$, the changes in $ \lambda^*_a$ under the full symmetry generated by $K_V + K_F + K_E$ can be obtained from the RHS of (\ref{auto2}), which is simply the identity transformation. 

In summary, over an intersection $U_e \cap U_f$, if $g_{ef}$ denotes the unique symmetry transformation of the fields implied by the automorphism relations of the sheaf of CDO's given by (4.4a)-(4.4b) of~\cite{GMS1}, and if $g^{tw}_{ef}$ denotes the unique symmetry transformation of the fields implied by the automorphism relations (\ref{auto1})-(\ref{autobeta}), we can write
\vspace{-0.0cm}\begin{eqnarray}
\label{TCDO tx 1}
& g^{tw}_{ef}\vert_{\CO_{U_e \cap U_f}}   =   g_{ef}\vert_{\CO_{U_e \cap U_f}}, &  \\
\label{TCDO tx 2}
&g^{tw}_{ef}(\lambda^*_a)  =  \lambda^*_a,& \\
\label{TCDO tx 3}
& g^{tw}_{ef}(\xi)  =   g_{ef}(\xi)  - \sum_a (\iota_\xi \lambda^{(1)}_a(U_e \cap U_f)) \lambda^*_a.&
\end{eqnarray}
Moreover, since the underlying theory is anomaly-free, we have, over triple intersections $U_e \cap U_f \cap U_g$, the relation
\be
\label{TCDO tx 4}
g^{tw}_{ef} = g^{tw}_{gf} \circ g^{tw}_{eg}.
\ee
As $\lambda^*_a$ lives in $H^1(X, \Omega^{1, cl}_X)$ (which was explained at the start of this subsection), (\ref{TCDO tx 1})-(\ref{TCDO tx 4}) is \emph{exactly} Lemma 4.4 of~\cite{Arakawa}; in other words, the underlying model describes, purely physically, the sheaf of TCDO's on $X$ defined by Arakawa et al. in~\cite{Arakawa}! 

\def\MA{{\mathscr A}}

\bigskip\noindent{\it The Chiral Algebra $\MA$}

With the above-observed connection to the sheaf of TCDO's, one can conclude, from our discussion in $\S$4.3, that the chiral algebra $\MA$ of the perturbative $T_C$-gauged twisted $(0,2)$ sigma model on $X$ can, as a vector space, be written as
\be
\MA =  \bigoplus_{q_R} H^{q_R}(X, {\widehat\Omega}^{ch,tw}_{X}). 
\label{A=H}
\ee
Here, $\widehat\Omega^{ch,tw}_{X}$ is the sheaf of TCDO's on $X$ generated by the fields $\beta$, $\gamma$, $A^a_z$ and their $z$-derivatives whose automorphism relations over any intersection $U_e \cap U_f \subset X$ are given by (\ref{auto1})-(\ref{autobeta}), and $q_R$ is the  $U(1)_R$ charge of the corresponding physical operator.

\bigskip\noindent{\it The Elliptic Genus}  

This brings us now to the elliptic genus of the sigma model. Consider the canonical quantization of the sigma model on an infinitely long cylinder ${\mathbb R} \times S^1$. Let $\cal V$ be the $Q$-cohomology of states which furnishes a module for the chiral algebra $\mathscr A$. Then, one can form a modular function 
\be
V( X, q) = q^{- {d / 12}}\sum_n \, {q}^{n} \, \textrm{Tr}_{{\cal V}_n} (-1)^F,
\label{elliptic genus}
\ee
called the elliptic genus. Here, $d = \textrm{dim}_{\mathbb C} X$; $q$ is some unit modulus complex parameter;  ${\cal V}_n$ are the states in $\cal V$ of energy level $n$; $F$ is the total fermion number of each state; and $\textrm{Tr}_{{\cal V}_n} (-1)^F$ is the Witten index that counts the difference between the number of bosonic and fermionic states of energy level $n$.

Notice that the elliptic genus in (\ref{elliptic genus}) involves the states but \emph{not} the local operators of the sigma model. When and how do the local operators come into the picture? If the Ricci tensor of $X$ vanishes, i.e., if $c_1(X) = 0$, the sigma model will be conformal (as explained around (\ref{tzzanomaly})). One can then proceed to employ the conformal field theory (CFT) state-operator isomorphism to relate the above states in (\ref{elliptic genus}) to the local operators in $\MA$ in a one-to-one manner. In this way, one can express the elliptic genus in terms of a difference between the number of bosonic and fermionic operators in the $Q$-cohomology, with the holomorphic (i.e.,~left-moving) dimension  of the operators $n$ corresponding to the energy level $n$ of the supersymmetric states that the operators are supposed to be isomorphic to. However, the reality is that $c_1(X) \neq 0$, and the state-operator correspondence will $\it{not}$ be an isomorphism. Nevertheless, one can -- via the description of $\MA$ in terms of the $\check{ \textrm C}$ech cohomology of ${\widehat{\Omega}}^{ch, tw}_X$ in (\ref{A=H}), and the fact that bosonic and fermionic operators have even and odd  values of $q_R$, respectively -- define the following expression in the $Q$-closed local operators which is analogous to $V(X, q)$: 
\be
A(X, q) = q^{-{d/12}}\sum_{{q_R}} \sum_{n=0}^{\infty} (-1)^{q_R} \, {\mathrm{dim}} {H}^{q_R}(X, {\widehat {\Omega}}^{ch, tw}_{X; n}) \, q^n.
\label{elliptic genus TCDO} 
\ee 
Here,  ${\widehat {\Omega}}^{ch, tw}_{X, n}$ is a sheaf of TCDO's on $X$ whose local sections correspond to the $\psi^{\bar i}$-independent  $Q$-cohomology classes with dimension $(n,0)$. Unlike in the case of $V(X, q)$ (which,  if we let $q = e^{2 \pi i \tau}$, can be physically interpret as a partition function of some sigma model with a toroidal worldsheet of modulus $\tau$), it is not clear if there exists a natural path integral proof that $A(X, q)$ must transform as a modular form under $SL(2, \mathbb Z)$. That being said, $\CA(X,q) = q^{{d/12}}A(X, q)$ has been computed purely mathematically in~\cite{chiral bott}, where it appears to possess modular properties.

Notice that $A(X, q)$ is  ${\mathbb Z_{\geq 0}} \times \mathbb Z_{\geq 0}$  graded by the holomorphic dimension $n$ and $U(1)_R$ charge $q_R$ of the $Q$-closed operators. The grading by dimension follows naturally from the scale invariance of the  correlation functions that define the chiral algebra $\MA$ of the sigma model. Note that $A(X, q)$ has no perturbative quantum corrections.\footnote{Absence of quantum corrections can be inferred from the fact that both the scaling dimension (or energy)  and the $(-1)^{q_R}$ (or $(-1)^F$) operator that distinguishes the bosonic and fermionic operators (or states), are conserved through to the quantum level.} However, since $c^{T_C}_1(X) \neq 0$, nonperturbative worldsheet twisted-instanton corrections can (i) destroy (via dimensional transmutation) the scale invariance of the correlation functions  and thus, violate the grading by dimension of the operators in $\MA$; (ii) ``connect'' operators with different values of $q_R$ through certain nonperturbative relations and thus, violate the grading by $U(1)_R$ charge of the operators in $\MA$. Consequently, $A(X, q)$ can vanish, as we will see in $\S$8.


\newsubsection{Examples Of Sheaves Of TCDO's}   

We will now study in detail, examples of sheaves of TCDO's and their $\check{\rm C}$ech cohomologies on smooth flag manifolds of \emph{simply-connected}, simple, complex Lie groups $G_{\mathbb C}$ (whose compact real form is $G$ with Cartan subgroup $T_C \subset G$). In the process, we will unravel an important connection between the chiral algebra $\MA$ of the underlying $T_C$-gauged twisted $(0,2)$ sigma model and a family of affine $G_{\mathbb C}$-algebras at the critical level parameterized by its center. In turn, this connection, together with the results of $\S$6, will allow us to furnish,  in $\S$7, a purely physical interpretation of the geometric Langlands correspondence for $G_{\mathbb C}$.

\newsubsubsection{The Sheaf of TCDO's on the Flag Manifold of $SL(2)$}

For our first example, we take $X$ to be the flag manifold of $G_{\mathbb C} = SL(2)$, i.e., $X = SL(2)/B = \mathbb{CP}^1$, where $B \subset SL(2)$ is a Borel subgroup.  In other words, we will be exploring and analyzing the chiral algebra $\MA$ of  local operators in the $T_C$-gauged twisted $(0,2)$ model on $\mathbb{CP}^1$. To this end, we will work locally on the worldsheet $\Sigma$, where $z$ is the local complex coordinate.   

As mentioned before, $\mathbb{CP}^1$ can be viewed as the complex $\gamma$-plane plus a point at infinity.  Thus, it can be covered using two open charts, $U_1$ and $U_2$, where $U_1$ is the complex $\gamma$-plane, $U_2$ is the complex $\tilde \gamma$-plane, and $\tilde\gamma=1/\gamma$. 

Since $U_1 \simeq \mathbb{C}$ and $\textrm{dim}({\frak h}) = 1$, where ${\frak h}$ is the Lie algebra of $T_C$, the sheaf of TCDO's in $U_1$ can be described by a single perturbed free $\beta\gamma$ system with action
\be
I={1\over 2\pi}\int|d^2z| \ \beta \partial_{\bar z} \gamma - c  A_z A_{\bar z},
\label{actionU1}
\ee 
where $c = |V|^2$ is some complex constant. The relevant fields $\beta$, $A_{z}$ and $\gamma$  are of dimension $(1,0)$, $(1,0)$ and $(0,0)$, respectively. They obey the usual free-field OPE's: there are no singularities in the operator products $\beta\cdot \beta$, $A_z \cdot A_{z'}$, $\gamma\cdot\gamma$, $A_z \cdot \gamma$ and $A_z \cdot \beta$,  while 
\be
\beta(z) \gamma(z')  \sim  -{1\over z-z'}.
\ee

Similarly, the sheaf of TCDO's in $U_2$ is described by a single perturbed free $\tilde\beta\tilde\gamma$ system with action 
\be
\tilde I= {1\over 2\pi}\int|d^2z| \ \tilde
\beta \partial_{\bar z} \tilde\gamma - c   {\tilde A}_z {\tilde A}_{\bar z}, 
\label{actionU2}
\ee
where the relevant fields $\tilde \beta$, $\tilde A_z$ and $\tilde \gamma$ obey the same OPE's as $\beta$, $A_z$ and $\gamma$. In other words, the only singular OPE is  
\be
\tilde \beta(z) \tilde \gamma(z')  \sim  -{1\over z-z'}.
\ee 

In order to describe a globally-defined sheaf of TCDO's, one will need to glue the free conformal field theories with actions (\ref{actionU1}) and (\ref{actionU2}) over pairwise intersections $U_1 \cap U_2$. To do so,  one must use the admissible automorphisms of the free conformal field theories defined in (\ref{auto1})-(\ref{autobeta}) to glue the free-fields together. In the case of $X = \mathbb{CP}^1$, the automorphisms will be given by\footnote{Note that in writing the following relations, we have -- in order to facilitate comparison of our results with those in the mathematical literature -- chosen the arbitrary local (1,0)-form $B(\gamma)d\gamma$ on $\mathbb{CP}^1$ (associated with the current $J_E$ of  $\S$4.5) to be such that $B(\gamma) = 2 \gamma$. For the same reason, we have also chosen the arbitrary local $\frak h$-valued $(1,0)$-form $F_a$ on $\mathbb{CP}^1$ (associated with the current $J_F$ of  $\S$4.5) to be $F_a = \gamma_a \, d \gamma$. (To arrive at our choice of $F_a$, we have taken the liberty to rescale $V_a \to \gamma V_a$ since (i) globally-defined holomorphic functions in $\gamma$ with no poles are -- on the compact Riemann surface $\mathbb{CP}^1$ -- equivalent to constants, (ii) $V_a$ is only defined up to scaling by a finite constant.)}   
 
\begin{eqnarray}
\label{autoCP1gamma}
{\tilde \gamma} & = & {1 \over \gamma},\\
\label{autoCP1A}
{\tilde A_z} & = & A_z, \\
\label{autoCP1beta}
{\tilde \beta} & = &   -  \beta \gamma^2  + 2 \partial_z\gamma - \gamma A_z
\end{eqnarray} 
As there is no obstruction to this gluing in a sigma model that is nonanomalous, a sheaf of TCDO's can be globally-defined on its $\mathbb{CP}^1$ target space.

\bigskip\noindent{\it Global Sections of the Sheaf}

Recall from (\ref{A=H}) that for a general manifold $X$ of complex dimension $n$, the chiral algebra $\MA$ will be given by ${\MA} = \bigoplus_{q_R = 0}^{q_R = n} H^{q_R}( X, {\widehat \Omega}^{ch,tw}_X)$ as a vector space. Since $\mathbb{CP}^1$ has complex dimension one, we will have, for $X=\mathbb{CP}^1$, the relation ${\MA} = \bigoplus_{q_R = 0}^{q_R = 1} H^{q_R}( \mathbb{CP}^1, {\widehat \Omega}^{ch,tw}_{\mathbb {CP}^1})$. Thus, in order to determine the chiral algebra of the sigma model, one needs only to ascertain the global sections  $H^0( \mathbb{CP}^1, {\widehat \Omega}^{ch,tw}_{\mathbb{CP}^1})$ of the sheaf ${\widehat \Omega}^{ch,tw}_{\mathbb{CP}^1}$, and its first $\check{ \textrm C}$ech cohomology $H^1( \mathbb{CP}^1, {\widehat \Omega}^{ch,tw}_{\mathbb{CP}^1})$. 

First, let us compute the global sections  $H^0( \mathbb{CP}^1, {\widehat \Omega}^{ch,tw}_{\mathbb{CP}^1})$. For brevity, we shall focus on the dimension 0 and 1  operators only; the higher-dimensional cases can be obtained in a similar manner. At dimension 0, the space of global sections $H^0( \mathbb{CP}^1, {\widehat \Omega}^{ch,tw}_{\mathbb{CP}^1; 0})$ must be spanned by regular polynomials in $\gamma$, i.e., ${\widehat \Omega}^{ch,tw}_{{\mathbb {CP}}^1; 0}$ is just the sheaf  $\cal O$ of regular holomorphic functions in $\gamma$ on $\mathbb{CP}^1$. Since all globally-defined regular holomorphic functions on a compact Riemann surface such as  $\mathbb{CP}^1$ are equivalent to constants, we find that  $H^0( \mathbb{CP}^1, {\widehat \Omega}^{ch,tw}_{\mathbb{CP}^1; 0})$ is one-dimensional and generated by 1.   

What about the space $H^0( \mathbb{CP}^1, {\widehat \Omega}^{ch,tw}_{\mathbb{CP}^1; 1})$ of global sections of dimension 1? In order to get a global section of ${\widehat \Omega}^{ch,tw}_{\mathbb{CP}^1}$ of dimension 1, we can act on a (gauge-invariant) global section of ${\widehat \Omega}^{ch,tw}_{\mathbb{CP}^1}$ of dimension 0 with the partial derivative $\partial_z$. Since $\partial_z 1 = 0$, this prescription will not apply here.

One could also consider dimension 1 operators of the form $f(\gamma) D_z \gamma = f(\gamma) \partial_z \gamma - f'(\gamma) A_z$, where $f(\gamma)$ and $f'(\gamma)$ are globally-defined holomorphic functions of $\gamma$. From (\ref{autoCP1A}), we immediately see that $A_z$ can be globally-defined over $\mathbb {CP}^1$; thus, the operator $f'(\gamma) A_z$ can be globally-defined over ${\mathbb{CP}^1}$ too. However, the operator $f(\gamma) \partial_z \gamma $,  by virtue of the way it transforms purely geometrically under (\ref{autoCP1gamma}), would correspond to a section of $\Omega^1_{\mathbb{CP}^1}$, the sheaf of holomorphic differential one-forms $f(\gamma) d\gamma$ on $\mathbb{CP}^1$; from the classical result $H^0(\mathbb{CP}^1, \Omega^1_{\mathbb{CP}^1}) = 0$, which continues to hold in the quantum theory, it is clear that $f(\gamma) \partial_z \gamma$ \emph{cannot }be a dimension 1 global section of ${\widehat \Omega}^{ch,tw}_{\mathbb{CP}^1}$.  As such, $f(\gamma) D_z \gamma \notin H^0( \mathbb{CP}^1, {\widehat \Omega}^{ch,tw}_{\mathbb{CP}^1; 1})$.

The remaining possibility at dimension 1 is to find an operator that contains $\beta$. In fact, from the automorphism relation of (\ref{autoCP1beta}), we  immediately find an example since the LHS, $\tilde \beta$, is regular in $U_2$ (by definition), while the RHS, being a polynomial in $\gamma$, $\beta$ and $A_z$, is manifestly regular in $U_1$. Their being equal means that they represent a dimension 1 global section of ${\widehat\Omega}^{ch,tw}_{\mathbb {CP}^1}$ that we will call $J_-$:
\be
J_-  = - :\beta \gamma^2:  + 2 \partial_z\gamma - \gamma A_z   = {\tilde \beta}.
\label{J_-}
\ee   
Notice that since $U_1$ and $U_2$ are on equal footing, one can also apply the above construction to the fields which are manifestly regular in $U_1$. In doing so, we obtain another dimension 1 global section $J_+$:
\be
J_+ =\beta = -:\tilde \beta \tilde\gamma^2: + 2\partial_z \tilde\gamma - \tilde \gamma \tilde A_z.
\label{J_+}
\ee
Hence,  $J_+$ and $J_-$  furnish us with two dimension 1 global sections of the sheaf ${\widehat\Omega}^{ch,tw}_{\mathbb{CP}^1}$.   Since these are global sections of a sheaf of chiral vertex operators, their nontrivial OPE's would generate other global sections. The $J_+ \cdot J_+$ and $J_-\cdot J_-$ operator products are trivial (i.e.  nonsingular), but 
\be
J_+ J_- \sim {2J_0\over z-z'} - {2 \over {(z-z')^2}},
\label{again}
\ee
where $J_0$ is another global section of dimension 1 given by 
\be
J_0 = :\beta \gamma: + {1 \over 2} A_z.
\label{J_0}
\ee

In short, $\{J_+, J_-, J_0 \}$ are bosonic dimension 1 operators that span $H^0(\mathbb{CP}^1, {\widehat\Omega}^{ch,tw}_{\mathbb{CP}^1 ; 1})$, and by a further computation, one can show that they satisfy the following closed nontrivial OPE algebra:
\begin{eqnarray}
\label{SL(2)-OPE}
 {J}_+ (z) {J}_- (z') \sim  {k \over {(z-z')^2}} + {{2{ J}_0 (z')} \over z-z'},  \quad   {J}_0 (z) {J}_+ (z') \sim  {{{ J}_+ (z')} \over z-z'},  \nonumber \\ \\
{J}_0 (z) {J}_- (z') \sim  {{-{ J}_- (z')} \over z-z'},  \quad  { J}_0 (z) { J}_0 (z') \sim {k /2 \over {(z-z')^2}}, \nonumber
\end{eqnarray}
where $k = -2$. Note that (\ref{SL(2)-OPE}) means that the $J$'s actually generate an affine algebra of $SL(2)$ at the \emph{critical} level of $-2$ in the Wakimoto free-field representation~\cite{CDO28}, in agreement with Lemma~4.6 of~\cite{Arakawa}. Since $\{J_+, J_-, J_0 \}$ are chiral vertex operators holomorphic in $z$, one can expand them in a Laurent series that allows an affinization of the $SL(2)$ Lie algebra generated by their resulting zero modes. Consequently, $H^0(\mathbb{CP}^1, {\widehat\Omega}^{ch,tw}_{\mathbb{CP}^1})$ would be a Wakimoto module for the affine algebra of $SL(2)$ at level $-2$, in agreement with Theorem~6.2 of~\cite{Arakawa}.  $\{J_+, J_-, J_0 \}$ also define a structure of a chiral algebra although not in the full physical sense -- it satisfies all the physical axioms of a chiral algebra, except invariance under arbitrary holomorphic reparameterizations of the coordinates on the worldsheet $\Sigma$. The fact that the chiral algebra is not coordinate reparameterization invariant  is -- as explained in $\S$4.7 for a general abelian model -- due to the fact that $\MA$ lacks a \emph{quantum} stress tensor $T(z)$; i.e.,  $T(z) \notin H^{0}( \mathbb{CP}^1, {\widehat \Omega}^{ch,tw}_{\mathbb{CP}^1})$ (see (\ref{fraktzanomaly})). 

\bigskip\noindent{\it The Center of a Chiral Algebra} 

The absence of $T(z)$ in $\MA$ can also be understood from a different albeit relevant viewpoint which we shall now elaborate upon. To this end, first note that for any affine algebra $\widehat {\frak g}_{\mathbb C}$ of $G_{\mathbb C}$ at level $k \neq
-h^{\vee}$, where $h^{\vee}$ is the dual Coxeter number of its Lie
algebra $ {\frak g}_{\mathbb C}$, one can construct the corresponding quantum stress
tensor out of the currents of $\widehat  {\frak g}_{\mathbb C}$ via a
Segal-Sugawara construction \cite{Ketov}. In our case at hand of
 $\widehat  {\frak g}_{\mathbb C} = \hat{\frak {sl}_2}$, where $h^{\vee}
=2$,  the quantum stress tensor can be constructed as
\be
 T(z) = \,  {{ :J_+ J_-  +  J_0^2: (z) \over {(k+2)}}}. 
\label{SS def T(z)}
 \ee
The shift in the level $k \to k + h^\vee$ in the denominator of (\ref{SS def T(z)}) as one transitions from the classical to quantum theory, is due to a renormalization effect. As required, for every $k \neq {-2}$, the modes of the
Laurent expansion of $T(z)$ will span a Virasoro algebra. In
particular, $T(z)$ will generate holomorphic reparameterizations of
the coordinates on the worldsheet $\Sigma$. Notice that this
definition of $T(z)$ in (\ref{SS def T(z)}) is ill-defined when
$k=-2$, i.e., at the critical level. Nevertheless,  notice from (\ref{SS def T(z)}) that one can write 
 \be
S(z) =  (k+2) T(z),
\label{S(z)} \ee 
where
\be S(z) =   \,  :J_+ J_-  +  J_0^2:  (z).
\label{s(z)} 
\ee 
$S(z)$ is commonly known as the Segal-Sugawara tensor. From (\ref{S(z)}), we see that in the quantum theory, $S(z)$ generates $(k+2)$ times the field transformations usually generated by the quantum stress tensor $T(z)$.
Therefore, at the critical level $k=-2$, $S(z)$ generates \emph{no}
field transformations at all -- its OPE with all other field operators ought to be regular at best. This is equivalent to saying that the quantum
stress tensor does not exist at $k =-2$, since $S(z)$ is the
only well-defined operator at this critical level which can generate field transformations under arbitrary
holomorphic reparameterizations of the worldsheet coordinates on
$\Sigma$.

$S(z)$ may or may not vanish identically at $k=-2$, and if it does not, it will merely play the role of a \emph{classical} field (i.e.,~$c$-number) that has no nontrivial interactions with itself and other fields at the quantum level. In fact, by substituting (\ref{J_-}), (\ref{J_+}) and (\ref{J_0}) in (\ref{s(z)}), a careful computation with the aid of (\ref{SL(2)-OPE}) would give 
\be
S(z) = {1 \over 4} A^2_z (z) - {1 \over 2} \partial_z A_z(z).
\label{S(z)-A}
\ee
(Note that  the above formula was also computed below (9.3) of~\cite{Frenkel}.) Since $A_z$ is a non-dynamical field which does not have any nontrivial propagators with itself or with the rest of the fields $\beta$ and $\gamma$ (and their $z$-derivatives) that define $\MA$,  $S(z)$ indeed plays the role of a purely classical field. 

The fact that $S(z)$ only has regular OPE's with all other relevant fields and with itself implies the following. If we denote by $V_{-2} (\frak {sl}_2)$ the chiral algebra generated by $\{J_{\pm}, J_0\}$ and their $z$-derivatives, and if we denote by $\frak z(V_{-2} (\frak {sl}_2))$ its center which is defined to be the set of fields which have \emph{regular }OPE's with all other fields within, we have $\frak z(V_{-2} (\frak {sl}_2)) = \mathbb C[\partial^m_z S(z)]_{m\geq 0}$, where $\mathbb C[\partial^m_z S(z)]_{m\geq 0}$ is the space of differential polynomials on $S(z)$ with complex coefficients. This result agrees with Theorem~7 of~\cite{Frenkel}. In turn, (\ref{S(z)-A}) means that $\frak z(V_{-2} (\frak {sl}_2)) \subset H_{\mathbb {CP}^1}$, where $ H_{\mathbb {CP}^1}$ is the space of differential polynomials on $A_z$ with complex coefficients. This result agrees with Lemma~4.6 of~\cite{Arakawa}. (Recall from $\S$5.2 that $A_z$ corresponds to $\lambda^*$ of \emph{loc.~cit.}.)

Moreover, notice that since $S(z)$ and the $J(z)$'s are holomorphic in $z$ and are of dimension 2 and 1, respectively,  one can Laurent expand them as 
\be S(z) = \sum_{n \in {\mathbb Z}} {S}_n z^{-n-2} \quad {\rm and} \quad J^\alpha(z) = \sum_{n \in \mathbb Z} J^\alpha_{n} z^{-n-1},
\label{S(z) expansion}
\ee 
where $\alpha = \{\pm, 0 \}$. Again, the fact that $S(z)$ only has regular OPE's with all other relevant fields and with itself implies that 
\be
[{S}_n, J^\alpha_m] = [{S}_n, {S}_m] =0.
\label{center}
\ee 
This means that the $S_n$'s generate the center ${\frak z}(\tilde U_{-2} (\hat{{\frak sl}_2}))$ of the completed enveloping algebra $\tilde U_{-2} (\hat{{\frak sl}_2})$ of $\hat{{\frak sl}_2}$ at critical level $-2$. 

The fact that ${\frak z}(\tilde U_{-2} (\hat{{\frak sl}_2}))$ ought to also furnish a Poisson algebra -- which is a claim made in (8.14) of~\cite{Frenkel} -- can be understood purely physically as follows. Firstly, (\ref{center}) means that one can define simultaneous eigenstates of the ${S}_n$ and $J^\alpha_n$ mode operators. In particular, one would be able to properly define a general state ${\Psi} = {S}_{l} {S}_{q} \dots {S}_{p} |0, \eta \rangle$, where $| 0, \eta \rangle$ is a vacuum state which is a representation of $\frak {sl}_2$ labelled by $\eta$, such that $J^\alpha_0 | 0, \eta \rangle = \eta^\alpha | 0, \eta \rangle$. However, note that any such $\Psi$ will correspond to a null-state, i.e., $\Psi$ decouples from the ultimately relevant Hilbert space of physically inequivalent quantum states spanned by the representations of $\frak {sl}_2$ \cite{lindstrom}. Therefore,  the ${S}_{m} $'s which generate  ${\frak z}(\tilde U_{-2} (\hat{{\frak sl}_2}))$,  can, in this regard, be interpreted as classical $c$-numbers which would then generate a Poisson algebra.

\bigskip\noindent{\it The First Cohomology}

Let us now proceed to ascertain the first cohomology group $H^1(\mathbb{CP}^1, {\widehat \Omega}^{ch,tw}_{\mathbb{CP}^1})$.  In dimension 0, we can again consider regular polynomials in $\gamma$. However, from ordinary algebraic geometry, we have the classical result $H^1(\mathbb{CP}^1, {\cal O}) = 0$, where $\cal O$ is the sheaf of regular functions over $\mathbb{CP}^1$ which are holomorphic in $\gamma$. Since a vanishing cohomology at the classical level continues to vanish at the quantum level, it would mean that we cannot have regular polynomials in $\gamma$; i.e., $H^1(\mathbb{CP}^1, {\widehat \Omega}^{ch,tw}_{\mathbb{CP}^1;0}) = \emptyset$.

In dimension 1, one could consider the operator $\Theta = (D_z \gamma) / \gamma = (\partial_z \gamma - A_z\gamma)/ \gamma$. Note that $(\partial_z \gamma) / \gamma = - (\partial_z \tilde \gamma) / \tilde\gamma$ and $A_z = \tilde A_z$. As such, $\Theta$ has a pole at $\gamma = 0$ and $\tilde \gamma = 0$, and it cannot be written as a difference of operators which are regular and  holomorphic in $U_1$ and $U_2$, respectively. In other words, $\Theta \in H^1(\mathbb{CP}^1, {\widehat\Omega}^{ch,tw}_{\mathbb{CP}^1;1})$. Indeed, by comparing $\Theta$ and the RHS's of (\ref{fraktzanomaly}) and (\ref{tanomaly}),\footnote{Because of footnote~27, one must rescaled $V_a \to \gamma V_a$ in (\ref{fraktzanomaly}) before making any comparison with the results in this subsection.}  we see that $\Theta$ corresponds to the dimension 1 fermionic sigma model operator $R_{1\bar 1} D_z \phi^1 \psi^{\bar 1}$ of $q=1$. 

What about in dimension 2? Let us try to differentiate $\Theta$, i.e., let us consider the operator $\partial_z \Theta$. Via (\ref{fraktzanomaly}), one can write $\partial_z\Theta = {\tilde T}(z) - T(z)$, where $\tilde T(z)$ and $T(z)$ are regular and holomorphic in $U_2$ and $U_1$, respectively. Thus, $\partial_z\Theta  \notin H^1(\mathbb{CP}^1, {\widehat\Omega}^{ch,tw}_{\mathbb{CP}^1;2})$. 

Nevertheless, since  we have the product formula $H^1(\mathbb {CP}^1, {\widehat\Omega}^{ch,tw}_{\mathbb{CP}^1; l}) \otimes H^0(\mathbb {CP}^1, {\widehat\Omega}^{ch,tw}_{\mathbb{CP}^1; m}) \to H^1(\mathbb {CP}^1, {\widehat\Omega}^{ch,tw}_{\mathbb{CP}^1; l+m})$, we can act $\Theta \in H^1(\mathbb{CP}^1, {\widehat\Omega}^{ch,tw}_{\mathbb{CP}^1;1})$ on every other element of  $H^0(\mathbb{CP}^1, {\widehat\Omega}^{ch,tw}_{\mathbb{CP}^1})$ to build $H^1(\mathbb{CP}^1, {\widehat\Omega}^{ch,tw}_{\mathbb{CP}^1})$ up: for example, the first element of $H^1(\mathbb{CP}^1, {\widehat\Omega}^{ch,tw}_{\mathbb{CP}^1})$ is $\Theta \cdot 1 = \Theta$  of dimension 1, the second set of elements of $H^1(\mathbb{CP}^1, {\widehat\Omega}^{ch,tw}_{\mathbb{CP}^1})$ is given by $\Theta \cdot \{J_+, J_-, J_0 \}$ of dimension 2, and so on. Thus, we have a one-to-one correspondence $H^0(\mathbb{CP}^1, {\widehat\Omega}^{ch,tw}_{\mathbb{CP}^1; m}) \leftrightarrow H^1(\mathbb{CP}^1, {\widehat\Omega}^{ch,tw}_{\mathbb{CP}^1; m+1})$, where $m \geq 0$. In turn, this implies that $H^0(\mathbb{CP}^1, {\widehat\Omega}^{ch,tw}_{\mathbb{CP}^1}) \cong  H^1(\mathbb{CP}^1, {\widehat\Omega}^{ch,tw}_{\mathbb{CP}^1})$. Hence, $H^1(\mathbb{CP}^1, {\widehat\Omega}^{ch,tw}_{\mathbb{CP}^1})$ would also be a Wakimoto module for the affine algebra of $SL(2)$ at level $-2$. This conclusion is in exact agreement with (6.18) and (6.20) of~\cite{Arakawa}. 

That being said, our isomorphism involves a shift of conformal weight 1:  $H^0(\mathbb{CP}^1, {\widehat\Omega}^{ch,tw}_{\mathbb{CP}^1})$ starts ``growing'' at dimension 0 with the operator $1$, while $H^1(\mathbb{CP}^1, {\widehat\Omega}^{ch,tw}_{\mathbb{CP}^1})$ starts ``growing'' at dimension 1 with the operator $\Theta$. On the other hand, the isomorphism proved in~\cite{Arakawa} involves a shift of conformal weight $n+1$, where $n \geq 0$. In other words, our model realizes the $n=0$ case only.  Why this is so, can be understood as follows. Firstly,  the dimension 1 gauge field $A_z(z)$  ought to be physically well-behaved and therefore regular over \emph{all} of $\Sigma$; in particular, it ought to be nonsingular at the origin $z=0$. In turn, this means that it will take the form $A_z (z) = \sum_{n < 0} a_n z^{-n -1} = a_0 z^{-1} + a_{-1} + a_{-2} z + \dots$, whereby $a_0 = 0$. Since $A_z(z)$ also corresponds to $\chi(z)$ of~\cite{Arakawa}, it would mean that we must set $\chi_0 = n = 0$ in \emph{loc.~cit.}. In short, physical consistency requires that $n =0$ in our case. Then, Theorem~6.2 of~\cite{Arakawa} tells us that $H^0$ and $H^1$ must be modules with highest weight 0. Indeed, the set $\{J_-, J_+, J_0 \}$ with formulas (\ref{J_-}), (\ref{J_+}) and (\ref{J_0}) which coincide with (9.3) of~\cite{Frenkel}, would furnish a module with highest weight 0 according to the discussion in $\S$9.6 of \emph{loc.~cit.}.  

\bigskip\noindent{\it About the Isomorphism Between $1$ and $\Theta$}

The fact that the isomorphism between the bosonic operator $1$ and fermionic operator $\Theta$ violates their grading by dimension, is a classical starting point for a nonperturbative phenomenon -- which we will discuss in detail in $\S$8 -- that gives rise to the relation $\{Q, \Theta\} \sim 1$ for a worldsheet that is a cylinder. Because of this relation, the picture changes radically -- as the identity operator itself is $Q$-exact, the $Q$-cohomology of local operators would be empty whence the chiral algebra vanishes! By canonically-quantizing the theory on the cylinder, one can also see that the relation $\{Q, \Theta\} \sim 1$ implies that the $Q$-cohomology of states (or the module $\CV$ of $\MA$ described in $\S$5.1) is empty too!

This radical phenomenon will hold not just for  $X = \mathbb {CP}^1$, but for $X$ being\emph{ any} flag manifold of $G_{\mathbb C}$. In turn, as we shall explain in $\S$8, this would imply certain delicate conditions for the existence of Beilinson-Drinfeld $\cal D$-modules of the geometric Langlands correspondence for $G_{\mathbb C}$. 

\newsubsubsection{The Sheaf of TCDO's on the Flag Manifold of $SL(3)$}

Let us move on to our second example and take $X$ to be a flag manifold of $G_{\mathbb C} = SL(3)$, i.e., $X = SL(3) /B$, where $B \subset SL(3)$  is a Borel subgroup with Lie algebra $\frak b$. In this case, $\textrm{dim}_{\mathbb C} ({\frak sl}_3) = 8$ and $\textrm{dim}_{\mathbb C} ({\frak b}) = 5$. Therefore, $\textrm{dim}_{\mathbb C} X = 3$, and one can cover $X$ with six open charts $U_w$, $w=1,2, \dots, 6$, such that
each open chart $U_w$ can be identified with the affine space $\mathbb {C}^3$. Consequently, since $\textrm{dim} (\frak h) =2$,   the sheaf of TCDO's in any $U_w \subset X$ can be described by the following perturbed free $\beta\gamma$ system with action
\be
I= {1\over 2\pi}\int|d^2z| \  \sum_{i=1}^{3} \left(
\beta_i \partial_{\bar z} \gamma^i -  \sum_{a, b =1}^2  {A}^a_z  V_ {i a} A^b_{\bar z}  V^i_b   \right),
\label{actionUSL(3)}
\ee
where the only singular OPE of this system is 
\be
\beta_i(z) \gamma^j(z')  \sim  -{\delta_i^j\over z-z'}.
\ee 

Similarly, the sheaf of TCDO's in a neighboring chart $U_{w+1}$ can be described by the following perturbed free $\tilde\beta\tilde\gamma$ system with action 
\be
\tilde I= {1\over 2\pi}\int|d^2z| \  \sum_{i=1}^{3} \left(
{\tilde\beta}_i \partial_{\bar z} {\tilde\gamma}^i -  \sum_{a, b =1}^2 {\tilde A}^a_z  V_ {i a} {\tilde A}^b_{\bar z}  V^i_b   \right),
\label{actionUSL(3) -2}
\ee
where the only singular OPE of this system is
\be
{\tilde \beta}_i(z) {\tilde \gamma}^j(z')  \sim  -{\delta_i^j\over z-z'}.
\ee 

In order to describe a globally-defined sheaf of TCDO's, one will need to glue the free conformal field theories with actions (\ref{actionUSL(3)}) and (\ref{actionUSL(3) -2})  over pairwise intersections $U_w \cap U_{w+1}$ for every $w = 1,2, \dots 6$, where $U_7 = U_1$. To do so,  one must use the admissible automorphisms of the free
conformal field theories defined in (\ref{auto1})-(\ref{autobeta}) to glue the free-fields
together. In the case of $X = SL(3)/B$, the relation between the
coordinates in $U_{w}$ and $U_{w+1}$ will mean that the $\tilde
\gamma^i$'s in $U_{w+1}$ will be related to the $\gamma^i$'s in
$U_{w}$ via the relation $[\tilde \gamma] = [V_{w+1}]^{-1}
[V_{w}][\gamma]$, where the $3 \times 3$ matrices $[V_{w+1}]$ and
$[V_w]$ are elements of the $S_3$ permutation subgroup of $GL(3)$
matrices associated with the open charts $U_{w+1}$ and $U_w$,
respectively, and $[\gamma]$ is a $3 \times1$ column matrix with
the $\gamma^i$'s as entries.  By substituting this relation
between the $\tilde \gamma^i$'s and $\gamma^i$'s in
(\ref{auto1})-(\ref{autobeta}), one will have the admissible automorphisms of the fields which one can then use to
glue together the local sheaves of TCDO's over pairwise intersections
$U_{w} \cap U_{w+1}$ for every $w=1,2,\dots,6$. The gluing
relations for the free fields are, in this case, given by
\begin{eqnarray}
\label{auto1SL(3)}
{\tilde \gamma}^i & = & [V^{-1}_{w+1}\cdot V_w] ^i{}_j \ \gamma^j ,\\
\label{auto2SL(3)}
{\tilde A}^a_z & = & A^a_z, \\
\label{autobetaSL(3)} 
{\tilde \beta}_i  & = &   D_i{}^k \beta_k 
+ \partial_z \gamma^j E_{i j}  -   F_{i, a} A^a_z,
\end{eqnarray}
where $i,j,k = 1, 2, 3$. Here, $D$ and $E$ are $3 \times 3$
matrices whereby $[D^{-1}]_i{}^k =  \partial_i
[V^{-1}_{w+1}\cdot V_w] ^k{}_j \ \gamma^j$ and $[E]_{ij} =
\partial_i B_j$; $B = B_l d \gamma^l$ is the non-closed holomorphic $(1,0)$-form on $X$ discussed in $\S$4.5; $F_{i,a}$ is the $i^{th}$-component of a (complex) $\frak h$-valued holomorphic one-form $F_a = F_{l, a} \, d\gamma^l = (\partial_l f_a)  d \gamma^l$ (where $f_a = V^k_a B_k$) on $X$.  Since the underlying sigma model is nonanomalous, one would be able to define the sheaf of TCDO's globally on its $SL(3)/B$ target space. In other words, if we let $R_w$ represent a transformation of the fields in going from $U_{w}$ to $U_{w+1}$, for appropriate choices of $B$ and $F_a$, one ought to be able to show that $(R_6R_5R_4R_3R_2R_1) \cdot \{\gamma^i, \beta_i, A^a_z\} =  \{\gamma^i, \beta_i, A^a_z\}$. 

\bigskip\noindent{\it Global Sections of the Sheaf}

Since $X = SL(3)/B$ has complex dimension three, we will have the relation ${\MA} = \bigoplus_{q_R = 0}^{q_R = 3} H^{q_R}(X, {\widehat \Omega}^{ch,tw}_{X})$. Thus, in order to determine the chiral algebra of the sigma model, one needs to ascertain the global sections  $H^0(X, {\widehat \Omega}^{ch,tw}_{X})$ of the sheaf ${\widehat \Omega}^{ch,tw}_{X}$, and its $\check{ \textrm C}$ech cohomology $H^p(X, {\widehat \Omega}^{ch,tw}_{X})$ for $p=1,2,3$. 

First, let us compute the global sections $H^0({X}, {\widehat \Omega}^{ch,tw}_{X})$. For brevity, we shall again focus on the dimension 0 and 1  operators only; the higher-dimensional cases can be obtained in a similar manner.  At dimension 0, the space of global sections $H^0({X}, {\widehat \Omega}^{ch,tw}_{X; 0})$ must be spanned by regular polynomials in the $\gamma^i$'s, i.e., ${\widehat \Omega}^{ch,tw}_{X; 0}$ is just the sheaf  $\CO_X$ of regular holomorphic functions in the $\gamma^i$'s on $X$. Since all globally-defined regular holomorphic functions on a compact,  connected, complex manifold such as  $X$ are equivalent to constants~\cite{griffith}, we find that  $H^0({X}, {\widehat \Omega}^{ch,tw}_{X; 0})$ is one-dimensional and generated by 1.   

What about the space $H^0(X, {\widehat \Omega}^{ch,tw}_{X; 1})$ of global sections of dimension 1? In order to get a global section of ${\widehat \Omega}^{ch,tw}_{X}$ of dimension 1, we can act on a (gauge-invariant) global section of ${\widehat \Omega}^{ch,tw}_{X}$ of dimension 0 with the partial derivative $\partial_z$. Since $\partial_z 1 = 0$, this prescription will not apply here. 

One could also consider dimension 1 operators of the form $f_i(\gamma) D_z \gamma^i = f_i(\gamma) \partial_z \gamma^i + A^a_z V^i_a(\gamma)$, where $f_i(\gamma)$ and $ V^i_a(\gamma)$ are globally-defined holomorphic functions of the $\gamma^k$'s. The operator $f_i(\gamma) \partial_z \gamma^i $,  by virtue of the way it transforms purely geometrically under (\ref{auto1SL(3)}), would correspond to a section of $\Omega^1_{X}$, the sheaf of holomorphic differential one-forms $f(\gamma) d\gamma$ on $X$; from the classical result $H^0(X, \Omega^1_{X}) = 0$, which continues to hold in the quantum theory, it is clear that $f_i(\gamma) D_z \gamma^i \notin H^0(X, {\widehat \Omega}^{ch,tw}_{X; 1})$.

As before, the remaining possibility at dimension 1 is to find operators that contain the $\beta_i$'s. In fact, from the automorphism relation of (\ref{autobetaSL(3)}), we  immediately find an example since the LHS, $\tilde \beta_i$, is by definition regular in $U_{w+1}$, while the RHS, being a polynomial in $\gamma^i$, $\beta_i$ and $A^a_z$, is manifestly regular in $U_w$. Their being equal means that the $\tilde \beta_i$'s represent dimension 1 global sections of ${\widehat\Omega}^{ch,tw}_{X}$. Since the construction is completely symmetric between $U_w$ and $U_{w+1}$, with $\gamma^i \leftrightarrow \tilde \gamma^i$, $\beta_i \leftrightarrow \tilde \beta_i$ and $A^a_z \leftrightarrow \tilde A^a_z$, a reciprocal formula would imply that the $\beta_i$'s also represent dimension 1 global sections of ${\widehat\Omega}^{ch,tw}_{X}$.  As in the previous $SL(2)$ example, one can generate more global sections by computing the OPE's between the $\tilde \beta_i$'s and $\beta_i$'s.  In the end, we get the following dimension 1 operators:
\begin{eqnarray}
\label{OPE-SL(3)}
J_{e_1} & = & \beta_1, \nonumber \\
J_{e_2} & =  & \beta_2 - \gamma^1 \beta_3, \nonumber \\
 J_{e_3} & =  & \beta_3, \nonumber \\
  J_{h_1} & =  &  V^1_a A^a_z + 2 \gamma^1 \beta_1 - \gamma^2 \beta_2  + \gamma^3\beta_3,  \\
   J_{h_2} & =  & V_a^2 A^a_z - \gamma^1\beta_1 + 2 \gamma^2\beta_2 + \gamma^3 \beta_3,  \nonumber \\
    J_{f_1} & =  &   - V_a^1A^a_z \gamma^1 + 3\partial_z \gamma^1 + \gamma^3 \beta_2 - (\gamma^1)^2 \beta_1 + \gamma^1 \gamma^2 \beta_2 - \gamma^1 \gamma^3 \beta_3,    \nonumber \\
     J_{f_2} & =  &   - V_a^2 A^a_z \gamma^2 +2 \partial_z\gamma^2 - \gamma^3 \beta_1 - (\gamma^2)^2 \beta_2,   \nonumber \\
      J_{f_3}  & =  &  - V^3_a A^a_z \gamma^3 - V_a^2 A^a_z \gamma^1 \gamma^2 + 3 \partial_z \gamma^3 + 2 \gamma^1 \partial_z \gamma^2 - \gamma^1 \gamma^3 \beta_1 - \gamma^2 \gamma^3 \beta_2  -  \gamma^3 \gamma^3 \beta_3 -  \gamma^1(\gamma^2)^2 \beta_2, \nonumber 
\end{eqnarray} 
where $V^1 = {1 \over \sqrt 2} (1, \sqrt 3)$, $V^2 =  {1 \over \sqrt 2} (1, - \sqrt 3)$, and $V^3 = V^1 + V^2$. As usual, normal-ordering is understood in the above formulas. One can verify that $\{J_{e_1}, J_{e_2}, J_{e_3}, J_{h_1}, J_{h_2},  J_{f_1}, J_{f_2}, J_{f_3} \}$ indeed satisfy a closed OPE algebra; they are thus dimension 1 operators which span $H^0(X, {\widehat\Omega}^{ch,tw}_{X; 1})$, as desired. Moreover, the closed OPE algebra in question is that of an affine $SL(3)$-algebra $\widehat {\frak {sl}_3}$ at the critical level of $-3$ in the Wakimoto representation; this is in perfect agreement with Lemma~4.6 of~\cite{Arakawa}. Since $\{J_{e_1}, J_{e_2}, J_{e_3}, J_{h_1}, J_{h_2},  J_{f_1}, J_{f_2}, J_{f_3} \}$  are chiral vertex operators holomorphic in $z$, one can expand them in a Laurent series that allows an affinization of the $SL(3)$ Lie algebra generated by their resulting zero modes. Consequently, $H^0(X, {\widehat\Omega}^{ch,tw}_{X})$ would be a Wakimoto module for the affine algebra of $SL(3)$ at level $-3$.

\bigskip\noindent{\it The Center of a Chiral Algebra}
         
As in the previous $SL(2)$ example, one can, in our case at hand of
 $\widehat  {\frak g}_{\mathbb C} = \hat{\frak {sl}_3}$ whence $h^{\vee}
=3$, define the quantum stress tensor at level $k \neq - h^{\vee}$ as
\be T(z) = {{: d^{\alpha \zeta}J_\alpha J_\zeta(z) :} \over {k+3}}, 
\label{SS def T(z) for SL(3)} 
\ee
where $d^{\alpha \zeta}$ is the inverse of the Cartan-Killing metric of $\frak {sl}_3$, and $\alpha, \zeta = 1, 2, \dots, {\rm dim}_{\mathbb C} (\frak {sl}_3) = 8$. As required, for
every $k \neq {-3}$, the modes of the Laurent expansion of $T(z)$
will span a Virasoro algebra. In particular, $T(z)$ will generate
holomorphic reparameterizations of the coordinates on the
worldsheet $\Sigma$. Notice that this definition of $T(z)$ in
(\ref{SS def T(z) for SL(3)}) is ill-defined when $k=-3$.
Nevertheless,  one can always associate $T(z)$ with the
Segal-Sugawara operator $S(z)$ that is well-defined at any level,
where
\be S(z) =  (k+3) T(z), 
\label{S(z) for SL(3)} 
\ee 
and 
\be
S(z) = {: d^{\alpha \zeta} J_\alpha J_\zeta:} (z). 
\label{s(z) for sl(3)} 
\ee 
From (\ref{S(z) for SL(3)}), we see that $S(z)$ generates, in its OPE
with other field operators, $(k+3)$ times the field transformations
usually generated by the stress tensor $T(z)$. Therefore, at the critical level $k= -3$, $S(z)$ generates \emph{no}
field transformations at all -- its OPE with all other field operators ought  to be regular at best. This is equivalent to saying that the quantum
stress tensor does not exist in $\MA$ at $k =-3$, since $S(z)$ is the
only well-defined operator at this critical level which can generate field transformations under arbitrary
holomorphic reparameterizations of the worldsheet coordinates on
$\Sigma$. This observation is consistent with (\ref{tzzanomaly}), as $R_{i \bar j} (X) \neq 0$. 

$S(z)$ may or may not vanish identically at $k=-3$, and if it does not, it will merely play the role of a \emph{classical} field (i.e.,~$c$-number) that has no nontrivial interactions with itself and other fields at the quantum level. In fact, by substituting (\ref{OPE-SL(3)}) in (\ref{s(z) for sl(3)}), a careful computation (with the aid of the relevant affine $SL(3)$ OPE algebra at level $-3$ obeyed by the $J$'s) would give 
\be
S(z) = {1 \over 2} \left[ (d_{ab}A^a_z A^b_z) (z) -  \sum_{i=1}^3 (V^i_a \partial_z A^a_z)(z) \right],
\label{S(z)-A for sl(3)}
\ee
where $a,b = 1, \dots, {\rm dim}_{\mathbb C} ({\frak h}_{\mathbb C}) = 2$; ${\frak h}_{\mathbb C}$ being the Cartan subalgebra of $\frak {sl}_3$. Since the $A^a_z$'s are non-dynamical fields which do not have any nontrivial propagators with themselves or with the rest of the other fields $\beta_i$ and $\gamma^i$ (and their $z$-derivatives) that define $\MA$, $S(z)$ indeed plays the role of a purely classical field.

For an affine $SL(N)$-algebra where $N > 2$,  one can generalize
the Sugawara formalism to construct higher-spin analogs of the
holomorphic stress tensor using the currents. These higher-spin
analogs have conformal weights $3, 4, \dots N$. These higher-spin
analogs are called Casimir operators, and were first constructed
in \cite{casimir operators}.

In the context of our affine $SL(3)$-algebra with a module that is
furnished by the global sections of the sheaf of TCDO's on
$X=SL(3)/B$, a spin-three analog of the holomorphic stress tensor
will be given by the 3rd-order Casimir operator~\cite{review}
 \be
T^{(3)} (z) = { {:{\tilde d}^{\alpha \zeta \xi } (k) (J_\alpha (J_\zeta J_\xi)):(z)} \over {k+3}}, 
\label{casimir of spin-three for SL(3)} 
\ee 
where ${\tilde d}^{\alpha \zeta \xi} (k)$ is a completely symmetric traceless
$\frak{sl}_3$-invariant tensor of rank 3 that depends on the level
$k$ of the affine $SL(3)$ algebra in question.  ${\tilde
d}^{abc}(k)$ is also well-defined and finite at $k=-3$. The
superscript on $T^{(3)}(z)$ just denotes that it is a spin-three
analog of $T(z)$.

As with $T(z)$ in (\ref{SS def T(z) for SL(3)}), $T^{(3)} (z)$ is
ill-defined when $k = -3$. Nevertheless, one can always make
reference to a higher-spin analog $S^{(3)}(z)$ of the Segal-Sugawara tensor that is well-defined for any finite value of $k$, where 
\be
 S^{(3)}(z) = (k+3) T^{(3)}(z), 
 \label{act}
\ee
whence
\be S^{(3)}(z) = {:{\tilde d}^{\alpha \zeta \xi } (k) (J_\alpha (J_\zeta J_\xi)):(z)}.
\label{S^{(3)}(z)} \ee From (\ref{act}), one can see that the operator $S^{(3)}(z)$
generates in its OPE with all other operators of the quantum
theory,  $(k+3)$ times the field transformations typically
generated by $T^{(3)}(z)$. Therefore, at the critical level $k=-3$, $S^{(3)}(z)$  generates \emph{no} field transformations at all -- its OPE with all other field operators ought  to be regular at best. This is equivalent to saying that $T^{(3)}(z)$ does not exist as a quantum operator in $\MA$ at $k = -3$, since $S^{(3)}(z)$ is the only well-defined operator at this critical level which can  generate the type of field transformations associated with $T^{(3)}(z)$.

$S^{(3)}(z)$ may or may not vanish identically at $k=-3$, and if it does not, it will merely play the role of a \emph{classical} field (i.e.,~$c$-number) that has no nontrivial interactions with itself and other fields at the quantum level. In fact, by substituting (\ref{OPE-SL(3)}) in (\ref{S^{(3)}(z)}), a careful computation (with the aid of the relevant affine $SL(3)$ OPE algebra at level $-3$ obeyed by the $J$'s) would give (cf.~\cite{review})
\begin{eqnarray}
S^{(3)}(z) &= &   \sqrt {- {6 \over 15}} \sum_{i < j < k}  \left( (\epsilon_i \cdot A_z) (\epsilon_j \cdot A_z) (\epsilon_k \cdot A_z) \right) (z) - \sum_{i < j} (i-1) \partial_z \left( (\epsilon_i \cdot A_z) (\epsilon_j \cdot A_z) \right) (z) \nonumber \\
&& -\sum_{i < j} (j - i-1) \partial_z \left( (\epsilon_i \cdot A_z) (\epsilon_j \cdot \partial_z A_z) \right) (z) + {1 \over 2} \sum_i (i-1) (i-2) (\epsilon_i \cdot \partial^2_z A_z) (z) \nonumber \\
&& - {1 \over 4} \partial_z (A_z \cdot A_z) (z) - {1 \over 2} \sum_i (i-1)\epsilon_i \cdot A_z (z),
\label{S^3(z)-A for sl(3)}
\end{eqnarray}
where the dot-product refers to an inner product with respect to the Cartan-Killing metric on ${\frak h}_{\mathbb C}$, and $\{ \epsilon_i, \, i = 1, 2, 3\}$ is a set of weights of the vector representation of $SL(3)$ normalized such that $\epsilon_i \cdot \epsilon_j = \delta_{ij} - {1 \over 3}$ and $\sum_i \epsilon_i = 0$. Since the $A^a_z$'s are non-dynamical fields which do not have any nontrivial propagators with themselves or with the rest of the other fields $\beta_i$ and $\gamma^i$ (and their $z$-derivatives) that define $\MA$, $S^{(3)}(z)$ indeed plays the role of a purely classical field.

The fact that $S(z)$ and $S^{(3)}(z)$ only have regular OPE's with all other relevant fields and with themselves implies the following. If we denote by $V_{-3} (\frak {sl}_3)$ the chiral algebra generated by the $J_e$'s, $J_h$'s, $J_f$'s and their $z$-derivatives, and if we denote by $\frak z(V_{-3} (\frak {sl}_3))$ its center which is defined to be the set of fields which have \emph{regular }OPE's with all other fields within, we have $\frak z(V_{-3} (\frak {sl}_3)) = \mathbb C[\partial^m_z S(z) + \partial^n_z S^{(3)}(z)]_{m, n\geq 0}$, where $\mathbb C[\partial^m_z S(z) + \partial^n_z S^{(3)}(z)]_{m, n\geq 0}$ is the space of differential polynomials on $S(z)$ and $S^{(3)}(z)$ with complex coefficients. This result agrees with Theorem~8 of~\cite{Frenkel}. In turn, (\ref{S(z)-A for sl(3)}) and (\ref{S^3(z)-A for sl(3)}) mean that $\frak z(V_{-3} (\frak {sl}_3)) \subset H_{X}$, where $ H_{X}$ is the space of differential polynomials on the $A^a_z$'s with complex coefficients. This result agrees with Lemma~4.6 of~\cite{Arakawa}. (Recall from $\S$5.2.1 that $A_z$ corresponds to $\lambda^*$ of\emph{ loc.~cit.}.)

Moreover, notice that since $S(z)$, $S^{(3)}(z)$ and the $J(z)$'s are holomorphic in $z$ and are of dimension 2, 3 and 1, respectively,  one can Laurent expand them as 
\be S(z) = \sum_{n \in {\mathbb Z}} {S}_n z^{-n-2}, \quad  S^{(3)}(z) = \sum_{n \in {\mathbb Z}} {S}^{(3)}_n z^{-n-3}, \quad J^\alpha(z) = \sum_{n \in \mathbb Z} J^\alpha_{n} z^{-n-1},
\label{S(z) and S^3(z) expansion}
\ee 
where $\alpha = \{e_i, h_j, f_i\}$. Again, the fact that $S(z)$ and $S^{(3)}(z)$ only have regular OPE's with all other relevant fields and with themselves implies that 
\be
[{S}_n, J^\alpha_m] = [{S}_n, {S}_m] =0 \qquad {\rm and} \qquad [{S}^{(3)}_n, J^\alpha_m] =  [{S}^{(3)}_n, {S}^{(3)}_m] = [{S}^{(3)}_n, {S}_m] = 0.
\label{center sl(3)}
\ee 
This means that the $S_n$'s and $S^{(3)}_n$'s generate the center ${\frak z}(\tilde U_{-3} (\hat{{\frak sl}_3}))$ of the completed enveloping algebra $\tilde U_{-3} (\hat{{\frak sl}_3})$ of $\hat{{\frak sl}_3}$ at critical level $-3$. As in the previous example,  ${\frak z}(\tilde U_{-3} (\hat{{\frak sl}_3}))$ would also furnish a Poisson algebra.

\bigskip\noindent{\it The Higher Cohomologies}

Let us proceed to determine the higher cohomology groups $H^p(X, {\widehat \Omega}^{ch,tw}_{X})$ where $p \geq 1$. In dimension 0, we can again consider regular polynomials in the $\gamma^i$'s. However, from geometric representation theory, we have the classical result $H^p(\mathscr X, {\cal O}) = 0$ when $p \geq 1$, where $\cal O$ is the sheaf of regular functions over an arbitrary flag manifold $\mathscr X$ of $G_{\mathbb C}$ that are holomorphic in the $\gamma^i$'s~\cite{nash}. Since a vanishing cohomology at the classical level continues to vanish at the quantum level, it would mean that we cannot have regular polynomials in the $\gamma^i$'s; i.e., $H^p(X, {\widehat \Omega}^{ch,tw}_{X;0}) = \emptyset$. In other words, the higher cohomologies start ``growing'' at dimension greater than zero, like in the previous $SL(2)$ example. 

Let us now compute the first cohomology $H^1(X, {\widehat \Omega}^{ch,tw}_{X;1})$ of operators of dimension 1. From the fact that $T_{zz}$ cannot be expressed as a total $z$-derivative (else $L_{-1} = 0$), and the relation (\ref{tzzanomaly}), we find that the dimension 1 fermionic sigma model operator $R_{i \bar j} D_z \phi^i \psi^{\bar j}$ with $q=1$ is not $Q$-exact. Moreover, from the nilpotency of $Q$, and  the relation (\ref{tzzanomaly}), we find that  $R_{i \bar j} D_z \phi^i \psi^{\bar j}$ is $Q$-closed. Hence, from our $Q$-$\check {\rm C}$ech cohomology dictionary, we find that $\mathscr R \in H^1(X, {\widehat \Omega}^{ch,tw}_{X;1})$, where  $\mathscr R$ is the $\check {\rm C}$ech cohomology counterpart of the sigma model operator $R_{i \bar j} D_z \phi^i \psi^{\bar j}$. ($\mathscr R$ is just the $SL(3)$ analog of $\Theta$ of the previous $SL(2)$ example.) What about  the space $H^1(X, {\widehat \Omega}^{ch,tw}_{X;2})$ of operators at dimension 2? Let us try to differentiate $\mathscr R$, i.e., let us consider the operator $\partial_z \mathscr R$. From (\ref{tzzanomaly}), one can see that $\partial_z \mathscr R$ would correspond to a $Q$-exact sigma model operator. Thus, from our $Q$-$\check {\rm C}$ech cohomology dictionary, we conclude that $\partial_z\mathscr R  \notin H^1(X, {\widehat\Omega}^{ch,tw}_{X;2})$.  Nevertheless, since  we have the product formula $H^1(X, {\widehat\Omega}^{ch,tw}_{X; l}) \otimes H^0(X, {\widehat\Omega}^{ch,tw}_{X; m}) \to H^1(X, {\widehat\Omega}^{ch,tw}_{X; l+m})$, we can act $\mathscr R \in H^1(X, {\widehat\Omega}^{ch,tw}_{X;1})$ on every other element of  $H^0(X, {\widehat\Omega}^{ch,tw}_{X})$ to generate $H^1(X, {\widehat\Omega}^{ch,tw}_{X})$: for example, $\mathscr R \cdot 1 = \mathscr R$ is an element of $H^1(X, {\widehat\Omega}^{ch,tw}_{X})$ of dimension 1, and $\mathscr R \cdot \{J_{e_1}, J_{e_2}, J_{e_3}, J_{h_1}, J_{h_2},  J_{f_1}, J_{f_2}, J_{f_3} \}$  is a set of elements of $H^1(X, {\widehat\Omega}^{ch,tw}_{X})$ of dimension 2, and so on. Hence, $H^1(X, {\widehat \Omega}^{ch,tw}_{X})$ would also be a Wakimoto  module for $\widehat{\frak {sl}_3}$ at the critical level.

Let us now compute the  second cohomology $H^2(X, {\widehat \Omega}^{ch,tw}_{X;1})$ of operators of dimension 1. One could consider operators of the form $f_i(\gamma) D_z \gamma^i = f_i(\gamma) \partial_z \gamma^i + A^a_z V^i_a(\gamma)$, where $f_i(\gamma)$ and $V^i_a(\gamma)$ are holomorphic functions of the $\gamma^k$'s. Note that by virtue of the way $A^a_z V^i_a(\gamma)$ transforms purely geometrically under (\ref{auto1SL(3)}) and (\ref{auto2SL(3)}), it would correspond to a section of the sheaf of holomorphic zero-forms $\CO_X$ on $X$. By virtue of the way $f_i(\gamma) \partial_z \gamma^i$ transforms purely geometrically under (\ref{auto1SL(3)}), it would correspond to a section of the sheaf $\Omega^1_{X}$ of holomorphic differential one-forms on $X$. From the classical result $H^2(X, \Omega^1_{X}) = 0 = H^2(X, \CO_X)$, which continues to hold in the quantum theory, it is clear that $f_i(\gamma) D_z \gamma^i \notin H^2(X, {\widehat \Omega}^{ch,tw}_{X;1})$. 

What about the  second cohomology $H^2(X, {\widehat \Omega}^{ch,tw}_{X;2})$ of operators of dimension 2? From footnote~24, we learn that we have a nonzero class $c_2^{T_C} (X) \in H^2(X, \Omega^{2,cl}_{X,T_C})$. This implies that one can consider dimension 2 gauge-invariant operators of the form $ \mathscr F = f_{kj} (\gamma) D_z \gamma^k D_z \gamma^j$ where $\partial_{[i}f_{kj]} = 0$. (By the way these operators transform purely geometrically under (\ref{auto1SL(3)}), one can see that they do indeed correspond to $\partial$-closed $T_C$-equivariant two-forms $ \Omega^{2,cl}_{X,T_C}$ on $X$.) Since there are no quantum relations analogous to (\ref{fraktzanomaly}) in the second cohomology, we conclude that $\mathscr F \in H^2(X, {\widehat \Omega}^{ch,tw}_{X;2})$. How about at dimension 3? Well, since we have the classical result $H^2(X, K_X) = 0$, where $K_X$ is the sheaf of $(3,0)$-forms on $X$, and since a vanishing cohomology at the classical level continues to vanish at the quantum level, we cannot have operators of the form $f_{ijk}(\gamma) D_z \gamma^i D_z \gamma^j D_z \gamma^k$. Nevertheless, since we have  the product formula $H^2(X, {\widehat\Omega}^{ch,tw}_{X; l}) \otimes H^0(X, {\widehat\Omega}^{ch,tw}_{X; m}) \to H^2(X, {\widehat\Omega}^{ch,tw}_{X; l+m})$, we can act $\mathscr F \in H^2(X, {\widehat\Omega}^{ch,tw}_{X;2})$ on every other element of  $H^0(X, {\widehat\Omega}^{ch,tw}_{X})$ to generate $H^2(X, {\widehat\Omega}^{ch,tw}_{X})$: for example,  $\mathscr F \cdot 1 = \mathscr F$  is an element of  $H^2(X, {\widehat\Omega}^{ch,tw}_{X})$ of dimension 2, $\mathscr F \cdot \{J_{e_1}, J_{e_2}, J_{e_3}, J_{h_1}, J_{h_2},  J_{f_1}, J_{f_2}, J_{f_3} \}$ is a  set of elements of $H^2(X, {\widehat\Omega}^{ch,tw}_{X})$ of dimension 3, and so on. Hence, $H^2(X, {\widehat \Omega}^{ch,tw}_{X})$ would also be a Wakimoto  module for $\widehat{\frak {sl}_3}$ at the critical level.

Similarly, for the third cohomologies $H^3(X, {\widehat \Omega}^{ch,tw}_{X; 1})$ and $H^3(X, {\widehat \Omega}^{ch,tw}_{X; 2})$ of operators of dimensions 1 and 2, respectively, the classical results $H^3(X, \Omega^1_{X})= 0$ and $H^3(X, \Omega^2_{X})= 0$ imply that they are empty. What about the third cohomology $H^3(X, {\widehat \Omega}^{ch,tw}_{X;3})$ of operators at dimension 3? From the classical result $H^3(X, K^{T_C}_X) \neq 0$, one could consider gauge-invariant operators of the form $\mathscr G = f_{ijk}(\gamma) D_z \gamma^i D_z \gamma^j D_z \gamma^k$. As there are no quantum relations analogous to (\ref{fraktzanomaly}) in the third cohomology, we conclude that $\mathscr G \in H^3(X, {\widehat \Omega}^{ch,tw}_{X;3})$. Since we have  the product formula $H^3(X, {\widehat\Omega}^{ch,tw}_{X; l}) \otimes H^0(X, {\widehat\Omega}^{ch,tw}_{X; m}) \to H^3(X, {\widehat\Omega}^{ch,tw}_{X; l+m})$, we can act $\mathscr G \in H^3(X, {\widehat\Omega}^{ch,tw}_{X;3})$ on every other element of  $H^0(X, {\widehat\Omega}^{ch,tw}_{X})$ to generate $H^3(X, {\widehat\Omega}^{ch,tw}_{X})$: for example,  $\mathscr G \cdot 1 = \mathscr G$  is an element of  $H^3(X, {\widehat\Omega}^{ch,tw}_{X})$ of dimension 3, $\mathscr G \cdot \{J_{e_1}, J_{e_2}, J_{e_3}, J_{h_1}, J_{h_2},  J_{f_1}, J_{f_2}, J_{f_3} \}$ is a  set of elements of $H^3(X, {\widehat\Omega}^{ch,tw}_{X})$ of dimension 4, and so on. Hence, $H^3(X, {\widehat \Omega}^{ch,tw}_{X})$ would also be a Wakimoto  module for $\widehat{\frak {sl}_3}$ at the critical level.

Last but not least, notice that the zeroth, first, second and third cohomologies -- all of which are Wakimoto modules for $\widehat{\frak {sl}_3}$ at the critical level -- start ``growing'' at dimensions 0, 1, 2 and 3, respectively. This observation is also consistent with the  representation-theoretic results of~\cite{chiral bott}.\footnote{To understand this claim, first note that the series $\CA(X, q) =\sum^{{\rm dim}_{\mathbb C}X}_{{i=0}}  (-1)^{i} [\sum_{n \geq 0}q^{n} {\mathrm{dim}} {H}^{i}(X, {\widehat {\Omega}}^{ch, tw}_{X; n})]$ just corresponds to $\chi(\CL^{ch}_{\nu(z)}) = \sum_{w \in W} (-1)^{l(w)}  {\rm ch} {\mathbb W}^w_{\nu(z)}$ of (4.6) of~\cite{chiral bott}, where $W$ is the Weyl group of $\frak{g}_{\mathbb C}$, $l(w) = i$ is the length of $w \in W$, and ${\rm ch}{\mathbb W}^w_{\nu(z)}$ is the character of the $w$-twisted Wakimoto module ${\mathbb W}^w_{\nu(z)}$ of critical level and central character $\nu(z) = \nu_0/z + \nu_{-1} + \nu_{-2} z + \dots$. For $y, v \in W$,  $l(y) = l(v) +1$ means that $y > v$; moreover, if  $y > v$, the conformal weight of the highest weight vector ${\bf 1}_{y \circ \nu_0} \in {\mathbb W}^y_{\nu(z)}$  is strictly greater than the conformal weight of the highest weight vector ${\bf 1}_{v \circ \nu_0} \in { \mathbb W}^v_{\nu(z)}$. Altogether, this implies that the lowest (scaling) dimension of operators in $H^i(X, {\widehat {\Omega}}^{ch, tw}_{X})$ increases with increasing cohomological degree $i$.} Furthermore, since $\nu(z)$ of \emph{loc.~cit. }can be identified with the holomorphic component $A_z(z)$ of the gauge field, according to our discussion in the third last paragraph of $\S$5.2.1, these modules would necessarily have highest weight 0.  

\newsubsubsection{The Sheaf of TCDO's on the Flag Manifold of $G_{\mathbb C}$}

What if $X$ is a flag manifold of an arbitrary $G_{\mathbb C}$ with Lie algebra ${\frak g}_{\mathbb C}$ of rank $l$? From a Cartan decomposition, we can write ${\frak g}_{\mathbb C} = {\frak n}_+ \oplus {\frak b}$, where $\frak b$ is a Borel subalgebra, and ${\frak n}_+$ is the subalgebra of upper triangular nilpotent matrices. Since $X = G_{\mathbb C} / B$, where $B \subset G_{\mathbb C}$ is the Borel subgroup with Lie algebra $\frak b$, we have ${\rm dim}_{\mathbb C} X = {\rm dim}_{\mathbb C}({\frak n}_+) = |\Delta_+|$, where $\Delta_+$ is the set of positive roots of ${\frak g}_{\mathbb C}$. Thus, each chart $U$ in $X$ can be identified with the affine space $\mathbb C^{ |\Delta_+|}$. Consequently, the sheaf of TCDO's in any $U \subset X$ can be described by the following perturbed free $\beta\gamma$ system with action
\be
I= {1\over 2\pi}\int|d^2z| \  \sum_{i=1}^{ |\Delta_+|} \left(
\beta_i \partial_{\bar z} \gamma^i -  \sum_{a, b =1}^l  {A}^a_z  V_ {i a} A^b_{\bar z}  V^i_b   \right),
\label{actionUG}
\ee
where the only singular OPE of this system is 
\be
\beta_i(z) \gamma^j(z')  \sim  -{\delta_i^j\over z-z'}.
\ee 

Similarly, the sheaf of TCDO's in a neighboring intersecting chart $\tilde U \subset X$ can be described by the following perturbed free $\tilde\beta\tilde\gamma$ system with action 
\be
\tilde I= {1\over 2\pi}\int|d^2z| \  \sum_{i=1}^{ |\Delta_+|} \left(
{\tilde\beta}_i \partial_{\bar z} {\tilde\gamma}^i -  \sum_{a, b =1}^l {\tilde A}^a_z  V_ {i a} {\tilde A}^b_{\bar z}  V^i_b   \right),
\label{actionUG -2}
\ee
where the only singular OPE of this system is
\be
{\tilde \beta}_i(z) {\tilde \gamma}^j(z')  \sim  -{\delta_i^j\over z-z'}.
\ee 

In order to describe a globally-defined sheaf of TCDO's, one will need to glue the free conformal field theories with actions (\ref{actionUG}) and (\ref{actionUG -2})  over all pairwise intersections $U \cap \tilde U$. To do so, one must again use the admissible automorphisms of the free
conformal field theories defined in (\ref{auto1})-(\ref{autobeta}) to glue the free-fields together. Since the underlying sigma model is nonanomalous, one ought to be able to define the sheaf of TCDO's globally on $X$. 

\bigskip\noindent{\it Global Sections of the Sheaf}

As $X = G_{\mathbb C} /B$ has complex dimension $ |\Delta_+|$, we have ${\MA} = \bigoplus_{q_R = 0}^{q_R =  |\Delta_+|} H^{q_R}(X, {\widehat \Omega}^{ch,tw}_{X})$. Thus, in order to determine the chiral algebra of the sigma model, one needs to ascertain the global sections  $H^0(X, {\widehat \Omega}^{ch,tw}_{X})$ of the sheaf ${\widehat \Omega}^{ch,tw}_{X}$, and its $\check{ \textrm C}$ech cohomology $H^p(X, {\widehat \Omega}^{ch,tw}_{X})$ for $p=1,2, \dots,  |\Delta_+|$. 

First, let us compute the global sections $H^0({X}, {\widehat \Omega}^{ch,tw}_{X})$. For brevity, we shall again focus on the dimension 0 and 1  operators only; the higher-dimensional cases can be obtained in a similar manner.  At dimension 0, the space of global sections $H^0({X}, {\widehat \Omega}^{ch,tw}_{X; 0})$ must be spanned by regular polynomials in the $\gamma^i$'s, i.e., ${\widehat \Omega}^{ch,tw}_{X; 0}$ is just the sheaf  $\CO_X$ of regular holomorphic functions in the $\gamma^i$'s on $X$. Since all globally-defined regular holomorphic functions on a compact, connected, complex manifold such as  $X$ are equivalent to constants~\cite{griffith}, we find that  $H^0({X}, {\widehat \Omega}^{ch,tw}_{X; 0})$ is one-dimensional and generated by 1.   

What about the space $H^0(X, {\widehat \Omega}^{ch,tw}_{X; 1})$ of global sections of dimension 1? In order to get a global section of ${\widehat \Omega}^{ch,tw}_{X}$ of dimension 1, we can act on a (gauge-invariant) global section of ${\widehat \Omega}^{ch,tw}_{X}$ of dimension 0 with the partial derivative $\partial_z$. Since $\partial_z 1 = 0$, this prescription will not apply here. 

One could also consider dimension 1 operators of the form $f_i(\gamma) D_z \gamma^i = f_i(\gamma) \partial_z \gamma^i + A^a_z V^i_a(\gamma)$, where $f_i(\gamma)$ and $ V^i_a(\gamma)$ are globally-defined holomorphic functions of the $\gamma^k$'s. The operator $f_i(\gamma) \partial_z \gamma^i $,  by virtue of the way it transforms purely geometrically under (\ref{auto1}), would correspond to a section of $\Omega^1_{X}$, the sheaf of holomorphic differential one-forms $f(\gamma) d\gamma$ on $X$; from the classical result $H^0(X, \Omega^1_{X}) = 0$, which continues to hold in the quantum theory, it is clear that $f_i(\gamma) D_z \gamma^i \notin H^0(X, {\widehat \Omega}^{ch,tw}_{X; 1})$.

As before, the remaining possibility at dimension 1 is to find operators that contain the $\beta_i$'s. To this end, consider the operators
\begin{eqnarray}
\label{current chevalley}
J_{e_{\alpha_i}} &= & \beta_{\alpha_i} + \sum_{j=1}^{|\Delta_+|}
: P^i_j(\gamma) \beta_j:, \quad i = 1, 2, \dots, l; \nonumber \\ 
J_{h_a} &= & - \sum_{j=1}^{|\Delta_+|} D_j :
\gamma^j \beta_j: + A^a_z, \quad a = 1, 2, \dots, l;  \\
J_{f_{\alpha_i}} &= & \sum_{j = 1}^{|\Delta_+|} :
Q^i_j(\gamma) \beta_j : + C_i \partial_z
\gamma^{\alpha_i} + A^a_z \gamma^{\alpha_i}, \quad a = i = 1, 2, \dots, l. \nonumber
\end{eqnarray}
Here, $\alpha_i \in \Delta_+$ are \emph{simple} roots, the $D_i$'s and $C_i$'s are complex constants, and for appropriate choices of the polynomials $ P^i_j(\gamma)$ and $Q^i_j(\gamma)$, the set $\{J_{e_{\alpha_i}}, J_{h_a}, J_{f_{\alpha_i}}\}$ furnishes an OPE algebra of an affine $G_{\mathbb C}$-algebra $\widehat {\frak {g}}_{\mathbb C}$ (in the Chevalley basis) at the critical level $-h^\vee$ in the  Wakimoto representation (see Theorem~4.7 of~\cite{waki-frenkel}). However, does the set  $\{J_{e_{\alpha_i}}, J_{h_a}, J_{f_{\alpha_i}}\}$  span $H^0(X, {\widehat\Omega}^{ch,tw}_{X; 1})$? The answer according to Lemma~4.6 of~\cite{Arakawa}, is ``yes''.  Moreover, since $\{J_{e_{\alpha_i}}, J_{h_a}, J_{f_{\alpha_i}}\}$  are chiral vertex operators holomorphic in $z$, one can expand them in a Laurent series that allows an affinization of the underlying Lie algebra ${\frak g}_{\mathbb C}$ generated by their resulting zero modes. Consequently, $H^0(X, {\widehat\Omega}^{ch,tw}_{X})$ would be a Wakimoto module for ${\widehat{\frak g}}_{\mathbb C}$ at level $-h^{\vee}$.

\bigskip\noindent{\it The Center of a Chiral Algebra}

As in the previous $SL(2)$ and $SL(3)$ examples, one can, in the general case at hand, define the quantum spin-2 stress tensor at level $k \neq - h^{\vee}$ as
\be T^{(2)}(z) = {{: d^{\alpha \zeta} J_{\alpha} J_{\zeta}:} \over {k+h^\vee}}, 
\label{SS def T(z) for G} 
\ee
where $d^{\alpha \zeta}$ is the inverse of the Cartan-Killing metric of $\frak g_{\mathbb C}$, and $J_{\alpha}$ and $J_{\zeta}$ are currents of $\widehat {\frak {g}}_{\mathbb C}$ expressed in the standard basis. As required, for
every $k \neq {-h^\vee}$, the modes of the Laurent expansion of $T^{(2)}(z)$
will span a Virasoro algebra. In particular, $T^{(2)}(z)$ will generate
holomorphic reparameterizations of the coordinates on the
worldsheet $\Sigma$. Notice that this definition of $T(z)$ in
(\ref{SS def T(z) for G}) is ill-defined when $k=-h^\vee$.
Nevertheless,  one can always associate $T^{(2)}(z)$ with the spin-2
Segal-Sugawara operator $S^{(2)}(z)$ that is well-defined at any level,
where
\be S^{(2)}(z) =  (k+h^\vee) T(z), 
\label{S(z) for G} 
\ee 
and 
\be
S^{(2)}(z) ={: d^{\alpha \zeta} J_{\alpha} J_{\zeta}:} (z). 
\label{s(z) for G} 
\ee 
From (\ref{S(z) for G}), we see that $S^{(2)}(z)$ generates, in its OPE
with other field operators, $(k+h^\vee)$ times the field transformations
usually generated by the stress tensor $T^{(2)}(z)$. Therefore, at the critical level $k= -h^\vee$, $S^{(2)}(z)$ generates \emph{no}
field transformations at all -- its OPE with all other field operators ought  to be regular. This is equivalent to saying that the quantum
stress tensor does not exist in $\MA$ at $k =-h^\vee$, since $S^{(2)}(z)$ is the
only well-defined operator at this critical level which can generate field transformations under arbitrary
holomorphic reparameterizations of the worldsheet coordinates on
$\Sigma$. This observation is consistent with (\ref{tzzanomaly}), as $R_{i \bar j} (X) \neq 0$. 

One can generalize the Sugawara formalism to construct higher-spin analogs of the
holomorphic stress tensor $ T^{(2)}(z)$ by using the currents of $\widehat {\frak {g}}_{\mathbb C}$.  These higher-spin
analogs are called Casimir operators. In particular, a spin-$s_i$ analog 
will be given by the ${s_i}^{th}$-order Casimir operator
\cite{review}
 \be
T^{(s_i)} (z) = {{:{\tilde d}^{\zeta_1 \zeta_2 \zeta_3 \dots \zeta_{s_i}} (k)
(J_{\zeta_1} J_{\zeta_2}\dots J_{\zeta_{s_i}})(z):} \over {k+h^{\vee}}},
\label{casimir of spin-three for G} \ee where ${\tilde
d}^{\zeta_1\zeta_2 \zeta_3 \dots \zeta_{s_i}}(k)$ is a completely symmetric
traceless ${\frak g}_{\mathbb C}$-invariant tensor of rank $s_i$ that
depends on the level $k$ of $\widehat {\frak {g}}_{\mathbb C}$. ${\tilde
d}^{\zeta_1\zeta_2 \zeta_3 \dots \zeta_{s_i}}(k)$  is also
well-defined and finite at $k=-h^{\vee}$. Note that $i = 1,2, \dots, l$,
and  the spins $s_i = 1+ e_i$, where the $e_i$'s are the exponents of $\frak g_{\mathbb C}$. Thus, one can have $l$ of
these Casimir operators, and the spin-2 Casimir operator is just
the holomorphic stress tensor $T^{(2)} (z) $ from the usual Sugawara
construction. For example, in the $SL(2)$ and $SL(3)$ cases of  $l= 1$ and $l = 2$ discussed earlier, the exponents are $e_1 = 1$ and $\{e_1, e_2\} = \{1, 2\}$, respectively. Consequently, we had, in the $SL(2)$ case, $T^{(2)}(z) = T(z)$ given in (\ref{SS def T(z)}); we also had, in the $SL(3)$ case, $T^{(2)}(z) = T(z)$ and $T^{(3)}(z)$ given in (\ref{SS def T(z) for SL(3)}) and (\ref{casimir of spin-three for SL(3)}).

As with $T^{(2)}(z)$ in (\ref{SS def T(z) for G}), $T^{(s_i)} (z)$
is ill-defined when $k = -h^{\vee}$. Nevertheless, one can always
make reference to a spin-$s_i$ analog $S^{(s_i)}(z)$ of the Segal-Sugawara tensor
 that is well-defined for any finite value of $k$,  where its relation to $T^{(s_i)}(z)$ is given by 
\be 
S^{(s_i)}(z) = (k+h^{\vee}) T^{(s_i)}(z), 
\label{act 2}
\ee 
so that 
\be S^{(s_i)}(z) = :{\tilde
d}^{\zeta_1 \zeta_2 \zeta_3 \dots \zeta_{s_i}}(k) (J_{\zeta_1} J_{\zeta_2}\dots
J_{\zeta_{s_i}})  (z):. 
\label{S^{(s_i)}(z)} 
\ee 
From (\ref{act 2}), one can see that the operator $S^{(s_i)}(z)$ generates in its OPE with all other operators of
the quantum theory,  $(k+h^{\vee})$ times the field transformations typically generated by $T^{(s_i)}(z)$. Therefore, at the critical level $k=-h^\vee$, the $S^{(s_i)}(z)$'s  generate \emph{no} field transformations at all -- their OPE's with all other field operators ought  to be regular. This is equivalent to saying that the $T^{(s_i)}(z)$'s do not exist as quantum operators in $\MA$ at $k = -h^\vee$, since the $S^{(s_i)}(z)$'s are the only well-defined operators at this critical level which can  generate the type of field transformations associated with the $T^{(s_i)}(z)$'s.


The fact that the $l$ number of $S^{(s_i)}(z)$'s only have regular OPE's with all other relevant fields and with themselves at $k=-h^\vee$, implies the following. If we denote by $V_{-h^\vee} (\frak {g}_{\mathbb C})$ the chiral algebra generated by the $J$'s in (\ref{current chevalley}) and their $z$-derivatives, and if we denote by $\frak z(V_{-h^\vee} (\frak {g}_{\mathbb C}))$ its center which is defined to be the set of fields which have \emph{regular }OPE's with all other fields within, we have $\frak z(V_{-h^\vee} (\frak {g}_{\mathbb C})) = \mathbb C[\partial^{m}_z S^{(s_i)}(z)]_{i = 1, \dots, l; \, m\geq 0}$, where  $\mathbb C[\partial^{m}_z S^{(s_i)}(z)]_{i = 1, \dots, l; \, m\geq 0}$ is the space of differential polynomials on the $S^{(s_i)}(z)$'s with complex coefficients. This result agrees with Theorem~8 of~\cite{Frenkel}. Since the $A^a_z$'s are non-dynamical fields which do not have any nontrivial propagators with themselves or with the rest of the fields $\beta$ and $\gamma$ (and their $z$-derivatives) that define $\MA$, it ought to be the true that  $\frak z(V_{-h^\vee} (\frak {g}_{\mathbb C})) \subset H_{X}$, where $ H_{X}$ is the space of differential polynomials on the $A^a_z$'s with complex coefficients. This last claim has indeed been proved as Lemma~4.6 in~\cite{Arakawa}. (Recall from $\S$5.2.1 that $A_z$ corresponds to $\lambda^*$ of \emph{loc.~cit.}.) Consequently, $\frak z(V_{-h^\vee} (\frak {g}_{\mathbb C}))$ would define a space of purely classical ($c$-number) fields.

Moreover, notice that since the $S^{(s_i)}(z)$'s and the $J(z)$'s are holomorphic in $z$ and are of dimensions $s_i$ and 1, respectively,  one can Laurent expand them as 
\be 
S^{(s_i)}(z) = \sum_{n \in {\mathbb Z}} {S}^{(s_i)}_n z^{-n-s_i}, \quad J^\alpha(z) = \sum_{n \in \mathbb Z} J^\alpha_{n} z^{-n-1},
\label{enveloping expansion}
\ee 
where $\alpha = \{e_{\alpha_i}, h_{a}, f_{\alpha_i}\}$. The fact that the $S^{(s_i)}(z)$'s only have regular OPE's with all other relevant fields and with themselves implies that 
\be
 [{S}^{(s_i)}_n, J^\alpha_m] =  [{S}^{(s_i)}_n, {S}^{(s_i)}_m] = 0.
\label{center G}
\ee 
This means that the $S^{(s_i)}_n$'s generate the center ${\frak z}(\tilde U_{-h^\vee} (\hat{{\frak g}}_\mathbb C))$ of the completed enveloping algebra $\tilde U_{-h^\vee} (\hat{{\frak g}}_\mathbb C)$ of $\hat{{\frak g}}_\mathbb C$ at critical level $k = -h^\vee$. As before,  ${\frak z}(\tilde U_{-h^\vee} (\hat{{\frak g}}_\mathbb C))$ would also furnish a Poisson algebra.

\bigskip\noindent{\it The Higher Cohomologies}

Let us proceed to determine the higher cohomology groups $H^p(X, {\widehat \Omega}^{ch,tw}_{X})$ where $p \geq 1$. In dimension 0, we can again consider regular polynomials in the $\gamma^i$'s. However, from geometric representation theory, we have the classical result $H^p(\mathscr X, {\cal O}) = 0$ when $p \geq 1$, where $\cal O$ is the sheaf of regular functions over an arbitrary flag manifold $\mathscr X$ of $G_{\mathbb C}$ that are holomorphic in the $\gamma^i$'s~\cite{nash}. Since a vanishing cohomology at the classical level continues to vanish at the quantum level, it would mean that we cannot have regular polynomials in the $\gamma^i$'s; i.e., $H^p(X, {\widehat \Omega}^{ch,tw}_{X;0}) = \emptyset$. In other words, the higher cohomologies start ``growing'' at dimension greater than zero. 

Let us now compute the first cohomology $H^1(X, {\widehat \Omega}^{ch,tw}_{X;1})$ of operators of dimension 1. From the fact that $T_{zz}$ cannot be expressed as a total $z$-derivative (else $L_{-1} = 0$), and the relation (\ref{tzzanomaly}), we find that the dimension 1 fermionic sigma model operator $R_{i \bar j} D_z \phi^i \psi^{\bar j}$ with $q=1$ is not $Q$-exact. Moreover, from the nilpotency of $Q$, and  the relation (\ref{tzzanomaly}), we find that  $R_{i \bar j} D_z \phi^i \psi^{\bar j}$ is $Q$-closed. Hence, from our $Q$-$\check {\rm C}$ech cohomology dictionary, we find that ${\cal R} \in H^1(X, {\widehat \Omega}^{ch,tw}_{X;1})$, where  $\cal R$ is the $\check {\rm C}$ech cohomology counterpart of the sigma model operator $R_{i \bar j} D_z \phi^i \psi^{\bar j}$. ($\cal R$ is just the $G_\mathbb C$ generalization of ${\mathscr R}$ and $\Theta$ of the $SL(3)$ and $SL(2)$ examples discussed earlier.) What about  the space $H^1(X, {\widehat \Omega}^{ch,tw}_{X;2})$ of operators at dimension 2? Let us try to differentiate ${\cal R}$, i.e., let us consider the operator $\partial_z {\cal R}$. From (\ref{tzzanomaly}), one can see that $\partial_z {\cal R}$ would correspond to a $Q$-exact sigma model operator. Thus, from our $Q$-$\check {\rm C}$ech cohomology dictionary, we conclude that $\partial_z{\cal R}  \notin H^1(X, {\widehat\Omega}^{ch,tw}_{X;2})$.  Nevertheless, since  we have the product formula $H^q(X, {\widehat\Omega}^{ch,tw}_{X; l}) \otimes H^p(X, {\widehat\Omega}^{ch,tw}_{X; m}) \to H^{q+p}(X, {\widehat\Omega}^{ch,tw}_{X; l+m})$, we can act ${\cal R} \in H^1(X, {\widehat\Omega}^{ch,tw}_{X;1})$ on every other element of  $H^0(X, {\widehat\Omega}^{ch,tw}_{X})$ to generate $H^1(X, {\widehat\Omega}^{ch,tw}_{X})$: for example, ${\cal R} \cdot 1 = {\cal R}$ is an element of $H^1(X, {\widehat\Omega}^{ch,tw}_{X})$ of dimension 1, and ${\cal R} \cdot \{J_{e_{\alpha_i}}, J_{h_a},  J_{f_{\alpha_i}}\}$  is a set of elements of $H^1(X, {\widehat\Omega}^{ch,tw}_{X})$ of dimension 2, and so on. Hence, $H^1(X, {\widehat \Omega}^{ch,tw}_{X})$ would also be a Wakimoto  module for $\widehat{\frak {g}}_\mathbb C$ at the critical level $-h^\vee$.

Let us now compute the  second cohomology $H^2(X, {\widehat \Omega}^{ch,tw}_{X;1})$ of operators of dimension 1. The arguments are similar to those employed in our earlier discussion of the $SL(3)$ case. In particular, since one has the classical result  $H^p(X, \Omega^q_X) = 0$ if $p \neq q$, and since a vanishing cohomology at the classical level continues to vanish at the quantum level, $H^2(X, {\widehat \Omega}^{ch,tw}_{X;1}) = \emptyset$.  What about the  second cohomology $H^2(X, {\widehat \Omega}^{ch,tw}_{X;2})$ of operators of dimension 2? From footnote~24, we learn that there is a nonzero class $c_2^{T_C} (X) \in H^2(X, \Omega^{2,cl}_{X,T_C})$. As in the $SL(3)$ case, since there are no quantum relations analogous to (\ref{fraktzanomaly}) in the second cohomology, the fact that $H^2(X, \Omega^{2,cl}_{X,T_C}) \neq 0$ implies that  $ \CF \in H^2(X, {\widehat \Omega}^{ch,tw}_{X;2})$,  where $\CF = f_{kj} (\gamma) D_z \gamma^k D_z \gamma^j$ and $\partial_{[i}f_{kj]} = 0$. How about at dimension 3? As in the case of the first cohomology, we can act $\CF \in H^2(X, {\widehat\Omega}^{ch,tw}_{X;2})$ on every other element of  $H^0(X, {\widehat\Omega}^{ch,tw}_{X})$ to generate $H^2(X, {\widehat\Omega}^{ch,tw}_{X})$: for example,  $\CF \cdot 1 = \CF$  is an element of  $H^2(X, {\widehat\Omega}^{ch,tw}_{X})$ of dimension 2, $\CF \cdot  \{J_{e_{\alpha_i}}, J_{h_a},  J_{f_{\alpha_i}}\}$ is a  set of elements of $H^2(X, {\widehat\Omega}^{ch,tw}_{X})$ of dimension 3, and so on. Hence, $H^2(X, {\widehat \Omega}^{ch,tw}_{X})$ would also be a Wakimoto  module for $\widehat{\frak {g}}_\mathbb C$ at the critical level $-h^\vee$.

Similarly, for the higher cohomologies $H^{q}(X, {\widehat \Omega}^{ch,tw}_{X; n})$ of operators of dimension $n$, where $q \geq 3$, the classical result $H^q(X, \Omega^p_X) = 0$ when $q \neq p$ implies that $H^q(X, {\widehat \Omega}^{ch,tw}_{X; n}) = \emptyset$ if $n \neq q$. Nevertheless,  we have the classical result $H^q(X,  \Omega^q_{X,T_C}) \neq 0$,  and since there are no quantum relations analogous to (\ref{fraktzanomaly}) in the higher cohomologies, we conclude that $\CG_q  \in H^q(X, {\widehat \Omega}^{ch,tw}_{X;q})$, where $\CG_q  =  f_{k_1 k_2 \dots k_q} (\gamma) D_z \gamma^{k_1} D_z \gamma^{k_2} \dots D_z \gamma^{k_q}$.  Again, we can act $\CG_q \in H^q(X, {\widehat\Omega}^{ch,tw}_{X;q})$ on every other element of  $H^0(X, {\widehat\Omega}^{ch,tw}_{X})$ to generate $H^n(X, {\widehat\Omega}^{ch,tw}_{X})$: for example,  $\CG_q \cdot 1 = \CG_q$  is an element of  $H^q(X, {\widehat\Omega}^{ch,tw}_{X})$ of dimension $q$, $\CG_q \cdot \{J_{e_{\alpha_i}}, J_{h_a},  J_{f_{\alpha_i}}\}$ is a  set of elements of $H^q(X, {\widehat\Omega}^{ch,tw}_{X})$ of dimension $q+1$, and so on. Hence, $H^q(X, {\widehat \Omega}^{ch,tw}_{X})$ would also be a Wakimoto  module for $\widehat{\frak {g}}_\mathbb C$ at the critical level $-h^\vee$. 

 Last but not least, notice that the zeroth, first, second and higher cohomologies -- all of which are Wakimoto modules for $\widehat{\frak {g}}_\mathbb C$ at the critical level $-h^\vee$ -- start ``growing'' at dimensions $0, 1, 2$ and so on. This observation is also consistent with the  representation-theoretic results of~\cite{chiral bott}. (See foonote~29.) Furthermore, since $\nu(z)$ of \emph{loc.~cit.} can be identified with the holomorphic component $A_z(z)$ of the gauge field, according to our discussion in the third last paragraph of $\S$5.2.1, these modules would necessarily have highest weight 0.

\newsection{$T$-Duality And The Appearance Of The Langlands Dual Group $^LG_{\mathbb C}$}

In this section, we will show how a generalized $T$-duality of the local gauged twisted sigma model on flag manifolds of $G_{\mathbb C}$ -- where $G_{\mathbb C}$ is any simply-connected, simple, complex Lie group -- leads naturally to an isomorphism of $\cal W$-algebras which involves the Langlands dual Lie algebra $^L{\frak g}_{\mathbb C}$. In the context of the \emph{global }sigma model with chiral algebra $\mathscr A$ however, only the classical limit of this isomorphism is found to be physically valid. Together with the results of the previous section, this means that $\mathscr A$ would furnish a \emph{family} of affine $G_{\mathbb C}$-algebras at the critical level parameterized by  $^LG_{\mathbb C}$-opers on the worldsheet, where  $^LG_{\mathbb C}$ is the Langlands dual of  $G_{\mathbb C}$. This crucial fact would then allow us to furnish, in the next section, a natural physical interpretation of the geometric Langlands correspondence for $G_{\mathbb C}$.

\newsubsection{$T$-Duality And An Isomorphism Of $\cal W$-Algebras}

\newsubsubsection{The $G_{\mathbb C} = SL(2)$ Case}

As a start, let us consider the local gauged twisted sigma model on the flag manifold $X$ of $SL(2)$ of rank $l=1$ and Lie algebra $\frak {sl}_2$. For convenience, let us, as was done in $\S$4.4, pick a hermitian metric that is flat when restricted to the underlying local patch $U \subset X$ over which the local model is defined. Since for any flag manifold $M$, we have  $H^1(M,\Omega^{2,cl}_{M,T}) = 0$, we can, according to our discussion at the beginning of $\S$4.4, write the action of the local sigma model as (cf.~(\ref{Su}))
\be
I_{local} = {1 \over 2 \pi} \int_{\Sigma} |d^2 z| \, \delta_{1 \bar 1} \left ( \partial_z \phi^{\bar 1} \partial_{\bar z}\phi^1 +   \psi ^1_{\bar z} \partial_z \psi^{\bar 1}  -  A_{\bar z} V^{1} A_z V^{\bar 1} \right).
\label{Su-SL(2)}
\ee 

Let the worldsheet $\Sigma$ be the complex plane with the origin $z=0$ removed. In this case, the condition stipulated in $\S$5.2.1 that $A_z(z)$ be regular over all of $\Sigma$ would still be met even if we expand $A_z(z)$ in all powers of $z$, i.e., $A_z(z) = \sum_m a_m z^{m-1}$.  Since $F = dA = 0$ in our case, one can express the (locally-defined) gauge field as $A = d Y(z, \bar z)$, where $Y(z, \bar z)$ is a zero-form in $\Sigma$. In turn, because $\partial_{\bar z} A_z = 0 = \partial_z A_{\bar z}$, it would mean that one can write $Y(z, \bar z) = Y_L(z) + Y_R(\bar z)$ such that $A_z = \partial_zY(z, \bar z)  = \partial_z Y_L(z) =  \sum_m a_m z^{m-1}$ and $A_{\bar z} = \partial_{\bar z} Y(z, \bar z)  = \partial_{\bar z} \bar Y_R(\bar z) = \sum_m {\bar a}_m {\bar z}^{m-1}$. Thus, if we let $\delta_{1 \bar 1}  \partial_z \phi^{\bar 1} = \beta$, $\phi^1 = \gamma$, $\psi_{\bar z \bar 1} = \bar \psi_{\bar z}$, $\psi^{\bar 1} = \bar \psi$, one can re-express (\ref{Su-SL(2)}) as 
\be
I_{local} =  {1 \over 2 \pi} \int_{\Sigma} |d^2 z| \,  \left ( \beta \partial_{\bar z}\gamma +   \bar\psi_{\bar z} \partial_z \bar \psi  -  V^1 V_1 \partial_{\bar z} Y \partial_z Y \right).
\label{Su-SL(2)-rewrite}
\ee 
Since $V^1,  V_1 \in \mathbb C$, we indeed have $\partial_{\bar z} (\partial_z Y) = \partial_{\bar z} A_z =  0$ and $\partial_z (\partial_{\bar z} Y) = \partial_z A_{\bar z} =  0$  from the equations of motion.  From $I_{local}$, the singular OPE's of the local model are found to be given by 
\be
\beta(z)\gamma(z') \sim - {1 \over z-z'},  
\label{beta-gamma OPE SL(2)}
\ee
\be
(\sqrt 2 \cdot Y_L) (z) (- \sqrt 2 \cdot Y_L) (z') \sim 2 \, {\rm ln} (z - z'),
\label{X-X OPE SL(2) - 1}
\ee
\be
(\sqrt 2 \cdot Y_R) (\bar z) ( -\sqrt 2 \cdot Y_R) (\bar z') \sim  2 \, {\rm ln} (\bar z - \bar z'),
\label{X-X OPE SL(2) - 2}
\ee
\be
\bar \psi_{\bar z} (\bar z) \bar \psi (\bar z') \sim  {1 \over \bar z - \bar z'},
\label{psi-psi OPE SL(2)}
\ee 
where we have chosen the normalization $V^1V_1 = -1/2$. (Recall that $V^1$ and $V_1$ are only defined up to scaling by a nonzero constant.) Clearly, $I_{local}$ represents an action of a \emph{conformal} field theory (CFT) that is a tensor product of the  $\beta\gamma$--$\bar \psi_{\bar z} \bar\psi$ CFT and the $Y$--$Y$ CFT. 

Note at this point that $\Sigma$ is conformally equivalent to a semi-infinite cylinder. This means that $I_{local}$ can be interpreted as an action of a closed string theory. Morever, notice that  $I_{local}$ is invariant under the translations $\gamma \to \gamma + {\rm const}$ and $Y \to Y + {\rm const}$; in particular, one is free to make the identifications $\gamma \cong \gamma + 2 \pi R$ and $Y \cong Y + 2\pi R$. Thus, $\gamma$ and $Y$ can also be interpreted as angular variables which characterize a compact direction in target space with radius $R$ that the closed string may then wind around.  However, as the target space of the  $\beta\gamma$--$\bar \psi_{\bar z} \bar\psi$ CFT is a local patch $U \subset X$, such a generalization is only possible in the  $Y$--$Y$ CFT (whose target space remains global even in the local sigma model). The most general boundary condition for the $Y$-field of the local sigma model can therefore be written as
\be
Y(\sigma + 2 \pi, t) = Y(\sigma, t) + 2 \pi w R,
\label{wind X}
\ee
where $\sigma$ and $t$ are the spatial and temporal directions along the semi-infinite cylinder, and $w$ is the winding number of the string. As $A_{z} = \partial_{z} Y$ is indexed in the (dual of the) Cartan subalgebra of ${\frak {sl}}_2$, one may therefore interpret the compactified target space of the $Y$--$Y$ CFT to be the maximal torus $S^1$ with radius $R$. 

Notice that (\ref{wind X}) implies the relation
\be
2 \pi w R = \oint  \, (dz\, \partial_z Y + d \bar z \, \partial_{\bar z} Y) = \oint A. 
\label{wind A}
\ee
Notice also that the RHS of (\ref{wind A}) defines -- via the holonomy ${\rm exp}(i\oint A)$ -- a map $\pi_1(\Sigma) \to U(1)$ which characterizes a nontrivial flat $U(1)$ connection over $\Sigma$. Thus, even though our above generalization to a compact target space of the $Y$--$Y$ CFT seemed rather \emph{ad-hoc}, it is clear that the condition $w \neq 0$ -- which implies that the target space ought to be given by $S^1$, as argued -- is necessary for the flat $U(1)$ gauge field $A$ to be nontrivial.

The underlying identification $Y \cong Y + 2\pi R$ also implies that the states of the $Y$--$Y$ CFT ought to map back to themselves under translations by $2 \pi R$ of the $Y$ field of the worldsheet theory; i.e., ${\rm exp}(2 \pi i \hat p) = 1$, where $\hat p$ is the momentum operator of the  $Y$--$Y$ CFT associated with the symmetry transformation $Y \to Y + {\rm const}$. Hence, if $p$ is the momentum of the string along $S^1$, we necessarily have $p = n / R$, where $n \in \mathbb Z$. Consequently, one can write $p = {1 \over 2} (p_L + p_R)$, where 
\be
p_L = {n \over R} + w R, \qquad p_R = {n \over R}  - wR.
\ee
As such, $wR = {1 \over 2} (p_L - p_R)$, and from (\ref{wind A}), we have $2 \pi p = \oint \, (dz \, \partial_z Y - d \bar z \, \partial_{\bar z} Y)$.

\bigskip\noindent{\it The Linear Dilation Theory Behind the $Y$--$Y$ CFT}

The $Y$--$Y$ theory, being a theory of a free, massless scalar field $Y$, has a huge underlying symmetry. For example, with the same $Y$--$Y$ action in (\ref{Su-SL(2)-rewrite}) and the OPE's it implies, one can construct -- independently in the holomorphic and antiholomorphic sectors -- a family of stress tensors which generate a family of Virasoro algebras with different central charges. In other words, one can associate  with the $Y$--$Y$ theory not one but a family of holomorphic and antiholomorphic conformal symmetries. 

Let us then consider the model whereby the holomorphic sector of the $Y$--$Y$ theory is described by a linear dilaton theory (see $\S$2.5 of~\cite{Polchinski 1}) with holomorphic conformal stress tensor
\be
T(z) = - {1 \over 2} : \partial_z Y_L \partial_z Y_L:   + \mathscr V  \partial_z^2 Y_L, 
\label{dilaton ST}
\ee
whence for different values of the $1$-vector $\mathscr V$, there are different (holomorphic) conformal symmetries which therefore define different CFT's behind the $Y$--$Y$ theory. Assuming that the antiholomorphic stress tensor is the usual one given by 
\be
\widetilde T(\bar z) =  - {1 \over 2} : \partial_{\bar z} Y_R \partial_{\bar z} Y_R:,
\ee
the $\mathscr V = 0$ case just gives us the standard CFT associated with the $Y$--$Y$ action.

\bigskip\noindent{\it $T$-Duality of the Linear Dilaton $Y$--$Y$ CFT}

Next, note that the Hamiltonian of the linear dilaton $Y$--$Y$ CFT can be written as 
\be
H = {1 \over 8}\left({n \over R} + w R\right)^2 + {1 \over 8}\left ({n \over R}  - wR\right)^2 +  \sum_{m=1}^\infty (\alpha_{-m} \alpha_m) +  \sum_{m=1}^\infty (\bar\alpha_{-m} \bar\alpha_m)  +  i  {\mathscr V \over 2}  ({n \over R} + wR)   + a^Y,
\label{spec}
\ee
where $(\alpha_m, \bar \alpha_m)  = i  (a_m, \bar a _m)$, and $a^Y$ is a normal-ordering constant.  Let us choose $\mathscr V = -  i ( R \rho - {1 \over R} \rho^\vee)$, where $R = 1 / \sqrt{k + 2}$ for some integer $k$ whose role will be understood shortly, and the constants $\rho = \rho^\vee = {1 / \sqrt 2}$. Notice that the spectrum (\ref{spec}) is invariant under $n \leftrightarrow w$, $R \to 1 / R$, $\rho \to - \rho^\vee$, and $\rho^\vee \to - \rho$; moreover, this exchange leaves $\mathscr V$ fixed but maps $p_L \to p_L$ and $p_R \to - p_R$, and consequently, $Y \to Y'$, where $Y' = Y_L(z) - Y_R (\bar z)$.  As $Y'$ has the same OPE's and stress tensors as $Y$, one can conclude that the linear dilaton $Y'$--$Y'$ CFT with target space radius $R' = 1/ R$, $p'_{L,R}$ parameters $(n', w') = (w,n)$, and $\mathscr V'$ parameters $(\rho', {\rho^\vee}') = (-\rho^\vee, -\rho)$, is \emph{dual} to the linear dilaton $Y$--$Y$ CFT with parameters $(R, n, w, \rho, \rho^\vee)$. This duality of the $Y$--$Y$ theory is also known as $T$-duality.  As mentioned earlier, since the local sigma model is a tensor product of the  $\beta\gamma$--$\bar \psi_{\bar z} \bar\psi$ CFT and the $Y$--$Y$ CFT, this $T$-duality is also a duality of the local sigma model itself. 

\bigskip\noindent{\it A $\cal W$-Algebra}

Let us now consider a relevant application of this $T$-duality. From (\ref{dilaton ST}), the underlying holomorphic stress tensor will be given by  
\be
T_R(z) = - {1 \over 2} : \partial_z Y_L \partial_z Y_L:   -   \left (R \rho - {1 \over R} \rho^\vee  \right) i \partial_z^2 Y_L. 
\ee
As the local sigma model is a tensor product of the  $\beta\gamma$--$\bar \psi_{\bar z} \bar\psi$ CFT and the $Y$--$Y$ CFT, its overall holomorphic stress tensor will be given by $T_{\sigma}(z) = T_{\beta\gamma}(z) + T_R(z)$, where $ T_{\beta\gamma}(z) =  - \beta \partial_z \gamma$ is the holomorphic stress tensor of the $\beta\gamma$--$\bar \psi_{\bar z} \bar\psi$ CFT. Thus, we have
\be
T_{\sigma}(z) =  - :\beta \partial_z \gamma:  - {1 \over 2} : \partial_z Y_L \partial_z Y_L:    -   \left (R \rho - {1 \over R} \rho^\vee  \right) i \partial_z^2 Y_L.
\label{Tsig}
\ee

Next, consider the local holomorphic operator  
\be
V_{R} (z) = \, : e^{- i 2 R \rho \cdot Y_L(z)}:.
\ee
By a short computation using (\ref{beta-gamma OPE SL(2)})--(\ref{X-X OPE SL(2) - 1}), we find that $V_{R} (z)$ has conformal dimension 1 with respect to  $T_{\sigma}(z)$. As such, 
\be
{\CQ} = \oint  dz \, V_{R} (z)  
\ee
would be a conformally-invariant, left-moving charge associated with the dimension 1 current $V_{R}(z)$. Since $\CQ$ is also conserved (a property which is implied by its conformal invariance), by Noether's theorem, $\CQ$ ought to generate a symmetry transformation of the local sigma model. In particular, its vacuum state $| 1 \rangle$ ought to be invariant under this symmetry, i.e., $\CQ | 1 \rangle = 0$. This statement is exact as there are no nonperturbative corrections to the local sigma model by worldsheet instantons that are necessarily global in $X$. 

What can we say about the $\CQ$-invariant spectrum $\CH_\CQ$ spanned by left-moving, non-vacuum states  $| \mathscr O \rangle$ which obey $\CQ |\mathscr O \rangle = 0$? Because the local sigma model is a CFT, we have a state-operator isomorphism, i.e., $ | \mathscr O \rangle \cong \mathscr O \cdot |1 \rangle$, where $\mathscr O$ is the corresponding local holomorphic operator. Hence, since $\CQ |1 \rangle = 0$, the condition that $\CQ | \mathscr O \rangle = 0$ is equivalent to the condition that $[\CQ, \mathscr O] | 1 \rangle  = 0$. Therefore,  the states in $\CH_\CQ$ are in one-to-one correspondence with the local holomorphic operators $\mathscr O$ that commute with $\CQ$. As the local sigma model is a tensor product of the  $\beta\gamma$--$\bar \psi_{\bar z} \bar\psi$ and the $Y$--$Y$ theories, $\CH_\CQ$ would be composed of three distinct sectors:
\be
\CH_\CQ = \CH_{\beta\gamma} \oplus \CH_{\beta\gamma \otimes Y_L} \oplus \CH_{Y_L},
\ee
where states in the sectors $\CH_{\beta\gamma}$, $\CH_{\beta\gamma\otimes Y_L}$ and $\CH_{Y_L}$ correspond to local holomorphic operators given by $z$-differential polynomials on the $(\beta,\gamma)$, $(\beta,\gamma, Y_L)$ and $Y_L$ fields, respectively, which all commute with $\CQ$. In particular, it is well-established that $\CH_{Y_L}$ can be identified with the local holomorphic operators whose Laurent expansion coeffcients generate  a $\cal W$-algebra ${\cal W}_k(\frak{sl}_2)$ at level $k$ (see~\cite{review} and references within).

\bigskip\noindent{\it $T$-Duality and an Isomorphism of $\cal W$-Algebras}

Let us now consider the $T$-\emph{dual} picture of things in the $(n,w) = (1,1)$ sector. $T$-duality  does not act on $\beta$ or $\gamma$, but maps $Y_L \to Y_L$, $R \to 1 /R$, and $(\rho, \rho^\vee) \to (-\rho^\vee, -\rho)$. Therefore, the $T$-dual holomorphic stress tensor will be given by
\be
T'_{\sigma}(z) =  - :\beta \partial_z \gamma:  - {1 \over 2} : \partial_z Y_L \partial_z Y_L:    -   \left (R \rho - {1 \over R} \rho^\vee  \right) i \partial_z^2 Y_L. 
\ee
This actually coincides with the original holomorphic stress tensor $T_{\sigma}(z)$. Also, the $T$-dual of the local holomorphic operator $V_R(z)$ will be given by
\be
V'_R(z) =   \, : e^{ i { 2 \over R}   \rho^\vee \cdot Y_L(z)}:.
\ee
Again, a short computation using (\ref{beta-gamma OPE SL(2)})--(\ref{X-X OPE SL(2) - 1}) would reveal that $V'_R(z)$ has conformal dimension 1 with respect to $T'_{\sigma} (z)$. As such,
\be
{\CQ}' = \oint  dz \, V'_{R} (z)  
\ee
would be a conformally-invariant, left-moving charge associated with the dimension 1 current $V'_{R}(z)$ that is the $T$-dual of $\CQ$. Likewise, the $\CQ'$-invariant spectrum $\CH_{\CQ'}$ spanned by left-moving, non-vacuum states  $| \mathscr O' \rangle$ which obey $\CQ' |\mathscr O' \rangle = 0$ can be written as
\be
\CH_{\CQ'} = \CH'_{\beta\gamma} \oplus \CH'_{\beta\gamma \otimes Y_L} \oplus \CH'_{Y_L},
\ee
where states in the sectors $\CH'_{\beta\gamma}$, $\CH'_{\beta\gamma\otimes Y_L}$ and $\CH'_{Y_L}$ correspond to local holomorphic operators given by $z$-differential polynomials on the $(\beta,\gamma)$, $(\beta,\gamma, Y_L)$ and $Y_L$ fields, respectively, which commute with $\CQ'$. 

Since $\rho = \rho^\vee$, we have $V'_R(z) = V_{^LR}(z)$ where $^LR = -1/R$; and since $\CQ'$ acts on the same fields ($\beta$, $\gamma$ and $Y_L$) as $\CQ$, the observations about $\CH_{Y_L}$ -- albeit with effective parameter $^LR = 1 / \sqrt{^Lk + 2}$ instead of $R$ -- apply exactly to $\CH'_{Y_L}$. Moreover, since $\frak{sl}_2$ is a simply-laced Lie algebra, we have an identification $\frak{sl}_2 \cong  {^L\frak{sl}_2}$ of Lie algebras, where $^L \frak{sl}_2$ is the Langlands dual of $\frak{sl}_2$. Altogether, this means that $\CH'_{Y_L}$ can be identified with the local homorphic operators whose Laurent expansion coeffcients generate  a $\cal W$-algebra ${\cal W}_{^Lk}(^L\frak{sl}_2)$ at level $^Lk$, where $1 / (^Lk+2) = (k +2)$. 

The $\CQ$-invariant spectrum of states ought to map back to itself under any duality of the local sigma model. Therefore, under the $T$-duality of the model, we ought to have an isomorphism $ \CH_{\CQ} \cong \CH_{\CQ'}$. As the distinct sectors of  both $\CH_{\CQ}$ and $\CH_{\CQ'}$ do not mix,  the isomorphism $ \CH_{\CQ} \cong \CH_{\CQ'}$ would imply the isomorphisms $\CH_{\beta\gamma} \cong \CH'_{\beta\gamma}$, $\CH_{\beta\gamma \otimes Y_L}  \cong \CH'_{\beta\gamma \otimes Y_L}$, and $ \CH_{Y_L} \cong  \CH'_{Y_L}$. In particular, we find that the isomorphism $ \CH_{Y_L} \cong  \CH'_{Y_L}$ -- taking into consideration our above analysis of $ \CH_{Y_L} $ and $ \CH'_{Y_L}$ and their relations to $\cal W$-algebras -- can also be expressed as 
\be
{\cal W}_{k}(\frak{sl}_2) \cong {\cal W}_{^Lk}(^L\frak{sl}_2) \qquad {\rm where}  \qquad (k + h^\vee) =  (^Lk + {^Lh}^\vee)^{-1}.
\label{iso of W-SL(2)}
\ee
Here, the dual Coxeter numbers $h^\vee$ and ${^Lh}^\vee$ of $\frak{sl}_2$ and $^L\frak{sl}_2$, respectively, are given by $h^\vee = {^Lh}^\vee = 2$.  This isomorphism of $\cal W$-algebras in (\ref{iso of W-SL(2)}) -- which has been shown here to be a direct consequence of $T$-duality -- has also been derived via a similar albeit abstract algebraic CFT approach in $\S$8.5 of~\cite{Frenkel}.

\newsubsubsection{The Arbitrary $G_{\mathbb C}$ Case}

\def\gc{{\frak {g}_{\mathbb C}}}
\def\GC{{G_{\mathbb C}}}

Now, let us consider the local gauged twisted sigma model on the flag manifold $X$ of an arbitrary $G_{\mathbb C}$ of rank $l$ and Lie algebra $\gc$. A generalization of (\ref{Su-SL(2)-rewrite}) to the present case gives us the action of the local sigma model as 
\be
I_{loc-gen} =  {1 \over 2 \pi} \int_{\Sigma} |d^2 z| \,  \sum_{i=1}^{|\Delta_+|} \left \{ \beta_i \partial_{\bar z}\gamma^i +  \psi^i_{\bar z} \partial_z \psi_i  -    \sum_{a,b = 1}^l V^i_a V_{ib} \partial_{\bar z} Y^a \partial_z Y^b \right \},
\label{Su-GC-rewrite}
\ee 
where $Y^a(z, \bar z) = Y^a_L (z) + Y^a_R(\bar z)$. From $I_{loc-gen}$ above, the singular OPE's of the local model are found to be given by 
\be
\beta_i(z)\gamma^j(z') \sim - {\delta_i^j \over z-z'},  
\label{beta-gamma OPE GC}
\ee
\be
(\alpha^i \cdot Y_L) (z) (\alpha_i \cdot Y_L) (z') \sim  - \kappa_c(\alpha^i, \alpha_i) \, {\rm ln} (z - z'), 
\label{X-X OPE GC - 1}
\ee
\be
(\alpha^i \cdot Y_R) (\bar z) (\alpha_i \cdot Y_R) (\bar z') \sim -  \kappa_c(\alpha^i, \alpha_i) \, {\rm ln} (\bar z - \bar z'),
\label{X-X OPE GC - 2}
\ee
\be
\psi^i_{\bar z} (\bar z) \psi_j (\bar z') \sim  {\delta^i_j \over \bar z - \bar z'},
\label{psi-psi OPE GC}
\ee 
where $(\alpha^i \cdot Y_{L,R}) = \sum_{a=1}^l  \alpha_a^i Y^a_{L,R}$, $(\alpha_i \cdot Y_{L,R}) = \sum_{b=1}^l  \alpha_{i b} Y^b_{L,R}$, $\alpha^i_a = k_1 V^i_a$, $\alpha_{i, b} = k_2 V_{i b}$, and $-2 \kappa_c(\alpha^i, \alpha_i) = k_1k_2$ for some $k_1, k_2 \in \mathbb R$. Since the $V^i_a$'s and $V_{ib}$'s are constants defined up to scaling only, with the appropriate scalings, one can interpret the $\alpha^i$'s and $\alpha_i$'s as the $|\Delta_+|$ positive and negative roots of $\gc$, respectively, with corresponding scalar product $\kappa_c(\alpha^i, \alpha_i)$, where $\kappa_c$ is the Killing form of $\gc$. As an example, consider the earlier $SL(2)$ case where $i, l =1$, $V^1V_1 = -1/2$, and there is just one positive and negative (simple) root $\alpha^1$ and $\alpha_1$; comparing (\ref{X-X OPE SL(2) - 1})--(\ref{X-X OPE SL(2) - 2}) with (\ref{X-X OPE GC - 1})--(\ref{X-X OPE GC - 2}), we have $\alpha^1\alpha_1 = k_1 k_2 V^1 V_1 = -k_1 k_2 / 2 =  -2$, which implies that $\kappa_c(\alpha^1, \alpha_1) = -2$, as it should.  At any rate, it is clear that $I_{loc-gen}$ represents an action of a \emph{conformal} field theory (CFT) that is a tensor product of the  $\beta\gamma$--$\psi_{\bar z} \psi$ CFT and the $Y$--$Y$ CFT. 

Also, as in the $SL(2)$ case, we have, in writing the above equations, assumed that the worldsheet $\Sigma$ is (conformally equivalent to) a semi-infinite cylinder. This means that $I_{loc-gen}$ can be interpreted as an action of a closed string theory. Morever, notice that  $I_{loc-gen}$ is invariant under the translations $\gamma^i \to \gamma^i + {\rm const}$ and $Y^a \to Y^a + {\rm const}$; in particular, one is free to make the identifications $\gamma^i \cong \gamma^i + 2 \pi R^i$ and $Y^a \cong Y^a + 2\pi R^a$. Thus, $\gamma^i$ and $Y^a$ can also be interpreted as angular variables which characterize the compact directions in target space with radii $R^i$ and $R^a$, respectively, that the closed string may then wind around.  However, as the target space of the  $\beta\gamma$--$ \psi_{\bar z} \psi$ CFT is a local patch $U \subset X$, such a generalization is only possible in the  $Y$--$Y$ CFT (whose target space remains global even in the local sigma model). The most general boundary condition for the $Y^a$-fields of the local sigma model can therefore be written as
\be
Y^a(\sigma + 2 \pi, t) = Y^a(\sigma, t) + 2 \pi w^a R^a,
\label{wind X^a}
\ee
where $\sigma$ and $t$ are the spatial and temporal directions along the semi-infinite cylinder, and $w^a$ is the winding number of the string along the $a^{th}$ compact direction. Since $a$ runs from 1 to $l$, one may therefore interpret the compactified target space of the $Y$--$Y$ CFT to be the maximal torus ${\cal T}^l = {S^1} \times \dots \times S^1$ with radii $\{R^1, \dots, R^l\}$. 

For our purpose, it suffices to consider the case where the radii $\{R^1, \dots, R^l\}$ are all the same and given by $R$, so that (\ref{wind X^a}) implies the relation
\be
2 \pi w^a R = \oint  \, (dz\, \partial_z Y^a + d \bar z \, \partial_{\bar z} Y^a) = \oint A^a. 
\label{wind A^a}
\ee
Notice also that the RHS of (\ref{wind A^a}) defines -- via the holonomy ${\rm exp}(i  \oint A)$ -- a map $\pi_1(\Sigma) \to U(1)^l$ which characterizes a nontrivial flat $U(1)^l$ connection over $\Sigma$, where $U(1)^l$ is the Cartan subgroup of $G$. Thus, even though our above generalization to a compact target space of the $Y$--$Y$ CFT seemed rather \emph{ad-hoc}, it is clear that the condition $w^a \neq 0$ -- which implies that the target space ought to be given by ${\cal T}^l$, as argued -- is necessary for the flat $U(1)^l$ gauge field $A$ to be nontrivial.

The underlying identification $Y^a \cong Y^a + 2\pi R^a$ also implies that the states of the $Y$--$Y$ CFT ought to map back to themselves under translations by $2 \pi R^a$ of the $Y^a$ fields of the worldsheet theory; i.e., ${\rm exp}(2 \pi i \hat {p^a}) = 1$, where $\hat {p^a}$ is the $a^{th}$-momentum operator of the  $Y$--$Y$ CFT associated with the symmetry transformation $Y^a \to Y^a + {\rm const}$. Hence, if $p^a$ is the momentum of the string along the $a^{th}$-direction, we necessarily have $p^a = n^a / R$, where $n^a \in \mathbb Z$. Consequently, one can write $p^a = {1 \over 2} (p^a_L + p^a_R)$, where 
\be
p^a_L = {n^a \over R} + w^a R, \qquad p^a_R = {n^a \over R}  - w^aR.
\ee
As such, $w^aR = {1 \over 2} (p^a_L - p^a_R)$, and from (\ref{wind A^a}), we have $2 \pi p^a = \oint \, (dz \, \partial_z Y^a - d \bar z \, \partial_{\bar z} Y^a)$.

\bigskip\noindent{\it The Linear Dilation Theory Behind the $Y$--$Y$ CFT}

The $Y$--$Y$ theory, being a theory of free, massless scalar fields $Y^a$, has a huge underlying symmetry, as discussed in the $SL(2)$ case. With the same $Y$--$Y$ action in (\ref{Su-GC-rewrite}) and the OPE's it implies, one can construct -- independently in the holomorphic and antiholomorphic sectors -- a family of stress tensors which generate a family of Virasoro algebras with different central charges. In other words, one can associate  with the $Y$--$Y$ theory not one but a family of  holomorphic and antiholomorphic conformal symmetries.

Let us then consider the model whereby the holomorphic sector of the $Y$--$Y$ theory is described by a linear dilaton theory (see $\S$2.5 of~\cite{Polchinski 1}) with holomorphic conformal stress tensor
\be
T(z) = - {1 \over 2} : \partial_z Y_L \cdot \partial_z Y_L:   + \mathscr V \cdot \partial_z^2 Y_L, 
\label{dilaton ST^a}
\ee
where $\partial_z Y \cdot \partial_z Y = \sum_{a=1}^l \partial_z Y^a \partial_z Y^a$ and $ \mathscr V \cdot \partial_z^2 Y =  \sum_{a =1}^l \mathscr V^a \partial_z^2 Y^a$. For different values of the $1$-vector $\mathscr V$, there are different (holomorphic) conformal symmetries which therefore define different CFT's behind the $Y$--$Y$ theory. Assuming that the antiholomorphic stress tensor is the usual one given by 
\be
\widetilde T(\bar z) =  - {1 \over 2} : \partial_{\bar z} Y_R \cdot \partial_{\bar z} Y_R:,
\label{dilaton ST^a-rightmoving}
\ee
the $\mathscr V = 0$ case just gives us the standard CFT associated with the $Y$--$Y$ action.

\bigskip\noindent{\it $T$-Duality of the $Y$--$Y$ CFT}

Next, note that the Hamiltonian of the linear dilaton $Y$--$Y$ CFT can be written (up to a normal ordering constant) as
\be
H = \sum_{a=1}^l \left[ {1 \over 8}\left({n^a \over R} + w^a R\right)^2 + {1 \over 8}\left ({n^a \over R}  - w^a R\right)^2 +  \sum_{m=1}^\infty (\alpha^a_{-m} \alpha^a_m) +  \sum_{m=1}^\infty ({\bar\alpha}^a_{-m} {\bar\alpha}^a_m)  +  i  {\mathscr V^a \over 2}  ({n^a \over R} + w^a R) \right],
\label{spec^a}
\ee
where $\partial_z Y_L^a = - i \sum_m {\alpha^a_m \over z^{m+1}}$ and $\partial_{\bar z} Y_R^a = - i \sum_m {{\bar\alpha}^a_m \over {\bar z}^{m+1}}$. Let us choose $\mathscr V = -  i (R \rho - {1 \over R} \rho^\vee)$, where $R = 1 / \sqrt{k + h^\vee}$ for some integer $k$ whose role will be understood shortly; $h^\vee$ is the dual Coxeter number of $\gc$; $\rho = {1 \over 2} \sum_{\alpha \in \Delta_+} \alpha$ and $\rho^\vee = {1 \over 2} \sum_{\alpha \in \Delta_+} \alpha^\vee$ are the Weyl vector of $\gc$ and its dual, respectively; and $\alpha^\vee = 2 \alpha / \kappa_c(\alpha, \alpha)$ is the coroot associated with the root $\alpha$. Notice that the spectrum (\ref{spec}) is invariant under $n^a \leftrightarrow w^a$, $R \to 1 / R$, $\rho \to - \rho^\vee$, and $\rho^\vee \to - \rho$; moreover, this exchange leaves $\mathscr V$ fixed but maps $p^a_L \to p^a_L$ and $p^a_R \to - p^a_R$, and consequently, $Y^a \to {Y^{a}}'$, where ${Y^a}' = Y^a_L(z) - Y^a_R (\bar z)$. As ${Y^a}'$ has the same OPE's and stress tensors as $Y^a$, one can conclude that the linear dilaton $Y'$--$Y'$ CFT with target space radius $R' = 1/ R$, $p^{a'}_{L,R}$ parameters $(n^{a'}, w^{a'}) = (w^a,n^a)$, and $\mathscr V'$ parameters $(\rho', {\rho^\vee}') = (-\rho^\vee, -\rho)$, is \emph{dual} to the linear dilaton $Y$--$Y$ CFT with parameters $(R, n^a, w^a, \rho, \rho^\vee)$. This is a $T$-duality of the $Y$--$Y$ theory.  As mentioned earlier, since the local sigma model is a tensor product of the  $\beta\gamma$--$\bar \psi_{\bar z} \bar\psi$ CFT and the $Y$--$Y$ CFT, this $T$-duality is also a duality of the local sigma model itself.

\bigskip\noindent{\it A $\cal W$-Algebra}

Let us now consider a relevant application of this $T$-duality.  From (\ref{dilaton ST^a}), the underlying holomorphic stress tensor will be given by  
\be
T_R(z) = - {1 \over 2} : \partial_z Y_L \cdot \partial_z Y_L:    -   (R \rho - {1 \over R} \rho^\vee) \cdot i \partial_z^2 Y_L. 
\label{T_R}
\ee
As the local sigma model is a tensor product of the  $\beta\gamma$--$\bar \psi_{\bar z} \bar\psi$ CFT and the $Y$--$Y$ CFT, its overall holomorphic stress tensor will be given by $T_{\sigma}(z) = T_{\beta\gamma}(z) + T_R(z)$, where $ T_{\beta\gamma}(z) =  - \sum_{i = 1}^{|\Delta_+|} \beta_i \partial_z \gamma^i$ is the holomorphic stress tensor of the $\beta\gamma$--$\bar \psi_{\bar z} \bar\psi$ CFT. Thus, we have
\be
T_{\sigma}(z) =  - \sum_{i = 1}^{|\Delta_+|} \, :\beta_i \partial_z \gamma^i:  - {1 \over 2} : \partial_z Y_L \cdot \partial_z Y_L:    -   (R \rho - {1 \over R} \rho^\vee) \cdot i \partial_z^2 Y_L.  
\label{Tsig-gen}
\ee

Next, consider the $l$ local holomorphic operators  
\be
V^s_{R} (z) = \, : e^{- i R \, \alpha^s \cdot Y_L(z)}:,  \quad {\rm where} \quad s = 1, \dots, l,
\label{exp-V}
\ee
and where $\alpha^1, \dots, \alpha^l$ are simple (positive) roots of $\gc$. By a computation using (\ref{beta-gamma OPE GC})--(\ref{X-X OPE GC - 1}), it can be shown that the $V^s_{R} (z)$'s have conformal dimension 1 with respect to  $T_{\sigma}(z)$. As such, 
\be
{\CQ}^s = \oint  dz \, V^s_{R} (z) \quad{\rm for} \quad s = 1, \dots, l,
\ee
would be conformally-invariant, left-moving charges associated with the dimension-one $V^s_{R}(z)$ currents. Since the $\CQ^s$'s are also conserved (a property which is implied by their conformal invariance), by Noether's theorem, each of the $\CQ^s$'s ought to generate a symmetry transformation of the local sigma model. In particular, its vacuum state $| 1 \rangle$ ought to be invariant under these symmetries, i.e., $\CQ^s | 1 \rangle = 0$ for all $s$. This statement is exact as there are no nonperturbative corrections to the local sigma model by worldsheet instantons that are necessarily global in $X$.

Note that all the $\CQ^{s}$'s  commute with one another.\footnote{From the OPE in (\ref{X-X OPE GC - 1}), and the explicit formula for $V^s_R(z)$ in (\ref{exp-V}), it is clear that $[\CQ^m, \CQ^n ] = \oint {dw \over 2\pi i} {\rm Res}_{z \to w} V^m_R(z) V^n_R(w) =  0$ if $m \neq n$. However, for $m=n$, we have the nontrivial OPE $V^m_R(z) V^m_R(w) \sim (z-w)^{R^2|\alpha^m|^2} V^m_{2R}(w)$; nevertheless, since $R^2 |\alpha^m|^2 > 0$, we have  $[\CQ^m, \CQ^m] = \oint {dw \over 2\pi i} {\rm Res}_{z \to w} V^m_R(z) V^m_R(w) =  0$. In short, all the $\CQ^s$'s commute with one another.} Simultaneous eigenstates of these charge operators therefore exist. In particular, there ought to be simultaneous left-moving zero-eigenstates $| \mathscr O \rangle$ which obey $\CQ^s |\mathscr O \rangle = 0$ for all $s$. These zero-eigenstates span a $\{\CQ^1, \dots, \CQ^l\}$-invariant spectrum which we shall denote as $\CH_{\widetilde \CQ}$. Because the local sigma model is a CFT, we have a state-operator isomorphism, i.e., $ | \mathscr O \rangle \cong \mathscr O \cdot |1 \rangle$, where $\mathscr O$ is the corresponding local holomorphic operator. Hence, since $\CQ^s |1 \rangle = 0$, the condition that $\CQ^s | \mathscr O \rangle = 0$ is equivalent to the condition that $[\CQ^s, \mathscr O] | 1 \rangle  = 0$. Thus,  the states in $\CH_{\widetilde \CQ}$ are in one-to-one correspondence with the local holomorphic operators $\mathscr O$ that commute with all the $\CQ^s$'s. As the local sigma model is a tensor product of the  $\beta\gamma$--$\bar \psi_{\bar z} \bar\psi$ and the $Y$--$Y$ theories, $\CH_{\widetilde \CQ}$ would be composed of three distinct sectors:
\be
\CH_{\widetilde \CQ} = \CH_{\widetilde {\beta\gamma}} \oplus \CH_{{\widetilde {\beta\gamma}} \otimes {\widetilde {Y_L}}} \oplus \CH_{\widetilde {Y_L}},
\ee
where states in the sectors $\CH_{\widetilde {\beta\gamma}}$, $\CH_{{\widetilde {\beta\gamma}} \otimes {\widetilde {Y_L}}}$, and  $\CH_{\widetilde {Y_L}}$ correspond to local holomorphic operators given by $z$-differential polynomials on the $(\beta_i,\gamma^i)$, $(\beta_i,\gamma^i, Y^a_L)$ and $Y^a_L$ fields, respectively, which all commute with the $\CQ^s$'s. In particular, it is well-established that $\CH_{\widetilde {Y_L}}$ can be identified with the local holomorphic operators whose Laurent expansion coefficients generate  a $\cal W$-algebra ${\cal W}_k(\gc)$ at level $k$ (see~\cite{review} and references within).

\bigskip\noindent{\it $T$-Duality and an Isomorphism of $\cal W$-Algebras}

Let us now consider the $T$-\emph{dual} picture of things in the $(n^a,w^a) = (1,1)$ sector. $T$-duality  does not act on $\beta$ or $\gamma$, but maps $Y^a_L \to Y^a_L$, $R \to 1 /R$, and $(\rho, \rho^\vee) \to (-\rho^\vee, - \rho)$. Therefore, the $T$-dual holomorphic stress tensor will be given by
\be
T'_{\sigma}(z) = - \sum_{i = 1}^{|\Delta_+|} \, :\beta_i \partial_z \gamma^i:  - {1 \over 2} : \partial_z Y_L \cdot \partial_z Y_L:    -   ( R \rho -  {1 \over R} \rho^\vee) \cdot i \partial_z^2 Y_L.
\ee
This actually coincides with the original holomorphic stress tensor $T_{\sigma}(z)$.  Also, the $T$-dual of the local holomorphic operator $V^s_R(z)$ will be given by
\be
V^{s'}_R(z)    \, : e^{{i \over R} {\alpha^{\vee^s}} \cdot Y_L(z)}:.
\ee
Again, a computation using (\ref{beta-gamma OPE GC})--(\ref{X-X OPE GC - 1}) would reveal that the $V^{s'}_R(z)$'s have conformal dimension 1 with respect to $T'_{\sigma} (z)$. As such,
\be
{\CQ}^{s'} = \oint  dz \, V^{s'}_{R} (z)  \quad {\rm for} \quad s = 1, \dots, l,
\ee
would be conformally-invariant, left-moving charges associated with the dimension one $V^{s'}_{R}(z)$ currents that are the $T$-duals of the $\CQ^s$'s. 

Now let us take a closer look at ${\CQ}^{s'}$ via its current $V^{s'}_R(z) = \, : e^{{i \over R} {\alpha^{\vee^s}} \cdot Y_L(z)}:$. To this end, note that $\alpha^{\vee^s} = \sqrt{r^\vee} \, {^L\alpha^s}$, where $r^\vee$ is the \emph{lacing number} of $\gc$ (equal to the maximal number of edges connecting two vertices of the Dynkin diagram of $\gc$), and $^L\alpha^s$ is a simple (positive) root of the Langlands dual Lie algebra $^L\gc$ corresponding to the simple (positive) root $\alpha^s$ of $\gc$. Thus,  one can write 
\be
V^{s'}_R(z) = {V}^{s}_{^LR}(z) =  \, : e^{- i ^LR \, {^L\alpha^s} \cdot  {Y}_L(z)}:,
\label{hat V}
\ee
where $^LR =   - {\sqrt{r^\vee} / R}$. 

\def\la{{ {^L\alpha}}}

Notice at this point that the $T$-dual of the original $Y$--$Y$ CFT action can be written as
\be
I'_{Y-Y} =  - {1 \over 2 \pi} \int_{\Sigma} |d^2 z| \,   \sum_{s=1}^{|\Delta_+|}  \, \partial_{\bar z} (V^s \cdot Y^{s'}) \partial_z (V_s \cdot Y^{s'}),
\ee 
where $Y^{s'} (z, \bar z) = Y'(z, \bar z) = Y_L (z) - Y_R(\bar z)$. Recall at this point that $\alpha^s_a = k_1 V^s_a$, $\alpha_{s, a} = k_2 V_{s a}$, and $-2 \kappa_c(\alpha^s, \alpha_s) = k_1k_2$; as a result, from $I'_{Y-Y}$ above, and the fact that $^L\kappa_c(^L\alpha^s, {^L\alpha_s}) = \kappa_c(^L\alpha^s, {^L\alpha_s})$, where $^L\kappa_c$ is the Killing form of $^L\gc$,\footnote{Why $^L\kappa_c(^L\alpha^s, {^L\alpha_s}) = \kappa_c(^L\alpha^s, {^L\alpha_s})$  can be seen as follows. First, note that one can write $\la^s = \sum_{a=1}^l \la^a V^s_a$, where the Cartan generators satisfy $[V^s_a, V^s_b] =0$ for all $s =1, \dots, l, \dots, |\Delta_+|$. Second, this means that $\kappa_c(^L\alpha^s, {^L\alpha_s}) = - \kappa_c(^L\alpha^s, {^L\alpha^s})  =  - \sum_{a,b =1}^l (\la^a \la^b) \, \kappa_c(V^s_a, V^s_b) = - \sum_{a =1}^l \la^a \la^a$, since $\kappa_c(V^s_a, V^s_b) = \delta_{ab}$ by definition. Therefore, as $ - \sum_{a =1}^l \la^a \la^a = {^L\kappa_c}(^L\alpha^s, {^L\alpha_s})$, we have   $^L\kappa_c(^L\alpha^s, {^L\alpha_s}) = \kappa_c(^L\alpha^s, {^L\alpha_s})$, as claimed.}  we compute the holomorphic $Y$--$Y$ OPE of the $T$-dual theory to be
\be
 (^L\alpha^s \cdot {Y}_L) (z) (^L\alpha_s \cdot {Y}_L) (z') \sim  - ^L\kappa_c(^L\alpha^s, {^L\alpha_s}) \, {\rm ln} (z - z'), \quad {\rm for} \quad s=1, \dots, l, \dots, |\Delta_+|.
\label{X-X OPE GC - 1a}
\ee

Repeating the analysis in footnote~30 here where we have (\ref{hat V}) and (\ref{X-X OPE GC - 1a}) instead, we find that all the  ${\CQ^{s'}}$'s commute with one another. Then, the $\{{\CQ}^{1'}, \dots, {\CQ}^{l'}\}$-invariant spectrum ${\CH}'_{\widetilde\CQ}$ spanned by left-moving, non-vacuum states  $| \mathscr O' \rangle$ which obey $\CQ^{s'} |\mathscr O' \rangle = 0$ for all $s$, can be written as
\be
{\CH}'_{\widetilde \CQ} = \CH'_{\widetilde{\beta\gamma}} \oplus \CH'_{\widetilde {\beta\gamma} \otimes \widetilde {{Y}_L}} \oplus \CH'_{\widetilde {{Y}_L}},
\ee
where states in the sectors $\CH'_{\widetilde{\beta\gamma}}$,  $\CH'_{\widetilde {\beta\gamma} \otimes \widetilde {{Y}_L}}$, and   $\CH'_{\widetilde {{Y}_L}}$ correspond to local holomorphic operators given by $z$-differential polynomials on the $(\beta_i,\gamma^i)$, $(\beta_i,\gamma^i, {Y}^a_L)$ and ${Y}^a_L$ fields, respectively, which all commute with the $\CQ^{s'}$'s.  According to our above discussion of $\CH_{\widetilde {{Y}_L}}$, and a comparison of (\ref{hat V}) with  (\ref{exp-V}), we find that $\CH'_{\widetilde {{Y}_L}}$ can be identified with the local holomorphic operators whose Laurent expansion coefficients generate  a $\cal W$-algebra ${\cal W}_{^Lk}(^L\gc)$ at level $^Lk$, where $ (^Lk + {^Lh^\vee})^{-1} = {^LR}^2 = {r^\vee} (k + h^\vee)$, and $^Lh^\vee$ is the dual Coxeter numbers of $^L\gc$.

The $\{\CQ^1, \dots, \CQ^l\}$-invariant spectrum of states ought to map back to itself under any duality of the local sigma model. Therefore, under the $T$-duality of the model, we ought to have an isomorphism $ \CH_{\widetilde \CQ} \cong \CH'_{\widetilde \CQ}$. As the distinct sectors of  both $\CH_{\widetilde \CQ}$ and $\CH'_{\widetilde \CQ}$ do not mix,  the isomorphism $ \CH_{\widetilde \CQ} \cong \CH'_{\widetilde \CQ}$ would imply the isomorphisms $\CH_{\widetilde{\beta\gamma}} \cong \CH'_{\widetilde {\beta\gamma}}$, $\CH_{\widetilde {\beta\gamma} \otimes \widetilde {Y_L}}  \cong \CH'_{\widetilde {\beta\gamma} \otimes \widetilde {{Y}_L}}$, and $ \CH_{\widetilde {Y_L}} \cong  \CH'_{\widetilde {{Y}_L}}$. In particular, we find that the isomorphism $ \CH_{\widetilde {Y_L}} \cong  \CH'_{\widetilde {{Y}_L}}$ -- in view of our above analysis of $ \CH_{\widetilde {Y_L}} $ and $ \CH'_{\widetilde {{Y}_L}}$ and their relations to $\cal W$-algebras -- can also be expressed as 
\be
{\cal W}_{k}(\gc) \cong {\cal W}_{^Lk}(^L\gc) \qquad {\rm where}  \qquad(^Lk + {^Lh^\vee})^{-1} = {r^\vee} (k + h^\vee).
\label{iso of W-GC}
\ee
The above isomorphism of $\cal W$-algebras -- which has been shown here to be a direct consequence of $T$-duality -- has also been derived via a similar albeit abstract algebraic CFT approach in $\S$8.6 of~\cite{Frenkel}. 

\bigskip\noindent{\it Validity of $\cal W$-Algebra Isomorphism on Arbitrary Worldsheets}

Last but not least, note that the explicit expressions of  the stress tensors $T(z)$ and $\widetilde T(\bar z)$ in (\ref{dilaton ST^a}) and  (\ref{dilaton ST^a-rightmoving}) imply that the $Y$--$Y$ CFT they describe actually leads to an underlying local theory whose overall action can also be written as
\be
\label{Iequiv}
I_{\rm equiv} =  {1 \over  \pi} \int_{\Sigma} |d^2 z| \, \sqrt {g}  \, e^{-2\sigma(z, \bar z)}  \, \left[\sum_{i=1}^{|\Delta_+|}    \{   \beta_i \partial_{\bar z}\gamma^i +   \psi^i_{\bar z} \partial_z \psi_i   + \partial_{\bar z} (V^i \cdot Y) \partial_z (V_i \cdot Y) \} +  {\cal R}_{\bar z z}  (\mathscr V \cdot Y_L)\right].
\ee
Here, $g^{z \bar z}  = g^{\bar z z} = 2 e^{-2\sigma(z, \bar z)}$ -- with $\sigma(z, \bar z)$ being a function of $z$ and $\bar z$ -- are the nonzero components of the metric on the arbitrary worldsheet $\Sigma$ whose determinant is $g$, and ${\cal R}_{\bar z z}$ is the Ricci curvature of $\Sigma$. As such, even though (\ref{iso of W-GC}) appears \emph{a priori }to have been derived  under the assumption that the worldsheet is a flat semi-infinite cylinder, it really holds for arbitrary (and thus possibly curved) worldsheets. This should come as no surprise since the intrinsic definition of a $\cal W$-algebra depends only on the \emph{local}  coordinate $z$ on $\Sigma$ -- i.e., it must be insensitive to the global topology of $\Sigma$. Moreover, $I_{\rm equiv}$ describes a local sigma model with target space a contractible patch $U \subset X$ over which any bundle can be trivialized; hence, as explained in $\S$4.6, one does not ``see'' the other anomaly $c_1(\Sigma)c_1^T(X)$ -- i.e., $I_{\rm equiv} $ is, as required, a physically valid action for \emph{any} $\Sigma$.

\newsubsection{Affine $G_{\mathbb C}$-Algebras At Critical Level Parameterized By  $^LG_{\mathbb C}$-Opers On The Worldsheet}

What we have discussed so far in this section pertains to the local sigma model over a local patch $U$ of the target space $X$. However, what is ultimately relevant is the \emph{global} sigma model over \emph{all} of $X$. The question then is whether our above result in (\ref{iso of W-GC}) -- on the isomorphism of $\cal W$-algebras -- would also hold in the global model. Given that the large symmetry of the free, massless $Y$--$Y$ theory of the local model would be partially or perhaps even fully broken upon ``lifting'' to the global model, one can, at best, expect (\ref{iso of W-GC}) to hold for certain values of the parameters $k$ and $^Lk$ only. 

\bigskip\noindent{\it The $\cal W$-Algebra Isomorphism in the Global Sigma Model}

For one, recall that $X = \GC /B$ -- where $B \subset \GC$ is a Borel subgroup -- is a homogeneous space whence there is a global $\GC$-action which serves as an automorphism that maps $X$ back to itself. This means that there necessarily is a $\GC$-symmetry in the global sigma model, and by Noether's theorem, there ought to be dimension-one currents whose conserved charges generate this symmetry. In particular, one should be able to find a set of bosonic dimension-one currents $\{J_{\GC}\}$ -- that would furnish an affine $\GC$-algebra at some level $k$ -- whereby the corresponding conserved charges $Q_{\GC}$ span a $\gc$ Lie algebra. Also, as the chiral algebra $\mathscr A$ of the global sigma model is expected to map back to itself under this symmetry, it should be true that $[Q, J_\GC] = 0$, where $Q$ is the scalar supercharge whose cohomology defines $\mathscr A$. Furthermore, since the $\GC$-symmetry is nontrivial and hence, would not act as an identity transformation on the elements of $\mathscr A$, it should also be true that $J_\GC \neq \{Q, \dots\}$.  These last two observations imply that  $J_\GC \in \mathscr A$. 

As $\mathscr A$ is a holomorphic chiral algebra, $\{J_\GC \}$ would be a set of \emph{holomorphic }currents whose conserved charges can then  be expressed as $Q_\GC = \oint dz \, J_\GC (z)$. In addition, from our $Q$-$\check{\rm C}$ech cohomology dictionary established in $\S$4,  one can conclude that $\{J_\GC \}$ should contain only an \emph{even} number of $\psi^{\bar i}$ fields. As there can only be at most \emph{one} $\psi^{\bar i}$ field in the operators that span $\mathscr A$ when $X = SL(2)/ B$, and since the expressions of $\{J_\GC\}$ for other higher-dimensional examples of $X$ (like the expressions of the holomorphic stress tensors $T_\sigma(z)$ which generate the conformal symmetries of the local model) would just be generalizations of the expression of $\{J_\GC\}$ for $X = SL(2)/B$, one can also conclude that the set $\{J_\GC\}$ would be $\psi^{\bar i}$-free, always. According to our $Q$-$\check{\rm C}$ech cohomology dictionary again, this last fact then means that $\{J_\GC\} \in  H^{0}( X, {\widehat \Omega}^{ch,tw}_X)$ for any $X$, where  ${\MA} = \bigoplus_{q_R = 0}^{q_R = |\Delta_+|} H^{q_R}( X, {\widehat \Omega}^{ch,tw}_X)$. In other words, the $J_\GC$ currents are global sections  of the sheaf $ {\widehat \Omega}^{ch,tw}_X$ of TCDO's which is described locally by the local sigma model over $U \subset X$. This implies that among the (continuous family of) CFT's (parameterized by the 1-vector $\mathscr V$ in (\ref{dilaton ST^a})) which underlie the local model, only those CFT's which allow for a construction of affine $\GC$ currents $\{ J_\GC \}$ out of the holomorphic bose fields $\{\beta_i, \gamma^i, Y^a_L\}$ and their $z$-derivatives, would survive the ``lift'' to the global model.   

A direct way to determine for what value(s) of $\mathscr V$ would the resulting CFT underlying the local model allow for a construction of $\{J_\GC\}$, is to check if the corresponding holomorphic stress tensor $T_\sigma (z)$ is amenable to a Sugawara construction involving a set of affine $\GC$ currents which we can naturally identify as $\{J_\GC\}$. With regard to this, note that out of the $(\beta_i, \gamma^i, Y^a_L)$ fields  (whose OPE's are given in (\ref{beta-gamma OPE GC})--(\ref{X-X OPE GC - 1})) and their $z$-derivatives, one can define a holomorphic stress tensor
\be
T_{sug} (z) = - \sum_{i = 1}^{|\Delta_+|} \, :\beta_i \partial_z \gamma^i:  - {1 \over 2} : \partial_z Y_L \cdot \partial_z Y_L:    -   R \rho \cdot i \partial_z^2 Y_L
\label{Tsug}
\ee  
which is amenable to a Sugawara construction at level $k$, where $R = 1 / \sqrt{k+ h^\vee}$~\cite{review}. Comparing $T_{sug}(z)$ with $T_\sigma(z)$ in (\ref{Tsig-gen}), one can see that the sought-after CFT which would survive (at least classically) the ``lift'' to the global model is the one in which the value of $\mathscr V$ gives $1/R = \sqrt{k+ h^\vee} =  0$, i.e., $k = -h^\vee$. Therefore, the particular $\cal W$-algebra isomorphism in (\ref{iso of W-GC}) that would also hold in the global model is the one whereby $k = - h^\vee$ and $^Lk = \infty$, i.e., 
\be
{\cal W}_{-h^\vee}(\gc) \cong {\cal W}_{\infty}(^L\gc). 
\label{iso classical W-algebras}
\ee   
It also means that the set $\{J_\GC\} \in \MA$ would furnish an affine $\GC$-algebra at the critical level $k = -h^\vee$ -- a conclusion which is consistent with our results in $\S$5.

\bigskip\noindent{\it Affine $G_{\mathbb C}$-Algebras At Critical Level Parameterized by  $^LG_{\mathbb C}$-Opers on the Worldsheet}

Now recall from the last subsection that ${\cal W}_{k}(\gc)$ is generated by the Laurent expansion coefficients of the holomorphic fields which define $\CH_{\widetilde {Y_L}}$ for arbitrary level $k$. Also, note that the minimum (conformal) dimension of the holomorphic fields whose Laurent expansion coefficients span ${\cal W}_{k}(\gc)$, is two; the dimension-two field in question is $T_R(z)$ (given in (\ref{T_R})), and its Laurent expansion coefficients generate a Virasoro subalgebra of  ${\cal W}_{k}(\gc)$~\cite{review}. These last two facts imply that the set of holomorphic fields which underlies ${\cal W}_{k}(\gc)$ must be spanned by certain $z$-differential polynomials on the $\partial_z Y^a_L$'s or rather, the $A^a_z$'s. At $k = - h^\vee$, recall from $\S$5 that the center $\frak z(V_{-h^\vee} (\frak {g}_{\mathbb C}))$ -- of the chiral algebra $V_{-h^\vee} (\frak {g}_{\mathbb C})$ generated by $\{J_\GC\}$ and their $z$-derivatives -- is spanned by certain $z$-differential polynomials on the $A^a_z$'s. The question then is whether  the set of holomorphic fields which underlies ${\cal W}_{-h^\vee}(\gc)$ can actually be identified with $\frak z(V_{-h^\vee} (\frak {g}_{\mathbb C}))$;  the answer according to~\cite{Frenkel}, is ``yes''. Also established in~\emph{loc.~cit.} is the fact that the set of holomorphic fields which underlies ${\cal W}_{\infty}(^L\gc)$ can be identified with the algebra ${\rm Fun} \, {\rm Op}_{^L\gc} (D)$ of functions on the space of $^L\GC$-opers on a disc $D \subset \Sigma$.\footnote{In general, ${\rm Fun} \, {\rm Op}_{^L\gc} (D)$ is the algebra of functions on the space of $^L{\frak g}_{\mathbb C}$-opers on a disc $D \subset \Sigma$. However, since $\GC$ is simply-connected and so, $^L\GC$ is of adjoint-type, an $^L {\frak g}_{\mathbb C}$-oper is the same as an $^L\GC$-oper~\cite{BD Langlands}.}  For all our purposes, an $^L\GC$-oper on $\Sigma$ can be understood to define the triple $(E, \nabla, E_{^LB})$, where $E$ is a principal $^L\GC$-bundle on $\Sigma$, $\nabla$ a holomorphic connection on $E$, and $E_{^LB}$ a reduction of $E$ to an $^LB$-bundle ($^LB$ being a Borel subgroup of $^L\GC$).  Altogether, one can conclude that (\ref{iso classical W-algebras}) implies that 
\be
\frak z(V_{-h^\vee} (\frak {g}_{\mathbb C})) \cong {\rm Fun} \, {\rm Op}_{^L\gc} (D). 
\label{iso-langlands duality}
\ee
This result is also proved as Theorem~9 in~\emph{loc.~cit.}. 

Note at this point that $^L\GC$-opers on $D$ can actually be represented explicitly by the operator (see $\S$9.3 of~\cite{Frenkel})
\be
\partial_z  + p_{-1} + \sum_{i =1}^l \langle S^{(s_i)}(z) {O}(w) \rangle \, p_i,
\label{oper}
\ee
where here, one recalls from $\S$5 that the space $\mathbb C[\partial^{m}_z S^{(s_i)}]_{i = 1, \dots, l; \, m\geq 0}$ (of differential polynomials on the  holomorphic spin-$s_i$ fields $S^{(s_i)}$ with complex coefficients) gives  $\frak z(V_{-h^\vee} (\frak {g}_{\mathbb C}))$;  $O(w) \in V_{-h^\vee} (\frak {g}_{\mathbb C})$; $w$ is an arbitrary point in $D$; and $p_{-1}$ and $p_i$ are constant matrices associated with the generators of $^L\gc$. As the $ S^{(s_i)}$'s have regular OPE's with all elements of $V_{-h^\vee} (\frak {g}_{\mathbb C})$,\footnote{The alert reader would have noticed that \emph{a priori}, the OPE's of $\partial_z Y^a_L = A^a_z$ with themselves are not regular, which contradicts our present claim that the $ S^{(s_i)}$'s -- which are $z$-differential polynomials in $A_z$ -- have regular OPE's with all fields, including themselves. There is actually no contradiction here, as we shall now explain. Firstly, as $R \to \infty$ when $k \to -h^\vee$, (\ref{wind A^a}) implies (since $w^a \neq 0$) that $A^a \to \infty$. However, this is unphysical, which means that the \emph{physically effective} gauge field at $k = -h^\vee$ must actually be a rescaled version of $A^a$ that is finite. An immediate example of such an effective gauge field would be given by $\tilde A = -A /R$. (According to our identification  of $A = dY$, such a rescaling can be understood as a trivial redefinition of the constants $V^i_a V_{ib}$ in the action (\ref{Su-GC-rewrite}).) Indeed, from (\ref{wind A^a}), we find that $\oint \tilde A^a = -2 \pi w^a$, which implies that $\tilde A$ is finite;  moreover, from (\ref{X-X OPE GC - 1}), we have $\tilde A(z) \tilde A(w) \sim {\rm regular}$, as required. Last but not least, note that the Sugawara tensor associated with $T_{sug}(z)$ is  $S_{sug}(z) = - {1 \over 2}  { \tilde A}_z \cdot {\tilde A}_z    +   \rho \cdot i \partial_z {\tilde A}_z$, and in both the $\GC = SL(2)$ and $SL(3)$ cases (where we have explicit formulas to compare with), the expressions of $S_{sug}(z)$ coincide (up to irrelevant overall constants) with (\ref{S(z)-A}) and (\ref{S(z)-A for sl(3)}), as they should. (Note that it is implicit in (\ref{S(z)-A}) and (\ref{S(z)-A for sl(3)})  that the gauge field in the formulas is the effective one.) In short, when $k = -h^\vee$, $A^a_z$ is implicitly the physically effective gauge field $\tilde A^a_z$ whence our present claim is in fact consistent.} the correlation function $\langle S^{(s_i)}(z) O(w) \rangle$ will be regular in $(z-w)$. Because regular, holomorphic functions on a compact Riemann surface $\Sigma$ are equivalent to constants which are thus globally-defined, and because  $\langle S^{(s_i)}(z) O(w) \rangle$ can always be regarded as a restriction  to $D$ of some regular holomorphic function on $\Sigma$,  it will mean that  the definition of $\langle S^{(s_i)}(z) O(w) \rangle$ can be extended to the\emph{ whole} of $\Sigma$. Consequently, an  $^L\GC$-oper on $D$ -- as given by (\ref{oper}) -- can, in our case, be extended to an $^L\GC$-oper on $\Sigma$. 

So, via (\ref{iso-langlands duality}) and our analysis in the previous paragraph, we can conclude that the set $\{J_{e_{\alpha_i}}, J_{h_a}, J_{f_{\alpha_i}}\}$ of local operators in (\ref{current chevalley}) -- through their dependence on the $A^a_z$'s and therefore the $S^{(s_i)}$'s -- are parameterized by $^L\GC$-opers on $\Sigma$. In turn, since the set $\{J_{e_{\alpha_i}}, J_{h_a}, J_{f_{\alpha_i}}\}$ furnishes -- for \emph{each} set of values of the $A^a_z$'s and therefore  $S^{(s_i)}$'s -- an OPE algebra of an affine $G_{\mathbb C}$-algebra  (in the Chevalley basis) at critical level (in the  Wakimoto representation), we effectively have a \emph{family} of affine $G_{\mathbb C}$-algebras at critical level parameterized by  $^LG_{\mathbb C}$-opers on the worldsheet $\Sigma$.


\newsection{Physical Interpretation Of The Geometric Langlands Correspondence For $G_{\mathbb C}$}

We are now ready to furnish in this section, a purely physical interpretation of the geometric Langlands correspondence for  $G_{\mathbb C}$, where $G_{\mathbb C}$ is any simply-connected, simple, complex Lie group. Firstly, we will show that our concluding result in $\S$6 -- that there is, within the context of the  gauged twisted sigma model over the flag manifold of $G_{\mathbb C}$, a family of affine $G_{\mathbb C}$-algebras at critical level parameterized by  $^LG_{\mathbb C}$-opers on the worldsheet $\Sigma$ -- can lead us to a natural correspondence between holomorphic $^L\GC$-bundles on $\Sigma$ and  Hecke eigensheaves on the moduli space $\textrm{Bun}_{G_{\mathbb C}}$ of holomorphic $G_{\mathbb C}$-bundles on $\Sigma$; in particular, we will argue that one can interpret the Hecke eigensheaves as the correlation functions of certain local operators which underlie the bosonic sector of the chiral algebra of the flag manifold model. Then, we will argue that one can interpret the Hecke operators as certain \emph{nonlocal} sigma model operators which are constructed out of the local affine  $G_{\mathbb C}$ currents that also span the chiral algebra of the flag manifold model. Along the way, we will also get to understand, from a purely physical perspective, the uniqueness or non-uniqueness property of the Hecke eigensheaves for various $\Sigma$ as established mathematically, from the anomaly-cancellation conditions of the model. 
     
\newsubsection{A Geometric Langlands Correspondence For $G_{\mathbb C}$}

Before we proceed any further, let us first state certain facts which have important implications for our analysis throughout this section. One, if the genus of $\Sigma$ is $g$, then the dimension of the space of (unramified) $^LG_{\mathbb C}$-opers on $\Sigma$ of $g =0,1$ and $\geq 2$ are, $1$, $\textrm{rank} \, ^L\GC$ and $\textrm{dim} ^L\GC (2g-1)$, respectively; this means that our result from $\S$6 - that there is a family of affine $G_{\mathbb C}$-algebras at critical level parameterized by  $^LG_{\mathbb C}$-opers on the worldsheet $\Sigma$ - holds true for any $g$. Two, recall from our analysis in $\S$3 and $\S$5 however, that the flag manifold model has an anomaly quantified by ${1\over 2} c_1(\Sigma)c^{T_C}_1(X)$, where $X$ is the flag manifold of $\GC$, and $T_C$ is the Cartan subgroup of the compact real form of $\GC$; thus, since $c^{T_C}_1(X) \neq 0$, the flag manifold model is actually anomalous when $g \neq 1$. Three, the last statement being made, notice from our analysis in $\S$3 that the  ${1\over 2} c_1(\Sigma)c^{T_C}_1(X)$ anomaly arises due to the nontriviality of the canonical bundle $K$ over $\Sigma$; hence, our model can still be physically well-defined for arbitrary $g$ if one introduces the right number of punctures on $\Sigma$ whence the resulting $K$ is effectively trivial. Finally, note that in the theory of Riemann surfaces, punctures are also known as marked points which can be realized by fixing the positions of local operators that may be defined over them.\footnote{The flag manifold model is not conformal and so, we do not have a CFT state-operator isomorphism whence we can replace an arbitrary puncture with a local operator. Nevertheless, since there are an infinite number of choices of punctures on $\Sigma$, let us, for convenience, choose those that \emph{can} be replaced by local operators.} Altogether therefore, let us assume that all correlation functions considered henceforth contain the right number of fixed-positioned local operators whence the model is nonanomalous and thus physically consistent.

\bigskip\noindent{\it Holomorphic $\GC$-bundles on $\Sigma$}

Let us now begin this subsection proper by explaining how holomorphic $\GC$-bundles on $\Sigma$ can be defined in the presence of an affine algebra $\widehat {\frak g}_{\mathbb C}$ of $\GC$ in the flag manifold model.  Recall that for the flag manifold model, we have (in the standard basis) the current-current OPE
\be 
J_a (z) J_b (w) \sim - {{ h^\vee d_{ab}} \over{(z-w)^2}} + \sum_c
f_{ab}{}^c {{J_c(w)}\over {(z-w)}},
\ee 
where $a,b = 1, \dots, \textrm{dim} (\gc)$. Here, $\gc$ is the Lie algebra of $\GC$; $f_{ab}{}^c$ are the structure constants of $\gc$; $d_{ab}$ is the
Cartan-Killing metric of $\gc$; and $h^\vee$ is the dual Coxeter number of $\gc$. Note also that since the
above dimension-one current operators are holomorphic over $\Sigma$,
they can be expanded in a Laurent expansion around the point $w$
in $\Sigma$ as 
\be J_a(z) = \sum_n{ {J^n_{a}(w)}
{(z-w)^{-n-1}}}.
\ee 
Consequently, from the above current-current OPE, we will get
the commuator relation 
\be 
[J^n_a (w), J^m_b (w)] = \sum_c
f_{ab}{}^c J^{n+m}_c (w) - ( h^\vee d_{ab})\  n \ \delta_{n+m, 0}.
\label{ramification 2}
\ee
As such, the Lie algebra $\gc$ generated by the zero-modes of the currents will be given by 
\be 
[J^0_a (w), J^0_b
(w)] = \sum_c f_{ab}{}^c J^{0}_c (w).
\label{zero-mode of currents}
\ee 
One can then exponentiate the above generators that span $\gc$ to define an element
of $\GC$;  since these generators depend on the point $w$
in $\Sigma$, it will mean that one can, via this exponential map,
consistently define a nontrivial principal $\GC$-bundle over
all of $\Sigma$. Moreover, this bundle will be holomorphic as the
underlying generators vary holomorphically in $w$ over $\Sigma$.

\bigskip\noindent{\it About the Moduli Space $\textrm{Bun}_\GC$ of Holomorphic $\GC$-bundles on $\Sigma$}

\def\cP{{\cal P}}

Now that we have seen how holomorphic $\GC$-bundles on $\Sigma$ can be naturally defined in the flag manifold model, let us review certain technical facts about their moduli space $\textrm{Bun}_\GC$ which will be essential to our forthcoming discussions. Firstly, note that we have the identification (see (7.8) of~\cite{Frenkel}) 
\be
\textrm{Bun}_\GC \cong   {\GC_{\rm out}}  \backslash  \GC((t)) / \GC[[t]], 
\label{id}
\ee
where $t$ is a local coordinate around a point $x \in \Sigma$; $\GC((t))$ is the corresponding loop group characterizing the space of continuous maps $S^1 \to \GC$;  ${\GC_{\rm out}}$ is the group of algebraic maps $\Sigma \backslash x \to \GC$; and $\GC[[t]]$ is the group of $\GC$ matrices whose entries are elements of the ring of formal power series in $t$. Secondly, as a consequence of (\ref{id}), the tangent space $T_{\cP} \textrm{Bun}_\GC$ to the point in $\textrm{Bun}_\GC$ corresponding to a $\GC$-bundle $\cP$ is isomorphic to the double quotient  ${\gc_{\rm out}}  \backslash  \gc((t)) / \gc[[t]]$; thus, any element $\eta(t) = \eta^a J_a(t)$ of the loop algebra $\gc((t))$ -- where one sums over the index $a$, and where $\eta^a$ is a $t$-dependent $c$-number -- would give rise to a tangent vector $\nu$ in $T_{\cP} \textrm{Bun}_\GC$; in turn, this means that the variation of some local holomorphic operator $S^{(s_i)}(x)$ (of say, scaling dimension $s_i$) at $x$ under an infinitesimal deformation of $\cP$ moving along $\textrm{Bun}_\GC$ would be given by 
\be
\delta_\nu S^{(s_i)}(x) = \oint_{\cal C} \eta^a(t) \{J_a (t) \cdot S^{(s_i)}(x) \} \, dt,
\label{deform operator}
\ee  
where $\cal C$ is a small loop around the point $x$, and $J_a (t) \cdot S^{(s_i)}(x)$ denotes the OPE between $J_a$ and $S^{(s_i)}$. Note that (\ref{deform operator}) applies to correlation functions of operators as well, i.e., for a one-point correlation function $\langle \Phi^0(x) \rangle$ of some local holomorphic operator  $\Phi^0(x)$ inserted at $x$, its variation under an infinitesimal deformation of $\cP$ moving along $\textrm{Bun}_\GC$ would be given by 
\be
\delta_\nu \big\langle \Phi^0(x) \big\rangle = \big\langle \oint_{\cal C} \eta^a(t) \{J_a (t) \cdot \Phi^0(x) \} \, dt \big\rangle.
\label{deform CF}
\ee
The above formula also has an obvious mutli-point generalization.    

\bigskip\noindent{\it Local Primary Field Operators From the Chiral Algebra}

As we will explain briefly, the sought-after Hecke eigensheaves on $\textrm{Bun}_{G_{\mathbb C}}$ can be interpreted as the correlation functions of certain local primary field operators of the flag manifold model. As such, it would be useful to describe these particular operators first. By definition, the holomorphic primary field operators $\Phi^{\lambda}_s(z)$ of any theory with a holomorphic affine
$\GC$ OPE algebra  obey (in the standard basis) the following OPE relations with the holomorphic currents \cite{CFT}:
 \be
J_a(z)\Phi^{\lambda}_r (z') \sim - \sum_{s}
{{(t_a^{\lambda})_{rs}\
\Phi^{\lambda}_s}(z')\over{z-z'}},
\label{primary field OPE's}
\ee
where $t_a^{\lambda}$ is a matrix in the representation of $\gc$ with highest weight $\lambda$; $r,s = 1, \dots, \textrm{dim}|\lambda|$; and $a=1,\dots, \textrm{dim}(\gc)$. 

Note that the scaling dimension $h_\lambda$ of the local operators  $\Phi^{\lambda}_s(z)$ obeys the formula~\cite{CFT}
\be
2 (k + h^\vee) h_\lambda = (\lambda, \lambda + 2 \rho),
\label{scaling dimension}
\ee
where $k$ is the level of the affine algebra  $\widehat {\frak g}_{\mathbb C}$ of $G_{\mathbb C}$; $h^\vee$ is the dual Coxeter number; and $\rho$ is the Weyl vector.  In the flag manifold model, the level is critical at $k = -h^\vee$; therefore, (\ref{scaling dimension}) will imply that $\lambda =0$ while $h_\lambda$  can be arbitrary. For $\lambda = 0$, we have $\textrm{dim}|\lambda| = 1$; thus, from (\ref{primary field OPE's}), we ought to have 
\be
J_a(z)\Phi^{0} (z') \sim -
{{(t_a^{0}) \
\Phi^{0}}(z')\over{z-z'}},
\label{primary field OPE's 1-dim}
\ee  
where $\Phi^{0} (z)$ is the sole, bosonic, holomorphic primary field operator of the flag manifold model, and the $t_a^0$'s are just constants.  

In $\S$5, we saw that the global sections $H^0(X, \Omega^{ch, tw}_X)$ of the sheaf of TCDO's furnish a module of  $\widehat {\frak g}_{\mathbb C}$ at  $k = -h^\vee$ with highest weight 0. Thus, a candidate for $\Phi^{0} (z)$ would be a (bosonic) local operator that corresponds to some element of $H^0(X, \Omega^{ch, tw}_X)$. Since  $\Phi^{0} (z)$ is the sole primary field operator, it should correspond to an element in $H^0(X, \Omega^{ch, tw}_{X,0})$ -- the one-dimensional subspace of  $H^0(X, \Omega^{ch, tw}_X)$ whose sole element corresponds to a scaling dimension 0 (bosonic) local operator in the $Q$-cohomology. Hence, one can regard $\Phi^{0} (z)$ as a scaling dimension 0 local operator in the chiral algebra of the flag manifold model.

\bigskip\noindent{\it A Sheaf of Correlation Functions Over $\textrm{Bun}_{G_{\mathbb C}}$}

Now consider the $n$-point correlation function $\big\langle \Phi^{0} (z_1) \dots \Phi^{0} (z_n) \big\rangle$, where the $z_i$'s are $n$ \emph{fixed} and \emph{distinct }points in $\Sigma$. This correlation function of purely bosonic operators with zero $U(1)_R$-charge is nonvanishing in the theory at hand: since the relevant set of (degree-one) twisted holomorphic maps is empty when there are $n$ marked points on $\Sigma$, $p$ in (\ref{p}), which is the $U(1)_R$-charge a correlation function is required to have in order for it to be nonvanishing, is zero.  From the multi-point generalization of (\ref{deform CF}) and the OPE relation (\ref{primary field OPE's 1-dim}), we find that the variation of the $n$-point correlation function under an infinitesimal deformation of $\cP$ would be given by  
\begin{eqnarray}
\delta_\nu \big\langle \Phi^{0} (z_1)\dots \Phi^{0} (z_n) \big\rangle & = & - \sum_{k=1}^n {1 \over {2 \pi i}} \oint_{{\cal C}_k} {dz \over {z - z_k}} \, \eta^a (z) \, t_a^0 \, \big\langle \Phi^{0} (z_1)\dots \Phi^{0} (z_n) \big\rangle \nonumber \\
& = & - \sum_{k=1}^n \, \eta^a(z_k) \, t_a^0 \, \big\langle \Phi^{0} (z_1)\dots \Phi^{0} (z_n) \big\rangle.
\end{eqnarray}
The second equality follows because ${\cal C}_k$ is a small loop around the point $z_k$. Since the $z_k$'s are pre-specified, the expression $\eta^a(z_k) t_a^0$ is just some nonzero number; in other words, the variation of the $n$-point correlation function is just some multiple of itself. Hence, as we move along $\textrm{Bun}_{G_{\mathbb C}}$ when we deform $\cP$ infinitesimally, the $n$-point correlation function $\Psi_n = \big\langle \Phi^{0} (z_1) \dots \Phi^{0} (z_n) \big\rangle$ changes as $\Psi_n \to \Psi_n'$, where $\Psi_n' = \alpha \Psi_n$ for some constant $\alpha$. In this sense, we have a one-dimensional sheaf of $n$-point correlation functions in the primary field operators $\Phi^0(z)$ over $\textrm{Bun}_{G_{\mathbb C}}$, and its section is given by $\Psi_n$.    

One can also make the following physical observation. Notice that $\delta_\nu \big\langle \Phi^{0} (z_1) \dots \Phi^{0} (z_n) \big\rangle =  \big\langle \oint_C dz \, \eta^a J_a (z) \cdot  \Phi^0 (z_1) \dots \Phi^0 (z_n) \big\rangle$ can also be interpreted (to lowest order in sigma-model perturbation theory) as the variation in the $n$-point correlation function due to a $\it{marginal}$ deformation of the sigma-model
action by the term $\oint dz \, \eta^a J_a (z)$. Since a deformation of
the action by the dimensionless term $\oint dz \, \eta^a J_a (z)$ is
tantamount to a displacement in the moduli space of the sigma-model itself, it will mean that $\delta_\nu \big\langle \Phi^{0} (z_1)\dots \Phi^{0} (z_n) \big\rangle$
is also the variation in the $n$-point correlation function due to a change in the moduli of the sigma-model. This implies that $\textrm{Bun}_{\GC}$ will at least correspond to a subspace of the entire moduli space of the flag manifold model. This last statement should come as no surprise since $\cal P$ is actually derived from the affine
$\GC$-algebra of the flag manifold model (as explained above) whose realization in turn does depend on the moduli. 

\bigskip\noindent{\it $\cal D$-Modules on $\textrm{Bun}_{G_{\mathbb C}}$}

\def\cD{{\cal D}}

At any rate, it can be shown that $\Psi_n$ actually represents a $\cal D$-module on $\textrm{Bun}_{G_{\mathbb C}}$ -- where $\cD$ is a free polynomial algebra in certain holomorphic differential operators on $\textrm{Bun}_{G_{\mathbb C}}$ -- as follows. For ease of illustration, let us first consider the case where $\GC = SL(2)$. An important point to note at this juncture is that the OPE in (\ref{primary field OPE's}) has an alternative representation as 
\be
J_a(z) \mathbb V_\lambda(x, z') \sim {{D}_a \mathbb V_\lambda(x, z')  \over {z-z'}},
\label{alt rep}
\ee
where $ \mathbb V_\lambda(x, z')$ is some polynomial function in a complex variable $x$ which represents the primary field operator $\Phi^\lambda_r(z')$ inserted at $z'$, and the ${D}_a$'s are differential operators given by 
\begin{eqnarray}
D_+ & = & - x^2 \partial_x + \sqrt 2 \lambda x, \\
D_0 & = & - x \partial_x +  {\lambda \over {\sqrt 2}}, \\
D_- & = & - \partial_x.  
\end{eqnarray}
For the flag manifold model, we have $\lambda = 0$; thus, $\Phi^0(z')$ would be represented by $ \mathbb V_0(x, z')$, and the corresponding differential operators would be given by
\begin{eqnarray}
\label{D1}
D_+ & = & - x^2 \partial_x, \\
D_0 & = & - x \partial_x, \\
\label{D3}
D_- & = & - \partial_x.  
\end{eqnarray}   
From (i) the expression of the spin-2 Segal-Sugawara stress tensor $S^{(2)}(z) = S(z)$ in (\ref{s(z)}), (ii) the relation (\ref{alt rep}), and (iii) the formulas (\ref{D1})--(\ref{D3}), it is clear that one can interpret the $n$-point correlation function $\Psi_n =  \big\langle \Phi^{0} (z_1) \dots \Phi^{0} (z_n) \big\rangle$ as some polynomial function $\Psi_n(x_1\vert z_1, \dots, x_n\vert z_n)$ in the variables  $x_i$ (and constants $z_i$) which $S(z)$ acts on as a second-order differential operator. Since $\Psi_n$ is also a section of a sheaf over $\textrm{Bun}_{SL(2)}$, one can regard the $x_i$'s as holomorphic coordinates on  $\textrm{Bun}_{SL(2)}$. This means that one can interpret $S(z)$ -- via its action on $\Psi_n(x_1\vert z_1, \dots, x_n\vert z_n)$ -- as a (second-order) holomorphic differential operator on (some line bundle on) $\textrm{Bun}_{SL(2)}$.\footnote{Why $S(z)$ in its action on $\Psi_n$ must be interpreted as a holomorphic differential operator on some line bundle on $\textrm{Bun}_{SL(2)}$ and not just on $\textrm{Bun}_{SL(2)}$, will be clarified shortly. For now, and similarly in the next few paragraphs, we shall just accept this to be true.} 

The arguments for groups other than $SL(2)$ are analogous; for example, see $\S$15.7.4 of~\cite{CFT} for the $SL(3)$ case. In short, for general $\GC$, the spin-$s_i$ tensors $S^{(s_i)}(z)$ -- whose expressions are given in (\ref{S^{(s_i)}(z)}) -- in their action on $\Psi_n (x^1_1\vert z_1, \dots, x^{|\Delta_+|}_1\vert z_1, \dots, x^1_n\vert z_n, \dots, x^{|\Delta_+|}_n\vert z_n)$ -- the corresponding $n$-point correlation functions for general $\GC$ with $|\Delta_+|$ number of positive roots -- can be interpreted as $s^{\rm th}_i$-order holomorphic differential operators on (some line bundle on) $\textrm{Bun}_{G_{\mathbb C}}$. 

That said, recall from our analysis in $\S$5 that because the level of the underlying affine $\GC$-algebra is critical, the $S^{(s_i)}(z)$'s are actually purely classical, $c$-number fields -- in other words, the $S^{(s_i)}(z)$'s effectively multiply the (correlation functions of) local operators by a $c$-number in their action on them. Since each $S^{(s_i)}(z)$ can either transform  as a projective connection or degree-$s_i$ differential on $\Sigma$ depending on whether $s_i =2$ or $s_i > 2$, respectively~\cite{Frenkel}, our discussions in this and the previous paragraph imply that 
\be
\cD^{s_i} \cdot \Psi_n (x^1_1\vert z_1, \dots, x^{|\Delta_+|}_n\vert z_n) = \Omega^{s_i}_\Sigma  \,  \Psi_n (x^1_1\vert z_1, \dots, x^{|\Delta_+|}_n\vert z_n),
\label{D-module}
\ee
where $\cD^{s_i}$ is an $s^{\rm th}_i$-order holomorphic differential operator on (some line bundle on) $\textrm{Bun}_{G_{\mathbb C}}$, and $\Omega^{s_i}_\Sigma$ is a $c$-number that is determined either by a projective connection or degree-$s_i$ differential on $\Sigma$ that is associated with $S^{(s_i)}(z)$. From (\ref{D-module}), it is clear that $\Psi_n$ is a $\cD$-module on  $\textrm{Bun}_{G_{\mathbb C}}$, where $\cD$ is a free polynomial algebra in the  $\cD^{s_i}$-operators. Alternatively, one can interpret $\Psi_n$ as a simultaneous eigenvector of the $\cD^{s_i}$-operators with eigenvalues $ \Omega^{s_i}_\Sigma$.

\bigskip\noindent{\it A Geometric Langlands Correspondence for $\GC$}

We are now finally ready to demonstrate, purely physically, a geometric Langlands correspondence for $\GC$. To this end, first note that since all the $S^{(s_i)}$'s have regular OPE's with the set $\{J_a\}$, the formula (\ref{deform operator}) tells us that the $S^{(s_i)}$'s are constant over all of $\textrm{Bun}_{G_{\mathbb C}}$; this means that the corresponding differential operators $\cD^{s_i}$ are globally well-defined. Second, since the  $S^{(s_i)}$'s  are effectively $c$-number fields, they and therefore the corresponding differential operators $\cD^{s_i}$, ought to commute with one another. Third, note that except for the constant functions, there are no global commuting differential operators on $\textrm{Bun}_{G_{\mathbb C}}$; thus, each  $\cD^{s_i}$ is necessarily a holomorphic differential operator $F$ acting on some line bundle $\cal L$ on  $\textrm{Bun}_{G_{\mathbb C}}$. To ascertain what $\cal L$ is, let $s$ and $u$ be compactly-supported sections of $\cal L$ and ${\cal L}^{-1} \otimes \omega$, respectively, where $\omega$ is the canonical bundle of (middle-dimensional forms on) $\textrm{Bun}_{G_{\mathbb C}}$. Then, $uFs$ is a section of $\omega$  which one can integrate over any real slice $Z$ of the complex manifold  $\textrm{Bun}_{G_{\mathbb C}}$ to define the inner product $\langle u \vert  F \vert s \rangle = \int_Z u F s$.  Via integration by parts, one can always introduce the transpose operator $F^t$ -- defined by $\langle u \vert  F \vert s \rangle = \langle s \vert  F^t \vert u \rangle$ -- via  the relation $\int_Z u F s = \int_Z (F^t u) s$. Clearly, $F^t$ ought to act on ${\cal L}^{-1} \otimes \omega$. At any rate, because the  differential operators  $\cD^{s_i}$ commute with one another, $F$ can be represented by a purely diagonal matrix whose transpose is itself, i.e., $F = F^t$. This means that $F$ and $F^t$ must act on bundles which are isomorphic, i.e., ${\cal L} \cong {\cal L}^{-1} \otimes \omega$; this implies that ${\cal L} = \omega^{1\over 2}$. Fourth, note that because  each $S^{(s_i)}$ either transforms as a projective connection or degree-$s_i$ differential on $\Sigma$, for $g>1$, the space of all $S^{(s_i)}$'s would be given by ${\rm Proj} (\Sigma) \times \oplus_{i=2}^l H^0(\Sigma, \Omega^{\otimes e_i +1})$ of dimension  $\sum_{i=1}^l (2e_i +1)(g-1) = {\rm dim} \, \GC (g-1)$, where ${\rm Proj} (\Sigma)$ is a projective connection on $\Sigma$; $s_i = e_i +1$;  $e_i$ is an exponent of $\GC$; and $l$ is the rank of $\GC$. In other words, there are altogether ${\rm dim} \, \GC (g-1)$ holomorphic differential operators  $\cD^{s_i}$ which obey (\ref{D-module}). Since ${\rm dim} \, \textrm{Bun}_{G_{\mathbb C}} = {\rm dim} \, \GC (g-1)$, it would mean that the system of differential equations  defined by (\ref{D-module}) and consequently $\Psi_n$, is holonomic.  In summary, for $g > 1$,  $\Psi_n$ is a holonomic $\cD$-module on $\textrm{Bun}_{G_{\mathbb C}}$ defined by the system of differential equations in  (\ref{D-module}), where $\cD$ is a polynomial algebra in the global commuting holomorphic differential operators $\cD^{s_i}$ on the line bundle $\omega^{1\over 2}$ on $\textrm{Bun}_{G_{\mathbb C}}$. Moreover, from (i) the isomorphism in (\ref{iso-langlands duality}), (ii) the observation made thereafter that $^L\GC$-opers on the disc $D \subset \Sigma$ can be extended to $^L\GC$-opers on $\Sigma$, and (iii) the fact that the LHS of (\ref{iso-langlands duality}) can be identified with the polynomial algebra in the $S^{(s_i)}$'s, we have the identification ${\rm Fun} \, {\rm Op}_{^L\gc} (X) \tilde {\longrightarrow} \cD$. As such, for $g > 1$, $\Psi_n$ can be identified with the Hecke eigensheaf of the geometric Langlands program defined by Beilinson and Drinfeld in~\cite{BD Langlands}. (In what sense is $\Psi_n$ an eigensheaf of a Hecke operator will be explained shortly.)  In addition, notice that since $\Psi_n$ is the $n$-point correlation function in the dimension 0 bosonic local operators $\Phi^0(z)$ of the chiral algebra of the flag manifold model, (\ref{primary field OPE's 1-dim}) would imply that the realization of $\Psi_n$ depends on the realization of the set $\{J_a\}$. Notice also that our concluding result in $\S$6 -- which states that we have a family of $\{J_a\}$'s parameterized by  $^LG_{\mathbb C}$-opers on $\Sigma$ -- implies that for each choice of a holomorphic  $^LG_{\mathbb C}$-bundle (with an oper structure) on $\Sigma$, we have a realization of the set $\{J_a\}$. Altogether therefore, our analysis in this paragraph tells us that for every holomorphic  $^LG_{\mathbb C}$-bundle on $\Sigma$ with $g>1$, there corresponds a Hecke eigensheaf $\Psi_n$ on $\textrm{Bun}_{G_{\mathbb C}}$. This is nothing but the statement by Beilinson and Drinfeld in~\cite{BD Langlands} of the geometric Langlands correspondence for $\GC$!

\bigskip\noindent{\it The Geometric Langlands Correspondence for $g \leq 1$}

So far, we have described the geometric Langlands correspondence for $g > 1$. What about for $g \leq 1$?  Well, for $g = 0$, if there are no punctures on $\Sigma$, because it is simply-connected (i.e.,~its fundamental group is trivial), all holomorphic (i.e.,~flat) $^L\GC$-bundles on $\Sigma$ are trivial; the correspondence is thus vacuous in this case. This observation is consistent with the fact that the flag manifold model is anomalous for $g =0$ with no punctures (as the resulting canonical bundle $K$ of $\Sigma$ is nontrivial).  What if we add punctures? The flag manifold model is certainly nonanomalous for $g=0$ with one or more punctures (as $K$ will be trivial).  On the other hand, it is a mathematically established fact that the relevant moduli spaces which support the Hecke eigensheaves that underlie the geometric Langlands correspondence for $g =0$, vanish when there are two or less punctures~\cite{Rubtsov}; i.e., the correspondence is also vacuous for $g =0$ unless one has three or more punctures. Note that there is no contradiction between the last two statements here:  as we shall elaborate in $\S$8, taking into account worldsheet twisted-instanton contributions in the full physical theory, we find that the chiral algebra will be trivial if $\Sigma$ has $g=0$ with two or less  punctures, and from the interpretation of $\Psi_n$ as a correlation function of the $\Phi^0(z)$ fields in the chiral algebra, this would mean that there are \emph{no} Hecke eigensheaves for such a $\Sigma$. In short, the physics of our flag manifold model implies that the geometric Langlands correspondence for general $\GC$ ought to hold for $g =0$ with three or more punctures. In the case where $\GC = SL(N)$, a geometric Langlands correspondence for $g=0$ with three or more punctures has been demonstrated purely mathematically  in~\cite{Rubtsov}.   

For $g=1$, $\Sigma$ is nonsimply-connected for any number of punctures (i.e.,~its fundamental group is never trivial). Thus, all holomorphic $^L\GC$-bundles on $\Sigma$ will be nontrivial, although they might be reducible. Moreover, the flag manifold model is always nonanomalous in this case. As such, from a purely physical standpoint of our flag manifold model, one would expect the correspondence to hold for $g=1$, always. Indeed, for certain Lie groups, a geometric Langlands correspondence for $g=1$ with and without punctures has been demonstrated purely mathematically in~\cite{genus 1 punctures, Nevin} and~\cite{genus 1 no punctures}, respectively.

\newsubsection{Uniqueness Or Non-Uniqueness Property Of The Hecke Eigensheaves}

Let us now discuss the uniqueness or non-uniqueness property of the Hecke eigensheaves. Recall that it was  mentioned at the start of $\S$7.1 that the flag manifold model is anomalous unless the correlation functions contain the right number of fixed-positioned local operators. This number is determined by the number of marked points required to trivialize the canonical bundle $K$ on $\Sigma$. For $g = 0, 1$ and $> 1$, this number is $\geq 1$, $\geq 0$ and $2g-2$, respectively.\footnote{I would like to thank C.W.~Chin for a detailed explanation of this mathematical fact.} Together with our analysis in $\S$4.2 -- which tells us that any correlation function purely in the local ground operators of the $Q$-cohomology such as $\Phi^0$ would be independent of their respective positions $z_i$ on $\Sigma$ -- this would mean that for $g>1$, for each choice of a holomorphic $^L\GC$-bundle on $\Sigma$, there is only \emph{one} (up to isomorphism) Hecke eigensheaf given by $\Psi_{2g-2}$. This claim is consistent with the mathematical results of Beilinson and Drinfeld in~\cite{BD Langlands}.  On the other hand, this also means that for $g=0$ (with three or more punctures) and $g = 1$ (with any number of punctures), for each choice of a holomorphic (parabolic) $^L\GC$-bundle on $\Sigma$, there are \emph{many} distinct Hecke eigensheaves given by $\Psi_1, \Psi_2, \dots$ and  $\Psi_0, \Psi_1, \dots$, respectively.  This second claim is also consistent with the mathematical results of~\cite{Rubtsov, genus 1 punctures,  Nevin, genus 1 no punctures}. In fact, for $g =1$ with no punctures, because the fundamental group of $\Sigma$ is abelian, all $^L\GC$-bundles on $\Sigma$ will be reducible; in the case where the bundle is reducible to its Cartan subgroup $^LT_{\mathbb C} \subset {^L\GC}$,  the corresponding Hecke eigensheaves have been shown to be given by the geometric Eisenstein series~\cite{genus 1 no punctures}, and in the best case scenario, they can be regarded as direct sums of infinitely many irreducible perverse  sheaves on ${\rm Bun}_\GC$.\footnote{I would like to thank A.~Beilinson for illuminating exchanges regarding this point.} In terms of our flag manifold model, these irreducible perverse sheaves on ${\rm Bun}_\GC$ have a physical interpretation as the (infinitely numerous set of) correlation functions in an \emph{arbitrary} number of $\Phi^0(z)$ fields with \emph{variable} positions $z$: such correlation functions, and their direct sums, also satisfy (\ref{D-module}) to be admissible as holonomic $\cD$-modules or Hecke eigensheaves on ${\rm Bun}_\GC$, and furthermore,  by the Riemann-Hilbert correspondence, these  holonomic $\cD$-modules are also perverse sheaves  on ${\rm Bun}_\GC$.

\newsubsection{Hecke Operators And The Hecke Modification And Correspondence}

So far, we have shown that $\Psi_n$ can be interpreted as a Hecke eigensheaf on ${\rm Bun}_\GC$. That being said, what, in the context of our flag manifold model, is the Hecke operator that acts on $\Psi_n$, and what is its corresponding eigenvalue, one might ask. Let us now, in this final subsection, address these questions, and expound on the related concepts of a Hecke modification and correspondence,  so as to complete our physical interpretation of the geometric Langlands correspondence for $\GC$. 

\bigskip\noindent{\it The Hecke Operator}

Consider an irreducible representation $V_{^Lw}$ of $^L\GC$ of highest (dominant integral) weight $^Lw = (^Lw^1, \dots. ^Lw^{r})$, where the integers $^Lw^i$ are the Dynkin labels and $r$ is ${\rm rank}(^L\GC) = {\rm rank}(\GC)$.  Consider the nonlocal operator 
\be
H_{^Lw, p} = \oint_{{\cal C}_p}  \rho^i_{^Lw}(z)J_i(z) \, dz
\label{H_w}
\ee
inserted near the local operator $\Phi^0(p)$ of position $p \in \Sigma$ that is encircled by the closed loop ${\cal C}_p$ in the $n$-point correlation function $\Psi_n = \big\langle \Phi^0(z_1), \dots, \Phi^0(p), \dots, \Phi^0(z_n) \big\rangle$, where one sums over the index $i = 1, \dots,  r$; here, the $J_i$'s are the subset of affine $\GC$ currents that are associated with the maximal torus $T_\mathbb C \subset \GC$, and $ \rho^i_{^Lw}(z) = {^Lw^i}  f^i(z)$ (no sum over index $i$), where the $f^i(z)$'s are certain nowhere-vanishing, regular, holomorphic functions in $z$ that are $c$-numbers. 

The dimension of the representation $V_{^Lw}$ is known to be given by the formula ${\rm dim} \, V_{^Lw} = \prod_{^L\alpha > 0}f_{^L\alpha} (^Lw^i)$~\cite{CFT}, where the $f_{^L\alpha} (^Lw^i)$'s are polynomial functions in the $^Lw^i$'s which are labeled by the roots $^L\alpha$ of $^L\gc$. Therefore, notice from (\ref{primary field OPE's 1-dim}) that for a choice of $z$-functions $ \rho^i_{^Lw}$ such that 
\be
 t^0_i \rho^i_{^Lw} (p) = {\rm dim} \, V_{^Lw}, 
 \label{dim V}
\ee  
(which should hold for any $p$ since the $\rho^i_{^Lw}$'s -- being nowhere-vanishing,  regular, and holomorphic in $z$ -- can be treated as global constants over $\Sigma$), the action of $H_{^Lw, p}$ on $\Psi_n$ at $p$ can be expressed as 
\be
H_{^Lw, p} \cdot \Psi_n = \underbrace{\Psi_n + \dots + \Psi_n}_{ {\rm dim} \, V_{^Lw} \ {\rm times}}.
\label{Hecke eqn}
\ee
Also, as mentioned in the last subsection, $\Psi_n$ does not depend on the insertion position $p$;  in other words, $p$ is irrelevant and if (\ref{Hecke eqn}) holds for some $p$, it would also hold for any other point in $\Sigma$. Therefore, given that we can interpret $\Psi_n$ as a sheaf on ${\rm Bun}_\GC$, one can also write (\ref{Hecke eqn}) as
\be
H_{^Lw} \cdot \Psi_n = E_{^Lw} \otimes \Psi_n,
\label{Hecke eigen-relation}
\ee  
where $E_{^Lw}$ is a vector space of dimension ${\rm dim} \, V_{^Lw}$ which can be associated with a holomorphic $^L\GC$-bundle on $\Sigma$.\footnote{For each representation $V_{^Lw}$ of $^L\GC$, one can always construct an associated vector bundle $P$ to an $^L\GC$-bundle $E$ on $\Sigma$, where $P = E \times_{^L\GC} V_{^Lw}$. Hence, since $p \in \Sigma$ is irrelevant -- i.e., $E_{^Lw}$ does not vary with $p$ -- one can always associate the vector space $E_{^Lw}$ with a holomorphic (i.e.,~flat) $^L\GC$-bundle on $\Sigma$.} It is in the sense of (\ref{Hecke eigen-relation}) that $\Psi_n$ is a Hecke eigensheaf of the Hecke operator $H_{^Lw}$ with eigenvalue  $E_{^Lw}$, as defined by Beilison and Drinfeld in~\cite{BD Langlands}. 

\bigskip\noindent{\it A Hecke Modification}

What else can one say about the action of  $H_{^Lw, p}$, in particular, with regard to the underlying $\GC$-bundle $\cal P$? Now that we have implicitly made a choice in the maximal torus $T_\mathbb C \subset \GC$ in defining, via (\ref{H_w}), the Hecke operator $H_{^Lw,p}$ inserted at position $p$, we can,  at $p$,  decompose the fiber of  ${\cal P}$ in some representation $R$ as ${\cal P}_R\vert_p = \oplus_w {\cal P}_{R, w}\vert_p $, where ${\cal P}_R = {\cal P} \times_\GC R$ and $R = \oplus_w R_w$ is the decomposition of $R$ in weight spaces $R_w$ (all but finitely many of which vanish). If we take $R_w$ to be the weight space associated with a simple (positive) root $w$ of $\GC$, we can write 
\be
{\cal P}\vert_p = \oplus_{i=1}^{r} {\cal L}_i
\ee
 in the vicinity of $p$, where ${\cal L}_i$ is a line bundle on $\Sigma$ corresponding to the $i^{\rm th}$-circle of $T_\mathbb C \cong (\mathbb C^\ast)^r$ with first Chern class $c_1({\cal L}_i)$ obeying $\int_\Sigma c_1({\cal L}_i) = q^i$.  

Notice from (\ref{H_w}) that the action of the Hecke operator $H_{^Lw, p}$ on $\Psi_n$ involves inserting a local operator  $\rho^i_{^Lw}J_i(z)$ in $\Sigma$ before taking a contour integral around it as $z \to p$.  Via the interpretation of the current zero modes  $J^0_i$ as the generators of the Lie algebra ${\frak t}_\mathbb C$ of $T_\mathbb C$ (see (\ref{zero-mode of currents})), and the fact that one can therefore exponentiate the $J^0_i$'s to generate $T_\mathbb C$, the insertion of the operator $\rho^i_{^Lw}J_i (p)$ in $\Sigma$ is --  according to how we first defined ${\cal P}$ in $\S$7.1 -- tantamount to the modification ${\cal L}_i \to {\cal L}_i \otimes {\cal O}_p^{(^Lw^i f^iq^i)}$ at $p$, where ${\cal O}_p$ is a line bundle  whose holomorphic sections are functions holomorphic away from $p$ with a possible single pole at $p$: the $i^{\rm th}$ generator of ${\frak t}_\mathbb C$ gets multiplied by a factor of $\left(1+ {^Lw^i} f^i\right)$ as the ``effective'' affine $T_\mathbb C$ current at $p$ becomes $(1+ \rho^i_{^Lw}) J_i(p)$ upon inserting the operator $\rho^i_{^Lw}J_i(p)$ in $\Sigma$, and this means that the ${\frak t}_\mathbb C$-valued curvature two-form of ${\cal L}_i$ and hence $\int_\Sigma c_1({\cal L}_i)$, gets modified from $q^i \to q^i + {^Lw^i} f^iq^i$, i.e., we effectively have a modification  from ${\cal L}_i \to {\cal L}_i \otimes {\cal O}_p^{(^Lw^i f^iq^i)}$ at $p$.     

Let us now make a relevant digression. Note at this point that we have a group homomorphism $^Lw: {\mathbb C}^\ast \to T_\mathbb C$;  however, it does not make sense to think of $^Lw$ as a homomorphism until a local coordinate, say $y$, on $\Sigma$, is chosen. Assuming that this has  been done,  we may write~\cite{Segal} 
\be
^Lw^i (y) = y^{^Lw^i}.
\label{lw(y)}
\ee 
If $z$ is some other choice of local coordinate, then 
\be
y = y(z) = z g(z)
\ee
for some nowhere-vanishing  holomorphic function $g(z)$. This means that we may also write
\be
^Lw^i(y(z)) = {^Lw^i(z)} {^Lw^i(g(z))}.
\label{lw(y(z))}
\ee
In sum, (\ref{lw(y)})-(\ref{lw(y(z))}) imply that for a choice of local coordinate $z$ whereby $g(z) = z^{(f^iq^i -1)}$, we have
\be
(^Lw^if^iq^i)(z) = {^Lw^i}(y).
\label{weight coord}
\ee

Coming back to our main discussion, the implication of (\ref{weight coord}) is  that if we adopt the local coordinate $y$ around the point $p$, we can re-express the modification of the line bundle ${\cal L}_i$ at $p$ -- discussed in the paragraph before the previous -- as
\be
{\cal L}_i \to {\cal L}_i \otimes {\cal O}_p^{\langle^Lw, w\rangle},
\label{Hecke mod}
\ee
where  $\langle^Lw, w\rangle = {^Lw^i}$ is the scalar product between the (dominant) coweight $^Lw$ of $\GC$ and the simple (positive) root $w$. Thus, the action of  $H_{^Lw, p}$ with regard to $\cP$, is to induce the modification (\ref{Hecke mod}). Note that (\ref{Hecke mod}) just coincides with the mathematical description~\cite{Lennox} of a Hecke modification of type $^Lw$ of  a holomorphic $\GC$-bundle $\cP$ at a point $p \in \Sigma$. 

\def\cP{{\cal P}}

\bigskip\noindent{\it A Hecke Correspondence}

Of the Hecke modification of the bundle $\cP$ which results from the insertion of the Hecke operator $H_{^Lw, p}$ at $p$, one can say more as follows. Firstly, let the bundle $\cal P$ ``before'' and ``after'' the insertion of $H_{^Lw, p}$ be ${\cP}_-$ and ${\cP}_+$, respectively; then, we have an isomorphism 
\be
\sigma: \cP_-\vert_{\Sigma \backslash p} \  \tilde\longrightarrow  \ \cP_+\vert_{\Sigma \backslash p}.
\ee 
Since each $\sigma$ corresponds to a way of obtaining $\cP_+$ from $\cP_-$ via a Hecke transformation of type $^Lw$ at $p$, the space of all possible $\sigma$'s is in one-to-one correspondence with the space of all possible Hecke modifications of $\cP$ at $p$.  

Secondly, recall that the Hecke operator  $H_{^Lw, p}$ is parameterized by $^Lw$ -- a highest weight of $^L\GC$ -- which classifies conjugacy classes of homomorphisms $\zeta_\mathbb C: \mathbb C^\ast \to \GC$. In turn, $\zeta_\mathbb C$ defines a point in the affine Grassmannian
\be
{\rm Gr}_{\GC} = \GC ((y)) / \GC[[y]],
\ee
where $\GC((y))$ is the corresponding loop group that characterizes the space of continuous maps $S^1 \to \GC$, and $\GC[[y]]$ is the group of $\GC$ matrices whose entries are elements of the ring of formal power series in $y$. Thus, the space of all possible Hecke operators   $H_{^Lw, p}$ would be given by ${\rm Gr}_\GC$. 

Alternatively, notice that the definition of  $H_{^Lw, p}$ in (\ref{H_w}) implies that it would correspond to an element of the Lie algebra $\gc((y))$ of $\GC((y))$; in turn, via the exponential map, it would correspond to an element of $\GC((y))$. Since it is the orbit in  $\GC((y))$  under the action of $\GC[[y]]$ that depends on the conjugacy class of $\zeta_\mathbb C$ that $^Lw$ -- which enters explicitly in the definition of  $H_{^Lw, p}$ in (\ref{H_w}) -- classifies, the space of all possible Hecke operators $H_{^Lw, p}$ would be given by  $ \GC ((y)) / \GC[[y]] = {\rm Gr}_{\GC}$. Consequently, the space of \emph{all} Hecke modifications at $p$ would also be given by ${\rm Gr}_\GC$.  

Finally, let us now consider the moduli space  of triples $(\cP_-, \cP_+, \sigma)$. Since $\cP_-, \cP_+ \in {\rm Bun}_\GC$, one can regard this moduli space as a variety $\textit{Hecke}_p$ that maps to ${\rm Bun}_\GC \times {\rm Bun}_\GC$ with fiber over ${\cP_- \times \cP_+} \in {\rm Bun}_\GC \times {\rm Bun}_\GC$ being the space of all $\sigma$'s. In other words, from our discussion in the previous three paragraphs, the variety  $\textit{Hecke}_p$ is just a fibration over ${\rm Bun}_\GC \times {\rm Bun}_\GC$ with fiber ${\rm Gr}_\GC$.  This description of $\textit{Hecke}_p$ coincides with the mathematical description~\cite{Frenkel} of a Hecke correspondence for $\GC$  at a point $p \in \Sigma$. Thus, we have demonstrated, purely physically, the Hecke correspondence of the geometric Langlands program for $\GC$.

\newsection{Nonperturbative Effects, Beilinson-Drinfeld $\cal D$-Modules, And The Geometry Of Loop Spaces Of Flag Manifolds Of $G_{\mathbb C}$}

In our analysis of the gauge twisted $(0,2)$ sigma model carried out hitherto, we have ignored nonperturbative effects which are nonetheless important for a complete study of the model. In this section, we shall take into account such nonperturbative contributions. In doing so, we will find that when the string described by the sigma model propagates freely without any interactions over a flag manifold of any simply-connected, simple, complex Lie group $\GC$, nonperturbative worldsheet twisted-instantons can radically alter the picture and trivialize the chiral algebra completely,  thereby resulting in a spontaneous breaking of supersymmetry. This trivialization also implies that (i) there can be no Beilinson-Drinfeld $\cal D$-modules when the underlying curve is rational with less than three punctures, in accordance with the mathematical literature~\cite{Rubtsov}; (ii) $A(X, q)$ of (\ref{elliptic genus TCDO}) must vanish nonperturbatively in such a situation. We then interpret this nonperturbative phenomenon in the context of supersymmetric gauged quantum mechanics on loop space. In doing so, we will find that (i) there can be no harmonic spinors on the loop space of flag manifolds of $G_\mathbb C$, (ii) the aforementioned condition on $\cal D$-modules is intimately related to  a conjecture by H\"ohn-Stolz~\cite{Stolz} which asserts that the Witten genus is zero on string manifolds with positive Ricci curvature.

\newsubsection{Nonperturbative Effects In The $\mathbb {CP}^1$ Model}

Let us now revisit the $\mathbb {CP}^1$ model whose perturbative aspects were discussed in detail in $\S$4.7 and $\S$5.2.1. In $\S$5.2.1, we saw  that there exists an isomorphism between the bosonic operator $1$ and fermionic operator $\Theta$ of the chiral algebra $\mathscr A$. The fact that the isomorphism between the operators $1$ and $\Theta$ violates their grading by dimension suggests that nonperturbative worldsheet twisted-instantons -- which through dimensional transmutation destroy the scale invariance of correlation functions by which operators are dimensionally graded -- can induce a relation of the form $\{Q, \Theta\} \sim 1$ that would then account for their apparent isomorphism. A further indication that nonperturbative worldsheet twisted-instantons can induce a relation of the form $\{Q, \Theta\} \sim 1$, is the fact that the isomorphism between $1$ and $\Theta$ also violates their grading by $U(1)_R$ charge: since $c^T_1(\mathbb {CP}^1) \neq 0$ (where $T = U(1)$ in this case), from our discussion in $\S$4.2 surrounding (\ref{p}), worldsheet twisted-instantons also destroy the continuous $U(1)_R$ symmetry by which operators are $U(1)_R$ charge graded. 

If we do have the relation $\{Q, \Theta\} \sim 1$, it would mean that the identity operator is $Q$-exact and therefore, all operators in $\mathscr A$  would also be $Q$-exact: for any operator ${\cal O} \in \mathscr A$ whence $\{Q, {\cal O}] = 0$, we would have ${\cal O} = {\cal O} \cdot 1 \sim {\cal O} \cdot \{Q, \Theta\} = \{Q, {\cal O}  \Theta]$, i.e., ${\cal O} = \{Q, \dots]$. Thus, if we do have the relation $\{Q, \Theta\} \sim 1$, the chiral algebra $\mathscr A$ of the  $\mathbb {CP}^1$ model would be completely trivialized! Such a radical phenomenon was first conjectured by Witten in~\cite{CDO}  for the \emph{ordinary} $\mathbb {CP}^1$ model. This conjecture was subsequently proved by the author and Yagi in~\cite{instantons 1,instantons 2}, where it was shown that a relation of the form $\{Q, \dots \} \sim 1$ indeed exists if worldsheet instanton contributions are taken into account. Shortly thereafter, the conjecture was also proved purely mathematically by Arakawa and Malikov using the technology of vertex algebras~\cite{A-M}.  

For the rest of this subsection, we shall prove that the \emph{operator relation} $\{Q, \Theta\} \sim 1$ can indeed be induced by worldsheet twisted-instantons in our \emph{gauged} $\mathbb {CP}^1$ model -- i.e., we shall now proceed to show, via an explicit path integral computation, that 
\be
\big\langle  \{Q, \Theta\}   \big\rangle \sim \big\langle  1  \big\rangle
\label{to be proved}
\ee 
holds to lowest order in sigma model perturbation theory around worldsheet twisted-instantons characterized by classical, $Q$-fixed, field configurations $ \psi^i_{\bar z, 0}$ and $ \phi^i_0$ which obey $\delta \psi^i_{\bar z; 0}  = - D_{\bar z} \phi^i_0 = 0$. (See (\ref{txgauged}), and recall that to lowest order in sigma model perturbation theory, the path integral will localize onto $Q$-fixed points since the action is $Q$-exact.)

\bigskip\noindent{\it Some Preliminary Remarks}

To this end, first let us consider our defining worldsheet to be $\Sigma = {\bf S}^2 - \{0, \infty\}$, i.e., the Riemann sphere with two punctures, one at each pole. In this case, the canonical bundle $K$ of $\Sigma$ is trivial, and from our analysis in $\S$3, we find that the model is nonanomalous even though $c^T_1(\mathbb {CP}^1) \neq 0$. In addition, since $K$ is trivial, the twisted model is the same as the untwisted model.   

Second, with the previous point about the twisted model being equal to the untwisted model in mind, let us recall from (\ref{action-gen}) that the action (in the absence of moduli deformation) can be written as 
\be
S_{gauged} =  \int_{\Sigma} |d^2z| \  g_{i{\bar j}}  ( D_{\bar z} \phi^i D_z\phi^{\bar j}
 + \psi^i {\widehat D}_z \psi^{\bar j}),
\label{action CP1}
\ee
where from (\ref{replacement 1}) and (\ref{replacement 2}), we have
\be 
D_\zb \phi^i = \partial_{\bar z} \phi^i - A^a_{\bar z}V^i_a,\quad D_z \phi^\jb = \partial_z \phi^{\bar j} + A^a_zV^{\bar j}_a, \quad {\widehat D}_z \psi^{\bar j} = \partial_z \psi ^{\bar j} + \Gamma^{\bar j}_{\bar l \bar k} D _z \phi^{\bar l} \psi^{\bar k},
\label{DO formulas}
\ee 
(since $\partial_{\bar k} V_a^{\bar j} = 0$ in the flag manifold model). Note that since the target space $X = \mathbb {CP}^1$, we actually have $i = 1$, $\bar i = \bar 1$, and $a =1$. As one can write the quantum fields $\phi$ and $\psi$ as $\phi = \phi_0 + \varphi$ and $\psi = \psi_0 + \eta$, where $(\phi_0, \psi_0)$ and $(\varphi, \eta)$ are associated with the zero and nonzero eigenfunctions of the Laplacians of the kinetic operators in (\ref{action CP1}), respectively, for an appropriately chosen coordinate, one can expand the target space metric as
\be
g_{i \bar j} (\phi^l, \phi^{\bar l}) = \delta_{i \bar j} +  R_{i \bar j k \bar m} (\phi_0) \varphi^k \varphi^{\bar m} + O(\varphi^3).  
\label{metric}
\ee 
Using (\ref{metric}), let us now expand (\ref{action CP1}) around worldsheet twisted-instantons $\phi^\jb_0$ defined by $D_z \phi^\lb_0 = 0$. Since the magnitudes of the fluctuating fields $\varphi$ and $\eta$ are themselves small by definition, in the large but finite $X$-volume limit -- i.e., to lowest order in sigma model perturbation theory -- whence the effective magnitude of $R_{i \bar j k \bar m}$ is also small, one can, to a good approximation, write the expanded action as
\be
S_{good} = {1 \over 2 \pi} \int_{\Sigma} |d^2 z| \,  \delta_{i \bar j} \left( \del_{\bar z}\varphi^i \del_z \varphi^{\bar j} +  \eta ^i \partial_z \eta^{\bar j} \right)  + I_{int},
\label{Sgood}
\ee
where 
\be
\begin{split}
I_{int} & =  \delta_{i \bar j} \left(\partial_{\bar z} \varphi^i A^a_z V_a^{\bar j}  - A^a_{\bar z} V^i_a \partial_z \varphi^{\bar j}  - A^a_{\bar z} V_a^{i} A^b_z V_b^{\bar j} + \psi^i_0 \del_z\eta^\jb + \psi^i_0 \Gamma^\jb_{\lb\kb}(\phi_0) \del_z\varphi^\lb \eta^\kb  \right. \\
&  \left. \hspace{1.5cm}  + \psi^i_0 \Gamma^\jb_{\lb\kb} (\phi_0)  A^a_zV^\lb_a \eta^\kb   +  \eta^i \Gamma^\jb_{\lb\kb} (\phi_0) A^a_z V^\lb_a \eta^\kb  \right)
\end{split} 
\label{Iint}
\ee
can be regarded as interaction terms in a quantum field theory of the dynamical $\phi$ and $\psi$ fields.

Third, notice that since one can bring down $I_{int}$ via an expansion of $e^{-S_{good}}$ in any path integral computation, an arbitrary correlation function in operators ${\cal O}_i, \, {\cal O}_j, \dots {\cal O}_k$ can be expressed as 
\be
\big\langle {\cal O}_i {\cal O}_j \dots {\cal O}_k\big\rangle = \big\langle {\cal O}_i {\cal O}_j \dots {\cal O}_k\big\rangle_{\rm CFT} + {\sum^\infty_{n =1}} \, {(-2 \pi)^{-n} \over{n!}} \, \big\langle {\cal O}_i {\cal O}_j \dots {\cal O}_k \cdot  (  \int_{\Sigma} |d^2 z| \,  I_{int})^n   \big\rangle_{\rm CFT},
\label{corr fn}
\ee
where the subscript ``CFT'' means that the correlation functions are to be computed with respect to the purely kinetic action
\be
S_{\rm CFT} =  {1 \over 2 \pi} \int_{\Sigma} |d^2 z| \,  \delta_{i \bar j} \left( \partial_{\bar z}\varphi^i \partial_z \varphi^{\bar j} +  \eta ^i \partial_z \eta^{\bar j} \right).
\label{SCFT}
\ee

Fourth, notice that $S_{\rm CFT}$, being a free-field action in the $\varphi$ and $\eta$ fields, is an action of a CFT. Thus, in computing $\big\langle {\cal O}_i {\cal O}_j \dots {\cal O}_k\big\rangle$ via (\ref{corr fn}), one is free to exploit arguments from CFT. In particular, by the state-operator correspondence in CFT, one can insert additional local operators over the two punctures in $\Sigma$ in (\ref{corr fn}), whereby the worldsheet is now effectively a complete sphere. 

Fifth, note that on a simply-connected Riemann surface such as a complete sphere, one can go to a pure gauge everywhere (see footnote~19). This means that  after inserting the additional local operators over the two punctures in $\Sigma$, one can set $A =0$ throughout our computation of  $\big\langle {\cal O}_i {\cal O}_j \dots {\cal O}_k\big\rangle$ in (\ref{corr fn}). Assuming that we adopt this particular gauge henceforth, $\phi^i_0$ -- which \emph{a priori} represents a twisted-instanton characterized by $D_{\bar z} \phi^i_0 = 0$ -- will now represent an ordinary instanton characterized by $\partial_{\bar z} \phi^i_0=0$. 
Furthermore, from (\ref{Iint}), we see that $I_{int}$ in (\ref{corr fn}) will vanish except for the term $\psi_{\jb,0} (\del_z\eta^\jb +  \Gamma^\jb_{\lb\kb}(\phi_0) \del_z\varphi^\lb \eta^\kb )$. Also,  the right-moving supercurrent of the supercharge $Q$ will now be given by $G(\bar z) = g_{i \bar j} \partial_{\bar z} \phi^i \psi^{\bar j}$, while  $\Theta (z) = R_{i \bar j} \partial_z \phi^i \psi^{\bar j}$. 

Last but not least, note that in the pure gauge $A =0$, the zero modes $\psi_{ \bar j, 0}$ and $\psi^{\bar j}_0$ -- which are intrinsically worldsheet spinors -- obey
\begin{equation}
 D_z \psi_{\bar j,0} = D_z \psi^{\bar j}_0 = 0,
 \label{fermi zero modes}
\end{equation}
where $D_z$ is the Dirac operator on $\Sigma$ twisted by the Levi-Civita connection on $X$. After taking the complex conjugate of (\ref{fermi zero modes}), one can see that the zero modes are respectively \emph{holomorphic} sections of ${\cal S} \otimes \phi_0^*T^*X$ and  ${\cal S} \otimes \phi_0^*TX$, where $\cal S$ is the spin bundle on the worldsheet.  If the worldsheet is effectively a complete sphere, ${\cal S} = \CO(-1)$;\footnote{The notation here is standard: ${\cal O}(n)$ refers to a holomorphic line bundle whose sections are functions homogeneous of degree $n$ in the homogeneous coordinates of the effective worldsheet $\hat \Sigma = \mathbb {CP}^1$. In particular, $\CO(0)= \CO$  is a trivial complex line bundle.} for degree-one instantons, $\phi_0^*TX = \CO(2)$ and $\phi_0^*T^*X = \CO(-2)$; since the relevant zeroth Hodge numbers are $h^0(\CO(1)) = 2$ and $h^0(\CO(-3)) = 0$, we have, in this instance, two $\psi^{\bar j}$ zero modes and no $\psi_{\bar j}$ zero modes. This means that $I_{int}$ in (\ref{corr fn}) actually vanishes completely, and the correlation function  $\big\langle {\cal O}_i {\cal O}_j \dots {\cal O}_k\big\rangle_{\rm CFT}$ in (\ref{corr fn}) can only be nonvanishing if it contains operators with \emph{exactly} two  $\psi^{\bar j}$ zero modes and no $\psi_{\bar j}$ zero modes.

\bigskip\noindent{\it The Explicit Computation}

We are now ready to perform an explicit computation of $\left\langle \{Q, \Theta\} \right\rangle$ on $\Sigma = {\bf S}^2 - \{0, \infty\}$. Altogether from the six points above, and the fact that $G (\bar z) \Theta (z)$ contains exactly two  $\psi^{\bar j}$ zero modes and no  $\psi_{\bar j}$ zero modes, we can write
\be
{ \big\langle \{Q, \Theta\} (1) \big\rangle_{\Sigma}   } =   {\left\langle 1 (\infty) \, \oint d\bar z \, G(\bar z) \Theta (1) \, 1 (0) \right\rangle_{\rm CFT; \, {\bf S}^2}},
\label{main computation}
\ee
where for convenience, we have put the operator $\{Q, \Theta\}$ at $z=1$. Here, $1(0)$ and $1(\infty)$ are the $Q$-closed unit operators  that represent  respectively the perturbative supersymmetric ground state in the far past and the perturbative supersymmetric de-excited state in the far future of the freely propagating string described by the sigma model.\footnote{The operators ${\mathscr V}_p (0)$ and $ {\mathscr V}_f (\infty)$ which actually represent the perturbative supersymmetric ground state in the far past and the perturbative supersymmetric de-excited state in the far future of the string, respectively, are supposed to be dimension-zero, $\psi$-free local operators in the perturbative chiral algebra $\mathscr A$. Hence, they ought to be given by regular, non-differential polynomials in $\phi$ which are holomorphic in the coordinate $z$ of ${\bf S}^2$. Since ${\bf S}^2$ is a compact Riemann surface, all regular holomorphic functions in $z$ over it are equivalent to constants. Thus, we can write (up to irrelevant constants) ${\mathscr V}_p (0) = 1(0)$ and $ {\mathscr V}_f (\infty) = 1(\infty)$.} As the subscripts indicate, the expression on the RHS of (\ref{main computation}) is to be computed  on ${\bf S}^2$ with respect to the action (\ref{SCFT}). The computation is also to be carried out to lowest order in sigma model perturbation theory around worldsheet instantons $\phi_0$ which satisfy $\partial_z \phi^{\bar i}_0 =0$ and $\partial_{\bar z} \phi^i_0 = 0$.

With these considerations in mind, first note that the automorphism group of ${\bf S}^2$ is given by the M\"obius group $PGL(2, \mathbb C)$. Thus, we can use a M\"obius transformation on ${\bf S}^2$ to map the three points $(0,1, \infty)$ to $(z_1, z_2, \infty)$, where $z_1$ and $z_2$ can take any desired value other than $\infty$. Let us therefore use a particular transformation on ${\bf S}^2$ whereby one can rewrite the RHS of (\ref{main computation}) as
\be
{\left\langle    1(\infty) \, 1(1)  \oint d\bar z \, G(\bar z) \Theta (0) \right\rangle_{\rm CFT; \, {\bf S}^2}}.
\label{proof 1}
\ee

Next, recall that we have two $\psi^{\bar j}$ zero modes. This means that one can write 
\be
\psi^{\bar j}_0 = c^1_0 u^{\bar j}_{0,1} + c^2_0 u^{\bar j}_{0, 2},   
\label{8.18}
\ee
where the $u_0$'s are the zero eigenfunctions of the Laplacian of the Dirac operator in (\ref{fermi zero modes}), and the $c$'s are anticommuting Grassmann coefficients. Therefore, by performing  in (\ref{proof 1}) the Gaussian integration over the nonzero modes $\varphi$ and $\eta$  with purely kinetic action, up to a ratio of fermionic to bosonic determinants, we are left with the following integral over the zero modes:  
\be
 \int d {\CM}_1 \, d c^1_0 \, d c^2_0    \, \oint d\bar z \, [G(\bar z) \Theta (0)]_{\rm OPE} \, 1(1) \, 1(\infty).
\label{proof 4}
\ee  
Here, $\CM_1$ is the one-instanton moduli space parameterized by the bosonic zero modes $\phi_0$, and the subscript ``OPE'' just means that one is to compute the OPE of $G(\bar z) \Theta (0)$ using the following propagator relations derived from the effective free-field action in (\ref{SCFT}):
\be
\varphi^i(z, \bar z) \varphi^{\bar j} (0) \sim - \delta^{i \bar j} \, {\rm ln} \, |z|^2 \qquad {\rm and} \qquad \eta^i (\bar z) \eta^{\bar j} (0) \sim {\delta^{i \bar j} \over \bar z}.
\label{prop relations}
\ee

Now observe that what we should be looking for in order to prove (\ref{to be proved}) is an antiholomorphic\emph{ single} pole in $ [G(\bar z) \Theta (0)]_{\rm OPE}$. Up to quadratic order in the fluctuating fields at the lowest order in sigma model perturbation theory, we have 
 \be
 G(\bar z) \cdot \Theta(0) =  \left(\delta_{i \bar j} \partial_{\bar z} \varphi^i (\psi^{\bar j}_0 + \eta^{\bar j}) \right)(\bar z) \cdot \left(R_{l \bar k} (\phi^m_0, \phi^{\bar m}_0) (\partial_z \phi^l _0 + \partial_z \varphi^l)  (\psi^{\bar k}_0 + \eta^{\bar k}) \right)(0).
 \label{G-Theta expansion}
 \ee 
However, as explained above, since only the part which contains exactly two $\psi^{\bar j}$ zero modes will contribute nonvanishingly to the calculation, in computing  $ [G(\bar z) \Theta (0)]_{\rm OPE}$,  we can take
\be
 G(\bar z) \cdot \Theta(0) =  \left(\delta_{i \bar j} \partial_{\bar z} \varphi^i \psi^{\bar j}_0 \right) (\bar z)  \cdot \left(R_{l \bar k} (\phi^m_0, \phi^{\bar m}_0) (\partial_z \phi^l_0 + \partial_z \varphi^l)  \psi^{\bar k}_0 \right) (0).
 \label{GT}
\ee 
We can further simplify (\ref{GT}) by expanding $\psi^{\bar j}_0 (\bar z)$ around $\bar z = 0$ to get
\be
G(\bar z) \cdot \Theta(0) =  \left(\delta_{i \bar j} \partial_{\bar z} \varphi^i  \right)(\bar z) \, \left(\psi^{\bar j}_0 (0) +  \bar z \, (\partial_{\bar z} \psi^{\bar j}_0) (0)  + \dots\right)   \cdot \left(R_{l \bar k}(\phi^m_0, \phi^{\bar m}_0)(\partial_z \phi^l_0 + \partial_z \varphi^l) \right) (0) \, \psi^{\bar k}_0 (0).
\label{GT2}
\ee
Due to the Grassmannian nature of the $\psi$ zero modes, we have $\psi^{\bar j}_0(0) \psi^{\bar k}_0(0) = 0$ (since $\bar j = \bar k =  \bar 1$). Therefore, to a good approximation at lowest order in $\bar z \ll 1$, (\ref{GT2}) is in fact
\be
G(\bar z) \cdot \Theta(0) =  \left( \bar z  \delta_{i \bar j}  \partial_{\bar z} \varphi^i \right)(\bar z)   \cdot \left(R_{l \bar k}(\phi^m_0, \phi^{\bar m}_0) (\partial_z \phi^l_0 + \partial_z \varphi^l)  \partial_{\bar z} \psi^{\bar j}_0 \psi^{\bar k}_0  \right) (0).
\label{GT3}
\ee
In light of the nontrivial propagator relations in (\ref{prop relations}), one can see that the $\varphi^i$ and $\varphi^l$ fields in (\ref{GT3}) will not interact. Hence, for the purpose of our computation, we can discard the $\partial_z \varphi^l$ term in (\ref{GT3}) (which will eventually contribute to higher orders in the fluctuations anyway), and reduce  (\ref{GT3}) to 
\be
G(\bar z) \cdot \Theta(0) =  \left( \bar z  \delta_{i \bar j}  \partial_{\bar z} \varphi^i \right)(\bar z)   \cdot \left(R_{l \bar k}(\phi^m_0, \phi^{\bar m}_0) \partial_z \phi^l_0  \partial_{\bar z} \psi^{\bar j}_0 \psi^{\bar k}_0  \right) (0).
\label{GT4}
\ee
Looking at (\ref{GT4}), we now know that because of the extra $\bar z$ variable in the first term, if there is to be a single antiholomorphic pole in $[G(\bar z) \Theta (0)]_{\rm OPE}$, there ought to be  a \emph{double} antiholomorphic pole coming from the contractions of the rest of the fields. From (\ref{GT4}), this appears to be impossible, or is it?

Note that the Laplacian of the Dirac operator in (\ref{fermi zero modes}) and therefore its  nontrivial eigenfunctions $u_n$, depend on the quantum fields $\phi^l$ and $\phi^\lb$; in turn,  this means that the mode expansion of $\psi^{\bar j}$ -- defined with respect to a spectral decomposition of this Laplacian -- would also depend on the quantum fields $\phi^l$ and $\phi^\lb$. Thus, one can, in general, write the zero modes of $\psi^{\bar j}$ as 
\be
\psi^{\bar j}_0 (z, \bar z; \phi)  = c^1_0 \, u^{\bar j}_{0,1} (z, \bar z, \phi^l_0, \phi^{\bar l}_0, \varphi^l,   \varphi^{\bar l}) + c^2_0 \, u^{\bar j}_{0, 2} (z, \bar z,  \phi^l_0, \phi^{\bar l}_0,  \varphi^l, \varphi^{\bar l}).
\ee
Consequently, $\psi_0$ can actually participate in a contraction with $\varphi$ in (\ref{GT4}). Hence, it is still possible that an antiholomorphic double pole can be found after contracting the aforementioned fields with each another. 


In order to ascertain whether the relevant contractions would yield an antiholomorphic double pole or not, we have to determine the explicit dependence of  $\psi^{\bar j}_0$ on the fluctuating bosonic field $\varphi$.   To this end, let us first expand $\psi^{\bar j}$  explicitly in the eigenfunctions $u_n$ of the Laplacian of the Dirac operator in (\ref{fermi zero modes}) as
\begin{equation}
\psi^{\bar j}(z,\zb; \phi) = \sum_{r =1}^2 c_0^r u^{\bar j}_{0,r}(z,\zb; \phi) + \sum_n c^n u^{\bar j}_n(z,\zb; \phi).
\label{exp psi}
\end{equation}
Next, note that for some point $\phi$ in the neighborhood $\CM'_1$ of the instanton moduli space $\CM_1 \subset \Map^1({\bf S}^2, X)$ -- where $\Map^1({\bf S}^2, X)$ is the space of all degree-one (anti)holomorphic maps from ${\bf S}^2$ to $X = \mathbb {CP}^1$ -- one can expand its fluctuation $\varphi$ near and \emph{orthogonal} to an instanton $\phi_0 \in \CM_1$ in the eigenfunctions of the Laplacian of the Dolbeault operator on $\Sigma$ as\footnote{Notice that the $\psi$'s and $\varphi$'s have the same nonzero eigenfunctions $u_n$. This is because (i)~the canonical bundle on $\Sigma$ is trivial and hence, the Dirac and Dolbeault operators are one and the same thing; (ii)~the fluctuations $\varphi$ define variations of the coordinates $\phi(z, \zb)$ of $X$ around $\phi_0(z, \zb)$ and therefore, can be viewed as tangent vectors in $X$, like the $\psi$'s.}

\begin{equation}
\varphi^i(z,\zb; \phi_0) = \sum_n a^n u^i_n(z,\zb;\phi_0), \qquad \varphi^{\bar j}(z,\zb; \phi_0) = \sum_n {\bar a}^n u^{\bar j}_n (z,\zb;\phi_0).
\end{equation}
Now, let the set of complex parameters $\{\zeta^{0, r}\}$ be holomorphic coordinates on $\CM_1$; let the set of complex parameters $\{\zeta^n_\perp\}$ be holomorphic coordinates in the space normal to $\CM_1 \subset \CM'_1$; then, we can write the point $\phi$ in $\CM'_1$ as $\phi^i(z,\zb; \zeta, \zeta_\perp)$.  Also, for an instanton $\phi_0$ that necessarily lies along $\CM_1$, we can (recalling that $\partial_{\bar z} \phi^i_0=0$) write $\phi_0^i(z;\zeta)$. Via the fact that (i) for a nearby instanton ${\tilde \phi}^{\jb}_0 = \phi^\jb_0 + \delta \phi^\jb_0$ which obeys $\del_z {\tilde \phi}^{\jb}_0 = 0$, we have $D_z (\delta \phi^\jb_0) = 0$ (to lowest order in $\delta \phi^\jb_0$), where $D_z$ is exactly the differential operator in (\ref{fermi zero modes}) which  defines the zero modes of $\psi^\jb$; (ii) the nonzero modes of any quantum field are orthogonal to the zero modes; one can view $\psi^\jb$ as an odd vector in $ \overline{T_\phi\CM'_1}$ with the  following general expansion
\begin{equation}
\psi^{\bar j}(z,\zb; \phi) = \sum_{r=1}^2 \tilde c_0^r \, {d\phi^\jb \over d{\bar \zeta}^{0,r}} + \sum_n \tilde c^n \, {d\phi^\jb \over d {\bar\zeta}^n_\perp},
\label{gen ex}
\end{equation}
where $(d/d{\bar \zeta}^{0, r}, d/d {\bar \zeta}^n_\perp) \in \overline{T_\phi\CM'_1}$. Again, the $\tilde c$'s are anticommuting Grassmannian coefficients.  By comparing (\ref{gen ex}) with (\ref{exp psi}), we see that in an appropriate basis whereby $\tilde c = c$, we necessarily have $d\phi^\jb/d{\bar \zeta}^{0,r} =  u^{\bar j}_{0,r}$ and $ d\phi^\jb/d {\bar \zeta}^n_\perp =  u^{\bar j}_n$.

Note at this juncture that a one-instanton maps the ${\bf S}^2$ effective worldsheet to the $\mathbb {CP}^1$ target space in a one-to-one manner; this means that the function $\phi^\jb_0(\zb; \bar \zeta)$ is invertible, i.e., we can write $\zb(\phi^\jb_0; \bar \zeta)$. In turn, we can express $\phi^\jb$ as
\be
\phi^\jb(z,\zb; \bar \zeta, {\bar \zeta}_\perp) = \hat\phi^\jb(\phi^i_0 (z; \zeta), \phi^{\bar i}_0 (\bar z; \bar \zeta); \bar \zeta, {\bar \zeta}_\perp). 
\label{re-express}
\ee
Using (\ref{re-express}), we can write (\ref{gen ex}) as
\be
\psi^{\bar j}(z,\zb; \phi)  =  \sum_{r=1}^2 \tilde c_0^r \,  \left({\del\hat\phi^\jb \over   \del\phi^\ib_0} { \del\phi^\ib_0 \over \del{\bar \zeta}^{0,r}}   + {\del \hat\phi^\jb \over \del  {\bar \zeta}^{0,r}} \right) + \sum_n \tilde c^n \, {d\hat \phi^\jb \over d {\bar \zeta}^n_\perp}.
\label{simplify psi}
\ee
Since $ \del_\zb \phi^\jb  = ( \del\hat\phi^\jb / \del\phi^\ib_0) \del_\zb \phi^\ib_0$,  assuming that we choose a basis whereby $\tilde c = c$, from (\ref{simplify psi}), the zero modes of $\psi^{\bar j}(z,\zb; \phi)$ are determined to be
\be
\psi^\jb_0 (z, \bar z, \phi) =  \sum_{r=1}^2 c_0^r \, \left( { \del_\zb \phi^\jb \over  \del_\zb \phi^\ib_0} u^{\bar i}_{0,r} (z, \zb, \phi_0)  +  {\del \hat\phi^\jb \over \del  {\bar \zeta}^{0,r}},\right),
\label{exp dep}
\ee
where we have used the fact that $ u^{\bar i}_{0,r} (z, \zb, \phi) = d\phi^\jb/d{\bar \zeta}^{0,r}$,  so $u^{\bar i}_{0,r} (z, \zb, \phi_0) = {d\phi^\ib_0 / d{\bar \zeta}^{0,r}} = {\del\phi^\ib_0 / \del{\bar \zeta}^{0,r}}$. 

Now that we have obtained the explicit dependence  of $\psi^\jb_0 (z, \bar z, \phi)$ on the fluctuating quantum field $\phi^\jb$ in (\ref{exp dep}), we are ready to compute  (\ref{GT4}). By inspecting (\ref{exp dep}) and (\ref{GT4}), it is not hard to see that the relevant term in (\ref{exp dep}) which can possibly contribute to an antiholomorphic double pole is
\be
\psi^\kb_0  =  \sum_{r=1}^2 c_0^r \, { \del_\zb \varphi^\kb \over  \del_\zb \phi^{\bar p}_0} u^{\bar p}_{0,r} (z, \zb, \phi_0).
\label{zero mode eff}
\ee
Indeed, by substituting (\ref{zero mode eff}) into (\ref{GT4}), we have a term involving $\del_{\zb} \varphi^i (\zb) \del_\zb \varphi^\kb (0)$ which upon using (\ref{prop relations}), gives us an antiholomorphic double pole. Consequently, up to quadratic order in fluctuations,  we have   
\be
[G(\zb) \Theta(0)]_{\rm OPE} = {1 \over \zb} \,  \left(R_{i \bar j}(\phi^m_0, \phi^{\bar m}_0) {\partial_z \phi^i_0 \over  \del_\zb \phi^{\bar p}_0}   \partial_{\bar z} \psi^{\bar j}_0 \psi^{\bar p}_0\right) (0) + \dots,
\label{final GT OPE}
\ee
where  the fermionic zero modes $\psi^\lb_0(0, 0; \phi^i)$ are to be evaluated at $\phi^i = \phi^i_0$, and where the ellipsis represents terms that are either regular or have poles of degree greater than one.

To complete the computation, we will need the explicit form of $\psi^\lb_0$. To this end, note that since our one-instanton $\phi^i_0$  is a biholomorphic map from the efffective worldsheet $\hat \Sigma = \mathbb {CP}^1$ to a target space $X = \mathbb {CP}^1$, it will be given by a M\"obius transformation:
\be
\phi^1_0(z) = {az + b \over cz + d}; \qquad ad-bc = 1. 
\label{Mobius tx}
\ee
(Recall that the indices $(i, \bar i)$ only take the values of $(1, \bar 1)$ since $X = \mathbb {CP}^1$.) As such,  we have
\be
\psi^{\bar 1}_0 = c^1_0 u^{\bar 1}_{0,1} + c^2_0 u^{\bar 1}_{0, 2},   
\label{final psi zero modes}
\ee
where $u^{\bar 1}_{0,1}$ and $ u^{\bar 1}_{0, 2}$ are~\cite{DG}
\be
u^{\bar 1}_{0,1} (\zb) = {1 \over {\cb \zb + \db}}, \qquad u^{\bar 1}_{0,2} (\zb) = {1 \over {\cb(\cb \zb + \db)^2}}.
\label{exp form of psi zero modes}
\ee
If we now substitute the expressions (\ref{final GT OPE}), (\ref{final psi zero modes}) and (\ref{exp form of psi zero modes}) in (\ref{proof 4}), and carry out the contour integral in the complex variable $\zb$ plus the integration over the Grassmannian variables $c^1_0$ and $c^2_0$, we will get (up to a constant)
\be
e^{-t(\mu)} \int d \CM_1 \, \left(i R_{1 \bar 1} (\phi^1_0, \phi^{\bar 1}_0) \partial_z \phi^1_0 \del_\zb \phi^{\bar 1}_0\right) (0)  \,   1(1) \,  1(\infty),
\label{final computation to do}
\ee  
where we have included the nonzero sigma model contribution 
\be
t(\mu) = \int_{\hat \Sigma} \phi^\ast_0 \, {\cal K} (\mu)
\label{t(u)}
\ee
due to worldsheet instantons. Here, ${\cal K}(\mu) = {\cal K}_0 + {\rm ln} (\mu / \Lambda) \, c_1(X)$  is the effective K\"ahler class at energy scale $\mu$; $\Lambda$ is a scale parameter; and ${\cal K}_0 \in H^2(X, \mathbb R)$ is the ``bare'' K\"ahler $(1,1)$-form on $X$. In terms of the points $X_0, X_1, X_\infty \in X$ which $0, 1, \infty \in \hat \Sigma$ are mapped to by one-instantons, a conformally invariant measure  on $\CM_1$  can also be written (up to an overall constant) as~\cite{DSWW} 
\be
d\CM_1 = {d^2X_0 d^2X_1 d^2X_\infty \over |X_0 - X_1|^2 |X_1 - X_\infty|^2 |X_\infty - X_0|^2}.
\label{mod measure}
\ee
The parameterization of $\CM_1$ by  $X_0, X_1, X_\infty$ allows us to identify $\CM_1$ with $(\mathbb {CP}^1)^3$.  Substituting (\ref{mod measure}) in (\ref{final computation to do}) while using the formula
\be
\del_z \phi^1_0 (0)= {(X_\infty - X_0) (X_0 - X_1) \over (X_1 - X_\infty)}, 
\ee
we then have 
\be
e^{-t(\mu)} \int_{\mathbb {CP}^1} d^2X_0 \, i R_{1 \bar 1} (X_0, {\bar X}_0) \int_{(\mathbb {CP}^1)^2} {d^2X_1 1(X_1, {\bar X}_1))   \over |X_1 - X_\infty|^4} \, d^2X_\infty \,1(X_\infty, {\bar X}_\infty).
\ee 
The $X_0$-integral is the evaluation of $2 \pi c_1(\mathbb {CP}^1) = i R_{1 \bar 1} \, d X_0 \wedge d \bar X_0$ on $\mathbb {CP}^1$; this just gives a constant of $4 \pi$. The $X_1$-integral diverges, reflecting the noncompactness of $\CM_1$; nevertheless, by imposing a lower bound on the distance between $X_1$ and $X_\infty$ measured with the target space metric, i.e., $g_{1 \bar 1} (X_\infty, {\bar X}_\infty)|X_1 - X_\infty|^2 \geq  l^2$ for some $l^2 > 0$, the integral can be regularized as
\be
\int_{\mathbb {CP}^1} \, {d^2X_1 1(X_1, {\bar X}_1)) \over  |X_1 - X_\infty|^4} \, \leq  \, {g_{1 \bar 1} (X_\infty, {\bar X}_\infty) \over l^2}. 
\label{integral}
\ee  
Since any flag manifold model can undergo RG flow whence the effective size of the target space $X$ varies proportionally with the  energy scale, one can interpret ${g_{1 \bar 1} (X_\infty, {\bar X}_\infty)/ l^2}$ on the RHS of (\ref{integral}) as the effective metric at an energy scale that differs from that implied by the large metric $g_{1 \bar 1} (X_\infty, {\bar X}_\infty)$ whenever $l^2 \neq 1$. That being said, since we ought to choose $l^2 \ll 1$ to best approximate the integral, keeping the integral nonetheless finite by asserting that $l > 0$ is tantamount to staying within the large but finite $X$-volume limit assumed throughout our computation. Indeed if one were to let $l \to 0$ so as to allow the integral \emph{per se} to diverge, because we will be in the infinite $X$-volume limit, $\phi_0$ would just be a constant map (i.e.,~$d\phi_0 = 0$) whence the integral over $\CM_1$ would actually be an integral over $X$ -- in other words, the question of a divergent result due to the $X_1$-integral would be irrelevant to begin with. Thus, it is clear that the regularization in (\ref{integral}) would be necessary if our instanton computation is to make any \emph{physical} sense. Since anything less than the upper bound of the $X_1$-integral is just a number (that is less than 1) times the RHS of (\ref{integral}), we finally have 
\be
{l^{-2} \,  \left({\Lambda \over \mu}\right)^2  \, e^{-t_0}} \int_{\mathbb {CP}^1}  d^2X_\infty \, g_{1 \bar 1} (X_\infty, {\bar X}_\infty) \,  1(X_\infty, {\bar X}_\infty)
\label{zero-instanton}
\ee 
up to a constant, where $t_0 = \int_{\hat \Sigma} \phi^\ast_0 ({\cal K}_0)$. (Note that we have made use of the fact that $\int_{\hat \Sigma} \phi^\ast_0 (c_1(X)) = 2$ in writing the above equation.)

A few remarks concerning (\ref{zero-instanton}) are in order. First, note that since a flag manifold has positive Ricci curvature, its sigma model is asymptotically-free, i.e., the effective size of $X$ increases or decreases as $\mu$ increases or decreases; because $l$ is inversely proportional to the effective size of $X$, and because $l$ and $\mu$ have dimensions of length and ${\rm length}^{-1}$, respectively, one can identify $l$ with $\mu^{-1}$. Second, notice that the $X_\infty$-integral is performed over $\CM_0 \cong X$ -- the zero-instanton moduli space -- with respect to the natural volume $(1,1)$-form $g_{1 \bar 1} \, dX_\infty \wedge d{\bar X}_\infty$. Third, recall that there is actually a ratio of fermionic to bosonic determinants in the overall (omitted) constant which accompanies (\ref{zero-instanton}); these determinants can be expressed as Gaussian integrals over the non-zero fermionic and bosonic modes $\eta$ and $\varphi$, respectively. Fourth, note that there are no fermionic zero modes for degree zero instantons; in particular, the Grassmannian variables $(c^1_{0}, c^2_0)$ will not appear in the path integral measure in the zero-instanton sector. In sum, this means that one can also write (\ref{zero-instanton}) as the following partition function on ${\bf S}^2$ with respect to the action (\ref{SCFT}) in the zero-instanton sector:
\be
{\Lambda^{2} \,  e^{-t_0}} \left\langle  1(\infty) \right \rangle_{{\rm CFT}; {\bf S}^2}.
\label{pf S2}
\ee
By inserting unit operators at $z =0, 1$ which would not change our result, we can also write (\ref{pf S2}) as
\be
{\Lambda^{2} \,  e^{-t_0}} \left\langle  1(\infty) \, 1(1) \, 1(0)  \right \rangle_{{\rm CFT}; {\bf S}^2}.
\label{second last}
\ee
Can more be said about (\ref{second last})? Most certainly. From the state-operator isomorphism in CFT which allows us to exchange a local operator on the worldsheet for a puncture which represents a \emph{corresponding} state, we find that (\ref{second last}) can actually be recast as
\be
{\Lambda^{2} \,  e^{-t_0}} \left\langle  1(1) \right \rangle_\Sigma,
\label{last}
\ee
where $\Sigma = {\bf S}^2 - \{0, \infty\}$ is the defining worldsheet of the freely propagating string that we started with. Since (\ref{last}) is, up to a constant, just the RHS of (\ref{main computation}), we conclude that in the large but finite $X$-volume limit -- i.e., to lowest order in sigma model perturbation theory -- whence $t_0 < \infty$ so (\ref{last}) does not vanish,  we have, in a one-instanton background,
\be
\left\langle \{Q, \Theta\}  \right \rangle_\Sigma    \sim \left\langle  1 \right \rangle_\Sigma.
\label{2 pts deleted}
\ee
This coincides with our claim in (\ref{to be proved}). 

\bigskip\noindent{\it Computation for Other Worldsheets}

What about for other choices of $\Sigma$?  Would the computation also hold? Let us consider another nonanomalous example:  $\Sigma = {\bf S}^2 - \{ \infty\}$. In this case, the computation is exactly the same as above except that in (\ref{main computation}), the unit operator $1(0)$ is now being inserted as a dummy operator while the unit operator $1(\infty)$ actually represents the de-excited state in the far future of the underlying string which is propagating freely from the unknown past. Consequently, we would still end up with (\ref{second last}) but without the $1(0)$ operator.  Thus, we again have $\left\langle \{Q, \Theta\}  \right \rangle_{\Sigma}    \sim \left\langle  1 \right \rangle_{\Sigma}$. Therefore, \emph{we have the operator relation $\{Q, \Theta\} \sim 1$ whence the chiral algebra $\mathscr A$ will be completely trivialized, if $\Sigma$ is a genus zero complex curve with two or less punctures}. 

As discussed in the previous section, the physical model would be nonanomalous as long as $\Sigma$  is an ${\bf S}^2$ with one or more punctures. Therefore, let us proceed to consider the computation in the case where there are three or more punctures. When there are three or more punctures, the RHS of (\ref{main computation}) will contain four or more operator insertions, including $1(0)$, $\oint d\zb \, G(\zb) \Theta(1)$ and $1(\infty)$. Note at this juncture that (i) $\CM_1$ (or rather, its compactified version) as expressed via the measure (\ref{mod measure}), is actually the same as the moduli space of (stable) maps $\CM_{g, n} (X, d)$ of genus $g =0$, degree $d =1$, with $n=3$ identified points (i.e.,~$0,1, \infty$); (ii) the complex dimension of $\CM_{g, n} (X, d)$ is given by $d_\CM = \int_{\hat \Sigma} \phi^\ast_0 c_1(X)  - 2(1-g) + n$;\footnote{Indeed, since $\int_{\hat \Sigma} \phi^\ast_0 c_1(X) =2$ when $d=1$ and $\hat \Sigma = {\bf S}^2$, the complex dimension of $\CM_1$ according to this formula will be $3$, consistent with (\ref{mod measure}).} (iii) since the number of $\psi$ zero modes here is the same as in the previous case with two or less punctures, in order to have a nonvanishing result, one cannot insert $\psi$-dependent local operators over any of the punctures; (iv)  from the fact that (\ref{last}) is just the RHS of (\ref{main computation}), we actually have, up to a dimensionless constant, $\left\langle \{Q, \Theta\} \right\rangle_\Sigma = {\Lambda^{2} \,  e^{-t_0}} \left\langle  1(1) \right \rangle_\Sigma$, where $\Lambda^2$, like the operator $ \{Q, \Theta\}$, has scaling dimension 2.\footnote{The local operator $[\{Q, \Theta\}]$ will have scaling dimension 2 because the underlying  model is effectively untwisted, i.e., $Q$ and $\Theta$ will have scaling dimensions $1/ 2$ and $3/2$, respectively.}    From (i)-(iv), we can deduce the following. First, from (i) and (ii), we know that the computation will eventually involve an integration over $d_M$ variables $X_1, \dots, X_{d_m}$, where $d_m = p +1$ and $p$ is the number of punctures. Second, from (iii), it will mean that we can only insert over the additional punctures either unit operators or operators which are composed of the $\phi$ fields and possibly their worldsheet derivatives. Third, if unit operators are inserted over the extra punctures at $p_i \neq 0, 1, \infty$, one would need to regularize the corresponding $X_i$-integrals as was done in (\ref{integral}), where each regularized integral would introduce an inverse length scale of scaling dimension 2; inserting operators that are worldsheet derivatives of the $\phi$ fields over the extra punctures $p_i \neq 0, 1, \infty$, would only serve to add on to the scaling dimension of the final answer.  Hence, from  (iv) -- which  tells us that our final answer ought to have scaling dimension exactly 2 -- and the aforementioned three points, it would mean that we can only consider operators which are regular, non-differential polynomials in $\phi$ over the additional punctures $p_i \neq 0, 1, \infty$. These local operators over the additional punctures ought to be from the perturbative chiral algebra $\mathscr A$ since they are supposed to represent perturbative supersymmetric states of the string. This means that they will be holomorphic in the coordinate $z$ of the effective worldsheet $\hat \Sigma = {\bf S}^2$. At any rate, since $\hat \Sigma$ is a compact Riemann surface, all regular holomorphic functions in $z$ over it must be constants; in other words, these local operators which are regular, non-differential polynomials in $\phi$ are actually equivalent (up to irrelevant constants) to the unit operator. In sum, due to a mismatch in scaling dimensions, we cannot have the operator relation $\{Q, \Theta\} \sim 1$ on a $\Sigma$ that is an ${\bf S}^2$ with three or more punctures. Hence, we conclude that \emph{the operator relation $\{Q, \Theta\} = 0$ will continue to hold if $\Sigma$ is a genus zero complex curve with three or more punctures}. 

Other choices of $\Sigma$ that lead to a nonanomalous physical model include the torus ${\bf T}^2$ with any number of punctures, and a genus $g > 1$ Riemann surface $\Sigma_g$ with $2g -2$ punctures. If $\Sigma$ is a ${\bf T}^2$ with no punctures, there are no degree-one holomorphic maps from $\Sigma$ to $X = {\mathbb {CP}}^1$, for if there are, $\Sigma$ would be isomorphic to $X$, which is not the case. Now a worldsheet twisted-instanton $\phi_0$ is defined by $\del_z\del_\zb \phi^l_0 = \del_z A^a_\zb V_a^l$ and $\del_\zb \del_z \phi^\lb_0 = - \del_\zb A^a_z V^\lb_a$. Because the previous point implies that $\del_z \del_\zb \phi^{l, \bar l}_0 \neq 0$ while a vanishing field strength implies that $\del_zA^a_\zb = \del_\zb A^a_z = 0$, it would mean that there can be no worldsheet twisted-instantons and therefore, one cannot have the operator relation $\{Q, \Theta\} \sim 1$ on a $\Sigma$ that is a ${\bf T}^2$ with no punctures.  If $\Sigma$ were to be a ${\bf T}^2$ with $p > 0$ number of punctures at positions $z_1, \dots, z_p$, we can first replace the $p$ punctures with certain state-representing local operators $\mathscr V_1(z_1), \dots, \mathscr V_p(z_p)$, as was done in the earlier case of the sphere with punctures. Because we now have a complete ${\bf T}^2$, the aforementioned arguments follow, whence we find that we cannot have the operator relation $\{Q, \Theta\} \sim 1$. Therefore, we conclude that\emph{  the operator relation $\{Q, \Theta\} = 0$ will continue to hold if $\Sigma$ is a genus one complex curve with any number of punctures}. 

\begin{figure}
  \centering
    \includegraphics[width=0.6\textwidth]{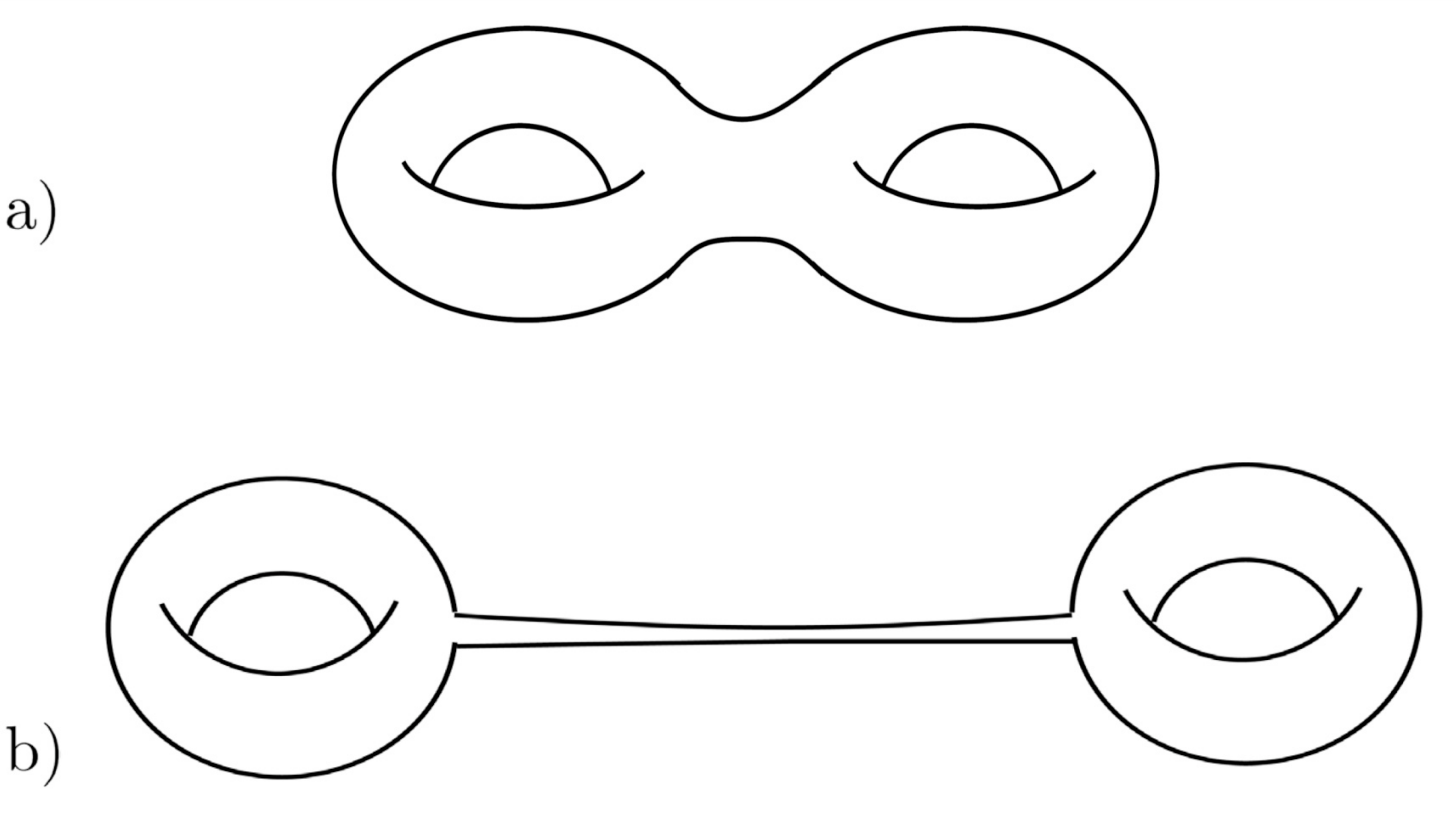}
  \caption{a) A genus 2 surface; b) Degeneration into a pair of ${\bf T}^2$'s}
\end{figure}

What about when $\Sigma$ is $\Sigma_g$ with $2g-2$ punctures? Before we answer this question, observe that $\Sigma_g$ is just $g$ number of ${\bf T}^2$'s connected pairwise by $g -1$ tubes. Under a conformal change of metric whereby these tubes degenerate and pinch off, we will have $g$ independent ${\bf T}^2$'s with punctures at the degeneration points. The $g=2$ case is illustrated in fig.~1. Conversely, one can construct $\Sigma_g$ by sewing in $(g-1)$ tubes at pairs of punctures on adjacent ${\bf T}^2$'s. This observation about Riemann surfaces has been exploited by string theorists to calculate higher genus amplitudes from lower genus ones via what is by now well-known as the factorization formula in CFT:
\be
\left \langle \dots_1 \dots_2 \dots \dots_g \right \rangle_{\Sigma_g}  = \sum_{a_1a_2 \dots a_{2g-2}} \left \langle \dots_1 \, {\mathscr A}^{(z_1)}_{a_1} \right \rangle_{{\bf T}^2_1} G^{a_1a_2} \left \langle {\mathscr A}^{(z_2)}_{a_2} \, \dots_2 \, {\mathscr A}^{(z_3)}_{a_3} \right \rangle_{{\bf T}^2_2} \dots \left\langle {\mathscr A}^{(z_{2g -2})}_{a_{2g-2}} \, \dots_g \right \rangle_{{\bf T}^2_g}.
\label{factorization}
\ee
Here, ${\bf T}^2_k$ denotes the $k$-th ${\bf T}^2$;  the ellipsis `$\dots_k$' represents the operators inserted in ${\bf T}^2_k$ away from the degeneration points;  the ${\mathscr A}^{(z_i)}_{a_i}$'s are the local operators placed over the degeneration points $z_i$ on the indicated two-torus; and $G^{a_i a_{i+1}}$ is an inverse metric defined by $G^{a_i a_{i+1}}  \left\langle {\mathscr A}^{(u)}_{a_{i+1}} \, {\mathscr A}^{(v)}_{a_m} \right \rangle_{{\bf S}^2} = \delta^{a_i}_{a_m}$. The factorization for a $g=2$ surface is illustrated in fig.~2. Since we are free to employ CFT arguments in our computation via (\ref{corr fn}), from (i) the CFT state-operator correspondence which allows us to trade a puncture for a local operator; (ii) the fact established in the last paragraph that $ \left \langle \{Q, \Theta\} \dots \right \rangle_{{\rm CFT}; {\bf T}^2} = 0$, where the ellipsis refers to arbitrary operator insertions, some of which are placed over  punctures of the defining worldsheet $\Sigma$; and (iii) the factorization formula of (\ref{factorization}); it is thus clear that \emph{the operator relation $\{Q, \Theta\} = 0$ will continue to hold if $\Sigma$ is a genus $g > 1$ complex curve with $2g-2$ punctures}. 

\begin{figure}
  \centering
    \includegraphics[width=0.6\textwidth]{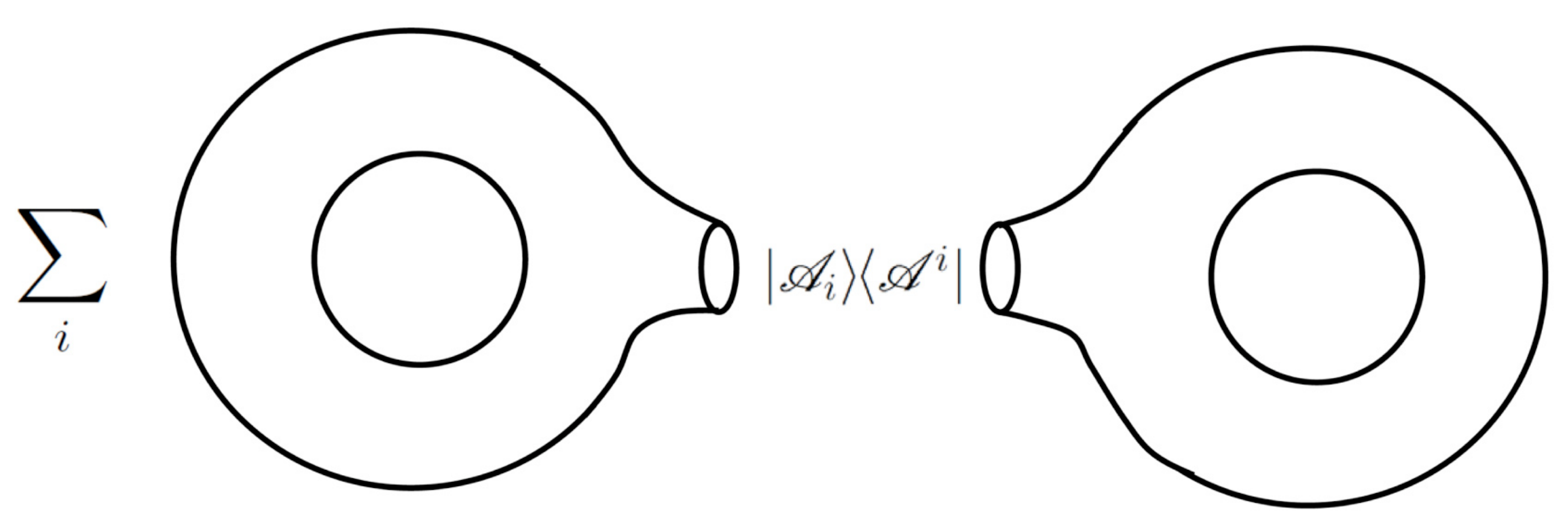}
  \caption{Factorization for genus 2 surface. The punctures at the left and right degeneration points $u$ and $v$ can be represent by the states $\vert {\mathscr A}_i \rangle$ and $\langle {\mathscr A}^i \vert$; one must sum over all pairings of $\vert {\mathscr A}_i \rangle$ and $\langle {\mathscr A}^i \vert$ since the underlying path integral is defined over the entire genus 2 surface; the state-operator isomorphism in CFT then allows us to express this sum as  $\sum_{ij} {\mathscr A}^{(u)}_{i} G^{ ij} {\mathscr A}^{(v)}_{j}$.}
\end{figure}

Alternatively, note that there are no degree-one holomorphic maps from $\Sigma_g$ to $X = {\mathbb {CP}}^1$, for if there are, $\Sigma_g$ would be isomorphic to $X$, which is not the case. Thus, as in the situation where $\Sigma$ is a torus with any number of punctures, there are no worldsheet twisted-instantons, and the conclusion in the previous paragraph follows.

\bigskip\noindent{\it Higher Degree Instanton Contributions}

What about contributions from higher degree instantons? How will it affect our above conclusions? To address these questions, first notice from (\ref{last}) that in the presence of degree-one instantons, when $\Sigma$ is a genus zero complex curve (with two or less punctures), we actually have the relation 
\be
[\{Q, \Theta\}] \sim \Lambda^2 e^{-t_0} \, [1],
\label{one-instanton relation}
\ee
where $[\{Q, \Theta\}]$ and $[1]$ are local operators; here, $\Lambda$ has dimensions of ${\rm length}^{-1}$, which means that $\Lambda^2$ would have scaling dimension 2, consistent with the fact that $[\{Q, \Theta\}]$ has scaling dimension 2 (see footnote~43). Second, recall that instanton-induced relations ought to violate the $U(1)_R$ charge and scaling dimension gradings of local operators; indeed, in the above degree-one instanton case, the local operator $[\{Q, \Theta\}]$ has $U(1)_R$ charge and scaling dimension equal to 2 while the identity operator $[1]$ has  $U(1)_R$ charge and scaling dimension equal to 0 and \emph{not} 2.  Third, note that the violation by degree-$k$ instantons $\Phi_0$ in the $U(1)_R$ charge grading is given by the index $\int_{\hat \Sigma} \Phi^\ast_0 (c_1(X)) = 2k$. Fourth, note that the violation by degree-$k$ instantons in the scaling dimension grading is captured by  a factor involving the dimensionful scale $\Lambda^{2k}$.  In order to determine what this factor involving $\Lambda^{2k}$ is, one just needs to observe that (i) the complex dimension  of the moduli space of instantons is $2k -1 + q$ (where $q \geq 2$ is related to the number of punctures on the defining worldsheet $\Sigma$) and hence, the integration over the moduli space in the computation of  the correlation function would entail an integration over $2k -1 + q$ complex variables which correspond to points on $X$; (ii) since the moduli space of instantons is noncompact, one would need to regularize at least one of the $2k -1 + q$ integrals; (iii) each regularization would introduce an overall factor of $l^{-2}$ to the final result, $l$ being a length scale.  Altogether therefore, recalling that one can identify $l^{-1}$ with the energy scale $\mu$ of (\ref{t(u)}), we can, in the presence of nonzero degree instantons, write the general relation 
\be
[\{Q, \Theta\}]  =   \sum_{k = 1}^\infty  l^{2 (k-n)} \Lambda^{2k} e^{-kt_0} \, [1],
\label{all-instanton relation}
\ee
where the integer $n > 0$ is the number of regularized integrals, and $[1]$  is supposed to have $U(1)_R$ charge and scaling dimension equal to $2 - 2k$ and $2-2n$, respectively. Notice that the identity operator $[1]$ cannot have any value other than 0 for its $U(1)_R$ charge and scaling dimension. Thus, only the $k=n=1$ part of the sum survives, giving us (\ref{one-instanton relation}). Hence, \emph{our above conclusions about whether $\{Q, \Theta\} \sim 1$ or $0$ when $\Sigma$ is a genus zero complex curve, holds, regardless of the presence of higher degree instantons}. 

What about the case where $\Sigma$ is a complex curve of genus one? Although there are no worldsheet twisted-instantons at degree-one, there will be worldsheet twisted-instantons at degrees two and higher. This is a consequence of the fact that holomorphic maps from genus one to zero Riemann surfaces exist only for degrees two and higher.\footnote{I would like to thank D.Q.~Zhang for a detailed explanation of this mathematical fact.}  What this means is that instead of $[\{Q, \Theta\}] = 0$,  one can potentially have a relation similar to (\ref{all-instanton relation}) with contributions coming from degree-two or higher instanton sectors. To ascertain the exact form of this relation, first note that in this case, the violation by degree $m \geq 2$ twisted-instantons $\phi_0$ in the  $U(1)_R$ charge grading is given by the index  
\be
\big\langle c^T_1(X), \phi_0 ({\bf T}^2) \big\rangle = \int_{{\bf T}^2} \Phi^\ast_0 (c_1(X)) = 2m,
\label{8.48}
\ee
where $\Phi_0: {\bf T}^2 \to X$ is a \emph{regular} holomorphic map of degree-$m$. To obtain the first equality in (\ref{8.48}), one simply needs to recall from our discussion in $\S$4.4 that the $T$-equivariant first Chern class $c^T_1(X)$ can be regarded as $c_1(X)$ in the flag manifold model; to obtain the second equality  in (\ref{8.48}), one just needs to know that $ \int_{{\bf T}^2} \Phi^\ast_0 (c_1(X))$ is equal to the degree $r$ of the pullback $\Phi^\ast_0 (c_1(X))$, and since $c_1(X) = 2H$, where $H$ is the hyperplane of $X$, we have $r = {\rm deg}(2H) \cdot {\rm deg} (\Phi_0) = 2m$. Second, note that the equivalent here of $t(\mu)$ in (\ref{t(u)})  is
\be
\big\langle {\cal K}^T(\mu), \phi_0 ({\bf T}^2) \big\rangle = \int_{{\bf T}^2} \Phi^\ast_0 ({\cal K}(\mu)), 
\label{8.47}
\ee
where ${\cal K}(\mu) = {\cal K}_0 + {\rm ln} (\mu / \Lambda) \, c_1(X)$ is the effective K\"ahler class at energy scale $\mu$; ${\cal K}_0 \in H^2(X, \mathbb R)$ is the ``bare'' K\"ahler $(1,1)$-form on $X$; and ${\cal K}_T(\mu)$ is the $T$-equivariant version of ${\cal K}(\mu)$. To obtain the equality in (\ref{8.47}), one simply needs to know that ${\cal K}(\mu)$, like $c_1(X)$, is a class in the de Rham cohomology group $H^2(X, \mathbb R)$, and according to our discussion in $\S$4.4, its $T$-equivariant version ${\cal K}_T(\mu)$, like $c^T_1(X)$, can be regarded as ${\cal K}(\mu)$ in the flag manifold model. Third, note that the complex dimension $d$ of the moduli space of twisted-instantons is given by $d = \big\langle c^T_1(X), \phi_0 ({\bf T}^2) \big\rangle - 1  + q = 2m -1 +q$ (\emph{c.f.}~\cite{mundet} and (\ref{8.48})), where $q$ is related to the number of punctures on the defining worldsheet $\Sigma$; hence, the integration over the moduli space in the computation of  the correlation function would entail an integration over $2m -1 +q$ complex variables which correspond to points on $X$.  Fourth, note that since the moduli space of twisted-instantons is noncompact, one would need to regularize at least one of the $2m -1 +q$ integrals;  each regularization would introduce an overall factor of $l^{-2}$ to the final result, $l$ here being a length scale identifiable with $\mu^{-1}$. In sum, one can write the equivalent of (\ref{all-instanton relation}) for genus one as 
\be
[\{Q, \Theta\}]  =   \sum_{m = 2}^\infty l^{2(m -s)} \Lambda^{2m} e^{-mt_0} \, [1],
\label{all-instanton relation for genus one}
\ee
where the integer $s > 0$ is the number of regularized integrals, and $[1]$  is supposed to have $U(1)_R$ charge and scaling dimension equal to $2 - 2m$ and $2 -2s$, respectively. As mentioned, since  the identity operator $[1]$ cannot have any value other than 0 for its $U(1)_R$ charge and scaling dimension, $m$ and $s$ can only be equal to $1$. Although it is possible to have $s=1$ in (\ref{all-instanton relation for genus one}), it is not possible to have $m=1$. As such, the relation (\ref{all-instanton relation for genus one}) cannot be true. Hence, \emph{our above conclusion that $\{Q, \Theta\}  =0$ if $\Sigma$ is a genus one complex curve with any number of punctures, holds, regardless of the presence of higher degree instantons}. 

The analysis for when $\Sigma$ is a complex curve of genus greater than one follows, via the discussion surrounding (\ref{factorization}), from the analysis in the previous paragraph; in particular, we find that  \emph{our above conclusion that $\{Q, \Theta\}  =0$ if $\Sigma$ is a genus $g > 1$ complex curve with $2g-2$ punctures, holds, regardless of the presence of higher degree instantons}.

\newsubsection{Nonperturbative Effects In The Flag Manifold Model And Beilinson-Drinfeld $\cal D$-Modules}

Let us now extend our analysis in the previous subsection to general flag manifold target spaces $X = \GC /B$, where $B$ is a Borel subgroup of $\GC$.  From $\S$5.2.3, we have an isomorphism between the perturbatively $Q$-closed local operators $1$ and $\cal R$, where $\CR = R_{i \bar j} D_z \phi^i \psi^\jb$ is the arbitrary $\GC$ generalization of $\Theta$ for $\GC = SL(2)$. If ${\rm dim}_{\mathbb C} X = d$ and ${\rm rank}(\gc) = r$, we have $i = 1, \dots, d$, $\jb = \bar 1, \dots, \bar d$, and $a = 1, \dots, r$. According to our explanations at the start of the previous subsection, which can be applied here since $c_1^T(X) \neq 0$, the isomorphism between $1$ and $\CR$ implies that one can expect a relation of the form $\{Q, \CR\} \sim 1$ to be induced by nonperturbative worldsheet twisted-instantons, at least in certain situations. We shall now show this to be true. 

\bigskip\noindent{\it Nonperturbative Effects at Genus Zero}

To this end, let us first assume that the defining worldsheet is $\Sigma = {\bf S}^2 - \{0, \infty\}$. In this case, recall that the key property of the $\mathbb {CP}^1$ model which allowed us to obtain the relation $\{Q, \Theta\} \sim 1$, is that there are \emph{precisely} two $\psi^{\bar 1}$ zero modes and no $\psi_{\bar 1}$ zero modes in the one-instanton sector. These numbers depend crucially on the fact that for one-instantons characterized by degree-one holomorphic maps $\phi_0: {\bf S}^2 \to \mathbb {CP}^1$, we have $\phi_0^*T{\mathbb {CP}^1} = \CO(2)$. Hence, since the only degree-one holomorphic maps from ${\bf S}^2$ are to genus zero rational curves, the aforementioned observation suggests that in order to have the instanton-induced relation $\{Q, \CR \} \sim 1$ in the $\GC / B$ model,  there ought to be in $X$ \emph{at least one} rational curve $L$ such that $\phi_0^*T{L} = \CO(2)$. If there exists such a rational curve $L \subset X$, then only the field components tangent to~$L$ -- since they are the only  ones related to instantons -- will contribute to $\{Q,\CR\}$.  Consequently, if there are no $\psi_\jb$ zero modes whence $I_{int}$ of (\ref{Iint}) vanishes (after going to pure gauge $A=0$ on the simply-connected ${\bf S}^2$ effective worldsheet), our computation will be the same as that in the $\mathbb {CP}^1$ case, and the integration over the two $\psi^{\bar b}$ zero modes will turn $\{Q, \CR\}$ into the pullback of the K\"ahler form $R_{a \bb} (\phi_0) \del_z \phi^a_0 \del_\zb \phi^\bb_0$, where the indices $a, \bb$ refer to directions along $L \subset X$.  The subsequent integration over the parameters of the instanton $\phi_0$ then becomes an integration over $L$. So, the critical question to ask is whether $X$ fulfills these requirements or not.

To shed light on the matter, we will now need to make a brief technical excursion. Note that if $M$ is a target K\"ahler manifold of complex dimension $d$ that contains at least one rational curve $L$, and $\phi_0$ is as before a holomorphic map from ${\bf S}^2$ to $L$,  we can decompose the tangent bundle of $M$ in the vicinity of $L$ as $TM = TL \oplus NL$, where $NL$ is the normal bundle of $L$ in $M$. For degree-one maps, the pullback bundle is $\phi_0^*TL = \CO(2)$, and $\phi_0^*NL$ further splits into a direct sum of line bundles; for example, if $\int_{{\bf S}^2} \phi_0^*c_1(M) = k$, we can write, in the vicinity of $L$,
\begin{equation}
\label{ST}
\phi_0^*TM
\iso \CO(2) \oplus \CO(p_1) \oplus \dotsb \oplus \CO(p_{d-1}),
\end{equation}
where $p_1 + \dotsb + p_{d-1} = k - 2$. For such a target manifold, the number of $\psi^\jb$ or $\psi_\jb$ zero modes can be found from the splitting type \eqref{ST}.  According to the formula
\begin{equation}
h^0\bigl(\CO(n)\bigr) =
\begin{cases}
n + 1 & \text{for $n \geq 0$}; \\
0     & \text{for $n < 0$},
\end{cases}
\end{equation}
and the fact that the relevant spin bundle $\cal S$ is given by $\CO(-1)$, each $\CO(n)$ with $n > 0$ in \eqref{ST} will contribute $h^0({\cal S} \otimes \CO(n)) = n$ number of $\psi^\jb$ zero modes (since $\psi^j \in {\cal S} \otimes \phi^\ast TM$).  We also know from the index theorem of the differential operators in (\ref{fermi zero modes}), that there are $\int_{{\bf S}^2} \phi_0^*c_1(M) = k$ more $\psi^\jb$ zero modes than $\psi_\jb$ zero modes.  

Coming back to our main discussion, note that it is well-known \cite{BH} that $c_1(\GC/B) = 2(x_1 + \dots + x_r)$, where each independent cohomology class $x_i$ is dual to a rational curve $L_i$; in other words, $X$ contains $r$ rational curves whose associated pullbacks of the first Chern class $\int_{{\bf S}^2} \phi_{0, i}^*c_1(M)$ -- where $\phi_{0, i}$ is a degree-one holomorphic map from ${\bf S}^2$ to $L_i$ -- are all equal to 2. As such, by the arguments of the previous paragraph, the splitting type of $\phi_{0,i}^*TX$ in the vicinity of each $L_i$ would be given by
\begin{equation}
\label{2000}
\phi_{0,i}^*TX \cong \CO(2) \oplus \CO(0) \oplus \dotsb \oplus \CO(0).
\end{equation}
Moreover, for each of the $r$ one-instantons $\phi_{0, i}$, we have two $\psi^\jb$ zero modes $\psi^\bb_{0, i}$ coming from the $\CO(2)$ factor along the $L_i$-directions, and no $\psi_\jb$ zero modes. Thus, $X$ fulfills the requirements spelt out in the paragraph before the previous, and the contribution from each one-instantons $\phi_{0, i}$ is the same as that in the $\mathbb {CP}^1$ case!

However, there is a caveat here, since $L_i$ is not a rigid curve if  the splitting type of $\phi_{0,i}^*TX$ around $L_i$ is as given in \eqref{2000}. An infinitesimal deformation of $L_i$ is given by a holomorphic section of~$\phi_{0,i}^*NL_i$;  in the case of the splitting type~\eqref{2000},  the normal bundle $NL_i$ is trivial and we have $d - 1$ independent deformations, one for each normal direction. Therefore, the instanton $\phi_{0, i}$ -- because it actually wraps around $L_i$ -- can be infinitesimally translated in every possible direction in the target space. This generates a family of instantons with $d-1$ complex parameters over which we still have to integrate after the integration over $L_i$ is done; as the instanton can be translated freely, it would mean that we would have to integrate over the target space as well. Nevertheless, as in the $\mathbb {CP}^1$ case, one would, in the end, still get the relation $\{Q, \CR \} = C_i$, where $C_i$ is some constant operator. 

In short, for each one-instanton $\phi_{0, i}$ wrapping $L_i$, we get a contribution $\{Q, \CR \} = C_i$ to the final answer. Summing up the $r$ contributions, we end up with $\{Q, \CR \} \sim 1$. The calculation for when the defining worldsheet is $\Sigma = {\bf S}^2 - {\infty}$ is similar, so we can conclude that \emph{we will have the operator relation $\{Q, \CR \} \sim 1$ if $\Sigma$ is a genus zero complex curve with two or less punctures}. 

As for when $\Sigma$ is a genus zero complex curve with three or more punctures, since the computation on $X$ eventually boils down to the computation on each rational curve $\mathbb {CP}^1 \subset X$, the arguments for the existence or lack thereof of the relation $\{Q, \CR \} \sim 1$ follow those given in the $\mathbb {CP}^1$ example. Hence, we can conclude that  \emph{the operator relation $\{Q, \CR \} = 0$ will continue to hold if $\Sigma$ is a genus zero complex curve with three or more punctures}.

\bigskip\noindent{\it Nonperturbative Effects at Genus One}

How about when $\Sigma$ is a ${\bf T}^2$ with any number of punctures? First, note that the only two-cycles in $X$ are the $r$ rational curves dual to the $x_i$'s, i.e., there are no two-cycles of nonzero genus in $X$. Second, as explained in the third paragraph after (\ref{2 pts deleted}), there are, in this case, no worldsheet twisted-instantons because the target of $\phi_0$ is a rational curve. Hence, the computation on each rational curve $\mathbb {CP}^1 \subset X$ would give a vanishing result. Therefore, we can conclude that  \emph{the operator relation $\{Q, \CR \} = 0$ will continue to hold if $\Sigma$ is a genus one complex curve with any number of punctures}.

\bigskip\noindent{\it Nonperturbative Effects at Genus $g > 1$}

And what if $\Sigma$ is a genus $g >1$ Riemann surface with $2g-2$ punctures? Once again, since there are no two-cycles of nonzero genus in $X$, according to our explanations in the paragraph before that which discusses (\ref{one-instanton relation}), there are, in this $g >1$ case, no worldsheet twisted-instantons either. Thus, we can conclude that \emph{the operator relation $\{Q, \CR \} = 0$ will continue to hold if $\Sigma$ is a genus $g > 1$ complex curve with $2g-2$ punctures}.

\bigskip\noindent{\it Higher Degree Instanton Contributions}

The analysis for the higher degree instanton contributions is the same as that for the $\mathbb {CP}^1$ case except for one subtlety. As pointed out above, because each instanton can be swept throughout $X$, one needs in this case to also integrate over $X$ in the computation. Nevertheless, since $X$ -- unlike the moduli space of instantons involved -- is compact, one would not need to regularize the integral associated with it. As such, no additional length scales would be introduced, and the relations (\ref{all-instanton relation}) and (\ref{all-instanton relation for genus one}), as well as the discussion surrounding them, would apply to the computation on each rational curve $\mathbb {CP}^1 \subset X$. Hence, we find that \emph{all our earlier conclusions hold regardless of the presence of higher degree instantons}.

\bigskip\noindent{\it Certain Delicate Conditions for the Existence of Beilinson-Drinfeld $\cal D$-Modules}

When we do have the relation $\{Q, \CR \} \sim 1$, the identity operator would be $Q$-exact and therefore, all operators in the chiral algebra $\mathscr A$ of the $\GC/B$ model  would also be $Q$-exact: for any operator ${\cal O} \in \mathscr A$ whence $\{Q, {\cal O}] = 0$, we would have ${\cal O} = {\cal O} \cdot 1 \sim {\cal O} \cdot \{Q, \CR\} = \{Q, {\cal O}  \CR]$, i.e., ${\cal O} = \{Q, \dots]$. Thus, from our above results, we find that $\mathscr A$  would be completely trivialized if $\Sigma$ is a genus zero complex curve with two or less punctures! In turn, this implies that $A(X, q)$ of (\ref{elliptic genus TCDO}) must vanish nonperturbatively, at least in these situations. 

Recall at this point from $\S$7.1 our interpretation of the Beilinson-Drinfeld $\cal D$-modules of the geometric Langlands program for $\GC$ as correlation functions $\Psi_n$ of the $\Phi^0(z)$ fields in the chiral algebra $\mathscr A$; what this trivialization of $\MA$ also means is that \emph{there can be no Beilinson-Drinfeld $\cal D$-modules when the underlying complex curve $\Sigma$ is rational with less than three punctures}. Hence, the physics of our flag manifold model implies, among other things, that the geometric Langlands correspondence for $\GC$ would be   non-vacuous at genus zero if and only if there are three or more punctures. This physically derived result is also consistent with the mathematical literature~\cite{Rubtsov}.

\newsubsection{Spontaneous Supersymmetry Breaking}

The trivialization of the chiral algebra $\mathscr A$ also leads to nontrivial consequences for the supersymmetric spectrum of the flag manifold model. Before we elucidate what these consequences are, first note that in any two-dimensional  theory with $(0,2)$ supersymmetry such as our flag manifold model, we have, from its supersymmetry algebra, the relation 
\be
\{Q, Q^\dagger \} = H_R,
\label{Q-Q}
\ee 
where the second supercharge $Q^\dagger$ is the hermitian conjugate of $Q$, and $H_R$ is the right-moving half of the Hamiltonian operator whose eigenvalue is the right-moving energy level of a state. Second, notice that any supersymmetric state $\ket{\Psi}$ must be annihilated by both $Q$ and $Q^\dagger$; this means that $\ket{\Psi}$ must be in the kernel of  $\{Q, Q^\dagger \}$, and that $H_R \ket{\Psi} = 0$.  Third, recall from $\S$4.1 that $Q$ will act as some differential operator  on the compact target space $X$; thus, one can regard $\{Q, Q^\dagger \}$ as a Laplace-Beltrami operator for $Q$, whence $\ket{\Psi}$ will be a harmonic element with respect to $Q$; in turn, this means that $\ket{\Psi}$ will span the $Q$-cohomology of states. 

Now consider the string described by the flag manifold model to be propagating freely either from the unknown past into the far future or from  the far past into the far future; since we are taking the large but finite $X$-volume limit, according to our analysis in $\S$8.1,  one can regard the string's worldsheet $\Sigma$ to be a sphere with either one or two punctures, respectively. Then, in the presence of worldsheet twisted-instantons, we have
\begin{equation}
\label{states trivial}
\ket{\Psi} = 1 \cdot \ket{\Psi} \sim \{Q,\CR\}\ket{\Psi}
   = Q\bigl(\CR\ket{\Psi}\bigr).
\end{equation}
Hence, $\ket{\Psi}$, though $Q$-closed, is now also $Q$-exact; i.e., $\ket{\Psi}$ is trivial in $Q$-cohomology. This means that there are actually\emph{ no} supersymmetric states in the full quantum theory! Therefore, since $\langle 0 \vert {\cal O}' \vert 0 \rangle = \langle 0 \vert  \{Q, \dots \}  \vert 0 \rangle \neq 0$ for some bosonic operator ${\cal O}' \in \mathscr A$, where $\vert 0\rangle$ is the vacuum state, one can conclude that supersymmetry is spontaneously broken! 

The above observation thus implies the following. From the defining condition $H_R \ket{\Psi} = 0$, supersymmetric states are necessarily right-moving ground states. Nevertheless, they can be nonperturbatively ``lifted'' in boson-fermion pairs by instantons such that eventually, $H_R \ket{\Psi} \neq 0$, thereby resulting in a spontaneous breaking of supersymmetry. We will now turn to the canonical quantization viewpoint involving supersymmetric gauged quantum mechanics on loop space, and investigate what this purely physical phenomenon implies for the geometry of loop spaces of flag manifolds of $\GC$, among other things.

\newsubsection{Supersymmetric Gauged Quantum Mechanics And Loop Space Geometry}

 The worldsheet $\Sigma$ of a string which propagates freely  over a K\"ahler manifold $X = \GC /B$ either from the unknown past into the far future or from the far past into the far future, can be modeled by an infinitely-long cylinder ${\bf S}^1 \times \mathbb R$ with coordinates $(\sigma, \tau)$, where $\sigma \sim \sigma + 2\pi$. Let us equip $\Sigma$ with a complex structure by setting $\del_z = \del_\sigma - i\del_\tau$; then $H$ and $P$ of $H_R = (H-P) /2$ in (\ref{Q-Q}) are the generators of translations in time $\tau$ and space $\sigma$, respectively.  Given that the $T$-equivariant K\"ahler form on $X$ is ${\cal K}_T = i g_{i\jb}D\phi^i \wedge D\phi^\jb /2$ (where we recall that $T$ is the Cartan subgroup of the compact real form of $\GC$), the action, including worldsheet twisted-instanton contributions, will be given by (\emph{c.f.}~(\ref{Sgauged})-(\ref{txgauged}))
\be
\begin{split}
 S_{gauged} &=  \frac{1}{2\pi} \int_\Sigma d\sigma d\tau
         \bigl\{Q, -g_{i\jb}  \psi^i (D_\tau + i D_\sigma)\phi^\jb  \bigr\}
         + \frac{1}{2\pi} \int_\Sigma \phi^*{\cal K}_T   \\
      &=  \frac{1}{2\pi} \int_\Sigma d\sigma d\tau
         \Bigl( g_{i\jb} (D_\tau\phi^i D_\tau\phi^\jb
                + D_\sigma\phi^i D_\sigma\phi^\jb)
         - i g_{i\jb} \psi^i ({\hat D}_\tau + i{\hat D}_\sigma) {\psi}^\jb\Bigr),
         \end{split}
         \ee
where the gauge-covariant derivatives $D$ and $\hat D$ are defined by 
\be
\begin{split}
D_{\tau, \sigma} \phi^i  & = \del_{\tau, \sigma} \phi^i \pm A^a_{\tau, \sigma} V^i_a,  \\
D_{\tau, \sigma} \phi^\jb  & = \del_{\tau, \sigma} \phi^\jb + A^a_{\tau, \sigma} V^i_a, \\
{\hat D}_{\tau, \sigma}  {\psi}^\jb & =   \del_{\tau, \sigma} {\psi}^\jb +  \Gamma_{\bar i \bar k}^{\bar j}\del_{\tau,\sigma} \phi^\ib \psi^\kb  +  A^a_{\tau, \sigma}  \Gamma_{\bar i \bar k}^{\bar j}V^{\bar i}_a {\psi}^\kb.
\end{split}
\ee

\bigskip\noindent{\it Supersymmetric Gauged Quantum Mechanics on Loop Space}

Working on ${\bf S}^1 \times \mathbb R$,  it is very natural to choose the temporal gauge $A^a_\tau = 0$. Doing so, the action can be simplified to
\be
\label{action S}
S =  \frac{1}{2\pi} \int_{{\bf S}^1 \times \mathbb R} d\sigma d\tau
         \Bigl( g_{i\jb} (\del_\tau\phi^i \del_\tau\phi^\jb
                + D_\sigma\phi^i D_\sigma\phi^\jb)
         - i g_{i\jb} \psi^i ({D}_\tau + i{\hat D}_\sigma) {\psi}^\jb\Bigr),
\ee
where $D_{\tau} {\psi}^\jb =  \del_{\tau} {\psi}^\jb +  \Gamma_{\bar i \bar k}^{\bar j}\del_{\tau} \phi^\ib \psi^\kb$. Notice that at each time $\tau$, the bosonic field $\phi: {\bf S}^1 \times \mathbb R \to X$ specifies a point $\phi_\tau$
in the loop space ${\cal L} X$ of smooth maps  ${\bf S}^1 \to X$ via $\phi_\tau(\sigma) = \phi(\sigma, \tau)$.
Similarly, the fermionic fields specify $\psi^i_{\tau}$ and
$\psi^\jb_{\tau}$, which, via
$\psi^i_{\tau}(\sigma) = \psi^i(\sigma, \tau)$ and
$\psi^\jb_\tau(\sigma) = \psi^\jb(\sigma, \tau)$, we may identify respectively with vectors in
$T_{\CL X}|_{\phi_\tau} \cong \Gamma(\phi_\tau^*T_X)$ and ${\overline T}_{\CL
  X}|_{\phi_\tau} \cong \Gamma(\phi_\tau^*{\overline T}_X)$.  In what follows, we
will fix a time $\tau$ and write these simply as $\phi$, $\psi^i$, and
$\psi^\jb$.  As is clear from this description, the theory may now be
viewed as supersymmetric gauged quantum mechanics on $\CL X$ with action 
\be
S_{QM} = \int_\mathbb R d\tau \, L (\phi, \psi, \del_\tau \phi, \del_\tau \psi, \tau),
\ee 
where 
\be
\label{L}
L (\phi, \psi, \del_\tau \phi, \del_\tau \psi, \tau) = {1 \over 2 \pi} \int_{{\bf S}^1} d\sigma  \Bigl( g_{i\jb} (\del_\tau\phi^i \del_\tau\phi^\jb
                + D_\sigma\phi^i D_\sigma\phi^\jb)
         - i g_{i\jb} \psi^i ({D}_\tau + i{\hat D}_\sigma) {\psi}^\jb\Bigr)
\ee
is the corresponding quantum mechanical Lagrangian. 

Let us now canonically quantize the theory on $\CL X$.  The canonical conjugate $\pi_\psi$ to $\psi$ will be given by $\delta L / \delta (\del_\tau \psi)$; from (\ref{L}), we have
\be
\pi_{\psi^\jb} = {\delta \over \delta \psi^\jb} =  {i \over 2 \pi} \int_{{\bf S}^1} d\sigma \,  g_{i \jb} \psi^i.
\ee
From the equal-time canonical commutation relation $\{\pi_{\psi^\kb_\tau}(\sigma'), \psi^\jb_\tau (\sigma)\} = i \delta^\jb_\kb \delta(\sigma - \sigma')$, we get
\be
 {1 \over \pi} \int_{{\bf S}^1} d\sigma \, \{  \psi^i_\tau (\sigma), \psi^\jb_\tau (\sigma')\} =   2 g^{i \jb} \delta(\sigma - \sigma').
\ee
This is the loop space version of the Clifford algebra $\{\Gamma^i,
\Gamma^\jb\} = 2 g^{i\jb}$, in which $\psi^i$, $\psi^\jb$ play the
roles of the gamma matrices $\Gamma^i$, $\Gamma^\jb$, with extra
continuous indices $\sigma$ and $\sigma'$ parametrizing the directions along the
loop. States furnish a representation of this algebra, so they are
spinors on $\CL X$. 

On the other hand, the canonical conjugate $\pi_\phi$ to $\phi$ will be given by $\delta L / \delta (\del_\tau \phi)$; from (\ref{L}), we have
\be
\label{pi psi}
\pi_{\phi^i} =  {\delta \over \delta \phi^i} = {1 \over 2 \pi} \int_{{\bf S}^1} d\sigma \, g_{i\jb} \del_\tau\phi^\jb \quad {\rm and} \quad \pi_{\phi^\jb} =  {\delta \over \delta \phi^\jb} = {1 \over 2 \pi} \int_{{\bf S}^1} d\sigma \, g_{i\jb} \del_\tau\phi^i -  { 2 i\over   \pi} \int_{{\bf S}^1} d\sigma \, \omega_\jb,
\ee
where $\omega_\jb =  {i \over 2} \Gamma_{i \jb \kb} \Sigma^{i \kb}$ and $\Sigma^{i \kb} = -i[\psi^i, \psi^\kb] / 4$. Given the interpretation of  $\psi^i$, $\psi^\jb$ as the gamma matrices $\Gamma^i$, $\Gamma^\jb$, it is thus clear that $\omega_\jb$ must be a spin connection on $X$. 
 
Now consider the covariant functional derivative on $\CL X$
\be
\label{Cov Fun}
{D \over \delta \phi^\jb}  = {\delta \over \delta \phi^\jb} + {2 i \over \pi} \int_{{\bf S}^1} d\sigma \, \omega_\jb.
\ee 
Notice from (\ref{pi psi}) that one can also write
\be
{D \over \delta \phi^\jb} = {1 \over 2 \pi} \int_{{\bf S}^1} d\sigma \, g_{i\jb} \del_\tau\phi^i.
\label{D/phi}
\ee
Thus, in terms of the (covariant) functional derivatives ${D / \delta \phi^\jb}$ on $\CL X$, the supercharges $Q$ can be expressed as
\be
\begin{split}
Q & = {1 \over 2 \pi} \int_{{\bf S}^1} d \sigma \, g_{i \jb} \psi^\jb (i \del_\tau + D_\sigma) \phi^i = Q_0 + {1 \over 2 \pi} \int_{{\bf S}^1} d \sigma \, g_{i \jb} \psi^\jb  D_\sigma \phi^i, 
\label{full Q}
\end{split}
\ee
where
\be
Q_0 = i \psi^\jb {D \over \delta \phi^\jb}.
\label{Q_0}
\ee
From (\ref{Cov Fun}), and the fact that $\omega_j$ must be a spin connection on $X$, it is clear that $Q$ can be interpreted as the antiholomorphic  half $Q_0$ of the Dirac operator on $\CL X$ twisted by the term ${1 / 2\pi} \int_{{\bf S}^1} d \sigma \, g_{i \jb}  \psi^\jb D_\sigma \phi^i$. 

An important point to note at this juncture is the following. Let $h$ be some function on $\CL X$. As $h$ is a scalar on $\CL X$, the covariant functional derivative will simply act as the functional derivative; in particular, $D h/ \delta \phi^\jb = \delta h / \delta \phi^\jb$. Bearing this in mind, it will mean that from (i) the Baker-Campbell-Hausdorff formula $e^{X} Y e^{-X} = Y - [Y, X] + \dots$ for operators $X$ and $Y$, and (ii) the action of $Q_0$ as indicated in (\ref{Q_0}), we can write 
\be
Q_h = e^{ih / 2\pi} Q_0 e^{-ih / 2\pi} = Q_0 + {1 \over 2 \pi}{\psi^\jb} {\delta h \over \delta \phi^\jb}. 
\label{similarity}
\ee 
Hence, for the right choice of $h$, we can have $Q_h = Q$. Moreover, since $Q_h$ and $Q_0$ are related by a similarity transformation, it will also mean that for the right choice of $h$, the $Q$-cohomology of states will just be the $Q_0$-cohomology of states.  Let us now determine what this choice of $h$ must be.

First, we pick a \emph{fixed}, base loop $\phi_0$ in each connected component of $\CL X$.
Then, for a given loop $\phi$ in that component, we choose a homotopy $\hat\phi$ that connects $\phi_0$ to $\phi$. Namely, $\hat \phi = \hat \phi(\sigma, \tau)$ is a map $\hat \phi \colon [0,1] \times S^1 \to X$ such that $\hat \phi(\sigma, 0) = \phi_0(\sigma)$ and $\hat \phi(\sigma, 1) = \phi(\sigma)$. Now consider the area
\begin{equation}
  \label{h}
  {\cal A}(\hat\phi) = -2 \int_{[0,1] \times {\bf S}^1} \hat\phi^*{\cal K}_T.
\end{equation}
Note that this area formula will make sense as long as we choose $\hat \phi$ such that  $\hat \phi_\ast ([0,1] \times {\bf S}^1)$ is a $T$-equivariant cycle in $X$. (An immediate example of such a two-cycle would be given by the dual of the non-vanishing $T$-equivariant first Chern class $c_1^T(X)$.) Since the gauge field strength $F$ is zero, i.e., $F = D^2 = 0$, we can write ${\hat\phi}^*{\cal K}_T =  i g_{i\jb}D ({\hat\phi}^i \wedge D{\hat\phi}^\jb) /4 - i g_{i\jb}D (D {\hat\phi}^i \wedge {\hat\phi}^\jb) /4$. As such, we have 
\begin{equation}
  \label{h}
  {\cal A}(\hat\phi) =   \int_{{\bf S}^1} d\sigma \,  g_{i \bar j}
 \left( D_\sigma {\hat \phi}^i
 {\hat \phi}^{\bar j} - {\hat \phi}^i
D_\sigma {\hat \phi}^{\bar j} \right) \vert^{\tau =1}_{\tau =0}. 
\end{equation}
Under a variation of $\hat \phi$, ${\cal A}(\hat \phi)$ will change by
\begin{equation}
  \label{deltah}
  \delta {\cal A(\hat\phi)}
  =  \int_{{\bf S}^1} d\sigma \,  g_{i \bar j}
 \left( D_\sigma {\phi}^i
 \delta {\phi}^{\bar j} - \delta {\phi}^i
D_\sigma { \phi}^{\bar j} \right),
 \end{equation}
where we have used the constraint that $\hat \phi\vert_{\tau =0}$ is fixed to be $\phi_0$ and thus, $\delta \hat \phi\vert_{\tau =0} = 0$. Notice that (\ref{deltah}) will imply that
\be
\label{8.75}
{\delta  {\cal A}(\hat\phi) \over \delta \phi^\jb} =  \int_{{\bf S}^1} d\sigma \,  g_{i \bar j} D_\sigma {\phi}^i.
\ee
If we  let $h =  {\cal A}(\hat\phi)$ and substitute (\ref{8.75}) in (\ref{similarity}), one can see by comparing with (\ref{full Q}) that indeed $Q_h = Q$.\footnote{In order to arrive at this result, we have used integration by parts to calculate that $\int_{{\bf S}^1} d \sigma \, \psi^\jb g_{i \jb} D_\sigma \phi^i =  \psi^\jb \int_{{\bf S}^1} d \sigma \, g_{i \jb} D_\sigma \phi^i -  \int_{{\bf S}^1} d\sigma \, \del_\sigma \psi^\jb ( \int d\sigma' \, g_{i \jb} D_{\sigma'} \phi^i) = \psi^\jb \int_{{\bf S}^1} d \sigma \, g_{i \jb} D_\sigma \phi^i$; the last equality was obtained using integration by parts again and the fact that $\int_{{\bf S}^1} d \sigma \, \del_\sigma\psi^\jb = 0$.}

\bigskip\noindent{\it Implications for the Geometry of Loop Spaces of Flag Manifolds of $\GC$}

Thus, the right choice of $h$ would be $ {\cal A}(\hat\phi)$ whence the $Q$-cohomology of states equals the $Q_0$-cohomology of states. Moreover, notice that a gauge transformation would change the second term in the second equality of (\ref{similarity}) (as this term is not gauge-invariant); hence, one can conclude that a gauge transformation would effect --  via a similarity transformation of $Q$ -- a change of basis in the $Q$-cohomology of operators, as deduced earlier in $\S$4.1.    

At any rate, note that $\CL X$, like $X$, is also K\"ahler; as such, the  $Q_0$-cohomology -- which is the cohomology of the antiholomorphic half of the Dirac operator on $\CL X$ -- actually coincides with the cohomology of the \emph{full} Dirac operator on $\CL X$.  In turn, this means that the $Q$-cohomology of states is simply the space of harmonic spinors on $\CL X$.  Therefore, the trivialization of the $Q$-cohomology of states by worldsheet twisted-instantons discussed in the previous subsection, implies that there can be \emph{no} harmonic spinors on $\CL X$! 

\bigskip\noindent{\it The H\"ohn-Stolz Conjecture and the Existence of Beilinson-Drinfeld $\cal D$-Modules}

Finally, we come to the intimate relation between the H\"ohn-Stolz conjecture in algebraic topology and the delicate conditions  unraveled in $\S$8.2 for the existence of  Beilinson-Drinfeld $\cal D$-modules in the geometric Langlands correspondence for $\GC$.  

The  H\"ohn-Stolz conjecture can be stated as follows~\cite{Stolz}: the Witten genus
\be
\Phi_W(M) = \eta(q)^d \, {V}(M, q)
\ee
of a closed manifold $M$ of real dimension $d$ and $p_1(M) /2 = 0$, where $\eta(q) = q^{1/24} \prod_{m=1}^\infty (1- q^m)$ is Dedekind's $\eta$-function and $V(M,q)$ is the elliptic genus  in (\ref{elliptic genus}), \emph{vanishes} if the Riemannian metric on $M$ admits a Ricci curvature that is \emph{positive}.\footnote{The conjecture is actually formulated for $M$ of real dimension $d = 4n$, where $n$ is an integer. However, since the Witten genus automatically vanishes if $d \neq 4n$, we have, for convenience, stated the conjecture for arbitrary $d$.}  An example of such an $M$ would be given by $X$, where $\Phi_W(X)$ indeed vanishes because the $Q$-cohomology of supersymmetric states counted by  $V(X, q)$ is empty. In fact, our above conclusion that there are no harmonic spinors on $\CL X$ affirms Stolz's heuristic ``proof'' (see $\S$4 of~\cite{Stolz}) which asserts that the H\"ohn-Stolz conjecture ought to be true because (i) the Witten genus depends on the (${\bf S}^1$-equivariant) index of the Dirac operator in $\CL M$; (ii) the Ricci scalar of $\CL M$ is positive as the Ricci curvature of $M$ is positive; (iii) one can apply Lichnerowicz' theorem in $\CL M$, whence the previous point would mean that there can be no harmonic spinors in $\CL M$, i.e., the (${\bf S}^1$-equivariant) index of the Dirac operator in $\CL M$ must be zero. 

As explained in $\S$8.3, the $Q$-cohomology of states is empty as a consequence of the trivialization of the perturbative chiral algebra $\MA$ by worldsheet twisted-instantons. Thus, from our ending remarks in $\S$8.2, and the discussion in the last paragraph, it is clear that  \emph{the fact that there can be no Beilinson-Drinfeld $\cal D$-modules at genus zero when there are two or less punctures  will imply the H\"ohn-Stolz conjecture for $\GC /B$!} One can also make the following statement -- \emph{there can be no harmonic spinors in the space of smooth maps ${\bf S}^1 \to \GC /B$ because there can be no Beilinson-Drinfeld $\cal D$-modules at genus zero when there are two or less punctures!}

\newsection{Chern-Simons Theory, Knot Invariants, And Langlands Duality}

In this final section, we will first elucidate the connection between our flag manifold model in the infinite-volume limit and a WZW model for compact, simply-connected, simply-laced Lie groups $G$. Exploiting the fact that states of a Chern-Simons theory on a three manifold $M$ with gauge group $G$ and Wilson lines in some representation of $G$, can be captured by the conformal blocks of a $G$-WZW model ``living'' on a certain Riemann surface $\hat \Sigma \subset M$, we will show -- via the interpretation of the conformal blocks as certain correlation functions of the flag manifold model in the infinite-volume limit -- how knot invariants of $M$ can be related to ``quantum'' ramified $\cal D$-modules. Next, we will specialize to the case whereby $G = SU(2)$ and $M = {\bf S}^3$, whence we will (i) see that the Jones polynomial of an arbitrary link~\cite{Jones} and its corresponding Khovanov homology~\cite{Khovanov} ought to be captured by various interesting features of  these aforementioned ``quantum'' ramified $\cal D$-modules, (ii) furnish a physical proof of a mathematical conjecture by Seidel-Smith~\cite{SS} which relates Lagrangian intersection Floer homology~\cite{Andreas} to Khovanov homology.  Lastly, we will demonstrate, via a generalized $T$-duality of the flag manifold model in the infinite-volume limit, (i) a ramified geometric Langlands correspondence for $\GC$ (the complexification of $G$); and (ii) a correspondence between representations of $^LG_{\mathbb C}$  and  ``classical'' ramified $\cal D$-modules on the moduli space of holomorphic parabolic $G_{\mathbb C}$-bundles on a rational curve, where $^LG_{\mathbb C}$ is the Langlands dual of $G_{\mathbb C}$; thereby proving physically a mathematical conjecture by Gaitsgory~\cite{Gaitsgory-summary, Gaitsgory-Whittaker}.

\newsubsection{The Infinite-Volume Limit Of The Flag Manifold Model And Chern-Simons Theory}

Consider the flag manifold model with target space $X = \GC/B$. (Recall that $\GC$ is a simple, simply-connected, complex Lie group, and $B \subset \GC$ is a Borel subgroup.)  Let us take the infinite $X$-volume limit; then,  the Ricci curvature $R_{i \bar j}$ of $X$ as ``seen'' by the sigma model, vanishes. In turn, this means that (i) one can regard the anomaly measured by $c_1(\Sigma)c_1^T(X)$ -- where $T$ is the Cartan subgroup of the compact real form $G$ of $\GC$ -- to be zero, i.e., one can consistently define the model on \emph{any} $\Sigma$, with \emph{any} number of punctures; (ii) (\ref{tzzanomaly}) ought to be replaced by $[Q, T_{zz}] = 0$ and thus, from our analysis in $\S$4.7 and our $Q$-$\check {\textrm C}$ech cohomology dictionary, the level $k$ that appears in $\S$5.2 and $\S$6.1 \emph{cannot} be equal to $h^\vee$, where $h^\vee$ is the dual Coxeter number of the Lie algebra $\gc$ of $\GC$. 

Note at this point that our aforementioned flag manifold model whose target space has infinite volume can also be viewed as the local flag manifold model introduced in $\S$6.1.  Bearing this in mind, now recall from our discussion following (\ref{spec^a}) that  $T$-duality of the model leaves $\mathscr V = -  i (R \rho - {1 \over R} \rho^\vee)$ invariant, where $R = 1 / \sqrt{k + h^\vee}$ and $\{\rho, \rho^\vee\}$ are the Weyl vector of the Lie algebra $\gc$ of $\GC$ and its dual, respectively; on the other hand, it maps $Y^a = Y^a_L(z) + Y^a_R(\bar z)$ to ${Y^a}' = Y^{a}_L(z) - Y^a_R(\bar z)$. Let $\mathscr V'$ be the $T$-dual of $\mathscr V$; then, since $T$-duality ought to leave the scalar product $\mathscr V \cdot Y$ invariant, i.e., $\mathscr V \cdot Y = \mathscr V' \cdot Y'$, from the previous point, we would have $\mathscr V \cdot Y_R = 0$. (This can be regarded as an \emph{a priori} account of the asymmetry between (\ref{dilaton ST^a}) and (\ref{dilaton ST^a-rightmoving}).) Consequently, if we henceforth restrict our analysis to \emph{simply-laced} $\GC$ so that $\rho = \rho^\vee$, from (\ref{Iequiv}), we can write the action of the flag manifold model in the infinite-volume limit as 
\be
\label{Iinfty}
I_{\infty} =  {1 \over  \pi} \int_{\Sigma} |d^2 z| \, \sqrt {g}  \, e^{-2\sigma(z, \bar z)}  \, \left[\sum_{i=1}^{|\Delta_+|}    \{   \beta_i \partial_{\bar z}\gamma^i +   \psi^i_{\bar z} \partial_z \psi_i   + \partial_{\bar z} (V^i \cdot Y) \partial_z (V_i \cdot Y) \} -   {i \, {\cal R}_{\bar z z} \over \sqrt{\hat k + h^\vee}}  (\rho \cdot Y)\right],
\ee
where
\be
\label{k-k}
{1 \over \sqrt {\hat k + h^\vee} }= R - {1 \over R}.
\ee

Let us express the field $Y$ in terms of its constant and fluctuating parts $Y_0$ and $\tilde Y$, i.e., let us write $Y = Y_0 + \tilde Y$. Then, notice that the last term in $I_{\infty}$ would weight the (Minkowskian) path integral by a factor of ${\rm exp} \left( -2 \pi(2-2g) \rho \cdot Y_0 / \sqrt{\hat k + h^\vee} \right)$, where $g$ is the genus of $\Sigma$.\footnote{One can see this by noting that (i) $|d^2 z| = i dz \wedge d{\bar z}$, $2 \pi c_1(\Sigma) = -i {\cal R}_{\bar z z} dz \wedge d{\bar z}$, $\int_{\Sigma} c_1 = 2 - 2g$, (ii) Wick rotating the action back to Minkowskian signature involves multiplying the action by $i$.} Thus, from the viewpoint of the string theory described by the flag manifold model, one can interpret this factor as $g_s^{2-2g}$, where 
\be
\label{g_s}
g_s =  {\rm exp}\left({-{2 \pi \rho \cdot Y_0\over \sqrt{\hat k + h^\vee}}}\right)
\ee 
is the string coupling. Hence, by tuning $g_s$, we can vary $\hat k$ and vice-versa. In the limit $k \to - h^\vee$ whence $R \to \infty$, we also have, from (\ref{k-k}), $\hat k \to -h^\vee$ and therefore $g_s \to 0$. One can therefore interpret the ``classical'' Langlands duality limit of $k \to -h^\vee$ as the  zero coupling, classical limit of the string. (The rationale for the word ``classical'' is that the isomorphism (\ref{iso of W-GC}) of $\cal W$-algebras  at $k \neq -h^\vee$ underlies -- according to~\cite{Frenkel-Ram} -- what is mathematically termed as the ``quantum'' geometric Langlands correspondence.)

At any rate, let us, for simplicity, consider $\Sigma = {\bf S}^2 - \{p_1, p_2, \dots, p_n\}$, where $p_i$ represents the $i$-th point deleted; in this case, the canonical bundle $K$ of $\Sigma$ will be trivial. Since we are in the infinite $X$-volume limit where there can be no worldsheet instanton contributions, the triviality of $K$ and (\ref{p}) will mean that there can be no $\psi$ zero modes either. One can then integrate out the $\psi$ fields that appear in (\ref{Iinfty}) by performing the Gaussian integral over its nonzero modes, and effectively write $I_\infty$ as
\be
\label{Iinfty-effective}
I_{\infty, {\rm eff}} =  {1 \over  \pi} \int_{\Sigma} |d^2 z| \, \sqrt {g}  \, e^{-2\sigma(z, \bar z)}  \, \left[\sum_{i=1}^{|\Delta_+|}    \{   \beta_i \partial_{\bar z}\gamma^i  + \partial_{\bar z} (V^i \cdot Y) \partial_z (V_i \cdot Y) \} -   i { {\cal R}_{\bar z z} \over \sqrt{\hat k + h^\vee}}  (\rho \cdot Y)\right].
\ee
If we consider values of $g_s$ such that $\hat k$ is an integer,\footnote{One could of course consider non-integer values of $\hat k$, but this would involve certain technical subtleties which would take us beyond the scope of the present paper.} according to~\cite{free-fields WZW}, $I_{\infty, {\rm eff}}$ is just the action of a WZW model for simply-connected, \emph{simply-laced} $G$ at level $\hat k$! (The reason why the gauge group is $G$ and not $\GC$ is because the topological term in the WZW model dictates that the gauge group be a compact, simple Lie group whenever the level $\hat k$ is an integer~\cite{Ketov}.)

\bigskip\noindent{\it Conformal Blocks of the WZW Model}

As there is an affine algebra $\widehat{\frak g}$ of $G$ at level $\hat k$ from our effective theory with action (\ref{Iinfty-effective}), we have the relation 
 \be
J_a(z)\Phi^{\lambda}_r (z') \sim - \sum_{s}
{{(t_a^{\lambda})_{rs}\
\Phi^{\lambda}_s}(z')\over{z-z'}},
\label{primary field OPE's WZW}
\ee
where the $J$'s are dimension-one currents of $\widehat{\frak g}$; the $\Phi$'s are local, holomorphic, bosonic primary field operators of $\widehat {\frak g}$; $t_a^{\lambda}$ is a matrix in the representation of the Lie algebra $\frak g$ of $G$ with highest weight $\lambda$; $r,s = 1, \dots, \textrm{dim}|\lambda|$; and $a=1,\dots, \textrm{dim}(\frak g)$. 

Note that the scaling dimension $h_\lambda$ of the operator  $\Phi^{\lambda}_s(z)$ obeys the formula~\cite{CFT}
\be
2 (\hat k + h^\vee) h_\lambda = (\lambda, \lambda + 2 \rho).
\label{scaling dimension WZW}
\ee
Since $k \neq - h^\vee$ in the infinite-volume limit of the flag manifold model and therefore, $\hat k \neq -h^\vee$ in the effective WZW model, if $\lambda = 0$, we have $h_\lambda = 0$ and $\textrm{dim}|\lambda| = 1$ -- i.e., we have a sole operator $\Phi^{0} (z)$, and the $t_a^0$'s in (\ref{primary field OPE's WZW}) are just constants.  If $\lambda \neq 0$, then we ought to have $h_\lambda \neq 0$, and because ${\rm dim}|\lambda| > 1$, we have not one but a set of operators $\Phi^{\lambda}_s(z)$ of \emph{positive} scaling dimension. As the model is now conformal, one can always employ the CFT state-operator isomorphism and replace the $n$ punctures on $\Sigma$ with these $\Phi$ operators to obtain an effective worldsheet $\hat \Sigma$ that is an ${\bf S}^2$.

\def\MC{{\mathscr C}}

Since $G$ is a simply-connected, simply-laced group, the corresponding WZW model is diagonal and factorizes into a holomorphic and an antiholomorphic sector. Because these sectors are identical, it suffices to focus only on the holomorphic half of the theory. Let us therefore focus on the holomorphic conformal blocks of the WZW model. According to the previous paragraph, we can express the holomorphic conformal blocks as
\be
\label{CB}
{\MC} = \left< \Phi (z_1) \dots \Phi(z_n) \right>_{\hat \Sigma},
\ee 
where the local operator $\Phi(z)$ can either be $\Phi^0(z)$ or $\Phi^{\lambda}_s(z)$. 

\begin{figure}
  \centering
    \includegraphics[width=0.3\textwidth]{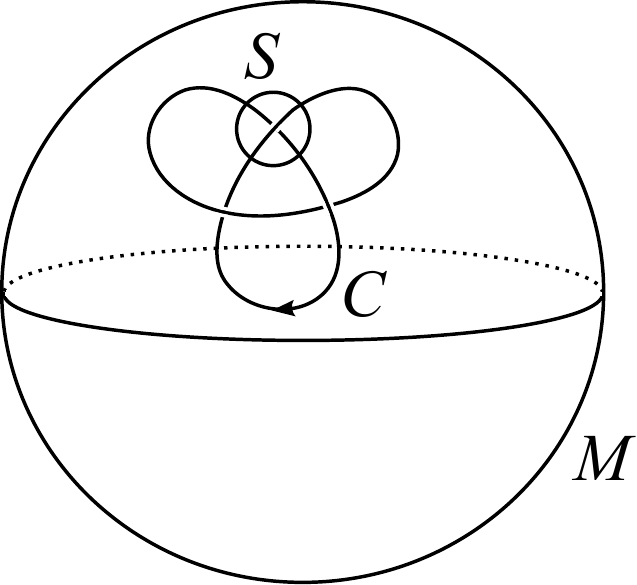}
  \caption{Path integral over whole of $M$}
\end{figure}

\bigskip\noindent{\it States of the Corresponding Chern-Simons Theory}

Observe that since $\Phi^0(z)$ ought to be given by a regular, holomorphic function in the coordinate $z$ on the compact Riemann surface $\hat \Sigma = {\bf S}^2$, it can be viewed as a constant; in particular, one can regard $\Phi^0(z)$ as the identity operator $1(z)$. Thus, if $\Phi(z_i) = \Phi^0(z_i)$ for all $i =1, \dots, n$ in (\ref{CB}),  we have
\be
\label{CB-unramified}
\MC = \MC_{\rm empty} =  \left< \Phi^0 (z_1) \dots \Phi^0(z_n) \right>_{\hat \Sigma} =   \left < 1 \right>_{\hat \Sigma}.
\ee

On the other hand, if say $p$ of the $n$ number of $\Phi$'s in (\ref{CB}) are given by the $\Phi^{\lambda}_s$ operators, because of what we said in the last paragraph, we have
\be
\label{CB-ramified}
\MC= \MC_{\rm knots} = \left< \Phi^{\lambda_1}_{s_1} (z_1) \dots \Phi^{\lambda_p}_{s_p}(z_p) \Phi^0(z_{p+1}) \dots \Phi^0(z_n) \right>_{\hat \Sigma} =  \left< \Phi^{\lambda_1}_{s_1} (z_1) \dots \Phi^{\lambda_p}_{s_p}(z_p) \right>_{\hat \Sigma}.
\ee 

We shall now elucidate the meaning of the subscripts ``empty'' and ``knots''. To this end, note that it was first established by Witten in~\cite{Jones-Witten} that if one can Heegaard split some three manifold $M$ along $\hat \Sigma \subset M$, then the states in the Hilbert space $\cal H$ of the Chern-Simons theory on $M$ with gauge group $G$ and inverse coupling $\hat k$, will correspond, in a one-to-one manner, to the conformal blocks  of the WZW model for $G$ at level $\hat k$ on $\hat \Sigma$. If there are no Wilson lines in $M$, the states of the Chern-Simons theory will be captured by $ \MC_{\rm empty}$ -- hence, the  subscript ``empty''. However, if there are Wilson lines in $M$ in various highest dominant weight representations that pierce through $\hat\Sigma$ at the points $z_1, \dots, z_p$  which, according to~\cite{Jones-Witten}, can also be interpreted as knots in $M$, the states of the Chern-Simons theory will be captured by $\MC_{\rm knots}$ -- hence, the subscript ``knots''. For example, in fig.~3, there is a Wilson line or knot $C$ in some highest dominant weight representation $\lambda$ of $G$ that pierces through $S = \hat \Sigma$ at four points, $z_1, z_2, z_3, z_4$; in this case, 
\be
\label{CB example}
\MC_{\rm knots} =  \MC^\lambda_{\rm knots} =  \left< \Phi^{\lambda}_{s_1} (z_1) \, \Phi^{\lambda}_{s_2} (z_2) \, \Phi^{\bar\lambda}_{s_3} (z_3) \, \Phi^{\bar\lambda}_{s_4}(z_4) \right>_{\hat \Sigma},
\ee
where $\bar \lambda$ is the representation dual to $\lambda$. The $\lambda$ and $\bar \lambda$ representations are associated with the points where the Wilson line or knot pierces into and out of $S$, respectively.

\newsubsection{``Quantum'' Ramified $\cal D$-Modules, Khovanov Homology, And Langrangian Intersection Floer Homology}

According to~\cite{Jones-Witten, Elitzur, Mehta},  the states in the Hilbert space $\cal H$ of the Chern-Simons theory on $M$ with gauge group $G$ in the presence of Wilson lines that pierce through $\hat \Sigma \subset M$ at points $z_1, \dots, z_p$ in the representations $\lambda_1, \dots, \lambda_p$, respectively, are -- once a complex structure on $\hat \Sigma$ is picked -- in one-to-one correspondence with the elements of $H^0(\CM_{G; z_1, \dots, z_p}, \mathscr L^{\hat k})$.  Here,  $\CM_{G; z_1, \dots, z_p}$ is the moduli space of flat $G$-bundles on $\hat \Sigma \backslash \{z_1, \dots, z_p\}$ whose connection has monodromy around the point $z_i$ given by 
\be
\label{mono}
g_{\lambda_i} = {\rm exp}\left(-{2 \pi  i \lambda^\ast_i \over \hat k}  \right),
\ee
where
\be
g_{\lambda_1} \dots g_{\lambda_p} = 1;
\ee
$\mathscr L$ is a line bundle whose first Chern class generates the second cohomology of $\CM_{G; z_1, \dots, z_p}$; and $\lambda^\ast_i$ is the dual of $\lambda_i$ in the following sense: using the quadratic form $-{\rm Tr}$ on $\frak g$ to identify the dual $\frak t^\ast$ of its Cartan subalgebra $\frak t$ as the Langlands-dual Cartan subalgebra ${^L\frak t}$, $\lambda_i \in {^L\frak t}$ maps to $\lambda^\ast_i \in \frak t$.


 So, if there are no Wilson lines or knots in $M$, in which case one would have to replace the operators $\Phi^{\lambda_1}_{s_1}(z_1), \dots, \Phi^{\lambda_1}_{s_1}(z_p)$ in (\ref{CB-ramified}) with $\Phi^{0}(z_1), \dots, \Phi^{0}(z_p)$,  we have $\MC_{\rm knots} \to \MC_{\rm empty}$ and
 $g_{\lambda_i} =  g_0 = 1$ for all $i=1, \dots, p$.

\bigskip\noindent{\it ``Quantum'' Ramified $\cal D$-Modules}

A useful fact to note at this juncture is that in the quantum geometric Langlands correspondence for $G$ with tame ramification, ``quantum'' ramified $\cal D$-modules would be given by sections of the line bundle ${\mathscr L}^{c- {h^\vee}}$ over $\CM_{G; z_1, \dots, z_p}$, where the nonzero integer $c = \hat k + h^\vee$~\cite{Kap-QGL}. Second,  based on the analysis in $\S$4 of~\cite{bq}, we find that $ \MC_{\rm knots} \subset \cal H \subset {\cal H}^\ast$, where ${\cal H}^\ast$ is the space of $({\cal B}_{cc}, {\cal B})$ strings in a topological $A$-model on the parabolic Hitchin fibration $\pi: Y \to {\bf B}$ with Lagrangian fiber $\bf F$ (along which a restriction $\omega_{J \vert {\bf F}}$ of a certain two-form $\omega_J$ is nonvanishing); ${\cal B}_{cc}$ is a space-filling canonical coisotropic brane (endowed with a unitary line bundle $\cal L$ with connection whose curvature is $\omega_J$); and $\cal B$ is a brane which wraps $\bf F$. In turn, noting the fact that $c$ is an integer, based on the analysis in $\S$11.3 of~\cite{KW} and that in $\S$4.4 of~\cite{GW} (with $\alpha \neq 0$; $\beta = \gamma = \eta = 0$; and $\theta \neq 0$ which ensures that $\omega_{J \vert {\bf F}}$ is nonvanishing),  we find that ${\cal H}^\ast$ would correspond to the space of ``quantum'' tamely-ramified $\cal D$-modules. Third,  just like there is supposed to be not one but a category of distinct ``quantum'' ramified $\cal D$-modules~\cite{Kap-QGL}, for each set of representations ${\lambda}_1, \dots, {\lambda}_p$, there is not one but several distinct $\MC_{\rm knots}$'s  due to the extra  $s_i$-index which runs from $0, \dots, {\rm dim}| {\lambda}_i| >1$. Fourth, in the $c \to 0$ limit whence the operators $\Phi(z) = \Phi^{0}(z)$ in (\ref{CB}) are, according to our analysis in $\S$5, $Q$-cohomology classes, according to our analysis in $\S$7, $\MC_{\rm empty} =  \left< \Phi^0 (z_{p+1}) \dots \Phi^0(z_n) \right>_{\hat \Sigma}$ would just be an ordinary ``classical'' $\cal D$-module. Consequently, this means that $\MC_{\rm knots}  =  \left< \Phi^{\lambda_1}_{s_1} (z_1) \dots \Phi^{\lambda_p}_{s_p}(z_p) \Phi^0(z_{p+1}) \dots \Phi^0(z_n) \right>_{\hat \Sigma}$ would actually generate -- according to the abstract algebraic CFT prescription in $\S$9.8 of~\cite{Frenkel} -- the category of ``classical'' $\cal D$-modules with tame ramification at points $z_1, \dots, z_p$. Last but not least, note that a connection between Chern-Simons theory  and a ``quantum'' geometric Langlands correspondence has also been unraveled recently by Witten in~\cite{5-branes and knots}, albeit via a gauge-theoretic approach; in \emph{loc.~cit.}, the geometric Langlands parameter and its dual are respectively $\Psi = \hat k + h^\vee$ and $^L\Psi = - 1 / \Psi$, whereby a ``classical'' geometric Langlands correspondence for $G$ is achieved in  the limits $^L\Psi \to \infty$ and  $\Psi \to 0$, i.e., $c \to 0$, in perfect agreement with our results obtained hitherto.  Altogether, the above five points imply that we can interpret $\MC_{\rm knots}$ as a ``quantum'' (tamely) ramified $\cal D$-module ${\cal D}_{mod}^{c}(\CM_{G; z_1, \dots, z_p})$ on $\CM_{G; z_1, \dots, z_p}$ with twist parameter $c$.

\begin{figure}
  \centering
    \includegraphics[width=0.35\textwidth]{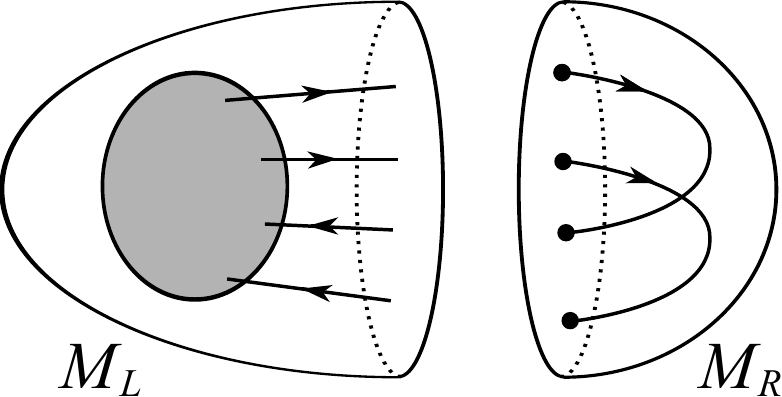}
  \caption{Path integral over $M_L$ and $M_R$}
\end{figure}

\bigskip\noindent{\it Relation to Knot Invariants of Three-Manifolds}

In~\cite{Jones-Witten}, it was argued that the states of the Chern-Simons theory with Wilson lines can be related to knot invariants of three-manifolds; in turn, this means that these knot-invariants can also be related to the above-mentioned ``quantum'' ramified $\cal D$-modules. For example, in fig.~3, the Chern-Simons path integral over the whole of $M$ with Wilson line or knot $C$ in some highest weight representation $\lambda$ of $G$, would be given, as shown in fig.~4, by a path integral over the interior three-ball $M_R$  whose boundary is  $S = \hat \Sigma$, followed by a path integral over the complicated exterior piece $M_L$ whose boundary is also $\hat \Sigma$ but with orientation reversed.  The former path integral would result in a state $ \ket \psi \in {\cal H}_R$, while the latter path integral would result in a state $ \ket \chi \in {\cal H}_L$. Since the Hilbert spaces ${\cal H}_{L,R}$ are determined by the corresponding boundary theories on $\hat \Sigma$, the fact that the orientations of $\hat \Sigma$ associated with ${\cal H}_L$ and ${\cal H}_R$ are opposite to each other means that ${\cal H}_L$ is canonically dual to ${\cal H}_R$. Hence, we can express the complete path integral over $M$ as the pairing of states
\be
Z^\lambda_{M}(C) = \langle \chi \vert \psi \rangle =    (\chi, \psi).
\label{knots 1}
\ee
The LHS of the above relation is a knot-invariant of $M$, while $\psi$ and ${\chi}$ on the RHS can be identified as ``quantum'' ramified $\cal D$-modules ${\cal D}_{mod}^{c}(\CM_{G; z_1, \dots, z_4})$ given by $\MC^\lambda_{\rm knots}$ of (\ref{CB example}). 

As $\chi$ and $\psi$ are vectors in a finite-dimensional space $\cal H$, we can expand them as
\be
\chi = \sum_{s=1}^{{\rm dim} {\cal H}} a^s_\chi \, \mathscr D_s \qquad {\rm and} \qquad \psi = \sum_{s=1}^{{\rm dim} {\cal H}} a^s_\psi \, \mathscr D_s, 
\ee
where the coefficients $a^s_\chi$ and $a^l_{\psi}$ are complex numbers (some of which may be zero), while $\mathscr D_s \in {\cal D}_{mod}^{c}(\CM_{G; z_1, \dots, z_4})$ spans an orthogonal basis in $\cal H$, i.e., $(\mathscr D_m, \mathscr D_n) = \delta_{mn}$.  Thus, we can also express (\ref{knots 1}) as
\be
Z^\lambda_{M}(C) =   \sum_{s =1}^{{\rm dim} \, \cal H} \,  a^s_\chi a^s_\psi.
\label{knots 2}
\ee
Notice that the $s^{\rm th}$ term on the RHS of (\ref{knots 2}) would vanish if the $s^{\rm th}$ component of either $\chi$ or $\psi$ were to be zero. Hence, we can interpret the RHS of (\ref{knots 2}) as a (weighted) count of the number of components of $\chi$ and $\psi$ that coincide.  

\bigskip\noindent{\it The Jones Polynomial and Khovanov Homology}

Let us now specialize to the case where  $M = {\bf S}^3$ and $G = SU(2)$; let $\lambda$ label the two-dimensional fundamental representation $\bf 2$ of $SU(2)$. Then,  $Z^{\bf 2}_{{\bf S}^3}(C)$ is simply the Jones polynomial of the knot $C$~\cite{Jones-Witten}. In turn, if the finite-dimensional vector space 
\be
{\cal K} (C) =  \oplus_{a,b} \, {\cal K}^{a,b}(C)
\ee
is the corresponding bi-graded Khovanov homology, we can,  according to~\cite{Khovanov},  rewrite (\ref{knots 2}) as
\be
\label{count1}
\sum_{a,b} \,  (-1)^a  q^b \, {\rm dim} \, {\cal K}^{a,b}(C) =   \sum_{c \in \mathscr C} \,  a^c_\chi a^c_\psi, 
\ee
where $\mathscr C$ is the set of components of $\chi$ and $\psi$ that coincide, and
\be
q = {\rm exp} \left ({{2 \pi i} \over {\hat k + h^\vee} } \right).
\ee
Thus, from (\ref{count1}), we learn that a (weighted) count of the Khovanov homology of the knot $C$ would be given by  a (weighted) count of the number of components of the ``quantum'' ramified $\cal D$-modules  $\chi$ and $\psi$ that coincide.

\bigskip\noindent{\it Relation to Lagrangian Intersection Floer Homology}

Now notice that we can also write 
\be
\sum_{n =1}^{{\rm dim} \cal H} \sum_{m =1}^{{\rm dim} \cal H}   \delta^n_{m} \langle \chi \vert \mathscr D_m \rangle \langle \mathscr D_n \vert \psi \rangle =  (\chi, \psi).
\ee 
In terms of the vectors $\varphi_n, \varphi_m \in \cal H$, where $\varphi_n = \sum_{n=1}^{{\rm dim} {\cal H}} \mathscr D_n$ and $\varphi_m = \sum_{m=1}^{{\rm dim} {\cal H}} \mathscr D_m$, this is  
\be
\label{hf1}
\delta^n_m \, (\chi, \varphi_m) (\varphi_n, \psi) = (\chi, \psi).
\ee
Compare this with the relation between Lagrangian intersection Floer homology groups  
\be
\label{hf2}
HF_{\rm symp}^\ast (L_0, L_1) \otimes HF_{\rm symp}^\ast (L_1, L_2) \rightarrow HF_{\rm symp}^\ast (L_0, L_2), 
\ee
where the pairwise $L_i$'s are intersecting Lagrangian submanifolds of some underlying symplectic manifold $\CM_{\rm symp}$, and the vector space $HF_{\rm symp}^\ast (L_i, L_j)$ is generated and counted by the intersection points of $L_i$ and $L_j$. Since $\varphi_m = \varphi_n$, and since for any $\zeta_1, \zeta_2 \in \cal H$, the pairing $(\zeta_1, \zeta_2)$ counts (with weights) the number of components of  $\zeta_1$ and $\zeta_2$ that coincide, the similarity between (\ref{hf1}) and (\ref{hf2}) suggests that we can interpret $\chi$ and $\psi$ as Lagrangian submanifolds $L_{\chi}$ and $L_{\psi}$ of $\CM_{\rm symp}$ that have nonzero intersection with each other. In turn, since the RHS of (\ref{count1}) is equal to $(\chi, \psi)$, it would mean that we can write 
\be
\label{count-symp}
\sum_{a,b} \,  (-1)^a  q^b \, {\rm dim} \, {\cal K}^{a,b}(C) =   \sum_{c \in \mathscr C} \,  \mathscr I_c, 
\ee
where $\mathscr C$ is now the set of intersection points of $L_{\chi}$ and $L_{\psi}$ in $\CM_{\rm symp}$, and $\mathscr I_c$ is some complex number whose value depends on the point $c$. In light of the fact that (i) $\chi$ and $\psi$ can be regarded as ``quantum'' ramified $\cal D$-modules, (ii) every Lagrangian brane of an $A$-model underlies a ``quantum'' ramified $\cal D$-module if the target is the parabolic Hitchin fibration $\CM_H$,\footnote{To see this, repeat the analysis in $\S$4.4 of~\cite{GW} using $\alpha \neq 0$, $\beta = \eta = \gamma = 0$, and $\theta \neq 0$.}  one can deduce that $\CM_{\rm symp} = \CM_H$.

\bigskip\noindent{\it A Gauge-Theoretic Approach}

We will now rederive (\ref{count-symp}) via four-dimensional gauge-theory, and in the process, obtain an explicit formula for $\mathscr I_c$. The relevant gauge theory for this purpose is GL-twisted ${\cal N} = 4$ SYM on the four-manifold $V = W \times \mathbb R_+$ with a surface operator, where $W = {\bf S}^3$ and $\mathbb R_+$ is the half-line $y \geq 0$ that one can interpret as the Euclidean ``time'' direction.  

According to $\S$4.2 of~\cite{5-branes and knots}, the LHS of (\ref{count-symp})  would  be given by the path integral
of this ${\cal N} = 4$ theory with (i) gauge group $SU(2)$; (ii) GL-twist parameter $t=1$; (iii) theta-angle $\theta \neq 0$; and (iv) a surface operator of the form $D= \mathbb R_+ \times C$ with (classical) monodromy parameter $\alpha =  {\lambda^\ast / \hat k}$, where $C \subset W$. This path integral counts (with appropriate weights) the number of solutions to the localization equations defined by setting the supersymmetric variation of the fermionic fields to zero. 

The above being said, note that it was also shown in $\S$2 of~\emph{loc.~cit.}, that the four-dimensional path integral is actually equivalent to a \emph{three-dimensional} path integral on $W$, i.e., it is in fact independent of ``time''. As such, the Hamiltonian of the four-dimensional path integral -- which generates field variations under ``time'' translations -- is effectively zero. Since each solution to the localization equations gives rise to a quantum supersymmetric state, the vanishing of the effective Hamiltonian would mean that the four-dimensional path integral really counts (with appropriate weights) the number of quantum supersymmetric \emph{ground} states of the ${\cal N} = 4$  theory.

\begin{figure}
  \centering
    \includegraphics[width=0.8\textwidth]{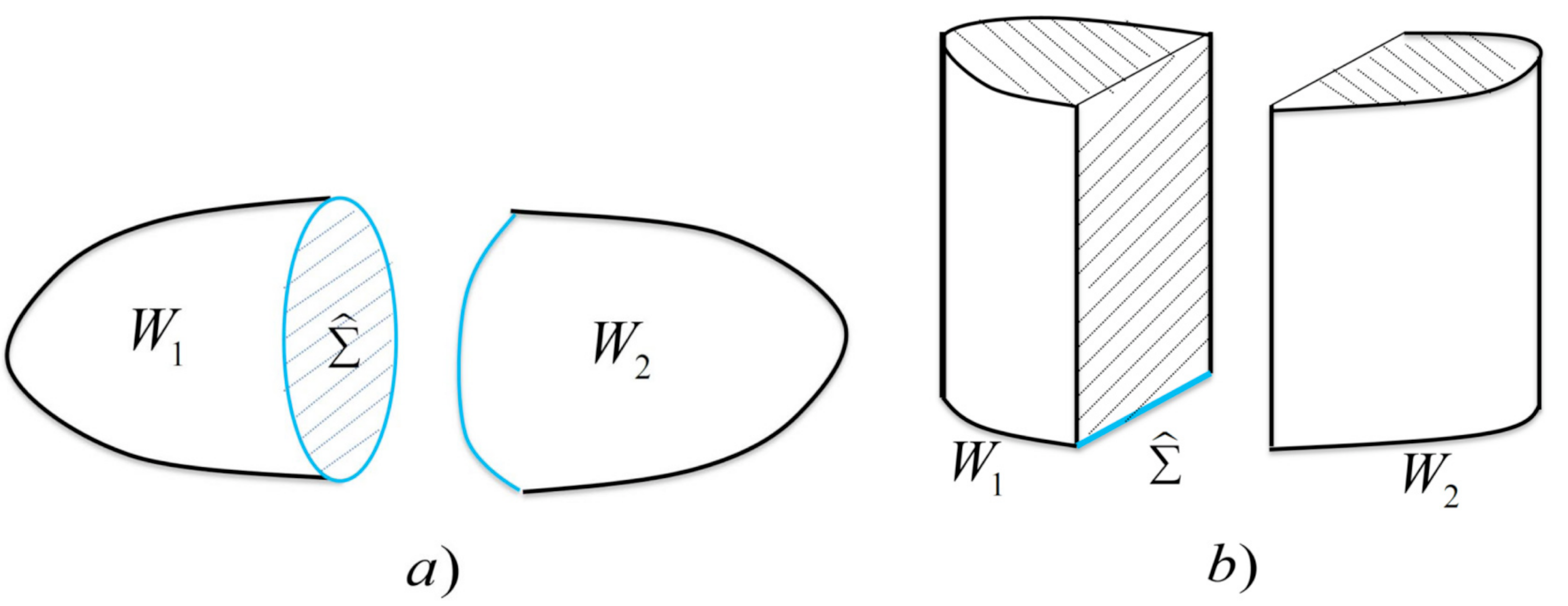}
  \caption{a) Heegaard Split of $W$; b) Heegaard Split of $V$}
\end{figure}

In order to describe these ground states, we will first need to make the following observation. Just as an $\bf S^2$ can be obtained by gluing a pair of two-discs ${\bf D}^2$ along their common ${\bf S}^1$ boundaries, one can also obtain an ${\bf S}^3$ by gluing a pair of three-discs ${\bf D}^3$ along their common ${\bf S}^2$ boundaries. In particular, this means that ${\bf S}^3$ can be Heegaard split into $W_1$ and $W_2$ along the two-surface $\hat \Sigma$, as shown in fig.~5a, where $\hat \Sigma = {\bf S}^2$ is just the equator of ${\bf S}^3$. Consequently, one can also Heegaard split $V$  into a pair of four-manifolds with corners $W_1 \times \mathbb R_+$ and $W_2 \times \mathbb R_+$ along the three-surface $\hat \Sigma \times \mathbb R_+$, as shown in fig.~5b.

We are now ready to describe the ground states. According to~\cite{gukov}, in the situation given by fig~5, the space of ground states would coincide with the space  of certain open strings that end on the branes ${\cal B}_1$ and ${\cal B}_2$ associated with the three-manifolds $W_1$ and $W_2$. These open strings are described by a topological $A$-model of type $K$ (since $t$=1) whose target is the parabolic Hitchin fibration $\pi: {\cal M}_H \to {\bf B}$ with Lagrangian fiber ${\bf F}$. The branes ${\cal B}_1$ and ${\cal B}_2$ are consequently $A$-branes of type $K$, and they are necessarily Lagrangian. 

For a single knot in $W$, we can write ${\cal B}_2 = {\tilde \phi_n}({\cal B}_1)$, where $\tilde \phi_n$ represents an autoequivalence action on the Fukaya category of ${\cal B}_1$ branes by some element of the braid group on $n$ letters; since $C$ in this case cuts $S = \hat \Sigma$ at four points as shown in fig.~3, we have  $n=4$.  For example, if $C$ is a $(2,k)$ torus knot and $\tilde \phi_4 = \phi_4$, where $\phi_4$ corresponds to a half-twist or an element of the braid group which exchanges the first two letters, we can view $C$ in $W$ as a union of the branes ${\cal B}_1$ and ${\cal B}_2 = \phi^k_4({\cal B}_1)$, as explained in fig.~6.  If $C$ is the unknot, one would have in fig.~6 a single twist instead of three half-twists; $C$ in $W$ can then be viewed as a union of the branes ${\cal B}_1$ and ${\cal B}_2 = \phi_4({\cal B}_1)$. 

\begin{figure}
  \centering
    \includegraphics[width=0.5\textwidth]{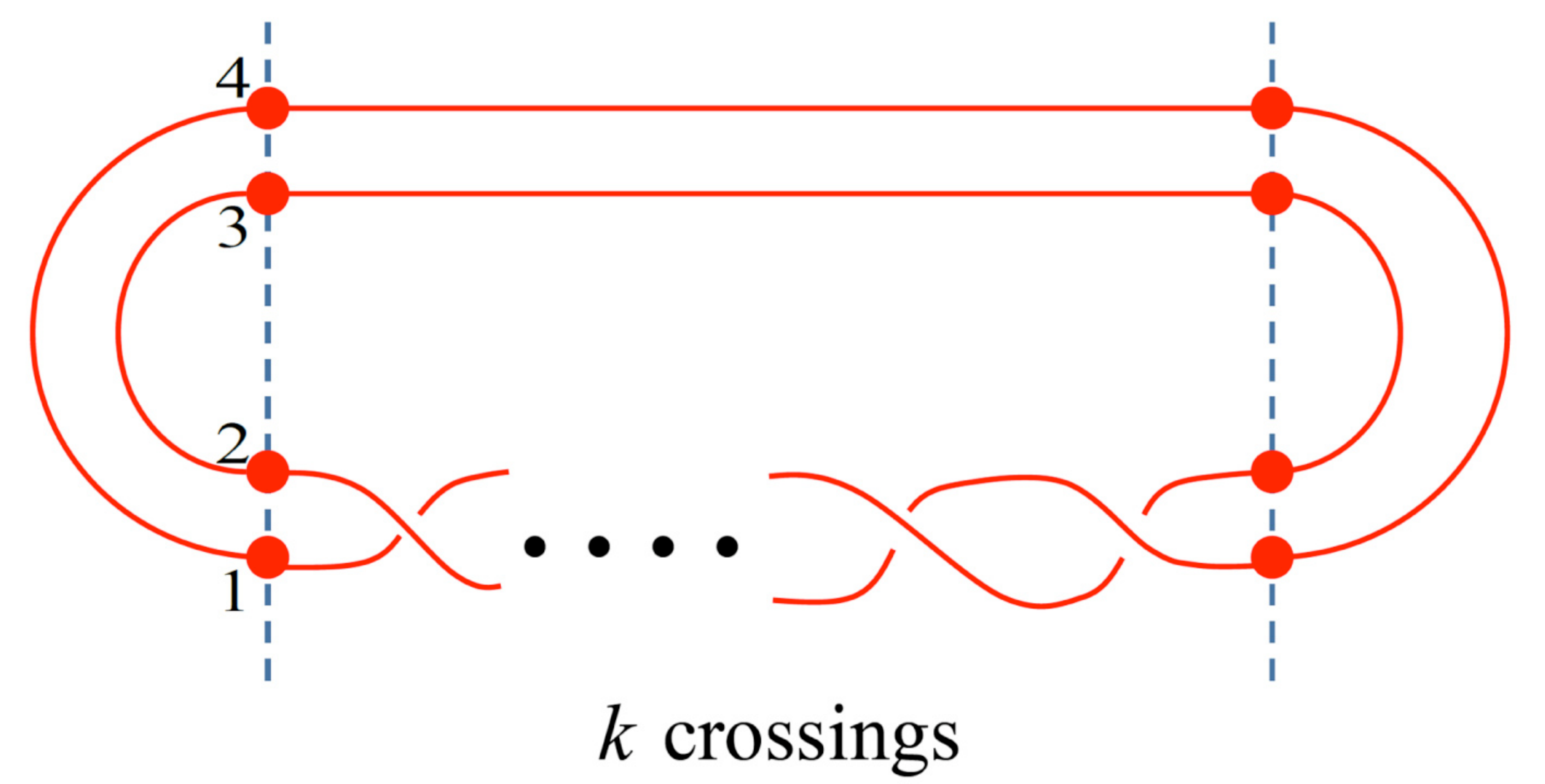}
  \caption{The $(2,k)$ torus knot in $W = {\bf S}^3$ obtained via a union of branes ${\cal B}_1$ and ${\cal B}_2 = \phi^k_4({\cal B}_1)$.  The vertical lines correspond to $\hat \Sigma$ which divides $W$ into $W_1$ and $W_2$. As shown, there are on $\hat \Sigma$ four distinct points where the knot or surface operator pierces through.}
\end{figure}

In short, the space of ground states would be given by the space of $({\cal B}_1, \tilde\phi_4({\cal B}_1))$ strings, where $\tilde\phi_4$ depends on the details of the knot $C$. A useful result~\cite{gukov} to state at this point is that the space of  $({\cal B}_1, \tilde\phi_4({\cal B}_1))$ strings is also  the Lagrangian intersection Floer homology $HF^\ast_{\rm symp}({\cal B}_1, \tilde\phi_4({\cal B}_1))$; in particular, the dimension of  $HF^\ast_{\rm symp}({\cal B}_1, \tilde\phi_4({\cal B}_1))$ is counted by the number of intersection points of ${\cal B}_1$ and  $\tilde\phi_4({\cal B}_1)$ in $\CM_H$.  In all, we can write (cf.~\cite{5-branes and knots})
\be
\label{count gauge-theoretic}
\sum_{a,b} \,  (-1)^a  q^b \, {\rm dim} \, {\cal K}^{a,b}(C) =   \sum_{c \in \mathscr C} \,  (-1)^{g_c} q^{n_c} , 
\ee
where $\mathscr C$ is the set of intersection points of ${\cal B}_1$ and  $\tilde\phi_4({\cal B}_1)$ in $\CM_H$, while $g_c$ and $n_c$ are integers whose values are $a$ and $b$. 

The relation (\ref{count gauge-theoretic}) is indeed consistent with (\ref{count-symp}) because (i) like $L_\psi$ and $L_\chi$,  the branes ${\cal B}_1$ and $ \tilde\phi_4({\cal B}_1)$ are associated with the three-spaces interior and exterior to $\hat \Sigma$ in ${\bf S}^3$; (ii) like $L_\psi$ and $L_\chi$, ${\cal B}_1$ and $ \tilde\phi_4({\cal B}_1)$ are Lagrangian $A$-branes of type $K$ in $\CM_H$ which consequently underlie ``quantum'' ramified $\cal D$-modules on $\CM_{SU(2); z_1, \dots, z_4}$; (iii)   like $\mathscr I_c$ of (\ref{count-symp})$, (-1)^{g_c} q^{n_c} $ of (\ref{count gauge-theoretic}) is a complex number. In sum, this implies that we can identify ${\cal B}_1$, $ \tilde\phi_4({\cal B}_1)$ and $(-1)^{g_c} q^{n_c} $ with $L_\psi$, $L_\chi$ and $\mathscr I_c$, respectively, whence (\ref{count gauge-theoretic}) would just coincide with (\ref{count-symp}). 

\bigskip\noindent{\it Explicit Check of Relation (\ref{count gauge-theoretic})}

Let us now subject (\ref{count gauge-theoretic}) and therefore (\ref{count-symp}) to an explicit check. As mentioned, $g_c$ and $n_c$ are integers whose values are $a$ and $b$; validity of (\ref{count gauge-theoretic}) would then imply that the total number of points spanning the set $\mathscr C$ is equal to $\sum_{a,b}{\rm dim} \, {\cal K}^{a,b}(C) = \sum_{a,b}{\rm rk} \, {\cal K}^{a,b}(C)$, where ${\rm rk} \, {\cal K}^{a,b}(C)$ is the rank of the group ${\cal K}^{a,b}(C)$.  Let us, for convenience, verify this for the $(2,k)$ torus knot, since its Khovanov homology is known. According to Prop.~35 of~\cite{Khovanov},  if $C_{2,k}$ is a $(2,k)$ torus knot (where $k$ is odd),  ${\cal K}^{\ast, \ast}(C_{2,k}) = {\mathbb Z}^{k +1} \oplus ({\mathbb Z / 2})^{(k-1) / 2}$. Thus, $ \sum_{a,b} {\rm rk} \, {\cal K}^{a, b}(C_{2,k}) = k+1$, and if  (\ref{count gauge-theoretic}) is to be true, ${\cal B}_1$ and $\phi^k_4({\cal B}_1)$ ought to intersect at $k+1$ points in $\CM_H$. 

In order to ascertain the number of intersection points of ${\cal B}_1$ and $\phi^k_4({\cal B}_1)$ in $\CM_H$, first note that hyperk\"ahler $\CM_H$ (in one of its three complex structures) can be described as the affine cubic~\cite{gukov}
\be
x_1 x_2 x_3 + \sum_{i=1}^3(x^2_i - \theta_i x_i)  + \theta_4 = 0,
\ee 
where $(x_1, x_2, x_3) \in \mathbb C^3$, and the $\theta_i$'s are constants that depend on the monodromy (\ref{mono}) associated with the knot. Second, notice from fig.~6 that the brane ${\cal B}_1$ on the left identifies the monodromies around points 1 and 4 (resp. 2 and 3). As such, ${\cal B}_1$ can be described as the degenerate quadric~\cite{gukov}
\be
(x_2 + x_3 - a^2)^2 = 0,
\ee
 where $a$ is also a constant that depends on the monodromy  (\ref{mono}) associated with the knot. Note that because of the double degeneracy of the quadric, ${\cal B}_1$ must be viewed as a stack of \emph{two }coincident branes supported along $x_2 + x_2 = a^2$.  Third, note that one can explicitly show~\cite{gukov} that there are $(k+1) /2$ distinct sets of triples $(x_1, x_2, x_3)$ which simultaneously solve the polynomial equations that describe  ${\cal B}_1$ and $\phi^k_4({\cal B}_1)$. Since  ${\cal B}_1$ is actually a stack of two coincident branes, the last statement means that  ${\cal B}_1$ and $\phi^k_4({\cal B}_1)$ effectively intersect at $k+1$ points in $\CM_H$, as anticipated. This completes our explicit check of relation (\ref{count gauge-theoretic}).

\bigskip\noindent{\it Some Common Examples}

Let us consider some common examples for illustration purposes. Take the unknot $C_{2,1}$. It is such that ${\cal K}^{a,b}(C_{2,1}) = \mathbb Z$ for $a =0$ and $b \pm 1$, and is zero otherwise; in other words, ${\cal K}^{\ast, \ast}(C_{2,1}) = {\mathbb Z}^{2}$, and $\sum_{a,b} {\rm rk} \, {\cal K}^{a, b}(C_{2,1}) = 2$. On the other hand, it can be verified~\cite{gukov} that  $(x_1, x_2, x_3) = (2,2,a^2 -2)$ simultaneously solves the polynomial equations that describe  ${\cal B}_1$ and $\phi_4({\cal B}_1)$. Hence, ${\cal B}_1$ and $\phi_4({\cal B}_1)$ effectively intersect at 2 points in $\CM_H$, in agreement with the fact that $\sum_{a,b} {\rm rk} \, {\cal K}^{a, b}(C_{2,1}) = 2$.

Next, take the trefoil knot $C_{2,3}$. In this case, ${\cal K}^{0,-1}(C_{2,3}) = {\cal K}^{0,-3}(C_{2,3}) = {\cal K}^{-2,-5}(C_{2,3}) = {\cal K}^{-3,-9}(C_{2,3}) =  \mathbb Z$, and ${\cal K}^{-2,-7}(C_{2,3}) = \mathbb Z / 2$, and  is zero otherwise; in other words, ${\cal K}^{\ast, \ast}(C_{2,3}) = {\mathbb Z}^{4} \oplus \mathbb Z/2$, and $\sum_{a,b}{\rm rk} \, {\cal K}^{a, b}(C_{2,3}) = 4$. On the other hand, it can be verified~\cite{gukov} that  $(x_1, x_2, x_3) = (2,2,a^2 -2)$ and $(x_1, x_2, x_3) = (2, a^2-1,1))$ simultaneously solve the polynomial equations that describe  ${\cal B}_1$ and $\phi_4({\cal B}_1)$. Hence, ${\cal B}_1$ and $\phi^3_4({\cal B}_1)$ effectively intersect at 4 points in $\CM_H$, in agreement with the fact that $\sum_{a,b} {\rm rk} \, {\cal K}^{a, b}(C_{2,3}) = 4$.

\bigskip\noindent{\it Relation to the Moduli Space of Hecke Modifications}

As explained in detail in  $\S$3.6.2 of~\cite{5-branes and knots},  the underlying localization equations  whose solutions (in the presence of a surface operator along $\mathbb R_+ \times C$)  are algebraically counted by the four-dimensional path integral on the RHS of (\ref{count gauge-theoretic}), are just the extended Bogomolny equations described in eqn.~(10.36) of~\cite{KW}. The moduli space of (singular) solutions to these equations is also the moduli space ${\rm Gr}_{SU(2)}$ of Hecke modifications of an $SU(2)$-bundle on $\hat \Sigma$~\cite{KW}. This means that we can replace $\CM_H$ in our above discussion with ${\rm Gr}_{SU(2)}$, in which case the Lagrangian $A$-brane ${\cal B}_1$ would just correspond to an element in its middle-dimensional cohomology $H^{\rm mid}({\rm Gr}_{SU(2)})$. Since ${\cal B}_1$ underlies the ``quantum'' ramified $\cal D$-module $\psi$ of (\ref{knots 1}), and since $\psi$ can be represented by $\mathscr C^\lambda_{\rm knots}$ of (\ref{CB example}), we have 
\be
\label{prop 2.8}
H^{\rm mid}({\rm Gr}_{^LSU(2)}) \cong \left< \Phi^{\bf 2}_{s_1} (z_1) \, \Phi^{\bf 2}_{s_2} (z_2) \, \Phi^{\bar{\bf 2}}_{s_3} (z_3) \, \Phi^{\bar{\bf 2}}_{s_4}(z_4) \right>_{\hat \Sigma}.
\ee
(We have made use of the fact that ${\rm Gr}_{SU(N)} = {\rm Gr}_{^LSU(N)}$ to arrive at the above expression.) Note that (\ref{prop 2.8}) is just a physical manifestation for $G = SU(2)$ of Proposition~2.8 of~\cite{Kam}!

\bigskip\noindent{\it Relation to a Conjecture by Seidel and Smith}

In~\cite{SS}, Seidel and Smith conjectured that 
\be
\label{SS}
\bigoplus_{l} HF^{l+ 2 +w}_{\rm symp}({\cal B}_1, \phi^k_4({\cal B}_1))  = \bigoplus_{l}  \bigoplus_{i- j = l} {\cal K}^{i,j} (C_{2,k}),
\ee
where $w$ is the number of positive minus the number of negative crossings of the knot $C_{2,k}$. (To arrive at the above expression, we have made use of the result in~\cite{Kam} that the symplectic manifold $M$ of~\cite{SS} coincides, in this case, with ${\rm Gr}_{SU(2)}$.) This conjecture was proved for the trefoil knot $C_{2,3}$ via Proposition~61 of~\cite{SS}. However, it remains an outstanding mathematical problem to prove it for all $k$, although a first effort in this direction recently appeared in~\cite{Waldron}. 

That said, note that (\ref{count gauge-theoretic}) asserts, at least on purely physical grounds,  that the conjecture ought to hold for all $k$: the derivation of (\ref{count gauge-theoretic}) depends squarely on the observation that the solutions to the four-dimensional localization equations -- which represent the time-invariant states of a five-dimensional Yang-Mills theory that generate ${\cal K}^{\ast, \ast} (C_{2,k})$ (see~\cite{5-branes and knots}, $\S$4.2) -- actually correspond to four-dimensional ground states which can be captured by $HF^{\ast}_{\rm symp}({\cal B}_1, \phi^k_4({\cal B}_1))$,  i.e., $\bigoplus_m HF^{m}_{\rm symp}({\cal B}_1, \phi^k_4({\cal B}_1)) = \bigoplus_{a,b} {\cal K}^{a, b} (C_{2,k})$ whence (\ref{SS}) immediately follows.  Thus, we have found a physical proof of the above conjecture by Seidel and Smith in relation (\ref{count gauge-theoretic}), or equivalently, in relation (\ref{count-symp})!

\begin{figure}
  \centering
    \includegraphics[width=0.3\textwidth]{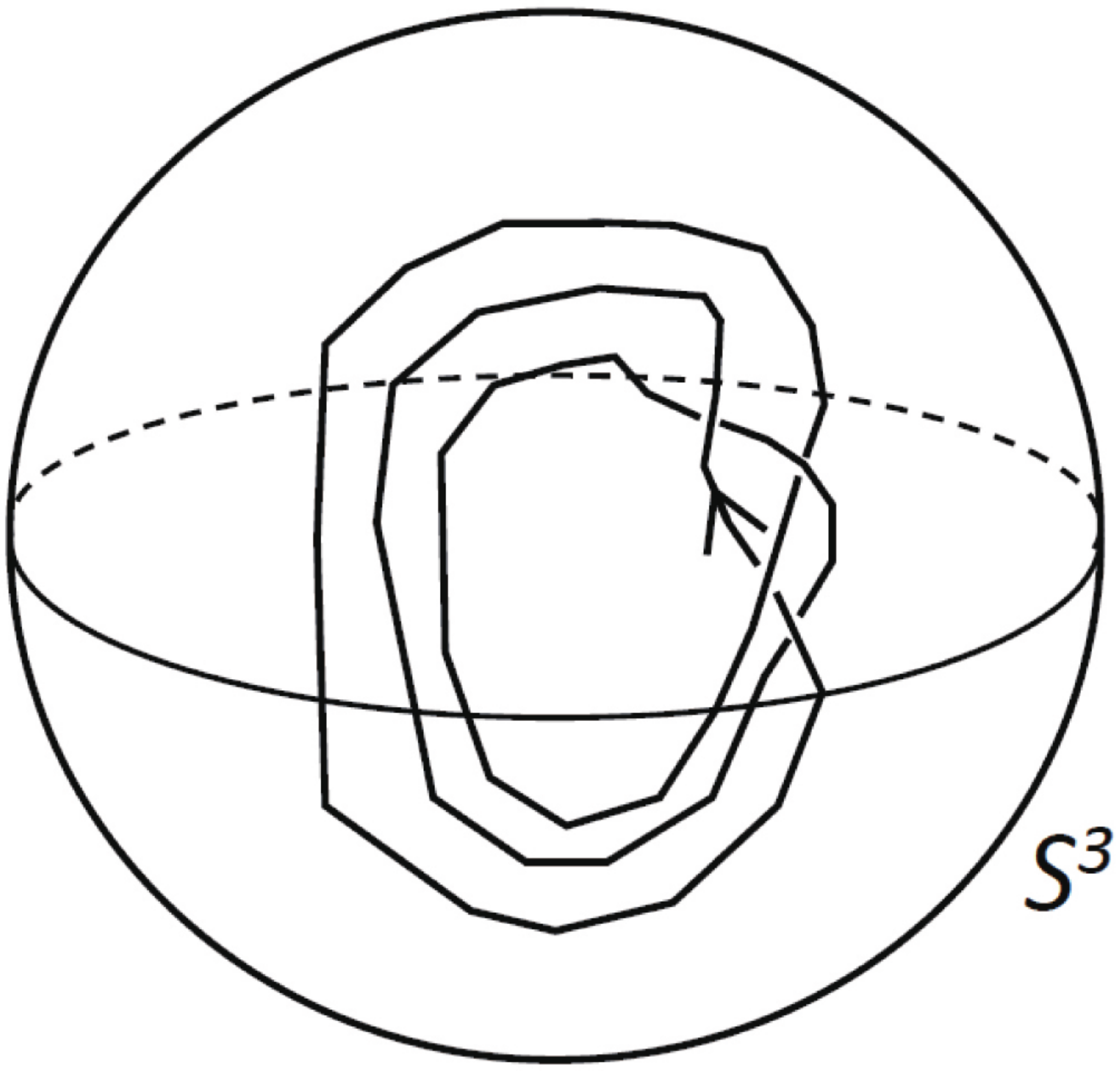}
  \caption{A link $L$ in ${\bf S}^3$}
\end{figure}

\bigskip\noindent{\it Generalization to an Arbitrary Link}

One can of course generalize $C$ in fig.~3 to an arbitrary link $L = \sum_{i=1}^{n-1} C_i$ composed of $n-1$ components $C_i$ each in the representation labeled by  $\lambda_i$. Then, according to the analysis in~\cite{Jones-Witten}, if $M = {\bf S}^3$ as shown in fig.~7, the corresponding link invariant would be given by
\be
\label{link}
Z^{\lambda_1 \dots \lambda_{n-1}}_{{\bf S}^3}(L) = \sum_j \, S_0{}^j    \tau_j(\hat B).
\ee
The above formula can be explained as follows. The index ``$j$'' labels the highest dominant weight representations of $G$; $\tau_j(\hat B)$ is the trace of the operator $\hat B$ in the Hlibert space ${\cal H}_j$ of states which can be represented by the ``quantum'' ramified $\cal D$-modules 
\be
\label{CB-ramified-braids}
\MC^j_{\rm knots} = \left< \Phi^{\lambda_1}_{s_1} (z_1) \dots \Phi^{\lambda_{n-1}}_{s_{n-1}}(z_{n-1}) \, \Phi^{\lambda_j}_{s_{n}} (z_{n}) \right>_{\hat \Sigma}
\ee 
on ${\CM}_{G; z_1, \dots, z_{n-1}, z_{n}}$; $\hat B$ is a representation of the subgroup $B$ of the mapping class group $M$ for ${\bf S}^2$ with $n$ marked points which leaves fixed the $n^{\rm th}$ point; and $S_0{}^j$ are certain functions in $\hat k$ which represent in the Hilbert space a modular transformation $S: \tau \to -1 /\tau$ of a two-torus embedded in ${\bf S}^3$ with complex structure $\tau$.

A useful result to quote at this point is the following. Elements of the mapping class group $M$ generate automorphisms of the line bundle ${\mathscr L}^{\hat k}$ on ${\CM}_{G; z_1, \dots, z_{n-1}, z_{n}}$~\cite{Charles}. Since the ``quantum'' ramified $\cal D$-modules $\MC^j_{\rm knots}$ correspond to sections of the aforementioned line bundle,  and since they are objects which span a category $\mathcal C$, it would mean that elements of $B \subset M$ generate autoequivalences of $\mathcal C$. This is also reflected in the fact that $\hat B: {\cal H}_j \to {\cal H}_j$. Hence, we can also write 
\be
\label{link2}
Z^{\lambda_1 \dots \lambda_{n-1}}_{{\bf S}^3}(L) =  \sum_j S_0{}^j  \sum_{s_j=1}^{{\rm dim} \, {\cal H}_j}   (\mathscr D^{s_j}_j,   \phi_{\hat B} (\mathscr D^{s_j}_j)),
\ee
where the $\mathscr D^{s_j}_j$'s  are ``quantum'' ramified $\cal D$-modules  ${\cal D}_{mod}^{c}(\CM_{G; z_1, \dots, z_{n-1}, z_n})$ that span an orthogonal basis in ${\cal H}_j$; $( \, \,  , \, \, )$ is the usual natural pairing in ${\cal H}_j$; and $ \phi_{\hat B}$ is the operator $\hat B$ -- with eigenvector $\mathscr D^{s_j}_j$ -- representing an autoequivalence of  $\mathcal C$. Thus, the link invariant $Z^{\lambda_1 \dots \lambda_{n-1}}_{{\bf S}^3}(L)$ just counts (with appropriate weights) the number of linearly-independent ``quantum'' ramified $\cal D$-modules $\mathscr D^{s_j}_j$.

Let us again specialize to the case where $G = SU(2)$; let all the $\lambda_i$'s label the two-dimensional fundamental representation $\bf 2$ of $SU(2)$. Then,  $Z^{\bf 2, \dots, \bf 2}_{{\bf S}^3}(L)$ is simply the Jones polynomial of the link $L$~\cite{Jones-Witten}. In turn, if the finite-dimensional vector space 
\be
{\cal K} (L) =  \oplus_{a,b} \, {\cal K}^{a,b}(L)
\ee
is the corresponding bi-graded Khovanov homology, we can,  according to~\cite{Khovanov},  rewrite (\ref{link2}) as
\be
\label{count}
\sum_{a,b} \,  (-1)^a  q^b \, {\rm dim} \, {\cal K}^{a,b}(L) = \sum_j S_0{}^j  \sum_{s_j=1}^{{\rm dim} \, {\cal H}_j}   (\mathscr D^{s_j}_j,   \phi_{\hat B} (\mathscr D^{s_j}_j)), 
\ee
where 
\be
q = {\rm exp} \left ({{2 \pi i} \over {\hat k + h^\vee} } \right).
\ee
Thus, from (\ref{count}), we learn that a (weighted) count of the Khovanov homology of the $n-1$ component link $L$ would be given by a (weighted) count of the number of linearly-independent ``quantum'' ramified $\cal D$-modules $\mathscr D^{s_j}_j$ on ${\CM}_{SU(2); z_1, \dots, z_{n-1}, z_{n}}$.

\newsubsection{Langlands Duality And Representations Of Complex Lie Groups}

Consider the flag manifold model of $\S$9.1. Furthermore, consider its $R \to 0$ limit.  From (\ref{k-k}) and (\ref{g_s}), one can see that its corresponding action ${I}^{R \to 0}_{\infty, {\rm eff}}$ describes the worldsheet theory of a string whose coupling strength is infinitely large, i.e.,  we are now considering the ultra-quantum limit of the string. From (\ref{Iinfty-effective}), (\ref{k-k}), and $R = 1 / \sqrt{k + h^\vee}$, we can write ${I}^{R \to 0}_{\infty, {\rm eff}}$ as\footnote{To arrive at the following expression, we have made the following trivial field redefinitions: $\beta \to i \beta$, $\gamma \to i\gamma$, and $Y \to -Y$.}  
\be
\label{Iinfty-effective-QG}
{I}^{R \to 0}_{\infty, {\rm eff}} =  - {1 \over  \pi} \int_{\Sigma} |d^2 z| \, \sqrt {g}  \, e^{-2\sigma(z, \bar z)}  \, \left[\sum_{i=1}^{|\Delta_+|}   \{ \beta_i \partial_{\bar z}\gamma^i  - \partial_{\bar z} (V^i \cdot Y) \partial_z (V_i \cdot Y) \} +  i { {\cal R}_{\bar z z} \over \sqrt{{\hat k} + {h}^\vee}}  (\rho \cdot Y)\right],
\ee
where
\be
{\hat k} + {h}^\vee =  {1 \over {k + h^\vee}}.
\ee

Recall that since the flag manifold model of $\S$9.1 can be viewed as the local flag manifold model of $\S$6.1, and since we consider $\GC$ to be simply-laced in this section whence $\rho = \rho^\vee$, the action ${I}^{R \to 0}_{\infty, {\rm eff}}$ enjoys a generalized $T$-duality which maps $R \to 1/R$, $\rho \to - \rho$ and $Y_R(\zb) \to - Y_R(\zb)$, where $Y = Y_L(z) + Y_R(\zb)$. Thus, since $\rho \cdot Y_R = 0$ (see discussion above (\ref{Iinfty})), and since $^L\rho = \rho^\vee = \rho$, where $^L\rho$ is the Weyl vector of $^L\GC$, it is clear that ${I}^{R \to 0}_{\infty, {\rm eff}} $ is the $T$-dual of the action 
\be
\label{Iinfty-effective-QG-T}
^L{I}^{R \to 0}_{\infty, {\rm eff}} =   - {1 \over  \pi} \int_{\Sigma} |d^2 z| \, \sqrt {g}  \, e^{-2\sigma(z, \bar z)}  \, \left[\sum_{i=1}^{|\Delta_+|}   \{ \beta_i \partial_{\bar z}\gamma^i  + \partial_{\bar z} (V^i \cdot Y) \partial_z (V_i \cdot Y)\}  -  i { {\cal R}_{ \bar z z} \over \sqrt{^L{\hat k} + {^L{h}}^\vee}}  (^L\rho \cdot Y)\right],
\ee
where
\be
\label{k-k l}
^L{\hat k} + {^Lh}^\vee  = k + h^\vee = {1 \over {\hat k} + {h}^\vee}.
\ee
Here, we have conveniently defined $^Lh^\vee$  to be the dual Coxeter number of the Lie algebra $^L\gc$ of $^L\GC$. 

Another consequence of $\GC$ being simply-laced is that the total number of positive roots $|\Delta_+|$ associated with $\GC$ and $^L\GC$ are the same. Moreover, $\GC$ and $^L\GC$ have the same rank, regardless. As such, restricting to integer values of $^L\hat k$, one can (up to an overall sign) also interpret $^L{I}^{R \to 0}_{\infty, {\rm eff}}$ as the action of a WZW model for $^LG$ (the real, compact form of $^L\GC$) at level $^L\hat k$. (See discussion following (\ref{Iinfty-effective}).) In short, $T$-duality of the flag manifold model in the infinite $X$-volume limit implies that when $R \to 0$, the WZW model for $G$ at level $\hat k$ that it describes can be regarded as a WZW model for $^LG$ at level $^L\hat k$, where $\hat k$ and $^L\hat k$ are related as shown in (\ref{k-k l}).

\bigskip\noindent{\it A Ramified Geometric Langlands Correspondence for $\GC$}

Now consider a general holomorphic conformal block of the $^LG$-WZW model at level $^L\hat k$ in the holomorphic primary field operators $\Phi^{{^L\lambda}}_s(z)$ associated with the highest dominant weight $^L\lambda$:
\be
\label{CB of LG}
{^L\MC_{\rm knots}} = \left< \Phi^{^L\lambda_1}_{s_1} (z_1) \dots \Phi^{^L\lambda_p}_{s_p}(z_p) \Phi^0(z_{p+1}) \dots \Phi^0(z_n) \right>_{\hat \Sigma} =  \left< \Phi^{^L\lambda_1}_{s_1} (z_1) \dots \Phi^{^L\lambda_p}_{s_p}(z_p) \right>_{\hat \Sigma}.
\ee 
(The reason for the second equality is given above (\ref{CB-unramified}).) By repeating our analysis in the first, second and third paragraphs of $\S$9.2 but with $G$ replaced by $^LG$,   and by noting a theorem of Mehta and Seshadri in~\cite{Mehta} regarding the one-to-one correspondence between $\CM_{^LG; z_1, \dots, z_p}$ and the moduli space ${\rm Bun}_{^L\GC; z_1, \dots, z_p}$ of (stable) holomorphic parabolic $^L\GC$-bundles on the rational curve $\hat \Sigma$ whose structure group reduces at the points $z_1, \dots, z_p$ to the Borel subgroup ${^LB} \subset {^L\GC}$, we learn that (i) ${^L\MC}_{\rm knots}$ must be given by elements of $H^0({\rm Bun}_{^L\GC; z_1, \dots, z_p}, \mathscr  L^{^Lc - {^Lh}^\vee})$, where $^Lc = {^L\hat k} + {^Lh^\vee}$, and $\mathscr L$ is a line bundle whose first Chern class generates the second cohomology of ${\rm Bun}_{^L\GC; z_1, \dots, z_p}$; (ii) we can interpret $^L\MC_{\rm knots}$ as a ``quantum'' (tamely) ramified $\cal D$-module ${\cal D}_{mod}^{^Lc}({\rm Bun}_{^L\GC; z_1, \dots, z_p})$ on ${\rm Bun}_{^L\GC; z_1, \dots, z_p}$ with twist parameter $^Lc$.

On the other hand, recall from (\ref{CB-ramified}) that a general holomorphic conformal block of the $T$-dual $G$-WZW model at level $\hat k$ can be written as
\be
\label{CB-ramified-QG}
\MC_{\rm knots} =   \left< \Phi^{\lambda_1}_{s_1} (z_1) \dots \Phi^{\lambda_p}_{s_p}(z_p) \right>_{\hat \Sigma}.
\ee 
$\MC_{\rm knots}$ can similarly be interpreted as a ``quantum'' (tamely) ramified $\cal D$-module ${\cal D}_{mod}^{c}({\rm Bun}_{\GC; z_1, \dots, z_p})$ on ${\rm Bun}_{\GC; z_1, \dots, z_p}$ with twist parameter $c$, where (\ref{k-k l}) means that 
\be
\label{c-c}
{^Lc} = {1 \over c}.
\ee

Note at this point that (\ref{iso of W-GC}) (which is valid for arbitrary values of $k$ denoted therein) tells us that $T$-duality leads to the identification of $\cal W$-algebras
\be
{\cal W}_{\hat k}(\gc) \cong {\cal W}_{^L\hat k}(^L\gc),
\ee
where the relation between $\hat k$ and $^L\hat k$ is as given in (\ref{k-k l}).  Note also that one can identify  the (generators of the) $\cal W$-algebras with $\mathbb C[\partial^{m}_z S^{(s_i)}(z)]_{i = 1, \dots, l; \, m\geq 0}$ -- the spaces of differential polynomials on the $S^{(s_i)}(z)$'s in (\ref{S^{(s_i)}(z)}) with complex coefficients. And, from our discussion leading up to (\ref{D-module}), we find that the differential polynomials spanning $\mathbb C[\partial^{m}_z S^{(s_i)}(z)]_{i = 1, \dots, l; \, m\geq 0}$ which correspond to ${\cal W}_{\hat k}(\gc)$ and ${\cal W}_{^L\hat k}(^L\gc)$, actually act as differential operators on $\MC_{\rm knots}$ and $^L\MC_{\rm knots}$, respectively. In sum, $T$-duality implies that we can identify $\MC_{\rm knots}$ with $^L\MC_{\rm knots}$. Since our discussion assumes the $R \to 0$ limit, and since we have $^Lc \to \infty$ and $c \to 0$ when $R \to 0$,\footnote{ Because $^Lh^\vee = h^\vee$ and thus, from (\ref{k-k l}), we have $^L\hat k = k$, when $R \to 0$, i.e., $k \to \infty$, we have  $^Lc \to \infty$ and $c \to 0$.} it would mean that we have the following correspondence:
\be
\label{QGL}
{\cal D}_{mod}^{0}({\rm Bun}_{\GC; z_1, \dots, z_p}) \longleftrightarrow {\cal D}_{mod}^{\infty}({\rm Bun}_{^L\GC; z_1, \dots, z_p}).
\ee
According to our earlier discussions, ${\cal D}_{mod}^{0}({\rm Bun}_{\GC; z_1, \dots, z_p})$ is a ``classical'' ramified $\cal D$-module on ${\rm Bun}_{\GC; z_1, \dots, z_p}$. On the other hand, ${\cal D}_{mod}^{\infty}({\rm Bun}_{^L\GC; z_1, \dots, z_p})$ is known to be a quasi-coherent sheaf on ${\rm Bun}_{^L\GC; z_1, \dots, z_p}$~\cite{Frenkel-Ram}, where the connection of the holomorphic parabolic $^L\GC$-bundle has unipotent monodromy around the points $z_1, \dots, z_p$.\footnote{Since the highest dominant weights ${^L\lambda_i}$ and hence their duals ${^L \lambda^\ast_i}$ are finite while $^L \hat k \to \infty$, the monodromy of the underlying $^L\GC$-connection around the point $z_i$ -- $g_{^L\lambda_i} = {\rm exp} (- 2 \pi i \, ^L{\lambda}^\ast_i / ^L\hat k)$ -- approaches 1, i.e., it is unipotent.} In other words,  $T$-duality in the $R \to 0$ limit implies the statement (\ref{QGL}) of the ``classical'' tamely-ramified  geometric Langlands correspondence for $\GC$!

\bigskip\noindent{\it Relation to Representations of $^L\GC$}

According to our analysis in $\S$9.2, $^L\MC_{\rm knots}$ on the RHS of (\ref{QGL}) can be related to knot invariants of three manifolds $M$, where $M$ can be Heegaard split along $\hat \Sigma \subset M$, and where the colored knots pierce through $\hat\Sigma$ at the points $z_1, \dots, z_p$ in the highest dominant weight representations labeled by $^L\lambda_1, \dots, {^L\lambda_p}$. This means that one can make the identification~\cite{Sawin}
\be
 {^L\MC}_{\rm knots} \longleftrightarrow {\rm Rep}_{^L\lambda_1}(U_q(^L\gc)) \otimes \cdots \otimes {\rm Rep}_{^L\lambda_p}(U_q(^L\gc)),
\ee
where ${\rm Rep}_{^L\lambda_i}(U_q(^L\gc))$ is an irreducible representation of the quantum group $U_q(^L\gc)$ that is associated with the ${^L\lambda}_i$-representation of $^LG$, and 
\be
\label{q}
q = {\rm exp} \left ({i \pi  \over {^L\hat k + {^Lh}^\vee}}\right) = {\rm exp} ({i \pi c}).
\ee
(The second equality is due to the fact that $^Lc = {^L\hat k} + {^Lh^\vee}$ and ${^Lc^{-1}} = c$.) Note that in the $R \to 0$ limit whence $c \to 0$ and $q \to 1$, we have $U_1(^L\gc) \to {^L\gc}$. Also, ${\rm Rep}_{^L\lambda_i}(^L\GC)$ is just the integrated form of  ${\rm Rep}_{^L\lambda_i}(^L\gc)$. Consequently, in addition to (\ref{QGL}), $T$-duality in the $R \to 0$ limit also implies that we have the following correspondence:
\be
\label{QG}
{\cal D}_{mod}^{0}({\rm Bun}_{\GC; z_1, \dots, z_p}) \longleftrightarrow {\rm Rep}_{^L\lambda_1}(^L\GC) \otimes \cdots \otimes {\rm Rep}_{^L\lambda_p}(^L\GC),
\ee
where  ${\rm Rep}_{^L\lambda_i}(^L\GC)$ is the $^L\lambda_i$-representation of $^L\GC$. This correspondence has also been mathematically conjectured by Gaitsgory in $\S$4.3 of~\cite{Gaitsgory-summary} (as a ``classical'' $c \to 0$ limit of the correspondence in Conjecture~0.13 of~\cite{Gaitsgory-Whittaker}). Thus, we have in the above a purely physical proof of Gaitsgory's conjecture for $\GC$.\footnote{Note that there is a slight deviation -- albeit a consistent one -- of our correspondence (\ref{QG}) from that in \emph{loc.~cit.}; in our case, we have $\cal D$-modules on  ${\rm Bun}_{\GC; z_1, \dots, z_p}$ and not ${\rm Bun}_{\GC}$. This is because $\hat \Sigma$ is a simply-connected rational curve and thus, ${\rm Bun}_{\GC}$ -- which is the moduli space  of holomorphic $\GC$-bundles on $\hat \Sigma$ -- is actually trivial.} 




\end{document}